\documentclass[12pt,a4paper,twoside]{book}
\usepackage{mysects}
\usepackage{amsmath}
\usepackage{amsfonts}
\usepackage{amssymb}
\usepackage{graphicx}
\usepackage{axodraw}
\usepackage{tabularx}
\usepackage{enumerate}
\usepackage{rangecite}
\usepackage{fancyhdr}
\long\def\ignore#1{}
\begin{document}
\renewcommand{\thepage}{\roman{page}}
\begin{titlepage}
  \begin{center}
    {\large TESIS DOCTORAL}
\vspace{3cm}
    
\textbf{\huge    Phenomenology of Bilinear Broken}

\vspace{0.2cm}

\textbf{\huge R--parity}
    
\vspace{3cm}
{\Large Departament de F{\'\i}sica Te{\`o}rica}
\vspace{1cm}
    
\includegraphics[scale=1.5]{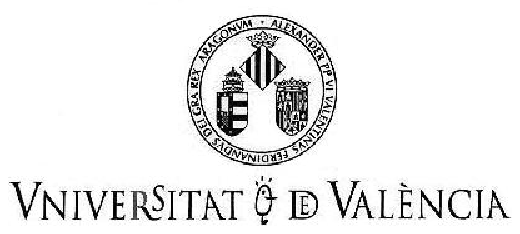} 

\vspace{4cm}
\textbf{\Large Diego Alejandro Restrepo Quintero}
    
  \end{center}
\end{titlepage}

\chapter*{Abstract}
\thispagestyle{empty}

The straightforward supersymmetrization of the Standard Model (SM)
results in a phenomenologically inconsistent theory in which Baryon
number ($B$) and Lepton number ($L$) are violated by dimension 4
operators, inducing fast proton decay.  Proton stability allows only
for separate $L$ or $B$ violation and, if neutrinos are massive
Majorana particles, $L$ violating terms must be present.  In this
thesis I will study a Supersymmetric Standard Model (SSM) realization
with $B$ conservation and minimal $L$ violation.  In this framework
$L$ is mildly violated only by super--renormalizable terms, allowing
for small neutrino Majorana masses. This model is more predictive than
the Baryon--Parity SSM. The induced dimension 4 $L$ violating couplings
are not arbitrary, and automatically satisfy all experimental
constraints.  After introducing the theoretical framework for
supersymmetric models without Lepton number, I will discuss the
phenomenology of the (unstable) lightest neutralino and of the
lightest stop. I will show that the leptonic decays of the stop can be
related to the neutrino parameters, and in particular their
measurement can indirectly probe the size of the solar neutrinos
mixing angle.

\chapter*{Resumen}
\thispagestyle{empty}

El Modelo Est{\'a}ndar Supersim{\'e}trico (MES) es el modelo supersim{\'e}trico
con m{\'\i}nimo n{\'u}mero de part{\'\i}culas correspondientes a aquellas del Modelo
Es\-t{\'a}n\-dar (ME), un doblete de Higgs adicional y todos los compa{\~n}eros
supersim{\'e}tricos de las part{\'\i}culas del ME.
El MES permite la aparici{\'o}n simultanea de t{\'e}rminos que violan n{\'u}mero
lept{\'o}nico y bari{\'o}nico y de este modo es excluido debido a que el
prot{\'o}n decae a trav{\'e}s de interacciones electrod{\'e}biles. Sin embargo es
bien conocido que la estabilidad del prot{\'o}n permite la presencia ya
sea de violaci{\'o}n de n{\'u}mero lept{\'o}nico o de la violaci{\'o}n de n{\'u}mero
bari{\'o}nico.  De hecho, si los neutrinos tienen masa como es sugerido
por las anomal{\'\i}as solar y atmosf{\'e}rica no existe una raz{\'o}n de peso para
asumir que los t{\'e}rminos que violan n{\'u}mero lept{\'o}nico est{\'e}n ausentes. 
En este trabajo estudiaremos la realizaci{\'o}n del MES con conservaci{\'o}n
de n{\'u}mero bari{\'o}nico y violaci{\'o}n \emph{m{\'\i}nima} de n{\'u}mero lept{\'o}nico. Los {\'u}nicos
t{\'e}rminos con violaci{\'o}n de n{\'u}mero lept{\'o}nico en este caso provienen del
t{\'e}rmino de masa m{\'a}s general que permite la invarianza gauge en el
superpotencial del MES. Al modelo correspondiente le llamaremos Modelo
Est{\'a}ndar Supersim{\'e}trico Superrenormalizable (MESS). El MESS es m{\'a}s
predictivo y te{\'o}ricamente m{\'a}s atractivo que el MES con conservaci{\'o}n de
n{\'u}mero bari{\'o}nico pero con t{\'e}rminos arbitrarios de violaci{\'o}n de n{\'u}mero
lept{\'o}nico y en general satisface todos las restricciones que surgen de
esta violaci{\'o}n.

En el MESS la Part{\'\i}cula Supersim{\'e}trica m{\'a}s Liviana (PSL) puede decaer. En
particular estudiaremos la fenomenolog{\'\i}a esperada del neutralino en
LEP2.
En el caso de part{\'\i}culas supersim{\'e}tricas con mayor masa que la PSL,
los decaimientos que violan n{\'u}mero lept{\'o}nico pueden llegar a ser
detectables en los aceleradores, incluso en el caso de masas de
neutrinos tan peque{\~n}as como las sugeridas por las anomal{\'\i}as de
neutrinos. Especialmente interesante en este caso son los decaimientos
que violan n{\'u}mero lept{\'o}nico del stop (compa{\~n}ero supersim{\'e}trico del
quark top), y que ser{\'a}n estudiados en este trabajo. Estos decaimientos
implican la posibilidad de probar el {\'a}ngulo de mezcla solar y as{\'\i} la
posibilidad de relacionar la f{\'\i}sica de colisionadores de alta energ{\'\i}a
con los experimentos sobre oscilaci{\'o}n de neutrinos solares.

\tableofcontents
\listoffigures
\listoftables
\newpage

\chapter*{Introduction\markboth{Introduction}{Introduction}}
\label{sec:introduction}

\addcontentsline{toc}{chapter}{Introduction} 
\renewcommand{\thepage}{\arabic{page}}
\setcounter{page}{1}

The Supersymmetric
Standard Model (SSM)%
\index{SSM: Supersymmetric Standard Model}
\cite{Nilles:1984ge,Haber:1985rc} is the Supersymmetric
(SUSY)\index{SUSY: Supersymmetric} model with minimal possible number
of particles, corresponding to those of the Standard Model
(SM)\index{SM: Standard Model}, one extra Higgs doublet and all the
superpartners of the SM particles.  We define the Minimal
Supersymmetric Standard Model (MSSM)\index{MSSM: Minimal
  Supersymmetric Standard Model} \cite{Haber:1985rc,Tata:1995zj} as
the SSM with minimum possible number of couplings.

Unlike the MSSM, the Supersymmetric Standard Model allows many gauge
invariant terms violating baryon number ($B$)%
\index{$B$: Baryon Number} and lepton number ($L$)\index{$L$: Lepton Number}.
Consequently, this most general case is excluded because the proton
would decay with a weak decay rate.  However, it is well known that
the stability of proton allows either $L$ or $B$ violation, because
the nucleon decays normally requires not only the baryon number
violation but also lepton number violation.
In the usual way to avoid proton decay, all such terms are forbidden
by imposing some $X$ symmetry~\cite{Carone:1996nd}. For example the
anomaly free gauge discrete symmetry known as $R$-parity
\index{$R$--parity}
\cite{Farrar:1978xj,Dimopoulos:1981zb,Ibanez:1992pr}, forbids all the
renormalizable $L$ and $B$ violating terms. It is worth to stress, in
that case it is possible have $L$ violation at the low energy theory.
For example, superrenormalizable (bilinear) $L$ violating terms in the
low energy SSM may appear from renormalizable (trilinear) terms
allowed at some high energy where the $R$--parity is still unbroken
\cite{Bento:1987mu}.

From a phenomenological point of view, it is most important to ensure
that there are no interaction terms in the lagrangian which lead to
rapid proton decay and, in this respect, other discrete symmetries can
be used which are even more effective than $R$--parity. For example
the anomaly free discrete symmetries equivalent to baryon parity
\cite{Ibanez:1992pr,Ibanez:1991hv,Ibanez:1993ji,Hinchliffe:1993ad,Eyal:1999gq}
or lepton parity \cite{Ibanez:1993ji,Choi:1999}. The former forbids
dimension--4 and 5 $B$ violating operators while the latter forbids
dimension--2, 4 and 5 $L$ violating operators.  In this way, general
models of $L$ or $B$ violation are expected to come from $X$
symmetries different from $R$ parity.

If neutrinos are massive Majorana particles, lepton number is
violated, and there is no compelling reason to assume that
$L$ violating terms are absent from the superpotential.
Indeed, there is considerable theoretical and phenomenological
interest in studying possible implications of alternative SUSY
scenarios in which $L$ is
broken~\cite{aul,ross:1985,Hall:1984id,Ellis:1985gi,Lee:1984kr,Dawson:1985vr,eigthies,Dimopoulos:1988jw,arca,beyond,Dreiner:1997uz,Barbier:1998fe}.  
This is especially so considering the fact that it
provides an appealing joint explanation of the solar and atmospheric
neutrino~\cite{Gonzalez-Garcia:2001sq} anomalies which has, in
addition, the virtue of being testable at present and future
accelerators like LEP~\cite{Abreu:2001nc,Abreu:2001mm,Abreu:2001ne},
Tevatron\cite{Abreu:2001nc,Barger:1989rk,Diaz:1999ge,Allanach:1999bf,Datta:2000xq},
LHC~\cite{Barbier:1998fe,Romao:2000up} or a linear $e^+ e^-$
collider~\cite{lastdiego}.
The effects of $L$ violation can be
large enough to be experimentally observable.

An special case of baryon parity SSM only contain dimension--2 $L$
violating terms in the superpotential. They have mass coefficients
protected by the non-renormalizable theorem in SUSY
\cite{Martin:1997ns} and are called superrenormalizable terms. In
fact, once mass parameters are introduced, the term
\begin{equation}
  \label{eq:1}
  W_\mu=\mu_\alpha\widehat L_\alpha\widehat H_u 
\end{equation}
(rather than only $\mu_0\widehat H_d\widehat H_u$) is the most natural
choice. Where $\widehat L_0\equiv\widehat H_d$ and $\alpha=0,1,2,3$.

The $L$ violation through only superrenormalizable terms could arise
explicitly as in 
\cite{Bento:1987mu,Hall:1984id,Smirnov:1996ey,Hempfling:1996wj,Tamvakis:1996dk,Choi:1997se,Nilles:1997ij,Benakli:1997iu,Giudice:1997wb,Binetruy:1998sm,Joshipura:1998sp,Chun:1999cq,kaplan,Chkareuli:2000at,Chkareuli:2000tp,Mira:2000gg} 
as a residual effect of some larger
theory.
Most of these models do not introduce new particles to the SSM and we
will call them Superrenormalizable Supersymmetric Standard Models
(SSSM)%
\index{SSSM: Superrenormalizable Supersymmetric Standard Model}.

Alternatively the $L$ violation could arise spontaneously, through
nonzero vacuum expectation values (vev's) of singlet
fields~\cite{Valle:1987sq,MASIpot3,Romao:1992vu,MASI,Chaichian:1993ra,Adhikari:1996bm,Frank:2001tr}
which break $R$--parity at low energies. Consequently these models
contain additional fields not present in the SSM. Although in this
work we will concentrate in the SSSM, most phenomenological features
of the spontaneous $L$ violating SUSY models are reproduced also for
the SSSM

SSSM is more predictive and theoretically more attractive than $L$
violating SSM through arbitrary dimension--4 operators.  Moreover,
unlike this baryon parity SSM, the induced dimension--4 $L$ violating
couplings in the SSSM are not arbitrary and in general automatically
satisfy all experimental constraints on $L$ violation. This renders a
systematic way to study $L$ violating signals
\cite{Smirnov:1996ey,ROMA,Hirsch:2000xe,Barbieri:1990vb,Mukhopadhyaya:1998xj,Navarro:1999tz,otros,Choi:1999tq}
and leads to effects that can be large enough to be experimentally
observable, even in the case where neutrino masses are as small as
indicated by the simplest interpretation of solar and atmospheric
neutrino data \cite[and references therein]{Romao:2000up}.  Moreover,
the SSSM follow a specific \emph{pattern } which can be easily
characterized. These features have been exploited in order to describe
the expected SSSM signals (see section~\ref{sec:superr-l-numb})

In chapter~\ref{sec:supersymm-lagr} we will study the SSM model for
one generation. This ``toy model'' turn to be very illustrative in the
discussion of key ingredients of the general baryon parity SSM, such
as the definition of basis independent parameters and the generation of
the tree level neutrino mass.

In chapter~\ref{sec:three-gener-supersym} we emphasize the problems
with the SSM and study in a systematic way the several alternative SSM
models. Instead of the usual approach of ad-hoc matter parities we use
theoretical best motivated anomaly free gauge discrete symmetries.
They allow an easier connection with other realistic ways of generate
the unknown $X$ symmetry responsible for the proton stability in the
SSM, such us $R$ symmetries, gauge symmetries, flavor symmetries,
Peccei-Quinn symmetries. Next we review how the SSSM can be generated
with all these $X$ symmetry possibilities. The SSSM is then presented
as the best motived and minimal realization of the SSM when the
experimental evidence on neutrino masses is taken into account.

In chapter \ref{sec:mass-spectrum-mssm} we build the mass spectrum of
the SUSY particles in the SSSM.
In the next chapters we study the phenomenology of the neutralino and
the stops in the SSSM. The neutralino and the two body stop decay
studies were performed in the supergravity version of the SSSM, while the
three body decay of the stop was studied with arbitrary low energy
parameters as inputs.

In the neutralino case we will present a detailed study of the
Lightest Supersymmetric Particle (LSP)\index{LSP: Lightest
  Supersymmetric Particle} decay properties and general features of
the corresponding signals expected at LEP2.
It is well known that in models with Gauge Mediated Supersymmetry
Breaking (GMSB) \index{GMSB: Gauge Mediated Supersymmetry Breaking}
the lightest neutralino decays \cite{GMSB1,Giudice:1999bp}, because in
this case the gravitino is the LSP. We therefore also discuss the
possibilities to distinguish between GMSB and the SSSM.

We will study the decay modes of the lightest top squark in
supergravity models where supersymmetry is realized with the SSSM. In
such models the lightest stop could even be the lightest
supersymmetric particle and be produced at LEP or Tevatron. Neither
$e^+ e^-$ collider
data~\cite{Abreu:2001nc,Abreu:2001mm,Abreu:2001ne,LEPSEARCH} nor
$p\bar{p}$ data from the Tevatron \cite{D0,Abbott:2000yu} preclude
this possibility. In contrast with ref.~\cite{Bartl:1996gz} here we
focus in the effective model where the $L$ violation is introduced
through an explicit superrenormalizable term. This is substantially
simpler than the full majoron version of the model considered
previously.

In order to discuss stop decays we also refine the work presented in
Ref.~\cite{Hikasa:1987db,baer,Porod:1999yp,Porod:1997at,Porod:1998kk,Boehm:1999tr}
by giving, for the first time, an exact numerical calculation for the
Flavor Changing Neutral Current (FCNC)\index{FCNC: Flavor Changing
  Neutral Current} process $\tilde t \to c\, \tilde\chi_1^0$. We also
compare the results obtained this way with those one gets by adopting
the usual one--step or leading logarithm approximation in the
Renormalization Group Equations (RGE)\index{RGE: Renormalization Group
  Equations}. In contrast with the MSSM such an approximation would be
rather poor for our purposes, since we will be interested in comparing
FCNC with $L$ violating stop decay modes.
Moreover, in contrast to ref. \cite{Bartl:1996gz}, where the magnitude
of the stop -- charm -- neutralino coupling was a phenomenological
parameter, here we assume a minimal supergravity scheme with
universality of soft terms at the unification scale in which this
coupling is induced radiatively and thus calculable. As we will see
this has important phenomenological implications, for example in the
behavior of the stop decays in the SSSM with respect to $\tan\beta$. We
calculate its magnitude using a set of RGE's in which the running of
the Yukawa couplings and soft breaking terms is taken into account.
Here we also provide the analysis of the relationship of the stop
decays in the SSSM with the magnitude of the heaviest neutrino mass.
Motivated by the simplest oscillation interpretation of the
Super-Kamiokande atmospheric neutrino data, we also generalize the
treatment of the $L$ violating decays by explicitly considering the
case of light $\nu_3$ masses, not previously discussed.

In the SSSM, the stop can have new decay modes such as
\begin{equation}
  \label{eq:2}
  \tilde{t}_1 \to b\,\tau
\end{equation}
due to mixing between charged leptons and charginos.  We show that
this decay may be dominant or at least comparable to the ordinary
$L$ conserving mode 
\begin{equation}
  \label{eq:3}
  \tilde{t}_1 \to c\,{\tilde{\chi}}^0_{1},
\end{equation}
where $\tilde{\chi}^{0}_{1}$ denotes the lightest neutralino.

Owing to the large top Yukawa coupling the stops have a quite
different phenomenology compared to those of the first two generations
of up--type squarks (see e.g.~\cite{Bartl97a,Bartl:1997wt} and
references therein).  The large Yukawa coupling implies a large mixing
between ${\tilde t}_L$ and ${\tilde t}_R$ \cite{Ellis83} and large
couplings to the higgsino components of neutralinos and charginos.
The large top quark mass also implies the existence of scenarios where
all MSSM two-body decay modes of ${\tilde t}_1$ are kinematically
forbidden at the tree-level (e.g. ${\tilde t}_1 \to t \, {\tilde
  \chi}^0_i, b \, {\tilde \chi}^+_j, t \, \tilde g$). In such case
higher order decays of ${\tilde t}_1$ become relevant
\cite{Hikasa:1987db,Porod:1999yp,Porod:1997at,Djouadi:2000bx}:
${\tilde t}_1 \to c \, {\tilde \chi}^0_{1,2}$, 
${\tilde t}_1 \to W^+ \, b \, {\tilde \chi}^0_1$,
${\tilde t}_1 \to H^+ \, b \, {\tilde \chi}^0_1$,
${\tilde t}_1 \to b \, {\tilde l}^+_i \, \nu_l$,
${\tilde t}_1 \to b \, {\tilde \nu}_l \, l^+$, 
%
where $l$ denotes $e,\mu,\tau$. Also 4-body decays may become important
if the 3-body decays are kinematically forbidden \cite{Boehm:2000tr}.
In \cite{Porod:1999yp,Porod:1997at,Djouadi:2000bx,Bartl:2000kw} it has
been shown that in the MSSM the three-body decay modes are in general
much more important than the two body FCNC decay mode.  Recently it
has been demonstrated that not only LSP decays but also the light stop
can be a good candidate for observing $L$ violation, even if its
magnitude is as small as indicated by the solutions to the present
neutrino anomalies
\cite{Diaz:1999ge,Allanach:1999bf,Bartl:1996gz,Datta:2000yc}.  In
particular in \cite{Diaz:1999ge} (see
section~\ref{sec:two-body-decays}) it has been demonstrated that there
exists a large parameter region where the SSSM decay
\begin{equation*}
{\tilde t}_1 \to b \, \tau
\end{equation*}
is much more important than the MSSM decays
\begin{equation*}
{\tilde t}_1 \to c \, {\tilde \chi}^0_{1,2} 
\end{equation*}
It is therefore
natural to ask if there exist scenarios where the decay ${\tilde t}_1
\to b \, \tau$ is as important as the three--body decays.  Note that in
the SSSM the neutral (charged) Higgs--bosons mix with the neutral
(charged) sleptons \cite{deCampos:1995av,Akeroyd:1998iq}.  These
states are denoted by $S^0_i$, $P^0_j$, and $S^\pm_k$ for the neutral
scalars, pseudoscalars and charged scalars, respectively.  Therefore
in the SSSM one has the following three-body decay modes:
\begin{eqnarray}
{\tilde t}_1 &\to& W^+ \, b \, {\tilde \chi}^0_1 \nonumber\\
{\tilde t}_1 &\to& S^+_k \, b \, {\tilde \chi}^0_1 \nonumber\\
{\tilde t}_1 &\to& S^+_k \, b \, \nu_l \nonumber\\
{\tilde t}_1 &\to& b \, S^0_i \, l^+ \,,\nonumber  \\
{\tilde t}_1 &\to& b \, P^0_j \, l^+ \,. \nonumber
\end{eqnarray}
We will show that there exist regions in parameter space where
${\tilde t}_1 \to b \, \tau^+$ is sizeable and even the most important
decay mode.  In particular we will consider a mass range of ${\tilde
  t}_1$, where it is difficult for the LHC to discover the light stop
within the MSSM due to the large top background \cite{ulrike}.
Contrary to the studies on LSP decay in the SSSM, the processes here
are very sensitive  to the value of the heaviest neutrino mass 

\chapter{One generation Supersymmetric Standard Model}
\label{sec:supersymm-lagr}

In the Standard Model (SM) the radiative corrections to the Higgs mass
are naturally of order of the Planck mass scale. However if there is a
symmetry relating bosons and fermions, called supersymmetry, such as
corrections turn to be at a controllable level. This symmetry at least
doubles the field content of the SM. In this work we will study SUSY
models with minimal possible number of particles, corresponding to
those the standard model, one extra Higgs doublet and all the
superpartners of the Standard Model (SM) particles.  The most general
renormalizable SUSY model respecting the gauge symmetries and with minimum
field content will be called \emph{Supersymmetric Standard Model}
(SSM)%
\index{SSM: Supersymmetric Standard Model}.

The field content of the SM together with the requirement of
$G_{SM}=SU(3)_C\times SU(2)_L\times U(1)_Y$ gauge invariance implies that the
most general Lagrangian is characterized by additional accidental
$U(1)$ symmetries implying Baryon number ($B$)%
\index{$B$: Baryon Number} and Lepton number ($L$)\index{$L$: Lepton Number}
conservation at the renormalizable level.  When the SM is
supersymmetrized, this nice feature is lost.  The introduction of the
superpartners allows for several new Lorentz invariant couplings which
are in conflict with present bounds in $B$ and/or $L$ violation.
However in the one generation SSM $B$ is automatically conserved (at
the perturbative level) while the $L$ violation may be compatible
with experimental constraints.  In this chapter we study the one
generation SSM and postpone the discussion of problems with the
three--generation SSM for the next chapter.

\section{Supersymmetric Lagrangian}
\label{sec:supersymm-lagr-1}

A supersymmetric transformation in a realistic supersymmetric model,
turns a bosonic state $\phi_i$ into its superpartner Weyl fermion $\psi_i$
state and vice versa. It also converts the gauge boson field $A^a_\mu$ into a
two component Weyl fermion gaugino $\lambda^a$ and vice versa

\begin{subequations}
  \label{eq:4}
  \begin{align}
    \delta\phi_i&=\epsilon\psi_i \\
    \delta(\psi_i)_\alpha&=i(\sigma^\mu\epsilon^{\dag})_\alpha D_\mu\phi_i+\epsilon_\alpha F_i \\
    \delta F_i&=i\epsilon^{\dag}\bar{\sigma}^\mu D_\mu\psi_i +\sqrt{2}g(T^a\phi)_i\epsilon^{\dag}\lambda^{{\dag}a}\\
    \delta A_\mu^a&=-\frac{1}{\sqrt{2}}
    \left(
      \epsilon^{\dag}\bar{\sigma}_\mu\lambda^a+\lambda^{{\dag}a}\bar{\sigma}_\mu\epsilon
    \right)\\
    \delta\lambda^a_\alpha&=-\frac{i}{2\sqrt{2}}
    \left(
      \sigma^\mu\bar{\sigma^\nu}\epsilon
    \right)_\alpha F^a_{\mu\nu}+\frac{1}{\sqrt{2}}\,\epsilon_aD^a\\
    \delta D^a&=\frac{1}{\sqrt{2}}
    \left(
      \epsilon^{\dag}\bar{\sigma}^\mu D_\mu\lambda-D_\mu\lambda^{\dag}\bar{\sigma}^\mu\epsilon
    \right)
  \end{align}
\end{subequations}
where the undotted (dotted in $\psi^{{\dag}i}_{\dot\alpha}$) greek indices are
used for the two components of the left--handed (right handed) Weyl
spinors\footnote{In general, $\alpha=1,2,\ldots,d=2^{2N}$ whit $N$ the number
  of supersymmetries.}. The $\sigma_\mu$ are $2\times2$ matrices with
$\sigma_0=\bar{\sigma}_i$ being the identity and $\sigma_i=-\bar{\sigma}_i$ the Pauli
matrices. The index $i$ runs over the  gauge and flavor
indices of the fermions (it is raised or lowered by hermitian
conjugation); $\epsilon^\alpha$ is an infinitesimal, anti commuting two-component
Weyl fermion object which parameterizes the supersymmetric
transformation. $F_i$ and $D_i$ are complex auxiliary fields which do
not propagate and can be eliminated using their classical equations of
motion.
The index $a$ runs over the adjoint representation of the gauge group
under which all the chiral fields transform in a representation with
hermitian matrices satisfying $[T^a,T^b]=if^{abc}T^c$. Finally the
gauge transformations are
\begin{subequations}
  \label{eq:5}
  \begin{align}
    D_\mu\phi_i&=\partial_u\phi_i+igA^a_\mu\left(T^a\phi\right)_i\\
    D_\mu\psi_i&=\partial_\mu\psi_i+igA^a_\mu\left(T^a\psi\right)_i\\
    D_\mu\lambda^a&=\partial_\mu\lambda^a-gf^{abc}A^b_\mu\lambda^c\\
    F^a_{\mu\nu}&=\partial_\mu A^a_\mu-\partial_\nu A^a_\mu-gf^{abc}A^b_\mu A^c_\nu
  \end{align}
\end{subequations}
where $g$ is the gauge coupling.  The gauge quantum numbers  of $\psi_i$ are the same as that its scalar superpartner
$\phi_i$.  For a complete discussion see~\cite{Martin:1997ns}.

As a result, in a renormalizable supersymmetric field theory, the
interactions and masses of all particles are determined just by their
gauge transformation properties and by the superpotential
$W$ (see for example \cite{Martin:1997ns})
\begin{equation}
  \label{eq:6}
  W=\frac12M^{ij}\widehat{\phi}_i\widehat{\phi}_j+\frac16y^{ijk}
\widehat{\phi}_i\widehat{\phi}_j\widehat{\phi}_k
\end{equation}
where the superfield $\widehat{\phi}$ is a singlet field which contains
as components all of the bosonic, fermionic and auxiliary field within
the corresponding supermultiplet, e.g.
$\widehat{\phi}_i\supset(\phi_i,\psi_i,F_i)$. $W$ determines the scalar
interactions of the theory as well the fermion masses and the Yukawa
couplings.

The superpotential, together with the $SU(3)_C\times SU(2)_L\times U(1)_Y$ gauge
symmetry, lead to the following generic lagrangian

\begin{multline}
  \label{eq:8}
  \mathcal{L}_{\mathrm{SUSY}}=-D^\mu\phi^{*i}D_\mu\phi_{i}
  -i\psi^\dagger\bar\sigma^\mu D_\mu\psi_i-\frac12M^{ij}\psi_i\psi_j
  -\frac12M^*_{ij}\psi^\dagger_i\psi^\dagger_j-
  \frac12y^{ijk}\phi_i\psi_j\psi_k\\
  -\frac{1}{2}y_{ijk}^{*}\phi^{*i}\psi^{\dagger j}\psi^{\dagger k}-
  \left[M^{*}_{ik}M^{kj}\phi^{*i}\phi_j+\frac{1}{2}
    M^{in}y^{*}_{jkn}\phi_i\phi^{*j}\phi^{*k}
  \right.\\
  \left.+\frac{1}{2}M^{*}_{in}y^{jkn}\phi^{*i}\phi_j\phi_k
    +\frac{1}{4}y^{ijn}y_{kln}^{*}
    \phi_i \phi_j \phi^{*ki} \phi^{* l}
  \right]\\
  +\sqrt{2} i g \lambda^a \Phi^{\dagger i} T^a \Psi_i +
  \frac{\sqrt{2}}{2} i g' \lambda ' \psi_i Y_{\phi_i} \phi^{* i} \\
  -
  \left[
    \frac{1}{2} g^2 \left( \Phi^{\dagger i} T^a \Phi_i
    \right) \left( \Phi^{\dagger j} T^a \Phi_j \right) +
    \frac{1}{4} {g'}^{2} \left( \phi^{* i} Y_{\phi_i} \phi_i \right)
    \left( 
      \phi^{* i} Y_{\phi_i} \phi_i 
    \right) 
  \right] \\
  -\frac14F^a_{\mu\nu}F^{\mu\nu a}-i\lambda^{{\dag}a}\bar{\sigma}^\mu D_\mu\lambda^a-i{\lambda'}^{\dag}\bar{\sigma}^\mu D_\mu\lambda'
\end{multline}
where $\Phi_i$ and $\Psi_i$ are the $SU(2)_L$ doublets (see eq.~(\ref{eq:13})).

Any realistic phenomenological model must contain
supersymmetry breaking. We use the usual approach of parameterized our
ignorance of the specific mechanism by just introducing extra terms
which break supersymmetry explicitly in the effective SUSY lagrangian.
The additional possible soft supersymmetry terms in the previous
lagrangian, assuming $SU(3)_C\times SU(2)_L\times U(1)_Y$ gauge symmetry is
\begin{equation}
  \label{eq:9}
  \begin{split}
    V_{\mathrm{soft}}=&-\frac12
    \left(
      M_1\lambda'\lambda'+M_2\lambda_2\lambda_2+M_3\lambda_3\lambda_3
      +\mathrm{h.c}
    \right)-m^2_{ij}\phi^{*j}\phi_i\\
    &\quad-
    \left(
      \frac12\, b^{ij}\phi_i\phi_j+\frac16\, a^{ijk}\phi_i\phi_j\phi_k +\mathrm{h.c}
    \right)
  \end{split}
\end{equation}

\section{One generation Supersymmetric Standard Model}
\label{sec:one-gener-supersymm}

Since the third generation fermions have the heaviest masses in the
Standard Model and the $L_\tau$ violating processes are less restricted,
it is often useful to make the approximation of keeping  only the third
family.

The most general renormalizable superpotential for one generation of
Standard Model quarks and leptons, imposing only gauge invariance
is~\cite{Nilles:1997ij,Navarro:1999tz,otros,Roy:1997bu,Joshipura:1995ib,Borzumati:1996hd,Diaz:1998xc,Diaz:1999wq,javi,Mukhopadhyaya:1999gy,Datta:2000yd,Diaz:1997xc}
\begin{equation} 
  \label{eq:10} 
  W=h_t\widehat Q_3\widehat U_3\widehat H_u
  +\lambda_0\widehat Q_3\widehat D_3\widehat L_0
  +h_{\tau}\widehat L_3\widehat E_3\widehat L_0
  -\mu_0\widehat L_0\widehat H_2
  -\mu_3\widehat L_3\widehat H_2
\end{equation}
where $L_0\equiv H_d$. Note that there is no possibility of $B$ violation
in the one generation case. Moreover we have chosen a basis where the
only allowed dimension--$4$ $L$ violating term, $\lambda'_{333}\widehat
L_3\widehat Q_3\widehat D_3$, is absent from the superpotential. This
minimizes the number of parameters in the superpotential because in
converse case, the $\mu_3$ terms should reappear when we evolve the
SUSY parameters to a different scale. The $\mu_0$ and $\mu_3$ are mass
terms that are protected by the supersymmetric non-renormalization
theorem (for a review see~\cite{Martin:1997ns}) . In particular, it
means that once we have a theory which can explain why $\mu_\alpha$
($\alpha=0,1,2,3$) is of order $10^2$ or $10^3\,$GeV at tree-level, we do
not have to worry about $\mu_\alpha$ made very large by radiative
corrections involving the masses of some very heavy unknown particles.
Thus, the $\mu_\alpha$ are called superrenormalizable terms to
differentiate from the bilinear soft mass terms which are not
protected by the supersymmetric non-renormalization theorem. We call
the models where the $L$ violation is induced only by $\mu_i$ terms and
have minimum content of fields ($i=1,2,3$), Superrenormalizable
Supersymmetric Standard Models (SSSM)%
\index{SSSM: Superrenormalizable Supersymmetric Standard Model}. In
particular, in the one generation case, the SSM is equivalent to the
SSSM.

The soft potential in this case is

\begin{multline}
  \label{eq:11}
  V_{\mathrm{soft}}=  \varepsilon_{ab}\left[ 
    A_t h_t\widetilde Q_3^a\widetilde U_3 H_u^b 
    +A_b \lambda_0 \widetilde Q_3^b\widetilde D_3 L_0^a 
    +A_\tau h_\tau\widetilde L_3^b\widetilde R_3 L_0^a\right.
  \\ 
  \left.-B_0\mu_0 L_0^a H_u^b-B_3\mu_3\widetilde L_3^a H_u^b\right] 
  \, + \mathrm{mass \,\, terms} .
\end{multline}

In this simple case it is easier the study of explicit calculations of $L$
violating process (section~\ref{sec:one-generation-ssm}), basis
independent $L$ violating parameters
(section~\ref{sec:basis-indep-param}), and tree level neutrino mass
generation (section~\ref{sec:orig-neutr-mass}).

\subsection{One generation SSM lagrangian}
\label{sec:one-generation-ssm}
The superpotential in eq.~(\ref{eq:10}), can be rewritten as
\begin{equation} 
  W=h_t \,\widehat{t^*_R} \widehat Q_3  \widehat H_2^0
  +\lambda_0\, \widehat{b^*_R} \widehat Q_3 \widehat L_0^0
  +h_{\tau}\, \widehat{\tau^*_R}\widehat L_3  \widehat L_0^0
  -\mu_0\, \widehat L_0^0 \widehat H_2^0
  -\mu_3\, \widehat L_3 \widehat H_2^0
  \label{eq:12}
\end{equation}
where
\begin{align*}
  \widehat Q_3&=
  \begin{pmatrix}
    \widehat{t_L} & \widehat{b_L}
  \end{pmatrix} &
  \widehat L_3&=
  \begin{pmatrix}
    \widehat{\nu_3} & \widehat{\tau_L}
  \end{pmatrix}
\end{align*}
which is of the form of the generic Lagrangian in eq.~\eqref{eq:8} if
we define
\begin{align*}
    (\phi_1,\psi_1)&=(\tilde t_L,t_L) & (\phi_5,\psi_5)&=(H_u^+,\tilde
    H_u^+) &(\phi_9,\psi_9)&=(\tilde\nu_L^\tau,\nu_L^\tau)\\
    (\phi_2,\psi_2)&=(\tilde b_L,b_L) & (\phi_6,\psi_6)&=(H_u^0,\tilde
    H_u^0) &(\phi_{(10)},\psi_{(10)})&=(\tilde\tau_L,\tau_L)\\
    (\phi_3,\psi_3)&=(\tilde t_R^*,\bar{t}_R) & (\phi_7,\psi_7)&=(L_0^0,\tilde
    L_0^0) &(\phi_{(11)},\psi_{(11)})&=(\tilde\tau_R^*,\bar{\tau}_R)\\
    (\phi_4,\psi_4)&=(\tilde b_R^*,\bar{b}_R) & (\phi_8,\psi_8)&=(L_0^-,\tilde
    L_0^-) 
\end{align*}
and with this notation we have
\begin{align}
  \Phi_i&=
  \begin{cases}
    \begin{pmatrix}
      \phi_i\\
      \phi_{i+1}
    \end{pmatrix}&\text{if $i$ is odd}, \\
    \quad0 &\text{if $i$ is even}
  \end{cases}&
  \Psi_i&=
  \begin{cases}
    \begin{pmatrix}
      \psi_i\\
      \psi_{i+1}
    \end{pmatrix}&\text{if $i$ is odd}, \\
    \quad0 &\text{if $i$ is even}
  \end{cases}
  \label{eq:13}
\end{align}

Comparing eqs.~(\ref{eq:6}) and \eqref{eq:12} we obtain the
non-vanishing couplings
\begin{align*}
  y^{136}&=h_t & y^{247}&=\lambda_0 & y^{(11)(10)7}&=h_\tau& M^{67}&=-\mu_0 &
  M^{69}&=-\mu_3\\
  y^{235}&=-h_t & y^{148}&=-\lambda_0 & y^{89(11)}&=-h_\tau& M^{58}&= \mu_0 &
  M^{5(10)}&=\mu_3
\end{align*}

In Appendix~\ref{sec:an-expl-calc} we show an explicit calculation by
using the previous formulas.

\subsection{Basis independent parameters}
\label{sec:basis-indep-param}
In the low energy superpotential of the one generation SSM in
eq.~(\ref{eq:10}), it is possible to rotate away the $\widehat
L_3\widehat H_u$ term from the superpotential by redefinition of
$\widehat H_d$ and $\widehat L_3$. It is worth stressing, however,
that such a redefinition does not leave the full lagrangian (including
soft breaking terms) invariant
\cite{Hall:1984id,Diaz:1999ge,Smirnov:1996ey,Joshipura:1995ib,Borzumati:1996hd,javi,Diaz:1997xc,Nardi:1997iy,Joshipura:1995wm,deCarlos:1996du,Nowakowski:1996dx,Banks:1995by,Diaz:1998vf}.
This generates a $\widehat L_3\widehat Q_3\widehat D_3$ term and also
affects the soft supersymmetry breaking terms. While the $\mu_3\widehat
L_3\widehat H_u$ term in the superpotential can be rotated away at a
fixed energy scale, the corresponding $\widetilde L_3 H_u$ would still
be present in the low energy theory.  Alternatively, if one chooses to
remove the $\mu_3$ term at the Planck scale, the terms $\widehat
L_3\widehat H_u$ will be radiatively generated due to the trilinear
$L$ violating term~\cite{Hall:1984id}. We
illustrate this statement below in the context of the neutrino mass
generation.

The two bilinear $L$ violating terms in eqs.~(\ref{eq:10}) and
(\ref{eq:11}) give rise to two different tree level contributions to
neutrino mass: (1) The term $\mu_3\widehat L_3\widehat H_u$ provides a
mixing between $\nu_\tau$ and $\tilde H_u$. (2) The term
$B_i\mu_i\widetilde L_3 H_u$ lead to a sneutrino vev $v_3$ which induce
a mixing between $\nu_\tau$ and the gauginos. Note that the contribution
to neutrino mass coming from the dimension--4 $L$ violating term was
rotated away. This is no longer possible in the full three--generation
SSM and may lead to too large tree level neutrino
mass \cite{Nardi:1997iy}.

The question is if all the bilinear terms can be rotated away from the
SSM lagrangian so that, for example, the neutrino mass can arise only
from the loop-level contribution from the dimension--4 $L$ violating
term.  We will see that this can only be done under very specific
conditions on the relevant soft parameters at the electroweak scale
which should look like unacceptable fine--tuning.  These conditions
can be determined from the minimization equations with $v_3=0$ in the
basis where already the superrenormalizable term have been rotated
away, inducing one trilinear violating term in the superpotential with
coupling $\lambda_3\equiv\lambda'_{333}$ .  Moreover, the conditions turn to be not
scale invariant because the Renormalization Group Equation (RGE) for
$\mu_3$ have a contribution proportional to $\mu_0$
\cite{Hall:1984id,Joshipura:1995ib,javi,Nardi:1997iy,Joshipura:1995wm,deCarlos:1996du,Nowakowski:1996dx}
\begin{equation}
  \label{eq:14}
  16\pi^2\frac{d\mu_3}{dt}=3\mu_0\lambda_0\lambda_3
\end{equation}
so that, even when $\mu_3$ is zero to one scale, it is radiatively
generated at another different scale.  However if the $\widehat
L_3\widehat Q_3\widehat D_3$ is absent at some energy, this will never
reappear in the superpotential because the RGE for $\lambda_3$ is
proportional to $\lambda_3$ itself.  Consequently the structure of the
superpotential in eq.~(\ref{eq:10}) is scale invariant.  The same
argument are valid in the SSSM with three generations.

To establish the conditions on the soft parameters we need first make
the rotation on the superfields that eliminates $\mu_3$ from the
superpotential \cite{Diaz:1999ge,javi}
\begin{equation}
  \label{eq:15}
  \begin{split}
    \widehat L_0'&=\frac{\mu_0}{\mu}\widehat L_0+\frac{\mu_3}{\mu}
    \widehat L_3\\
    \widehat L_3'&=\frac{\mu_0}{\mu}\widehat L_3-\frac{\mu_3}{\mu}
    \widehat L_0
  \end{split}
\end{equation}
where $\mu^2=\mu_0^2+\mu_3^2$. The sneutrino vev in this basis is just
\begin{equation}
  \label{eq:16}
  v_3'=\frac{\mu_0v_3-\mu_3v_0}{\mu}
\end{equation}
where $v_\alpha=\langle\tilde L_\alpha\rangle$, $\alpha=0,3$.  The bilinear term cannot be
rotated away from the SUSY lagrangian unless that $\mu_0/v_0=\mu_3/v_3$.
In this case $\mu_\alpha$ is aligned with $v_\alpha$, the $L$ violating
bilinears terms can be rotated away from the SSM lagrangian and the
neutrino acquires their mass only at the loop--level through the
induced trilinear $L$ violating term.  However in general $v_3\neq0$.
The misalignment can be quantified by means of an angle $\xi$ defined
as
\cite{Joshipura:1995ib,Borzumati:1996hd,Nardi:1997iy,Nowakowski:1996dx,Banks:1995by},
\begin{equation}
  \label{eq:17}
  \sin\xi=\frac{v_3'}{v_d}=\frac{\mu_0v_3-\mu_3v_0}{\mu v_d}
\end{equation}
where $v_d^2=v_0^2+v_3^2$. Using the minimization equations $\sin\xi$
can be written in terms $\Delta m^2=m^2_{L_0}-m^2_{L_3}$ and $\Delta
B=B_3-B_0$~\cite{Diaz:1999ge,Smirnov:1996ey,Diaz:1998vf} as
\begin{equation}
  \sin\xi=-\frac{\mu_3\mu_0}{\mu^2}
  \left(\frac{v_0}{v_d}\frac{\Delta m^2}{{m'}_{\tilde\nu^0_{\tau}}^2}+
    \frac{v_u}{v_d}\frac{\mu\Delta B}{{m'}_{\tilde\nu^0_{\tau}}^2}\right)
  \label{eq:18}
\end{equation}
where ${m'}_{\tilde\nu^0_{\tau}}^2$ is the tau sneutrino mass in the
MSSM. Consequently the necessary conditions in the soft terms are $\Delta
m^2=0$ and $\Delta B=0$. It is worth to stress that they correspond to
universality conditions in Supergravity (SUGRA)\index{SUGRA:
  Supergravity} scenarios naturally realized only at the unification
scale ($\sim10^{16}\,$GeV).  In fact, to supress $\sin\xi$ we need to
require universality only on lepton and Higgs soft masses.
Universality will be effectively broken at the weak scale due to
calculable renormalization effects.  For definiteness and simplicity
we will adopt this assumption throughout this work, unless otherwise
stated.

Consequently, the presence of the bilinear term $\widetilde L_3H_u$ in
the scalar potential in the low energy superpotential is unavoidable
if one assumes that some $L$ violation was produced in the
superpotential of the theory at some large scale (such as Planck or
grand unification scale).  In particular, the effect of $\widetilde
L_3H_u$ in the generation of the neutrino mass can be combined either
with the effect of $\widehat L_3\widehat H_u$ or $\widehat L_3\widehat
Q_3\widehat D_3$ term in the superpotential, depending on the basis
choice.

Therefore, one could argue that models which break explicitly $L$, in
which $\mu_i$ are neglected, may be considered to be intrinsically
incomplete, if not inconsistent.

It is worth noting also, in contrast with spontaneous $R$--parity
violation (\cite{Romao:2000up} and references therein) that doublet
sneutrino vev in the bilinear model is much more loosely constrained
because it is not subject to constraints from astrophysics
\cite{Comelli:1994nt}

We now turn to the calculation of the neutrino mass in an basis
independent way.

As observed before, the one generation SSM can be described in
various equivalent bases, for example
\begin{enumerate}
\item
  one in which bilinear term and sneutrino vev are non-zero, $\mu_3^I
  \neq 0$ and $v_3^I \neq 0$~\cite{beyond,Diaz:1999ge,Navarro:1999tz,Diaz:1999wq,javi,Diaz:1997xc,Bartl2000rp,Hirsch:1999kc,Porod:2000pw,Restrepo:2001me,Akeroyd:1998sv,Diaz:2000wm}
\item one in which trilinear term and sneutrino vev are non-zero,
  $\lambda_3^{II} \neq 0$ and $v^{II}_3 \neq 0$~\cite{Borzumati:1999th}
\item the vev-less basis in which $\mu_3^{III}$ and
  $\lambda_3^{III}$ are non-zero but
  $v_3^{III}=0$~\cite{Grossman:1998py,Bisset:1999nw}
\end{enumerate}
where the $L$ violating parameters can be expressed in terms of
dimension-less basis-independent alignment parameters $ \sin \xi$
$\sin \xi'$ and $ \sin \xi''$~\cite{Diaz:1999ge,javi,sacha} ($X=I,II$ or $III$) as
follows:
\begin{eqnarray}
  \label{eq:19}
  \sin\xi&=&\frac{\mu_0^X v_3^X-\mu_3^Xv_0^X}{\mu{v_d}}=
  \frac{v_3^{II}}{{v_d}}=-\frac{\mu_3^{III}}{\mu}\\
  \label{eq:20}
  \sin\xi'&=&\frac{\mu_0^X \lambda_3^X-\mu_3^X\lambda_0^X}{\mu h_b}=
  \frac{\lambda_3^{II}}{h_b}=-\frac{\mu_3^I}{\mu}\\
  \label{eq:21}
  \sin\xi''&=&\frac{-v_0^X\lambda_3^X+v_3^X\lambda_0^X}{{v_d}h_b}=
  \frac{v_3^I}{{v_d}}=-\frac{\lambda_3^{III}}{h_b}
\end{eqnarray}
where
\begin{equation} 
  h_b=\sqrt{{\lambda_0^X}^2+{\lambda_3^X}^2}\qquad
  \mu=\sqrt{{\mu_0^X}^2+{\mu_3^X}^2}\qquad
  {v_d}=\sqrt{{v_0^X}^2+{v_3^X}^2}, \quad X=I,II,\hbox{ or }III
  \label{eq:22}
\end{equation}
%

Note that, in the notation of eqs.~(\ref{eq:19})--(\ref{eq:21}), the
parameters $\mu_3$ and $\mu_0$ appearing in eq.~(\ref{eq:12})
should bear the superscript I.

Of these parameters only two are independent because they satisfy
\begin{equation}
  \sin\xi''=\cos\xi'\sin\xi-\sin\xi'\cos\xi
  \label{eq:23}
\end{equation}

$\sin\xi$ is easily generalized to three--generation case
\cite{Joshipura:1995ib,Borzumati:1996hd,Nardi:1997iy,Nowakowski:1996dx,Banks:1995by},
\begin{equation}
  \label{eq:24}
  \sin^2\xi=\frac12\,\frac{\Sigma_{\alpha,\beta}
    \left(
      \mu_\alpha v_\beta-\mu_\beta v_\alpha
    \right)^2}{\mu^2v_d^2}=\frac12\,\frac{\Sigma_{\alpha,\beta}\Lambda_{\alpha\beta}^2}{\mu^2v_d^2},\qquad
  \left(
    \mu^2\equiv\Sigma_\alpha\mu_\alpha\mu_\alpha,\quad v_d^2\equiv\Sigma_\alpha v_\alpha v_\alpha
  \right)
\end{equation}
and  $\alpha=0,1,2,3$.

From now on we will work in the $\lambda_3^I=0$--basis, unless otherwise
stated.  As a result we will omit the label $I$ in all the parameters
associated with this basis.  Note we can use $h_b$ instead $\lambda_0$. One
of the advantages in working in this basis is that the RGE's evolution
does not induce the trilinear $L$ violating terms neither in the
superpotential nor in the scalar potential \cite{javi}.

We define also the basis independent parameter
\begin{equation}
  \label{eq:25}
  \tan\beta'=\frac{v_u}{v_0^2+v_3^2}
\end{equation}

It makes sense in the $\lambda_3^I=0$--basis where the usual MSSM relation
\begin{equation}
  m_b=
  \frac1{\sqrt2}h_bv_0
  \label{eq:26}
\end{equation}
to introduce the following notation in spherical
coordinates for the vacuum expectation values (vev):
\begin{eqnarray} 
  v_0&=&v\sin\theta\cos\beta\cr 
  v_u&=&v\sin\theta\sin\beta\cr 
  v_3&=&v\cos\theta 
  \label{eq:27}
\end{eqnarray} 
which preserves the standard definition $\tan\beta=v_u/v_0$. 

\subsection{Origin of Neutrino Masses}
\label{sec:orig-neutr-mass}

In this model the presence of bilinear $L$ violating term in both the
superpotential and the soft potential induces a mass for the
$\nu_3$ at the tree level~\cite{Hall:1984id,Ellis:1985gi,arca}.
In order to study the $\nu_3$ mass it is convenient to have an
analytical expression for $m_{\nu_3}$ in this limit. The tree level
$\nu_3$ mass may be expressed in an basis independent way as
\cite{Diaz:1999ge,Joshipura:1995ib,javi,Nardi:1997iy,Nowakowski:1996dx}
\begin{equation}
  m_{\nu_{3}} \approx 
  -\frac{(g^2M_1+{g'}^2M_2){\mu}^2}{
    4 M_2 M_1{\mu}^2-2(g^2M_1+{g'}^2M_2){\mu}v_u{v_d} \cos\xi}{v_d}^2\sin^2\xi
  \label{eq:28}
\end{equation}
in terms of basis-independent parameters $\mu$, $v_d$ and $\sin \xi$
defined in eq.~\eqref{eq:22} and \eqref{eq:19}.  The formula was given
first in specific basis in for example
\cite{Hall:1984id,Dimopoulos:1988jw,Barger:1989rk,expl}.
The second term in the denominator may be neglected if $M_2,\mu\gtrsim m_Z$,
as often happens in minimal supergravity models with universal soft
SUSY breaking terms~\cite{Martin:1997ns,bartl:1996snow}.  Thus one may
obtain an estimate of the neutrino mass by keeping only the first term
in the denominator.
\begin{equation}
  m_{\nu_3}\approx\frac{g^2}{2M_2}{v_d}^2\sin^2\xi\,,
  \label{eq:29}
\end{equation}
where we have used $M_1=M_2{g'}^2/g^2$.  For $\sin \xi \approx 1$ one can
easily check that $m_{\nu_3}$ could be as large as the direct
experimental upper bound of 18 Mev~\cite{Barate:1998zg}. However in
SUGRA models with universality (in fact, we need to require
universality only of lepton and Higgs soft masses) one may obtain
naturally small $\sin \xi $ values, calculable from the RGE evolution
from the unification scale down to the weak scale
\cite{Smirnov:1996ey,Hempfling:1996wj,Nilles:1997ij}.  Indeed, using
the minimization equations $\sin\xi$ was written in terms $\Delta
m^2=m^2_{L_0}-m^2_{L_3}$ and $\Delta B=B_3-B_0$ in eq.~\eqref{eq:18}. In
term of the basis invariant quantities we have
\begin{equation}
  \sin\xi=-\cos\xi'\sin\xi'
  \left(\cos\xi\frac{\Delta m^2}{{m'}_{\tilde\nu^0_{\tau}}^2}+
    \frac{v_u}{v_d}\frac{\mu\Delta B}{{m'}_{\tilde\nu^0_{\tau}}^2}\right)
  \label{eq:30}
\end{equation}
One may give a simplified approximate analytical expression for the
$\nu_3$ in this model by solving the renormalization group
equations for the soft mass parameters $m^2_{L_0}$, $m^2_{L_3}$,
$B_0$, and $B_3$ in the one--step approximation.  This gives
\cite{Diaz:1999ge,Smirnov:1996ey,Hempfling:1996wj,javi,Nardi:1997iy,Diaz:1998vf}
\begin{eqnarray}
  \sin\xi\left|_{\Delta m^2}\right.&\approx&-\cos\xi'\sin\xi'\cos\xi
  h_b^2\left[\frac{m_{L_0}^2+M_Q^2+M_D^2+A_D^2}
    {{m'}_{\tilde\nu^0_{\tau}}^2}\right]\left(\frac3{8\pi^2}
    \ln\frac{M_{U}}{m_t}\right)\nonumber\\
  &\sim&-\cos\xi'\sin\xi'\cos\xi
  h_b^2\left(\frac3{8\pi^2}\ln\frac{M_{U}}{m_t}\right)
  \label{eq:31}
\end{eqnarray}
and
\begin{equation}
  \sin\xi\left|_{\Delta B}\right.\approx\cos\xi'\sin\xi'
  \tan\beta'
  h_b^2\left[\frac{\mu A_D}{{m'}_{\tilde\nu^0_{\tau}}^2}\right]
  \left(\frac3{8\pi^2}\ln\frac{M_{U}}{m_t}\right)
  \label{eq:32}
\end{equation} 
where we have denoted by the symbols $\sin\xi|_{\Delta m^2}$ and
$\sin\xi|_{\Delta B}$ the two terms contributing to $\sin\xi$ in
eq.~(\ref{eq:30}). In section~\ref{sec:lightest-stop-two} we compare
these formulas with the full numerical calculation and explore
non-universality effects. Using these expressions and assuming no
strong cancellation between these terms one finds that the minimum
neutrino mass is controlled by the $\sin\xi|_{\Delta m^2}$. As a result one
finds,
\cite{Diaz:1999ge,Smirnov:1996ey,Hempfling:1996wj,javi,Nardi:1997iy}
\begin{equation}
  m_{\nu_3}\left|_{\mathrm{min}}\right.\sim
  \frac{g^2m_b^2}{M_2}\left(\sin^2\xi'
    h_b^2\right)\left(\frac3{8\pi^2}
    \ln\frac{M_{U}}{m_t}\right)^2
  \label{eq:33}
\end{equation}

The above approximate analytical form of the $\nu_3$ mass is
useful, as we will see later (e.g. eq.~\eqref{eq:118}) in order to
display explicitly the degree of correlation between the $L$ violating
decays, such as $\tilde t_1\to b\tau$, with the $\nu_3$ mass.

The minimum value for $\sin \xi' h_b$ is determined by the value $\sin
\xi'$ and that of $\tan \beta$. For $\sin \xi' \sim 1$ and relatively
small 
$\tan \beta$ so that $h_t$ is perturbative, one has
\begin{equation}
  m_{\nu_3}\gtrsim 10 \hbox{KeV}
  \label{eq:34}
\end{equation}
for $M_2 \sim 1$ TeV. In order to get smaller $\nu_3$ masses one
needs to suppress $\sin^2 \xi'$ additionally, for example to reach one
electron-volt the required $L$ violating parameters are given in
Table~\ref{tab:1}.  These order-of-magnitude estimates are given in
terms of the basis--independent angles $\xi$ and $\xi'$, and in the
relevant parameters for the three bases defined before.

Note that whenever the parameter has two values, the first correspond
to $\tan\beta=2$ (the lower perturbativity limit) and the second to
$\tan\beta=35$. In Table~\ref{tab:1}, $\sin\xi$ was estimated from
eq.~(\ref{eq:29}) and $\sin\xi'$ from eq.~(\ref{eq:33}).

Note also that the RGE-induced suppression depends basically in the
$h_b^2$ factor in eq.~(\ref{eq:31}) which is $\sim 10^{-3}$ ($\sim 1$) for
small (large) $\tan\beta$. As a result the bigger the value of $\tan\beta$,
the smaller $\sin\xi'$ will have to be for a fixed $\nu_3$ mass.
The $L$ violating parameters in the several bases were estimated from
eqs.~\eqref{eq:19}, \eqref{eq:20} and \eqref{eq:23}.

\begin{table}
  \begin{center}
    \small
    \begin{tabular}{|c|c|c|c|c|c|c|c|c|c|c|c|c|c|c|c|}\hline
      &\multicolumn{4}{c|}{basis--independent}&
      \multicolumn{4}{c|}{Basis I: $\lambda_3^I=0$}&
      \multicolumn{3}{c|}{Basis II: $\mu_3^{II}=0$}&
      \multicolumn{4}{c|}{Basis III: $v_3^{III}=0$ }\\ \cline{2-16}
      &\multicolumn{2}{c|}{$\sin\xi$}&
      \multicolumn{2}{c|}{$\sin\xi'$}&
      \multicolumn{2}{c|}{$\!\!\mu_3^{\dag} \!\!$}&
      \multicolumn{2}{c|}{$\!\!v_3^{\dag} \!\!$}&
      \multicolumn{2}{c|}{$\lambda_3^{II}$}&$\!\!{v_3^{II}}^{\dag} \!\!$&
      \multicolumn{2}{c|}{$\lambda_3^{III}$}&
      \multicolumn{2}{c|}{$\!\!{\mu_3^{III}}^{\dag}\!\!$}\\ \hline
      $\!\!$(a)$\!\!$&$\!\!10^{-5}\!\!$&$\!\!10^{-4}\!\!$&$\!\!10^{-2}\!\!$&
      $\!\!10^{-4}\!\!$&1&$\!\!10^{-2}\!\!$&1&$\!\!10^{-3}\!\!$&
      \multicolumn{2}{c|}{$\!\!10^{-4}\!\!$}&$\!\!10^{-3}\!\!$&
      \multicolumn{2}{c|}{$\!\!10^{-4}\!\!$}&
      $\!\!10^{-3}\!\!$&$\!\!10^{-2}\!\!$\\ \hline
      $\!\!$(b)$\!\!$&$\!\!10^{-5}\!\!$&$\!\!10^{-4}\!\!$&$\!\!10^{-5}\!\!$&
      $\!\!10^{-4}\!\!$&$\!\!10^{-3}\!\!$&$\!\!10^{-2}\!\!$&
      $\!\!10^{-4}\!\!$&$\!\!10^{-4}\!\!$&
      $\!\!10^{-7}\!\!$&$\!\!10^{-4}\!\!$&$\!\!10^{-3}\!\!$&$\!\!10^{-7}\!\!$&
      $\!\!10^{-5}\!\!$&$\!\!10^{-3}\!\!$&$\!\!10^{-2}\!\!$\\ \hline
      \multicolumn{16}{l}{\small $^{\dag}$ in GeV}
    \end{tabular}
  \end{center}
  \caption[Estimated magnitude of $L$ violating parameters.]{\small
    Estimated magnitude of $L$ violating parameters required for a
    $\nu_3$ mass in the eV range in the three bases defined
    before for (a) SUGRA  (b) Horizontal symmetry. }
  \label{tab:1}
\end{table}

In eq.~\eqref{eq:33} we have neglected $\Delta B$ contribution with
respect to the one coming from $\Delta m^2$. It is possible, however, that
the $\Delta B$ term may be sizeable. In the large $\Delta B$ case then it may
cancel the $\Delta m^2$ contribution in $\sin\xi$, leading to an
additionally suppressed neutrino mass. As we will see, however, in
SUGRA models with universal soft terms at the unification scale we do
not need any substantial cancellation in order to obtain $\nu_3$
masses below the electron-volt scale.

In Horizontal models to be discussed in
section~\ref{sec:flavour-symmetries}, it is  possible make a
prediction for $\sin\xi$.  Instead of eq.~(\ref{eq:31}) that is roughly
of order $h_b^2\mu_3/\mu_0$, the Horizontal models predict, see
eq.~(\ref{eq:69}) and
\cite{Mira:2000gg,Borzumati:1996hd,Banks:1995by,Binetruy:1996xk}
\begin{equation}
  \label{eq:35}
  \sin\xi\sim\frac{\mu_3}{\mu_0}
\end{equation}
corresponding to no cancellation at all in the two terms contributing
to $\sin\xi$ in the first equality of eq.~(\ref{eq:19}).
Consequently, for a fixed neutrino mass, $\mu_i$ is required to be
much lower in horizontal models than in the SUGRA case. In
Table~\ref{tab:1} we compare all the parameters in the various basis
for the two cases.  This is the reason why the limits on $L$ violating
parameters are stronger in horizontal models
\cite{Mira:2000gg,Borzumati:1996hd,Banks:1995by}. And as a result, the
magnitude of $L$ violating processes is correspondingly suppressed in
these horizontal models.  In this way the strength of $L$ violation
effects at colliders may give light on the misalignment origin of the
SSSM.

In summary for one generation, one consistent SUSY model with minimum
field content based only in gauge principles can be constructed. This
model give rise to a tree level neutrino mass through $\Delta L=1$
violation and in general satisfies the bounds on $L$ violation and as
well as conserves $B$ number. 


\chapter{Supersymmetric Standard Models}
\label{sec:three-gener-supersym}

The field content of the Standard Model (SM) together with the
requirement of $G_{SM}=SU(3)_c\times SU(2)_L\times U(1)_Y$ gauge invariance
implies that the most general Lagrangian is characterized by
additional accidental $U(1)$ symmetries implying Baryon ($B$) and
Lepton flavor number ($L_i\,$, $i=e\,,\mu\,,\tau$) conservation at the
renormalizable level.  When the SM is supersymmetrized, this nice
feature is lost.  The introduction of the superpartners allows for
several new Lorentz invariant couplings.  The most general
superpotential respecting the gauge symmetries and with minimum field
content reads
\begin{subequations}
  \label{eq:36}
  \begin{align}
    W_1=&\varepsilon_{ab}
    \left[ 
      h_U^{ij}\widehat Q_i^a\widehat U_j\widehat H_u^b 
      +\lambda'_{ij0}\widehat Q_i^b\widehat D_j\widehat L_0^a 
      +\lambda_{0jk}\widehat L_0^a\widehat L_j^b \widehat E_k
    \right]\label{eq:5a}\\
    W_2=&-\varepsilon_{ab}\mu_0 \widehat L_0^a \widehat H_u^b \label{eq:5b}\\
    W_3=&- \varepsilon_{ab}\mu_i\widehat L_i^a\widehat H_u^b\label{eq:5c}\\
    W_4=&\varepsilon_{ab}
    \left[
      \lambda_{ijk} \widehat L_i^a \widehat L_j^b \widehat E_k +
      \lambda'_{i j k} \widehat L_i^a \widehat Q_j^b\widehat D_k 
    \right] \label{eq:5d}\\
    W_5=&\lambda''_{i j k} \widehat U_i \widehat D_j \widehat D_k &\label{eq:5e}\\
    W_6=&\varepsilon_{\mu\nu\rho}\varepsilon_{ab}\varepsilon_{cd}\frac{\kappa'_{ijkl}}{M_p}\,\widehat
    Q_i^{a\mu}\widehat Q_j^{b\nu}\widehat Q_k^{c\rho}
    \widehat L_l^d+\cdots
  \end{align}
\end{subequations}
where $i,j,k,l=1,2,3$ are generation indices, $a,b,c,d=1,2$ are
$SU(2)$ indices, $\mu,\nu,\rho$ are $SU(3)$ indices, and $\varepsilon$ is a
completely antisymmetric matrix, with $\varepsilon_{12}=1$.  For later use, it
is convenient to generalize the $L$ violating superrenormalizable term
to include the usual $\mu$--term: $\mu_\alpha\widehat L_\alpha\widehat H_u$,
where $\alpha=0,\ldots,3$, $\widehat L_0\equiv\widehat H_d$, $\lambda'_{ij0}\equiv
h^{ij}_D$, and $\lambda_{0jk}\equiv h^{ij}_E$.
The symbol ``hat'' over each letter indicates a superfield, with
$\widehat Q_i$, $\widehat L_i$, $\widehat L_0$, and $\widehat H_u$
being $SU(2)$ doublets with hypercharges $\frac13$, $-1$, $-1$, and
$1$ respectively, and $\widehat U$, $\widehat D$, and $\widehat E$
being $SU(2)$ singlets with hypercharges $-\frac43$, $\frac23$, and
$2$ respectively. The couplings $\mathbf{h_U}$, $\boldsymbol{\lambda'_{0}}$ and
$\boldsymbol{\lambda_{0}}$ are $3\times 3$ Yukawa matrices, and $\mu_0$ and
$\mu_i$ are parameters with units of mass.

We call this model Supersymmetric Standard Model (SSM) %
\index{SSM: Supersymmetric Standard Model} but we will see below at
least one initial description before ``Supersymmetric'' is necessary
in order to construct  viable phenomenological models. We will then
suggest the Superrenormalizable Supersymmetric Standard Model as the
minimal realization of SSM at low energies when the neutrino anomalies
are taken into account.
As it stands, eq.~(\ref{eq:36}) has potentially dangerous
phenomenological consequences
\begin{enumerate}[i)]
\item The simultaneous presence of the Dimension--4 ($D=4$)
  \index{Dimension--4}\index{$D=4$: Dimension--4} terms in
  eqs.~(\ref{eq:5d}) and (\ref{eq:5e}) that violates Baryon Number
  ($B$)\index{$B$: Baryon Number} and Lepton Number
  ($L$)\index{$L$: Lepton Number} give rise to a very fast proton
  decay.
  \label{item:1}

\item Flavor problem. The $D=4$ Yukawa couplings in
  eq.~(\ref{eq:5a}) are expected to be of order unity, suggesting that
  all the fermion masses should be close to the electroweak breaking
  scale.
  \label{item:2}

\item If the dimension--4 $B$ (or $L$) violating terms are not absent from the
  superpotential,  the trilinear couplings
  $\lambda_{ijk}\,,\lambda'_{ijk}$  (or $\lambda''_{ijk}$) are also expected to be of order
  unity, implying unsuppressed $L$ (or $B$) violating processes.
  \label{item:3}

\item The Dimension--5 ($D=5$)\index{Dimension--5}\index{$D=5$:
    Dimension--5} $B$/$L$ non-renormalizable violating couplings are
  also expected to be order unity, implying a too fast proton decay.
\label{item:4}
  
\item $\mu$ problem. The superrenormalizable\index{Superrenormalizable}
  parameters $\mu_\alpha$ are gauge and supersymmetric invariant, and thus
  their natural value is expected to be much larger than the
  electroweak and supersymmetry breaking scales.  A large value of
  $\mu_0$ would result in too large Higgsino mixing term (this is the
  supersymmetric $\mu$ problem). The parameters $\mu_i$ are required to
  be further suppressed by the smallness of the neutrino Masses 
  \label{item:5}

\item Large neutrino masses: the presence of general $D=4$, $L$
  violating terms expected to be order unity imply a potentially large
  tree level neutrino mass.
  \label{item:6}
\end{enumerate}
All these puzzles strongly indicate that SUSY models should be
restricted by some additional symmetry other than SUSY and $SU(3)\times
SU(2)\times U(1)$ gauge symmetry. In general we label such symmetry as $X$
symmetry.

The lack of explanation for the order unity couplings in
eq.~(\ref{eq:36}) is a problem common both SM and SSM. But we know SM
Yukawa couplings are of order $h_{ij}\sim\sqrt{m_im_j/v^2}$. In this way
and following~\cite{Hinchliffe:1993ad} we call natural conditions on
the dimension--4 and 5 $L$ and $B$ violating couplings to require them
to be order $\lambda_{ijk}\sim\sqrt{m_im_jm_k/v^3}$ and
$\kappa'_{ijkl}\sim\sqrt{m_im_jm_km_l/v^4}$ respectively, so that
$\lambda_{ijk}\gtrsim10^{-5}$ and $\kappa'_{ijk3}\gtrsim10^{-5}$. In this case the
problems in \ref{item:3})~\cite{Hinchliffe:1993ad,Nilles:1997ij} and
\ref{item:6})~\cite{Nardi:1997iy} disappear. However, proton stability
forces $\lambda'$ and $\lambda''$ to be much smaller
\cite{Dreiner:1997uz,Weinberg:1982wj}
\begin{equation}
  \label{eq:37}
  \lambda'_{11k}\lambda''_{11k}\leq10^{-27},\qquad k\neq1
\end{equation}
and \cite{Hinchliffe:1993ad}
\begin{equation}
  \label{eq:38}
  \lambda'_{ijk}\lambda''_{11k}\leq10^{-24},\qquad j=1,2
\end{equation}
When higher generations are involved, weaker constraints apply
\cite{Smirnov:1996ey}
\begin{equation}
  \label{eq:39}
  \lambda'_{ijk}\lambda''_{lmn}\leq10^{-10}
\end{equation}

\begin{figure}[htbp]
\label{fig:1}
  \def\shift{5} \def\alto{120} \def\altomitad{60} \def\altoshift{115}
  \def\altotercer{46.6} \def\altotercershift{51.6} \def\altotercermitad{23.3}
  \def\largo{300} \def\largomitad{150}
  \def\pl{10} \def\plshift{15} \def\kl{290} \def\klshift{285}
  \def\ul{20} 
  \def\ulend{280} 
  \def\dl{196.6}
  \def\vertice{78.2} \def\punto{2}
  \def\tshift{80pt} \def\raisep{16pt} \def\raisem{-20pt}
  \def\sul{103.3}
  \begin{center}
    \begin{picture}(320,\alto)(0,0)
      \Line(\pl,0)(\pl,\alto)
      \Line(\pl,0)(\plshift,0)
      \Line(\pl,\alto)(\plshift,\alto)
      \Text(0,\altomitad)[]{$p$}
      \Line(\kl,0)(\kl,\altotercershift)
      \Line(\kl,0)(\klshift,0)
      \Line(\kl,\altotercershift)(\klshift,\altotercershift)
      \Text(\kl,\altotercermitad)[]{\hspace{50pt}$\pi^+,K^+$}
      \ArrowLine(\ul,\shift)(\ulend,\shift)
      \Text(\largomitad,\shift)[]{\raisebox{15pt}{$u$}}
      \ArrowLine(\ul,\altotercer)(\sul,\vertice)
      \Text(\ul,\altotercer)[]{\hspace{\tshift}\raisebox{\raisep}{$d$}}
      \ArrowLine(\ul,\altoshift)(\sul,\vertice)
      \Text(\ul,\altoshift)[]{\hspace{\tshift}\raisebox{\raisem}{$u$}}
      \DashLine(\sul,\vertice)(\dl,\altoshift){6}
      \Text(\sul,\altoshift)[]{\hspace{\tshift}\raisebox{\raisem}{$\tilde e$}}
      \DashLine(\sul,\vertice)(\dl,\altotercer){6}
      \Text(\sul,\altotercer)[]{\hspace{\tshift}\raisebox{\raisep}{$\tilde u$}}
      \ArrowLine(\dl,\altotercer)(\ulend,\altotercer)
      \Text(\dl,\altotercer)[]{\hspace{\tshift}\raisebox{\raisep}{$\bar d$}}
      \ArrowLine(\dl,\altoshift)(\ulend,\altoshift)
      \Text(\dl,\altoshift)[]{\hspace{\tshift}\raisebox{\raisem}{$\bar \nu$}}
      \Line(\dl,\altotercer)(\dl,\altoshift)
      \Text(\dl,\vertice)[]{\hspace{20pt}{$\tilde \chi_i^+$}}
      \Vertex(\sul,\vertice){\punto}
    \end{picture}
    \label{fig:2}
  \end{center}
  \caption{Possible contribution to proton decay, involving $\kappa'_{ijkl}$}
\end{figure}
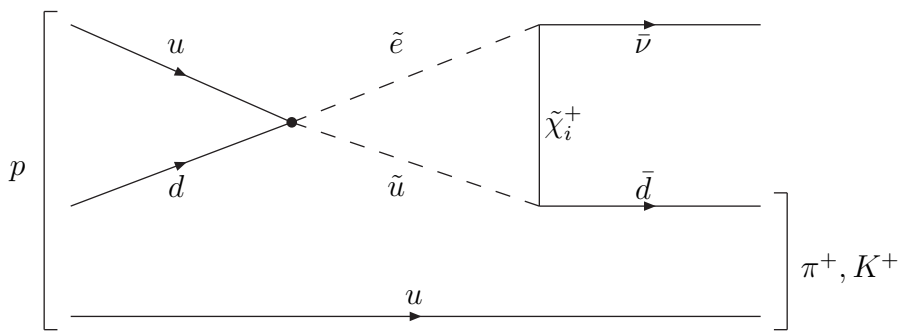

Concerning problem \ref{item:4}), SUSY models are sensitive to flavor
physics through $B$ violation suppressed by Planck scale. For
instance, the operator $(1/M_p)(Q_1Q_1)(Q_2L_i)$ gives a proton life
time shorter than the experimental bound by about 14 orders of
magnitude \cite{Murayama:1994tc}. An example of a contributing Feynman
diagrams shown in Fig.~\ref{fig:1}. In general the present upper
bound on the proton decay rate can be translated on $\kappa'_{ijkl}$ in
the following way \cite{Ben-Hamo:1994bq}
\begin{equation}
  \label{eq:40}
  \frac{\kappa'_{112l}}{M_{\mathrm{SUSY}}M}\leq10^{-29}\,\mathrm{GeV}^{-2}
\end{equation}
If, for example, we take $M\sim M_P/\sqrt{8\pi}=2.4\times10^{18}\,$GeV and
$M_{\mathrm{SUSY}}\sim 10^{3}\,$GeV, then eq.~\eqref{eq:40} become
\begin{equation}
  \label{eq:41}
  \kappa'_{112l}\leq10^{-8}
\end{equation}

Thus the simplest supersymmetric extension of the SM is excluded: an
extra $X$ symmetry is required to protect the proton.  Moreover
stability of proton allows either $L$ or $B$ violation, because the
nucleon decays normally requires not only baryon number violation
but also lepton number violation. This implies at least the absence
of either dimension--4 $L$ or $B$ violating terms and the absence of
the dimension--5 $B/L$ violating terms.  The most popular example of
$X$ symmetry is $R$--parity which forbids all the $\mathrm{dimension}\leq4$ $B$
and $L$ violating terms. Consequently it avoids the most dangerous
constraints in eq.~(\ref{eq:37}). However it was promptly realized
\cite{Weinberg:1982wj} that such a symmetry emerging for example from
minimal $SU(5)$ SUSY \cite{Dimopoulos:1981zb}, is in conflict with
proton decay because of the limit in eq.~\eqref{eq:41}. 
In fact, the minimal $SU(5)$ SUSY realization of $R_p$
\cite{Dimopoulos:1981zb} is nearly excluded by a combination of the
$\alpha_3(m_Z)$ prediction and the proton life time, and will be severely
tested at Superkamiokande and LEP2  \cite{Murayama:1996qe}.
Since the Planck scale operators give proton decay rates which are too
large, one needs another suppression mechanism. 

In this way the effect of the dimension--5 $B$/$L$ violating operators
cannot be neglected neither in strings nor Grand Unification Theories
(GUT)\index{GUT: Grand Unification Theory},  even after assuming that
they have natural couplings~\cite{Hinchliffe:1993ad}. This suggests
that $B$ and $L$ conservation in SUSY models may not be a consequence
of $R$ parity, but of a \emph{different $X$ symmetry}.

\section{Possible $X$ symmetries}
\label{sec:x-symmetry-1}

There is a wide variety of possibilities for the $X$ symmetry (see
\cite{Carone:1996nd} and references therein). One key ingredient of a
such symmetry is to have a justified origin. The $X$ symmetries have
their origin in one of the three categories of symmetries which occur in
field theory models of particle physics: space time symmetries, gauge
symmetries or flavor symmetries.  The $X$ symmetry  most frequently
used is $R$--parity. Since it acts in the anti-commuting
coordinate of superspace it can be viewed as a superspace analogue of the
familiar discrete space-time symmetries.  However the experience with
space time symmetries such as $P$ and $CP$, that in the real world are
broken, suggests that broken $R$ parity models are a likely
possibility.
                                                    
Assuming that the low energy discrete symmetries come somehow from a
larger gauge symmetry gives a rationale for the very existence of
discrete symmetries. Otherwise, the existence of discrete
symmetries seem unmotivated from a fundamental point of view.

We start considering the possibility for $X$ to be a discrete subgroup
of an enlarged anomaly free gauge symmetry. Such a symmetry will be an
anomaly free discrete symmetry that could arise naturally in string
theories
\cite{Ibanez:1992pr,Bento:1987mu,Ibanez:1991hv,Ibanez:1993ji,Hinchliffe:1993ad}.
This kind of symmetries include all previously discussed symmetries
used in this context such as matter parities \cite{Moreau:2000hz}

\section{Gauge Discrete symmetries}
\label{sec:gauge-discr-symm}

In $N=1$ supersymmetric theories there are two types Abelian internal
symmetries: ordinary symmetries and $R$--symmetries. The first commutes
with the SUSY generator and the second does not because the superspace
grassman variable $\theta$ is charged under the $R$--symmetry . They lead
to two types of discrete symmetries. They are (a) discrete $Z_N$ symmetries
\cite{Ibanez:1992pr,Bento:1987mu,Ibanez:1993ji,Hinchliffe:1993ad,Choi:1999,Choi:1997se,Ben-Hamo:1994bq,Pati:1996fn}
or (b) discrete $R$ symmetries
\cite{Ibanez:1993ji,Nilles:1997ij,Choi:1997fr,Kurosawa:2001iq,Chamseddine:1996gb}
  
We will study in detail only the $Z_N$ (case (a)) because the analysis for
the other is rather similar\cite{Ibanez:1993ji,Choi:1997fr}.
Specifically we consider a $Z_N$ symmetry under which each chiral
superfield transforms as
\begin{equation}
  \label{eq:42}
  \widehat\Phi\to\exp(i2\pi\alpha_\Phi/N)\widehat\Phi
\end{equation}
where the $\alpha_\Phi$ are additive $Z_N$ charges. An operator is allowed
only if the sum of its $Z_N$ charges is 0 (mod$\,N$).  We will assume
that the $Z_N$ charges are not family dependent.  A global discrete
symmetry is not protected against violation by Planck-scale and other
non-perturbative effects (see \cite{Ibanez:1993ji}). Moreover the
origin of discrete symmetries is arbitrary unless they are
\emph{gauge} discrete symmetries (see \cite{Ibanez:1993ji} and
references therein).  Furthermore it was pointed in
ref.~\cite{Ibanez:1991hv} that discrete gauge symmetries are
restricted by certain anomaly cancellation conditions.  Thus many
candidate gauge discrete symmetries may be ruled out on the basis of
these conditions. One way \cite{Eyal:1999gq,Choi:1997se} to obtain a
gauged discrete symmetry is to break a gauged $U(1)$ symmetry with an
order parameter (vev) whose charge is normalized to $q$, where the smallest
non-zero $U(1)$ charge assignment in the theory is
$q$~\cite{Eyal:1999gq,Choi:1999,Chun:1996xv}. For a complete
discussion on the possible origin of these gauge discrete symmetries
see \cite{Ibanez:1993ji}.

An analysis of the anomaly cancellation of discrete symmetries was
given in
\cite{Ibanez:1992pr,Ibanez:1991hv,Ibanez:1993ji,Hinchliffe:1993ad,Choi:1997fr,Berezinsky:1998pb}.
We will follow the notation of \cite{Hinchliffe:1993ad} and write down
the charges of the superfield as a vector of the form
\begin{equation}
  \label{eq:43}
  \alpha=
  \left(
    \alpha_Q,\alpha_u,\alpha_d,\alpha_L,\alpha_e,\alpha_{H_d},\alpha_{H_u}
  \right).
\end{equation}
The presence of the Yukawa couplings of eq.~(\ref{eq:36})
and invariance under hypercharge, reduce the number of independent
couplings to just three \cite{Hinchliffe:1993ad}. Thus we can choose a
convenient basis in which the charge of any field is given in terms of
three integers ($m$,$n$, and $p$)
\begin{subequations}
  \label{eq:44}
  \begin{align}
    \alpha_R&=(0,-1,1,0,1,-1,1),\\
    \alpha_A&=(0,0,-1,-1,0,1,0),\\
    \alpha_L&=(0,0,0,-1,1,0,0).
  \end{align}
\end{subequations}
So the total charge can be written as
\begin{equation}
  \label{eq:45}
  \alpha=m\alpha_R+n\alpha_A+p\alpha_L=(0,-m,m-n,-n-p,m+p,-m+n,m)
\end{equation}
The elements of the basis in eq.~(\ref{eq:44}) can be considered as the
three independent $Z_N$ generators discussed in
refs. \cite{Ibanez:1992pr,Ibanez:1993ji}
\begin{equation}
  \label{eq:46}
  R_N=\exp(i2\pi\alpha_R/N)\quad,\quad A_N=\exp(i2\pi\alpha_A/N)\quad,\quad L_N=\exp(i2\pi\alpha_L/N),
\end{equation}
In terms of these generators one can write the Flavor--independent
discrete symmetry as \cite{Ibanez:1992pr,Ibanez:1993ji}
\begin{equation}
  \label{eq:47}
  Z_N=R^m_N\times A^n_N\times L^p_N
\end{equation}

The conditions to have the various terms in  the
superpotential of eq.~(\ref{eq:36}) are
\begin{subequations}
  \label{eq:48}
  \begin{align}
    W_2&&\Rightarrow&&n=0 \quad (\mathrm{mod}\,N)\\
    W_3&&\Rightarrow&&m-n-p=0 \quad (\mathrm{mod}\,N)\\
    W_4&&\Rightarrow&&m-2n-p=0 \quad (\mathrm{mod}\,N)\\
    W_5&&\Rightarrow&&m-2n=0 \quad (\mathrm{mod}\,N)\\
    W_6&&\Rightarrow&&-n-p=0 \quad (\mathrm{mod}\,N)
    \label{eq:39e}
  \end{align}
\end{subequations}
where we have corrected a misprint in the sign of eq.~(\ref{eq:39e})
in \cite{Hinchliffe:1993ad}, and the $W_i$ were defined in
eq.~(\ref{eq:36}). They correspond to terms with couplings $\mu_0$,
$\mu_i$, $\lambda'_{ijk}$ (and $\lambda_{ijk}$), $\lambda''_{ijk}$, and $\kappa'_{ijkl}$
respectively.

If the $\mu_0$ term is allowed in the superpotential ($n=0$ mod$\,N$),
the possible models are quite restricted. We show them in the upper
part of Table~\ref{tab:2}. We also show there the equivalent matter
parity names \cite{Moreau:2000hz}. The generalized $Z_N$--type matter
parity is also given according to the definition given in
eq.~\eqref{eq:47}. Finally we display the conditions on $n$, $m$ and
$p$ that need be satisfied in order to obtain the corresponding set of
operators. All these conditions are mod$\,N$. The column for
$\lambda'_{ijk}$, actually includes also the $\lambda_{ijk}$ coupling.

\begin{table}[htbp]
  \begin{center}
    \small
    \begin{tabular}{|>{$}c<{$}|>{$}c<{$}|>{$}c<{$}|>{$}c<{$}|>{$}c<{$}|%
        r@{$\,$:$\,$}l|>{$}c<{$}|>{$}c<{$}|>{$}c<{$}|>{$}c<{$}|}
      \hline\hline
      \mu_0&\mu_i&\lambda'_{ijk}&\lambda''_{ijk}&\kappa'_{ijkl}&Name&Type&n&m&p&\\[0.5ex]\hline\hline
      \surd&\surd&\surd&\surd&\surd&SSM& &0&0&0&-\\[0.5ex]\hline
      \surd&\mathsf{X}&\mathsf{X}&\mathsf{X}&\mathsf{X}&MSSM&$R_N^mL_N^p$&%
      0&\neq0&\neq0&m\neq p\\[0.5ex]\hline
      \surd&\mathsf{X}&\mathsf{X}&\mathsf{X}&\surd&$R_p$SSM&$R_N^m$&0&\neq0&0&-\\[0.5ex]\hline
      \surd&\mathsf{X}&\mathsf{X}&\surd&\mathsf{X}&$L_p$SSM&$L_N^p$&0&0&\neq0&-\\[0.5ex]\hline
      \surd&\surd&\surd&\mathsf{X}&\mathsf{X}&$B_p$SSM&$R_N^mL_N^m$&0&\neq0&\neq0&m=p^{\dag}\\[0.5ex]\hline\hline
      \mathsf{X}&\mathsf{X}&\mathsf{X}&\mathsf{X}&\mathsf{X}&SSSM&$R_N^mA_N^nL_N^p$&\neq0&
      \neq n+p&\neq m-2n&p\neq -n^*\\[0.5ex]\hline
      \mathsf{X}&\multicolumn{2}{l|}{$\surd\Rightarrow\mathsf{X}$}&\mathsf{X}&%
      \mathsf{X}&SSSM&$R_N^{n+p}A_N^nL_N^p$&\neq0&n+p&\neq n&p\neq-n\\[0.5ex]\hline
      \mathsf{X}&\multicolumn{2}{l|}{$\mathsf{X}\Leftarrow\surd$}&\mathsf{X}&%
      \mathsf{X}&$B_p$SSM&$R_N^{2n+p}A_N^nL_N^p$&\neq0&2n+p&\neq0&p\neq-n^*\\[0.5ex]\hline
      \mathsf{X}&\mathsf{X}&\mathsf{X}&\surd&\mathsf{X}&&$R_N^{2n}A_N^nL_N^p$&\neq0&
      2n&\neq0&p\neq n^\star \\[0.5ex]\hline
      \mathsf{X}&\mathsf{X}&\mathsf{X}&\mathsf{X}&\surd&&$R_N^mA_N^nL_N^{-n}$&\neq0&
      \neq0&-n&m\neq n^\bullet\\[0.5ex]\hline
      &\multicolumn{2}{c|}{}&\surd&\mathsf{X}&&$R_N^{2n}A_N^nL_N^n$&\neq0&2n&n&2n\neq0\\
      \multicolumn{1}{|c|}{\raisebox{1.9ex}[0pt]{$\mathsf{X}$}}&%
      \multicolumn{2}{c|}{\raisebox{1.9ex}[0pt]{$\surd\Rightarrow\mathsf{X}\;$}}&%
      \surd&\surd&&$A_N^nL_N^n$&\neq0&2n&n&2n=0\\[0.5ex]\hline
      &\multicolumn{2}{c|}{}&\mathsf{X}&\surd&&$A_N^nL_N^{-n}$&\neq0&0&-n&2n\neq0\\
      \multicolumn{1}{|c|}{\raisebox{1.9ex}[0pt]{$\mathsf{X}$}}&%
      \multicolumn{2}{c|}{\raisebox{1.9ex}[0pt]{$\surd\Rightarrow\mathsf{X}\;$}}&%
      \surd&\surd&&$A_N^nL_N^{-n}$&\neq0&0&-n&2n=0\\[0.5ex]\hline
      \mathsf{X}&\multicolumn{2}{c|}{$\mathsf{X}\Rightarrow\surd$}&\multicolumn{2}{c|}%
      {$\surd\;\Rightarrow\mathsf{X}$}&&$R_N^{2n}A_N^n$&\neq0&2n&0&-\\[0.5ex]\hline
      \mathsf{X}&\multicolumn{2}{c|}{$\mathsf{X}\Rightarrow\surd$}&\multicolumn{2}{c|}%
      {$\mathsf{X}\Leftarrow\;\surd$}&&$R_N^nA_N^nL_N^{-n}$&\neq0&n&-n&-\\[0.5ex]\hline
      &\mathsf{X}& & & & &$R_N^{2n}A_N^nL_N^{-n}$&\neq0&2n&-n&2n\neq0\\
      \multicolumn{1}{|c|}{\raisebox{1.9ex}[0pt]{$\mathsf{X}$}}&%
      \surd&\multicolumn{1}{c|}{\raisebox{1.9ex}[0pt]{$\mathsf{X}$}}&%
      \multicolumn{1}{c|}{\raisebox{1.9ex}[0pt]{$\surd$}}&%
      \multicolumn{1}{c|}{\raisebox{1.9ex}[0pt]{$\surd$}}&&$A_N^nL_N^{-n}$&%
      \neq0&2n&-n&2n=0\\[0.5ex]\hline\hline
      \multicolumn{11}{l}{\small $^{\dag}$ This is a simplifying notation. %
        The real condition is $p-m=0$ (mod$\,N$). Consider for}\\[-0.8ex]
      \multicolumn{11}{l}{\small $^{}$\quad example: e.g, $N=3$, $n=-6$, %
        $m=-2$, $p=4$.}\\
      \multicolumn{10}{l}{\small $^*$Additional condition: $m\neq2n$}\\
      \multicolumn{10}{l}{\small $^\star$Additional condition: $p\neq-n$}\\
      \multicolumn{10}{l}{\small $^\bullet$Additional condition: $m\neq2n$}\\
    \end{tabular}
    \caption[Possible models arising for discrete symmetries]{Possible 
      models arising for discrete symmetries. All the relations in
      $n$, $m$, $p$ shown in the table are also mod$\,N$.  Note that
      the $n=0$ case is very restrictive in the type of possible
      models.  For the case of $n\neq0$ (mod$\,N$) the simultaneous
      presence of superrenormalizable and dimension--4 $L$ violating
      operators is not allowed.}
    \label{tab:2}
  \end{center}
\end{table}

Note that all models in $n=0$ case forbid proton decay avoiding the
simultaneous presence of renormalizable $L$ and $B$ violation.
Moreover, the anomaly cancellation conditions impose restrictions on
$N$, $m$, $n$, $p$. The corresponding possibilities are the symmetries
\cite{Ibanez:1993ji}: $R_3L_3^2$ (MSSM, $n=0$, $m=1$, $p=2$ in
Table~\ref{tab:2}); $R_2$, $R_3$ ($R_p$--SSM)\footnote{The $R_3$
  symmetry does essentially the same job as the standard $R_p$, at
  least with respect to the operators displayed in Table~\ref{tab:2}};
$L_3$ ($L_p$--SSM)\index{$L_p$--SSM: Lepton parity Supersymmetric
  Standard Model}; $R_3L_3$ ($B_p$--SSM)\index{$B_p$--SSM: Baryon
  parity Supersymmetric Standard Model}; and $R_3^2L_3$
(MSSM)~\cite{Castano:1994ec}

As the $\mu_0$ is likely to have a different origin than the Yukawas in
$W_1$ we also will consider the possibility that the $\mu_0$ can be
absent from the high energy superpotential. This corresponds to the
extra lines with $n\neq0$ in Table~\ref{tab:2}. In this case it will be
necessary to use a discrete version of the Green-Schwartz mechanism
(GS)\index{GS: Green-Schwartz}\cite{Green:1984sg}, and the number of
potentially anomaly free solutions substantially increases. Therefore,
this kind of symmetries may be well suited for the solution of the
$\mu$ problem. In fact, once mass parameters are introduced, the term
$\mu_\alpha\widehat L_\alpha\widehat H_u$ (rather than only $\mu_0\widehat
L_d\widehat H_u$) is the most natural choice.  Consequently the SSM
model with minimal set of couplings in this case, is likely to include
general superrenormalizable terms but not dimension--4 ones. We call
it Superrenormalizable Supersymmetric Standard Model (SSSM). The main
difference of the SSSM with the general $B_p$--SSM is that in the
former case the induced dimension--4 $L$ violating couplings are
naturally suppressed and automatically satisfy the experimental
constraints on $L$ violation.  Examples of anomaly free (through GS)
discrete symmetries in this case are \cite{Ibanez:1993ji}: $Z_2$
symmetries $A_2L_2$ ($n=1$, $m=2$, $p=1$) corresponding to
$A_N^nL_N^n$ case ($m=2n$, $p=n$, $2n=0$) in Table~\ref{tab:2}.  and
$R_2A_2L_2$ ($R_N^{2n+p}A^nL^p$ in Table~\ref{tab:2}). $Z_N$
symmetries with $N>3$ may be made anomaly free now.  Examples are
\cite{Ibanez:1993ji}: $A_4^{-1}L_4$ that correspond to $A_N^nL_N^{-n}$
case in the table (SSSM plus
$\widehat{Q}\widehat{Q}\widehat{Q}\widehat{L}$); and $A_5L_5^2$ that
correspond to $R^m_NA^n_NL^p_N$ case in Table~\ref{tab:2}
(SSSM)

The possibilities for anomaly free gauge discrete symmetries
stabilizing the proton further increases if one allows for gauged
$R$--symmetries \cite{Ibanez:1993ji}

The resulting models compatible with proton decay are listed in
Table~\ref{tab:3}.
 \begin{table}[htbp]
  \begin{center}
    \begin{tabular}{|c|c|r@{$\,:\,$}l|c|}
      \hline
      &&\multicolumn{2}{c|}{Symmetry}&\\[-0.6ex]
      \raisebox{2ex}[0pt]{Dim. $\leq5$ $L$ or $B$ violating terms}& \raisebox{2ex}[0pt]{Example}%
      & Name&type  &\raisebox{2ex}[0pt]{Origin} \\ \hline
      {\Large $ \blacktriangle$} &\cite{Castano:1994ec}&MSSM&$R_3^2L_3$ &Strings\\ \hline
      $\mu_i\widehat{L}_i\widehat{H}_u$ &\cite{Bento:1987mu,MASIpot3} & SSSM&$A_5L_5^2 $ &Strings\\\hline
      $\lambda''_{ijk}\widehat{U}_i\widehat{D}_j\widehat{D}_k$&\cite{Ben-Hamo:1994bq}&$L_p$SSM&$L_3$  &Flavor\\\hline
      $\lambda_{ijk}\widehat{L}_i\widehat{L}_j\widehat{E}_k%
      +\mu_i\widehat{L}_i\widehat{H}_u$               &\cite{Brahm:1989iy}  & --&--  &GUT\\ \hline
      $\lambda'_{ijk}\widehat{L}_i\widehat{Q}_j\widehat{D}_k%
      +\mu_i\widehat{L}_i\widehat{H}_u$                &\cite{Giudice:1997wb} & --&--        &GUT\\ \hline
      $\lambda_{ijk}\widehat{L}_i\widehat{L}_j\widehat{E}_k%
      +\lambda'_{ijk}\widehat{L}_i\widehat{Q}_j\widehat{D}_k%
      +\mu_i\widehat{L}_i\widehat{H}_u$              &\cite{Ibanez:1992pr} & $B_p$SSM&$R_3L_3$ %
      &Flavor\\ \hline
    \end{tabular}
    \caption[Possibilities of low energy $L$ or $B$
    violating superpotential renormalizable terms]{Possibilities of
      low energy Lepton or Baryon number violating superpotential
      renormalizable terms where the proton should be enough stable.
      For each case one specific model is cited with their
      correspondent discrete symmetry if possible.}
    \label{tab:3}
  \end{center}
\end{table}

The triangle in Table~\ref{tab:3} is just to illustrate that when we roll
down the ``hill'' the number of parameters in the superpotential 
increases: 0,3,9,12,30 and 39 respectively. This is the main reason
why most of the studies were initially performed in the context of the MSSM
(really in the context of $R_p$-SSM). When
the evidence of neutrino masses and mixings is taken into account
clearly the  minimum model which emerges is just the Superrenormalizable
SSM.  We now review all the possibilities with emphasis in the SSSM
alternative.

We define the superpotential with minimum content of couplings as the
one without any dimension $\leq5$ $L$/$B$ violating operators
\cite{Ibanez:1993ji,Castano:1994ec}
\begin{equation}
  \label{eq:49}
  W_{\mathrm{MSSM}}=W_1+W_2
\end{equation}
On the other hand, $R$--parity is
equivalent to the $R_2$  anomaly free gauge
discrete symmetry ($m=1$, $n=p=0$)
\begin{equation}
  \label{eq:50}
  R_2=R_3\Rightarrow\left(
    \alpha_Q,\alpha_u,\alpha_d,\alpha_L,\alpha_e,\alpha_{H_d},\alpha_{H_u}
  \right)=
  \left(
    0,-1,1,0,1,-1,1
  \right).
\end{equation}
In the $R_2$ case the high energy superpotential
contains~\cite{Castano:1994ec}
\begin{equation}
  \label{eq:51}
  W_{R_p}=W_{MSSM}+\frac{\kappa'_{ijkl}}{M}\widehat{Q}_i\widehat{Q}_j\widehat{Q}_k\widehat{L}_l+
  h_\nu\widehat{L}\widehat{H}_u\widehat{S}+M_S\widehat{S}\widehat{S}
\end{equation}
The latter allows both proton decay and neutrino masses with $R_p$
conservation.  For $R_3$ we should have $\lambda_S\widehat S\widehat S
\widehat S$ instead of the last term.  The introduced massive Majorana
superfield $\widehat S$ with $R_2$ charge $\alpha_S=-1$, is
\emph{necessary} to cancel the mixed gravitational anomaly
\cite{Ibanez:1993ji} and also leads to a $\Delta L=2$ Majorana neutrino
mass.

The presence of the $D=5$ term may give to rise to fast proton decay
\cite{Carone:1996nd,Ibanez:1992pr,Ibanez:1993ji,Hinchliffe:1993ad,Eyal:1999gq,Weinberg:1982wj,Ben-Hamo:1994bq,Choi:1997fr}.

\subsection{Superrenormalizable Supersymmetric Standard Model
  (Bilinear Model)}
\label{sec:super-supersymm-stan}

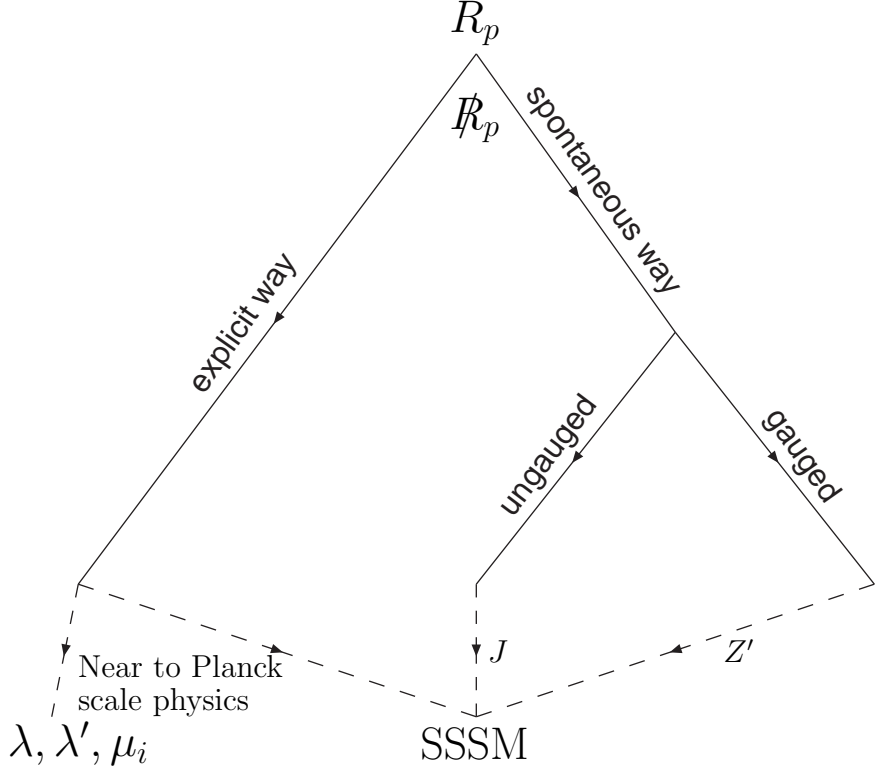
\begin{figure}[htbp]
\def\alto{260} \def\altoshift{60} \def\altomitad{155}
\def\ancho{310} \def\anchomitad{160} \def\anchodostercios{235}
\def\shift{10} \def\anchofull{320}
  \begin{center}
    \begin{picture}(\anchofull,270)(0,0)
      \SetPFont{Helvetica}{12}
      \ArrowLine(\anchomitad,\alto)(\shift,\altoshift)
      \ArrowLine(\anchomitad,\alto)(\anchodostercios,\altomitad)
      \ArrowLine(\anchodostercios,\altomitad)(\anchomitad,\altoshift)
      \ArrowLine(\anchodostercios,\altomitad)(\ancho,\altoshift)
      \DashArrowLine(\shift,\altoshift)(\anchomitad,\shift){6}
      \DashArrowLine(\anchomitad,\altoshift)(\anchomitad,\shift){6}
      \DashArrowLine(\ancho,\altoshift)(\anchomitad,\shift){6}
      \DashArrowLine(\shift,\altoshift)(0,\shift){6}
      \Text(\anchomitad,\alto)[]{\raisebox{30pt}{\Large $R_p$}}
      \Text(\anchomitad,\alto)[]{\raisebox{-60pt}{\Large $R_p$}}
      \Text(\anchomitad,\alto)[]{\hspace{-5pt}\raisebox{-58pt}{\Large $/$}}
      \Text(\anchomitad,0)[]{\Large SSSM}
      \Text(\anchomitad,35)[l]{\hspace{5pt}$J$}
      \Text(\anchomitad,35)[]{\hspace{200pt}$Z'$}
      \Text(\shift,0)[]{\Large $\lambda,\lambda',\mu_i$}
      \Text(\shift,35)[l]{\raisebox{-20pt}{Near to Planck}}
      \Text(\shift,35)[l]{\raisebox{-47pt}{scale physics}}
      \PText(70,\anchomitad)(55)[]{explicit way}
      \PText(185,110)(53)[]{ungauged}
      \PText(210,210)(-55)[]{spontaneous way}
      \PText(285,110)(-55)[]{gauged}
    \end{picture}
    \caption{Possibilities of $R_p$ breaking}
    \label{fig:99}
  \end{center}
\end{figure}

From now on we focus on the $R_p$--SSM superpotential defined in
eq.~(\ref{eq:51}). There are two main approaches to break $R_p$ in the bilinear way,
explicitly and spontaneously. See Fig.~\ref{fig:99}.

In the spontaneous approach, the breaking of $R_p$ occurs by
minimizing the scalar potential and the most relevant parameter is the
vacuum expectation value of the sneutrino which breaks lepton number
\cite{aul,ross:1985,Ellis:1985gi,Abreu:2001nc,MASIpot3,Romao:1992vu,MASI,Adhikari:1996bm,ROMA,Hirsch:2000xe,Barbieri:1990vb,Gato:1985jq,Chaichian:1996wr}.

For the ungauged $L$ case (see Fig.~\ref{fig:99}) there is a Goldstone
boson (Majoron) so that the theory must contain a singlet right handed
neutrino superfield \cite{MASIpot3,Romao:1992vu} with a vev which
typically lies at the weak scale. For the gauged case the Goldstone
boson is absorbed by a new gauge boson
\cite{Valle:1987sq,Gonzalez-Garcia:1991qf}. Either way one arrives to
a superrenormalizable supersymmetric model with a relatively small
$\mu_i$, as indicated by present anomalies in $\nu$--physics.

Here I will concentrate mainly on the explicit $R_p$ violating case in
which the superrenormalizable nature of the $R_p$ violation mechanism
must come from new physics near the Planck scale (see section
\ref{sec:altern-sssm-orig}). Note that in an explicit $R_p$ violating
model (like the SSSM) one can have also a sneutrino vev. Conversely a
spontaneous model may be accompanied by small explicit $R_p$ breaking
terms.

The resulting renormalizable superpotential which emerges is of the
form
\begin{equation}  
  \label{eq:54}
  W_{\mathrm{SSSM}}=\varepsilon_{ab}\left[ 
    h_U^{ij}\widehat Q_i^a\widehat U_j\widehat H_u^b 
    +h_D^{ij}\widehat Q_i^b\widehat D_j\widehat L_0^a 
    +h_E^{ij}\widehat L_i^b\widehat E_j\widehat L_0^a 
    -\mu_\alpha\widehat L_\alpha^a\widehat H_u^b 
  \right]\,,
\end{equation}
the soft superpotential is expected to satisfy the same symmetry of the
superpotential and therefore we have 
\begin{eqnarray} 
  V_{\mathrm{soft}}&=& 
  M_Q^{ij2}\widetilde Q^{a*}_i\widetilde Q^a_j+M_U^{ij2} 
  \widetilde U^*_i\widetilde U_j+M_D^{ij2}\widetilde D^*_i 
  \widetilde D_j+M_L^{\alpha\beta2}\widetilde L^{a*}_\alpha\widetilde L^a_\beta+ 
  M_E^{ij2}\widetilde E^*_i\widetilde E_j \nonumber\\ 
  &&\!\!\!\!+m_{H_u}^2 H^{a*}_u H^a_u\nonumber\\
  &&\!\!\!\!- \left[{\textstyle\frac{1}{2}} M_3\lambda_3\lambda_3
    +{\textstyle\frac{1}{2}} M\lambda_2\lambda_2
    + {\textstyle\frac{1}{2}}M_1\lambda_1\lambda_1+h.c.\right] 
  \nonumber\\ 
  &&\!\!\!\!+\varepsilon_{ab}\left[ 
    A_U^{ij}h_U^{ij}\widetilde Q_i^a\widetilde U_j H_u^b 
    +A_D^{ij}h_D^{ij}\widetilde Q_i^b\widetilde D_j L_0^a 
    +A_E^{ij}h_E^{ij}\widetilde L_i^b\widetilde R_j L_0^a\right.
  \nonumber\\ 
  &&\!\!\!\!\left.-B^{\alpha\beta}\mu_\beta L_\alpha^a H_u^b\right] \,,
  \label{eq:55}
\end{eqnarray} 
If the non-diagonal entries of $M_L^{\alpha\beta}$ and $B_{\alpha\beta}$ turn to be
negligible we can simply define: $m_{H_d}=M_{L}^{00}$,
$M_{L_i}=M_L^{ii}$, $B_0=B_{00}$ and $B_i=B_{ii}$.

In SSSM the appearance of misalignment between the $v_\alpha\equiv\langle\tilde
L_\alpha\rangle$ and $\mu_\alpha$, parameterized by $\sin\xi$ in eq.~(\ref{eq:24}),
is unavoidable with the possible exception of one point characterized
by very specific relations between some of the soft parameters. In the
one generation case they were simply $\Delta m^2=0$ and $\Delta B=0$ in
eq.~(\ref{eq:30}).  In the three--generation case these conditions are
\cite{Hall:1984id,Borzumati:1996hd,javi,Nardi:1997iy,Banks:1995by}.

\begin{enumerate}[(A)] 
\item $\mu_\alpha$ is an eigenvector of $m^2_{L_{\alpha\beta}}$, e.g, 
$m^2_{L_{\alpha\beta}}\mu_\alpha=m_0^2\mu_\alpha$
\label{item:7}
\item $B_{\alpha\beta}\mu_\beta/\mu_\alpha=B\,\delta_{\alpha\beta}$ for all $\alpha$. 
\label{item:8}
\end{enumerate}
and are naturally realized by the universal conditions at the
unification scale.  However, once $\sin\xi\neq0$, radiatively induced for
example, the term proportional to $\widetilde L_iH_u$ cannot be rotated
away of the SSSM lagrangian. They give rise to a plethora of new
effects without counterpart in a model with only dimension-4 $L$
violating terms (absent if only dimension--4 $L$ violating terms were
considered) like the emerging a tree level neutrino mass, additional
contribution to radiative neutrino masses, new decay modes for
superparticles, sneutrino splitting, and so on. See
section~\ref{sec:superr-l-numb} for a review on the phenomenology of
SSSM. Unless otherwise stated, we will assume universality of at least
lepton and Higgs soft masses at the unification scale in the SSSM. In
this way we may obtain naturally small tree level neutrino mass,
induced by the RGE running from the unification scale down to the weak
scale as explained in section~\ref{sec:orig-neutr-mass}.

We consider the model defined in eq.~\eqref{eq:54} and
eq.~\eqref{eq:55} as our reference Superrenormalizable Supersymmetric
Standard Model.

Note that $L$ violation can be rotated away from the superrenormalizable $L$
violating terms (but not from bilinear soft terms) into $L$ violating
dimension--4 terms
\begin{equation}
  \label{eq:56}
  W_{D=4}^{\not L}(\Lambda)=\frac{\mu_i}{\mu}h^D_{jk}\widehat{L}_i\widehat{Q}_j\widehat{D}_k+
\frac{\mu_i}{\mu}h_E^{jk}\widehat{L}_i\widehat{L}_j\widehat{E}_k
\end{equation}
where $\mu^2=\mu_0^2+\mu_i^2$. The new $L$ violating couplings are not
arbitrary and lead the same phenomenology of the general $B_p$--SSM
(see \cite{Roy:1997bu} for a detailed discussion on this).  Note also
that the rotation leading to \eqref{eq:56} is not scale invariant and
that $W_{D=4}^{\not L}$ will regenerate $L$ violation in
superrenormalizable terms via renormalization at scale $\Lambda'$. So that
a model without bilinear $L$ violating terms is incomplete and
inconsistent.

Before closing this section let us comment that another way to
generate the SSSM is provided by the first row with $n\neq0$ in
Table~\ref{tab:2}
\cite{Choi:1997se,Nilles:1997ij,Mira:2000gg,Choi:1997fr}. We will
study one specific example \cite{Mira:2000gg} in the next section.

\section{Alternative SSSM origin}
\label{sec:altern-sssm-orig}

Models about the origin of SSSM (with explicit $\mu_i$ terms) have been
constructed in the context of Strings
\cite{Bento:1987mu,Benakli:1997iu}; GUTs
\cite{Hall:1984id,Smirnov:1996ey,Hempfling:1996wj,Chkareuli:2000at,Chkareuli:2000tp,Brahm:1989iy,Hall:1990dg,Smirnov:1996jt,Suematsu:2001sp};
$R$--symmetries \cite{Nilles:1997ij,kaplan,Shafi:2000gw};
Peccei--Quinn symmetries
\cite{Tamvakis:1996dk,Joshipura:1998sp,Chun:1999cq,Chun:1999kd}; and
flavor symmetries \cite{Choi:1997se,Binetruy:1998sm,Mira:2000gg}. Next
we will consider the latter possibility.

\subsection{Flavor symmetries}
\label{sec:flavour-symmetries}
The emergence of string theories as a universal theory encompassing
all known fundamental interactions including gravity provides a
framework which allows to relate features of the effective low energy
theory which seemed otherwise uncorrelated. Of special interest for
the problems that we are discussing here is the presence of
non-renormalizable interactions, coming from a large number of
horizontal gauge symmetries, especially Abelian, which are
spontaneously broken at scales that may vary between the electroweak
scale and the Planck scale.

All these properties lied as to reconsider the original idea of
Froggatt and Nielsen (FN)\index{FN, Froggatt and Nielsen}
\cite{Froggatt:1979nt}.  The approach originally suggested by FN to
solve the flavor problem and account for the fermion mass hierarchy,
turns out to be quite powerful in the context of the SSM to solve also
the $\mu $ problem.  FN postulated an horizontal $U(1)_H$ symmetry that
forbids most of the fermion Yukawa couplings.  The symmetry is
spontaneously broken by the vacuum expectation value (vev) of a SM
singlet field $\chi\,$ and a small parameter of the order of the Cabibbo
angle $\theta =\langle\chi\rangle/M\simeq 0.22$ (where $M$ is some large mass scale) is
introduced.  The breaking of the symmetry induces a set of effective
operators coupling the SM fermions to the electroweak Higgs fields,
which involve enough powers of $\theta$ to ensure an overall vanishing
horizontal charge.  Then the observed hierarchy of fermion masses
results from the dimensional hierarchy among the various higher order
operators.  When the FN idea is implemented within the SSM, it is
often assumed that the breaking of the horizontal symmetry is
triggered by a single vev, for example the vev of the scalar component
of a chiral supermultiplet $\chi$ with horizontal charge $H(\chi)=-1\,$.

More recently it has been realized that the FN mechanism can play a
crucial role also in keeping under control the trilinear $B$ and $L$
violating terms in (\ref{eq:36}) without the need of introducing an ad
hoc R-parity quantum
number~\cite{Eyal:1999gq,Choi:1999,Barbier:1998fe,Choi:1997se,Binetruy:1998sm,Mira:2000gg,Binetruy:1996xk,Chun:1996xv,Joshipura:2000sn,Mira:2000fx}.
There are three ways to naturally suppress the $B$ and $L$ violating
couplings in models with horizontal symmetries without require too
large horizontal charges for the SSM fields
\begin{enumerate}[(a)]
\item Since $H(\chi)=-1\,$, then because the superpotential is
  holomorphic all the operators carrying a negative charge are
  forbidden in the supersymmetric limit.~\cite{Choi:1999,Binetruy:1998sm,Mira:2000gg,Binetruy:1996xk,Chun:1996xv,Mira:2000fx}
  \label{item:9}
  
\item If the operator carry a fractional charge only can appear
  at very high dimensionality~\cite{Choi:1997se,Binetruy:1998sm,Mira:2000gg,Binetruy:1996xk,Chun:1996xv}
  \label{item:10}
  
\item If the horizontal symmetry is not completely broken: a residual
  symmetry remain that forbids the dangerous operators
  \cite{Ibanez:1992pr,Eyal:1999gq,Choi:1999,Choi:1997se,Binetruy:1998sm}.
  In fact a simple way to obtain either $B_p$~\cite{Choi:1999} or
  $L_p$~\cite{Eyal:1999gq} is by the spontaneous breaking $U(1)\to
  Z_N$ which arises if the field which breaks $U(1)$ has a charge $N$
  normalized to the smallest charge of the theory
  \label{item:11}
\end{enumerate}

The mechanism (\ref{item:9}) is very related to the solution of the
$\mu_0$--problem.  If under $U(1)_H$ the superrenormalizable terms $L_0
H_u$ has a charge $n_0<0\,$, a $\mu_0$ term can only arise from the
(non-holomorphic) K{\"a}hler potential, suppressed with respect the
supersymmetry breaking scale $m_{3/2}$ as~\cite{Giudice:1988yz}
\begin{equation}
  \label{eq:66}
  \mu_0\simeq m_{3/2}\>\theta^{|n_0|}\,.
\end{equation}
A too large suppression ($|n_0| > 1$) would result in unacceptably
light Higgsinos, so that in practice on phenomenological grounds $n_0
= -1$ is by far the preferred value.

For example in~\cite{Mira:2000fx} it
was argued that under a set of mild phenomenological assumptions about
the size of neutrino mixings a non-anomalous $U(1)_H$ symmetry
together with the holomorphy conditions implies the vanishing of all
the superpotential $B$ and $L$ violating couplings.  A systematic
analysis on the restrictions on trilinear $L$ violating couplings
in the framework of $U(1)_H$ horizontal symmetries was also recently
presented in~\cite{Joshipura:2000sn}.
 
In this work we argue that if the $\mu_0$ problem is solved by the
horizontal symmetry in the way outlined above, and if the additional
bilinear terms $\mu_i$ are also generated from the K{\"a}hler potential and
satisfy the requirement of inducing a neutrino mass below the eV
scale, as indicated by data on atmospheric
neutrinos~\cite{Gonzalez-Garcia:2001sq,Fukuda:1998mi,atm99}, then in the basis
where the horizontal charges are well defined, all the trilinear
$L$ violating couplings are automatically absent.  This hints at
a self-consistent theoretical framework in which $L$ is violated
only by bilinear terms that induce a tree level neutrino mass in the
range suggested by the atmospheric neutrino anomaly, $L$ and $B$
violating processes are strongly suppressed, and the radiative
contributions to neutrino masses are safely small so that $m_\nu^{\rm
  loop}\approx 10^{-4}\,$eV, which barely allows for the LOW or
quasi-vacuum solutions to the solar neutrino
problem~\cite{Gonzalez-Garcia:2001sq,qva,MSW99} with $\tan\beta$ in the range $\approx10$--40.

In the following we will denote a field and its horizontal
charge with the same symbol, e.g.  $H(e_i)= e_i$ for the lepton
singlets, $H(Q_i)= Q_i$ for the quark doublets, etc.  It is also
useful to introduce the notation $f_{ij} =f_i-f_j$ to denote the
difference between the charges of two fields.  For example $L_{i0}$
denotes the difference between the charges of the $L_i$ `lepton
doublet' and the $L_0$ `Higgs field'.  On phenomenological grounds we
will assume that the charge of the $\mu_0$ term is $n_0 = -1\,$ and we
will also assume negative charges $n_i=L_i+H_u < n_0$ for the other
three bilinear terms $L_iH_u\,$.  It is worth stressing that the
theoretical constraints from the cancellation of the mixed $G_{SM}\times
U(1)_H$ anomalies hint at the same value $n_0=-1$ both in the
anomalous~\cite{Nir:1995bu} and in the non-anomalous
\cite{Mira:2000fx} $U(1)_H$ models.  With the previous assumptions the
four components of the vector $\mu_\alpha$ in (\ref{eq:36}) read
\begin{equation*}
  \mu_\alpha \simeq m_{3/2}\,
  (\theta^{|n_0|},\theta^{|n_1|},\theta^{|n_2|},\theta^{|n_3|})\,,
\end{equation*}
where coefficients of order unity multiplying each entry have been
left understood.  It is well known that if $\mu_\alpha$ and the vector of
the hypercharge $Y=-1/2$ vevs $v_\alpha \equiv \langle{L_\alpha}\rangle$ are not aligned
\cite{Hall:1984id,Banks:1995by} ($\sin\xi\neq0$ in eq.~(\ref{eq:24}))
the neutrinos mix with the neutralinos \cite{Hall:1984id,arca}, and
one neutrino mass is induced at the tree level giving by
eq.~(\ref{eq:67}) which can be rewritten as
\begin{equation}
  \label{eq:67}
  m_\nu^{\rm tree} \simeq \frac{\mu
    \cos^2\!\beta\,}{\sin2\beta\,\cos\xi-\frac{\mu
      M_1M_2}{M_Z^2M_\gamma}}\,\sin^2\!\xi \,,
\end{equation}
where $M_\gamma=M_1\cos^2\theta_W+M_2\sin^2\theta_W\,$, $M_1$ and $M_2$ are the
$U(1)_Y$ and $SU(2)_L$ gaugino masses, and $\tan\beta =
\langle{H_u}\rangle/\langle{H_d}\rangle\,$.
Since $m_b/\langle{H_u}\rangle \tan\beta \approx \theta^{2.7} \tan\beta$ (with $m_b(m_t)\sim
2.9\,$GeV \cite{Fusaoka:1998vc}) in the following we will use the
parameterization $\tan\beta=\theta^{\>x-3}$ that ranges between 90 and 1 for
$x$ between 0 and 3.  Keeping in mind that we are always neglecting
coefficients of order unity, we can approximate $\cos^2\beta
=(1+\tan^2\beta)^{-1}\approx \theta^{\>2\, (3-x)}$. Taking also $M_1\simeq M_\gamma\,$,
$\mu M_2 / M_Z^2\gg \sin2\beta\,\cos\xi\,$ and $100\,$GeV$\lesssim M_2\lesssim$
$500\,$GeV we obtain from (\ref{eq:67})
\begin{equation}
\label{eq:68}
  m_\nu^{\rm tree} 
  \approx \left[\theta^{\,-(5+x)}\> \sin\xi\right]^2 \,{\rm eV}\,.
\end{equation}
The magnitude of the tree-level neutrino mass as a function of
$\log_\theta\sin\xi \approx L_{30}$ for different 
values of $x$ (which in our notations parameterizes $\tan\beta$) 
is illustrated in fig. 1.  The
grey bands correspond to equation (\ref{eq:67}) with $M_2$ ranging
between $100\,$GeV and $500\,$GeV, while the dashed lines correspond
to the approximate expression (\ref{eq:68}).

In general, the two conditions (\ref{item:7}) and (\ref{item:8}) in
section \ref{sec:super-supersymm-stan} have to be satisfied to ensure
exact $\mu_\alpha$--$v_\alpha$ alignment and $m_\nu^{\rm tree}=0\,$.  In our
case the goodness of the alignment between $\mu_\alpha$ and $v_\alpha$ is
controlled by the horizontal symmetry, and in particular there is no
need of assuming universality of the soft breaking terms to suppress
$m_\nu^{\rm tree}$ to an acceptable level.  This is because the
previous two conditions are automatically satisfied in an approximate
way up to corrections of the order $\theta^{|L_{i0}|}\,$, where the
minimum charge difference between $L_0$ and the $L_i$ `lepton' fields
is responsible for the leading effects.  Thus we can estimate
\begin{equation}
\label{eq:69}
\sin\xi \approx  
\theta^{|L_{i0}|} =
\theta^{|n_i-n_0|} 
\simeq \frac{\mu_i}{\mu_0}\,.
\end{equation}
Confronting (\ref{eq:69}) with (\ref{eq:68}) it follows that in
order to ensure that $m_\nu^{\rm tree}$ is {\it parametrically}
suppressed below the eV scale we need
\begin{equation*} 
|n_i-n_0| > 5+x \qquad (i=1,2,3)\,.
\end{equation*}
\begin{figure}
\centerline{\includegraphics[height=9cm]{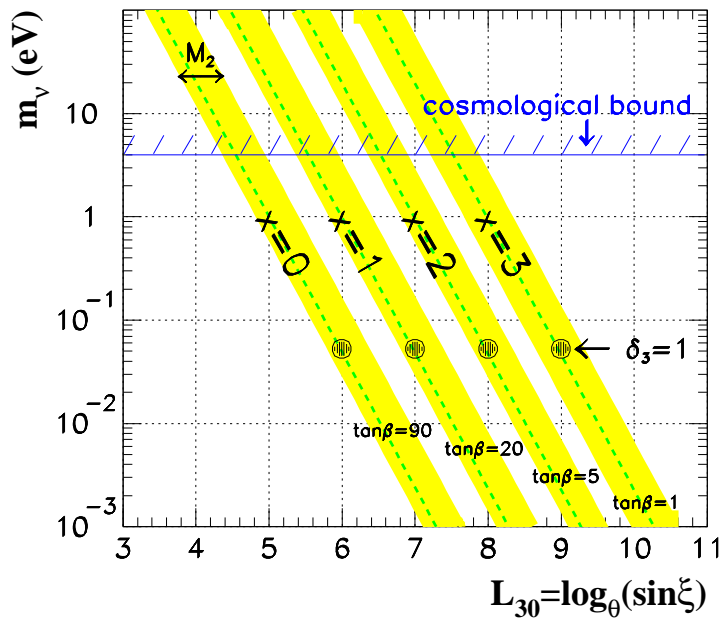}}
\caption[Tree-level neutrino mass dependence]{Tree-level neutrino
  mass dependence on $\log_\theta\sin\xi \approx L_{30}$ for different
  assignments of the charge difference $L_{30}=L_3-L_0$ and for
  different values of $\tan\beta$.  Details in the text.}
\label{fig:3}
\end{figure}

Let us introduce the parameterization
\begin{equation*} 
|n_i-n_0| - (5+x)= \delta_{i} \,.
\end{equation*}
Without loss of generality, we can also assume $n_1\leq n_2\leq n_3$
which implies
\begin{equation*}
m_\nu^{\rm tree} \approx  \theta^{\,2\delta_3}\>{\rm eV}\,. 
\end{equation*}
It is worth stressing that the parameter that controls the scaling of
$m_\nu^{\rm tree}$ with respect to changes in the values of the
horizontal charges is $\theta^{\,2}\simeq 0.05\,$, and thus neutrino
masses are much more sensitive to the horizontal symmetry than the
other fermion masses that scale with $\theta$.  For example
$\delta_3=-1$ yields $m_\nu^{\rm tree} \sim 20\,$eV; $\delta_3=0$ yields
$m_\nu^{\rm tree}\sim 1\,$eV; for $\delta_3=1$ all the
trilinear $L$ violating couplings are forbidden, and at the same
time $m_\nu^{\rm tree}\sim 5\times 10^{-2}\,$ eV (see Fig.~1) is in
the correct range for a solution to the atmospheric neutrino
problem~\cite{Gonzalez-Garcia:2001sq,Fukuda:1998mi,atm99}; finally, $\delta_3=2$
would suppress $m_\nu^{\rm tree}$ too much to allow for such a
solution.

Let us now write the down-quarks and lepton Yukawa matrices as
\begin{eqnarray}
h^D_{jk}&\simeq&\theta^{L_0+Q_j+d_k}=\theta^{Q_{j3}+d_{k3}+x}\,,
\qquad \nonumber   \\
h^E_{jk}&\simeq&\theta^{L_0+L_j+e_k}=\theta^{L_{j3}+e_{k3}+x}\,,\nonumber
\end{eqnarray}
where $x=L_0+Q_3+d_3=L_0+L_3+e_3\,$ consistently with our
parameterization of $\tan\beta$ and 
with the approximate equality between
the bottom and tau masses at sufficiently high energies (which in
particular allows for $b$--$\tau$ 
Yukawa unification).  The order of
magnitude of the trilinear $L$ violating couplings is then:
\begin{eqnarray}
\lambda'_{ijk}&\simeq&\theta^{n_i-n_0}\,h^d_{jk} \simeq
\theta^{Q_{j3}+d_{k3}-(5+\delta_i)}\,, \nonumber \\
\lambda_{ijk}&\simeq&\theta^{n_i-n_0}\,h^l_{jk} \simeq 
\theta^{L_{j3}+e_{k3}-(5+\delta_i)}\,.
\label{eq:70}
\end{eqnarray}
One can show that the phenomenological information on the charged
fermion mass ratios and quark mixing angles 
can be re-expressed in terms of the sets of eight charge
differences given in Table~\ref{tab:4}
\cite{Choi:1997se,Binetruy:1996xk,Leurer:1993wg,Binetruy:1995ru,Dudas:1995yu}.
Consequently gives rise to eight conditions on the fermion charges

\begin{table}[htbp]
  \begin{center}
    \begin{tabular}{|c|cccccc|c|cc|}
      \hline
      \rm model  & $Q_{13}$ & $Q_{23}$ & $d_{13}$ & $d_{23}$ & $u_{13}$ & $u_{23}$
      &\rm model & $L_{13}+e_{13}$ & $L_{23}+e_{23}$  \\ 
      \hline
      \strut MQ1: \ &\  3  & 2  &  1 & 0 &  5 & 2 & \ ML1: \ & 5 &\  2 \\
      \strut MQ2: \ & --3  & 2  &  7 & 0 & 11 & 2 & \ ML2: \ & 9 & --2 \\
      \hline
    \end{tabular}
    
    \caption{Models of horizontal charge differences}
    \label{tab:4}    
  \end{center}
\end{table}

 The phenomenological analysis leading to
 these sets of charge differences has been extensively
 discussed in the literature
 \cite{Choi:1997se,Binetruy:1996xk,Leurer:1993wg,Binetruy:1995ru,Dudas:1995yu}.

In~\cite{Mira:2000gg} was shown that in the framework of models of
Abelian horizontal symmetries, the phenomenological information on the
charged fermion mass ratios and quark mixing angles expressed in terms
of the eight horizontal charge differences in Tab.~(\ref{tab:4}), when
complemented with the requirement that $m_\nu^{\rm tree}$ is adequately
suppressed below the eV scale ($\delta_i \geq 1$) hints at one
self-consistent model (MQ1+ML1) where all the $\lambda$ and $\lambda'$ couplings
vanish.  It is interesting to note that $\delta_3=1$ which yields
$m_\nu^{\rm tree}\approx \theta^2$~eV in the correct range required by the
atmospheric neutrino problem is also the minimum value that ensures
$\lambda=0\,$, $\lambda'=0$ and, as we will below, $\lambda''=0\,$.

Concerning MQ2 either the neutrino masses are uninterestingly small
there, or the $\lambda'$ conflicts with existing experimental limits. MQ2
is also excluded by the requirement that the $\lambda''$ couplings vanish
  
In ML2, once we set $L_{23}=0$ to allow for maximal
  $\nu_\mu$--$\nu_\tau$ mixing, the
  lepton mass ratios can be correctly reproduced
  only if $L_{12}\geq 4\,$, which would again exclude the possibility
  of explaining the solar neutrinos deficit through 
  $\nu_e$--$\nu_\mu$ oscillations.

In conclusion, our analysis results in the  set of 
fields charge differences and of $n_\alpha=L_\alpha+H_u$ 
charge sums displayed in Tab.~\ref{tab:5}

\begin{table}[htbp]
  \begin{center}
    \begin{tabular}{|cccccc|cccc|cc|}
      \hline
      $Q_{13}$ & $Q_{23}$ & $d_{13}$ & $d_{23}$ & $u_{13}$ & $u_{23}$
      & $L_{13}$ & $L_{23}$ & $e_{13}$ & $e_{23}$ & $n_i   $ & $n_0 $\\ 
      \hline
      \strut  3  & 2  &  1 & 0 &  5 & 2  
      & 0  & 0  &  5 & 2 & $-7-x$& $-1$ \\
      \hline
    \end{tabular}
    \caption{Final charge differences}
    \label{tab:5}
  \end{center}
\end{table}
If we further use the analysis of the neutrino loop effects we would
have in addition the value $x=1$ (corresponding to $\tan\beta \approx
10$--$40\,$). Of course, $x=0$ implies that the value of $\tan\beta$ is
very large ($\sim60$) and therefore this case is phenomenological
disfavored \cite{Barger:1993ac,Grossman:1994ax}. Finally $x=2$
($\tan\beta\sim5$) would yield a too large suppression to the loop neutrino
mass to be interesting for the solar neutrinos.

Models based on a single $U(1)_H$ Abelian factor are completely
specified in terms of the horizontal charges of the SM fields.  There
are five charges for each fermion family plus two charges for the
Higgs doublets, for a total of 17 charges that {\it a priori} can be
considered as free parameters (the charge of the $U(1)_H$ breaking
parameter $\theta$ is just a normalization factor). The 17 horizontal
charges are constrained by eleven phenomenological conditions
corresponding to the eight constraints in Table~\ref{tab:4}, the $m_t$
value, the condition $m_b\approx m_\tau$ at high energy and the preferred
value for the solution of the $\mu_0$ problem $n_0=-1$. Also by two
theoretical conditions from anomaly cancellation through the GS
mechanism. This leaves us with four free parameters, and we can chose
them to be the charges $n_i$ ($i=1,2,3$) of the bilinear terms $\mu_i$,
and $x=Q_3+d_3+L_0$ that fixes the value of $\tan\beta\,$.  The
expressions of the horizontal charges for all the SM fields as a
function of these four parameters is given in the
Appendix~\ref{cha:constr-horiz-charg}. Then the self-consistent
solution given in Table~\ref{tab:5} is used to find the 17 individual
charges presented in Table~\ref{tab:10}. In the present case the
K{\"a}hler contributions to the trilinear couplings are of order
$m_{3/2}/M_p$ and powers of $\theta$ depending on the horizontal charges
of the various dimension--4 terms. Moreover, for this set of
individual charges the horizontal charges of $\lambda''$ are in addition
fractional and consequently zero.  In summary the resulting Dimension
4 and 5 $L$ and $B$ violating couplings in the basis where the
horizontal charges are well defined are
\begin{equation}
  \label{eq:71}
  \begin{split}
    \mu_i&=m_{3/2}\theta^8,\qquad \lambda'_{ijk}\approx
    (m_{3/2}/M_p)\theta^{|L_i+Q_j+d_k|},\qquad
    \lambda_{ijk}\approx (m_{3/2}/M_p)\theta^{|L_i+L_j+e_k|},\\
    \lambda''_{ijk}&=0,\qquad\qquad \kappa'_{ijkl}=0.
  \end{split}
\end{equation}
This clearly define one SSSM.

In summary we have obtained one SSSM originating from an anomalous
horizontal symmetry where the anomalies are canceled through a GS
mechanism.  We have assumed that all the superrenormalizable terms
coupling the up-type Higgs doublet with the four hypercharge $-1/2$
doublets carry negative horizontal charges, and hence are forbidden by
holomorphy.  We have constrained the value of these charges by several
theoretical and phenomenological requirements, such as having an
acceptable Higgsino mass ($\mu$ problem) and neutrino masses suppressed
below the electron-volt scale, as suggested by present neutrino data.
We have found that under these conditions all the trilinear $L$
violating superpotential couplings vanish, yielding a SSSM which is
defined by the charge differences in Tab.~(\ref{tab:5}), where lepton
number is mildly violated only by small bilinear terms.  The model
allows for neutrino masses in the correct ranges suggested by the
atmospheric neutrino problem and by the LOW and quasi-vacuum solutions
to the solar neutrino problem.  However, no precise theoretical
information can be obtained about the neutrino mixing angles except
for the fact that, unlike the quark mixings, there is no parametric
suppression of their values and thus they can be naturally large. Note
that this model solves all the problems mentioned at the beginning of
this chapter.

A similar mechanism but without the solution to the $\mu$ problem was
implemented in~\cite{Binetruy:1998sm}. In the example considered there
the dimension--4 $L$ and $B$ violating operators are absent by
holomorphy before the SUSY breaking.  The dimension--4 $L$ and $B$
violating terms are generated from the K{\"a}hler potential and one get
$\lambda'\lambda''\sim10^{-26}$.  However the strength of the superrenormalizable
term $\theta^{|-13|}$ is too small to be relevant in neutrino physics.
The resulting individual charges for $\mu_i\approx m_{3/2}\theta^{|-6|}$ are
shown in Table~\ref{tab:11} .  The dimension--4 $L$ and $B$ violating
terms satisfy now $\lambda'\lambda''\lesssim10^{-19}$.

\section{Phenomenology of SSSM}
\label{sec:superr-l-numb}
The phenomenology of the SSSM has been extensively studied in
literature. In particular:

\begin{itemize}
\item Neutralino-neutrino mixing:
The tree level neutrino mass has been studied in
\cite{Hall:1984id,Lee:1984kr,Dawson:1985vr,Banks:1995by,Barbieri:1990qj,Davidson:2000uc}.
Analysis of one-loop neutrino masses and/or Solution to the neutrino
anomalies can be found in \cite{Hall:1984id,Joshipura:1995ib,Borzumati:1996hd,Datta:2000yd,Takayama:2000pc,Bednyakov:1998cx,Chun:1999gp,Bhattacharyya:1999tv,Haug:2000kr,Mukhopadhyaya:2000wn,Chun:2000bq,Davidson:2000ne,Abada:2000xr,Abada:2001ma}.
Numerical calculation with the full one-loop corrections:
\cite{Romao:2000up,Hempfling:1996wj}.
An analytical study including the effect of arbitrary dimension--4 $L$
violating terms was performed in \cite{Nardi:1997iy}. 
The neutrino spectrum in horizontal models have been estimated in
\cite{Choi:1997se,Binetruy:1998sm,Mira:2000gg,Borzumati:1996hd,Banks:1995by,Chun:1996xv,Joshipura:2000sn,Chun:1999gp}

Neutralino decays: \cite{Hall:1984id,Dawson:1985vr,Smirnov:1996ey,Nilles:1997ij,Roy:1997bu,Joshipura:1995ib,Bartl2000rp,Hempfling:1997je,Porod:2000hv}.
Tevatron signatures of neutralino decay in \cite{Datta:2000xq}.
Neutralino decay processes based on the assumption that the
atmospheric neutrino masses and mixings are mainly due to the bilinear
terms in \cite{Choi:1999tq}. In \cite{Hall:1984id,kaplan} is concluded
that the dominant decay of the neutralino is into $b\,\bar b$ quark
jets and neutrino (or $b\,\bar t l^+$ if energetically allowed). 
In~\cite{Datta:2000xq,Smirnov:1996ey,Mukhopadhyaya:1998xj} it is shown that
the decay of the lightest neutralino $\tilde\chi_1^0$ will produce
comparable numbers of muons and taus as a result of the large mixing
implied by the atmospheric anomaly. This result is also obtained in
\cite{Romao:2000up,Porod:2000hv}.
Some of the neutralino decays channels for the one generation case
were studied in~\cite{Godbole:1993fb}.
The full numerical calculation of the neutralino decay in the one
generation case was presented in~\cite{Bartl2000rp}, while the full
numerical calculation in the three--generation case was presented
in~\cite{Porod:2000hv}.
A systematic study of the SSSM including approximate formula for the
neutralino--neutrino mixing was performed in
\cite{Romao:2000up,Nowakowski:1996dx,Hirsch:1999kc,Porod:2000hv,Hirsch:2000jt}.
In \cite{Chun:1999ub} it is shown that neutralino mix only with the
heaviest neutrino in SSSM.

\item Neutral Higgs -- sneutrino mixing.
Sneutrino production and/or decays have been studied in 
\cite{Nilles:1997ij,deCampos:1995av,Roy:1997bu} and 
recently in~\cite{Chun:2001mm}.
Sneutrino mass splitting have been analyzed in 
\cite{Grossman:1998py,Chun:2001mm,Grossman:1999hc}.
An basis--independent analysis of this
problem was performed in 
\cite{Grossman:2001ex}.
In \cite{Joshipura:1995wm} is studied the effect of $\widetilde L_3
H_u$ term of the low energy scalar potential in the generation of
spontaneous CP violation.
Scalar resonance enhancement in $e^+e^-\to Z^0Z^0\,W^+W^-$  and in
$e^+e^-\to t \bar{t}$ due two the sneutrino-Higgs mixing, have been
recently studied in \cite{Bar-Shalom:2001ew}.

\item Charged Higgs -- stau mixing.
Charged scalar decays have been studied in~\cite{Akeroyd:1998iq}

\item Stop decays have been studied in
  \cite{Diaz:1999ge,Bartl:1996gz,Porod:2000pw,Restrepo:2001me}
  
\item Rare decays: $b\to s\gamma$~\cite{Diaz:1999wq}; $Z^0$ decays as
  $Z^0\to\bar{\nu_3}\tilde\chi_1^0$ and $Z^0\to\tau^+\tilde\chi_1^-$ in
  \cite{Barbieri:1990vb,Akeroyd:1998sv}; flavor changing coupling of the $W$
  and $Z$ \cite{Roy:1997bu}; top decays \cite{Navarro:1999tz}.  The double beta
  decay have been studied in
  \cite{Hirsch:1999kc,Hirsch:2000jt,Faessler:1998db}.  Proton decay in
  bilinear models were studied in
  \cite{Chun:1999ub,Bhattacharyya:1998dt,Bhattacharyya:1999bx}.
In particular in \cite{Bhattacharyya:1998dt,Bhattacharyya:1999bx}
the limit in eq.~\eqref{eq:38} translate into
\begin{equation}
  \label{eq:74}
  \frac{\mu_i}{\mu_0}\lambda''_{112}\lesssim10^{-21}
\end{equation}
which is important in SSSM where the dimension--4 $B$ violating
couplings are generated from the K{\"a}hler potential without further
suppression, $\lambda''_{ijk}\lesssim\Lambda_{ijk}m_{3/2}/M_p$ ($j=1,2$).
$\mu\to e\gamma$ and the numerical study of misalignment
\cite{deCarlos:1996du}.
$\mu\to e\gamma$ and other rare processes 
\cite{Lee:1984kr,Dawson:1985vr,Frank:2000gs}.
Leptonic phenomenology and constraints \cite{Bisset:1999hz}.

\item Others:
Effects with large neutrino mass: \cite{Diaz:2000wm}.
SSSM in gauge mediated SUSY:
\cite{Chun:1999cq,Choi:1999tq,Chun:1999gp,Joshipura:1999fn}.
Fermion dipole moment:  \cite{Choi:2001bg}.
Magnetic moment of the neutrino \cite{Babu:1990px}.

\item Spontaneous breaking of R-parity. In addition to the papers
  quoted in Section~\ref{sec:super-supersymm-stan}, the phenomenology
  of this models have been studied in
  \cite{Chaichian:1993ra,Adhikari:1996bm,Frank:2001tr,Bartl:1996gz,Huitu:1999qu,deCampos:1999mf}

\end{itemize}


\chapter{The mass spectrum of the SSSM}
\label{sec:mass-spectrum-mssm}
Here we discuss the tree level structure of the supersymmetric masses
and mixing of SSSM, discussing the fermion  (chargino, neutralino)
and scalar (squarks, sleptons and Higgs) mass matrices 

\section{Chargino mass matrix}
\label{sec:chargino-mass-matrix}

In the SSSM the charginos mix with the charged leptons forming a set
of five charged fermions $F_i^{\pm}$, $i=1,\ldots,5\,$ in two component
spinor notation. In a basis where ${\psi^+}^T=[-i\lambda^+,\widetilde
H_u^+,e_R^+,\mu_R^+,\tau_R^+ ]$, the mass terms in the lagrangian are
\begin{equation*}
  \mathcal{L}_{SSSM}=-\frac12
  \begin{bmatrix}
    {\psi^+}^T&{\psi^-}^T
  \end{bmatrix}
  \begin{bmatrix}
    \mathbf{0}     &  \mathbf{M_C}^T\\
    \mathbf{M_C} &  \mathbf{0}
  \end{bmatrix}
  \begin{bmatrix}
    \psi^+\\
    \psi^-
  \end{bmatrix}+\textrm{h.c}+\cdots
\end{equation*}
where the chargino/lepton mass matrix is given by \cite{Romao:2000up}
\begin{equation}
  \mathbf{M_C}=
  \begin{bmatrix}
    M_2 & {\textstyle\frac1{\sqrt{2}}}gv_u & 0  \\
    {\textstyle\frac1{\sqrt{2}}}gv_0 & \mu_0 & 
    -{\textstyle\frac1{\sqrt{2}}}\sum_{j}h^E_{ij}v_j \\
    {\textstyle\frac1{\sqrt{2}}}gv_k & \mu_k &
    {\textstyle\frac1{\sqrt{2}}}\mathbf{h^E}v_0
  \end{bmatrix}
  \label{eq:75}
\end{equation}
where the index $i$ expands over the columns and $k$ labels the
rows.  We note that the chargino sector decouples from the leptonic
sector in the limit $\mu_i=v_i=0$.  As in the MSSM, the chargino mass
matrix is diagonalized by two rotation matrices $\mathbf{U}$ and
$\mathbf{V}$
\begin{equation}
  {\bf U}^*{\bf M_C}{\bf V}^{-1}=\mathrm{diag}
  \left(
    m_{\tilde\chi^{\pm}_i} ,m_{e_i}
  \right)=\mathrm{diag}
  \left(
    m_{\tilde F^{\pm}_a} 
  \right)\qquad a=1,\ldots,5
  \label{eq:76}
\end{equation}
where $i=1,2$ for the charginos, and $j=1,2,3$ for the neutrinos.

In the one generation case, the composition of the tau is given by
\begin{equation}
  \tau^+_R=V_{3j}\psi_j^+, \qquad \tau^-_L=U_{3j}\psi_j^- 
  \label{eq:77}
\end{equation}
where $\psi^{+T}=(-i\lambda^+,\widetilde H_u^1,\tau_R^{0+})$ and
$\psi^{-T}=(-i\lambda^-,\widetilde L_0^2,\tau_L^{0-})$. The
two-component Weyl spinors $\tau_R^{0-}$ and $\tau_L^{0+}$ are weak
eigenstates, while $\tau^+_R$ and $\tau^-_L$ are the mass eigenstates.
It follows easily from eq.~(\ref{eq:76}) that the matrix
${\bf{M_CM_C^T}}$ is diagonalized by ${\bf U}$ and the matrix
${\bf{M_C^TM_C}}$ is diagonalized by ${\bf V}$.

\section{Neutralino mass matrix}
\label{sec:neutr-mass-matr}
In the basis ${\psi^0}^T=[-i\lambda',-i\lambda^3,\widetilde H_u^0,\widetilde
H_d^0,\nu_e,\nu_\mu,\nu_\tau]\,$ the neutral fermion mass matrix
$\mathbf{M_N}$ is given by \cite{Romao:2000up}:

\begin{equation}
\mathbf{M_N}=
\begin{bmatrix}
  M_1       &0                  & \frac12\,g'\,v_u  & -\frac12\,g'\,\mathbf{v}\\
  0              &M_2              & -\frac12\,g\,v_u & \frac12\,g\,\mathbf{v}\\
\frac12\,g'\,v_u &-\frac12\,g\,v_u&      0         &-\boldsymbol{\mu} \\
-\frac12\,g'\,\mathbf{v}^T &\frac12\,g\,\mathbf{v}^T &   -\boldsymbol{\mu}^T  &   0 \\
\end{bmatrix}
\label{eq:78}
\end{equation} 
where $\mathbf{v}$ ($\boldsymbol{\mu}$) is the vector defined by $v_\alpha$
($\mu_\alpha$).  This matrix is diagonalized by a $7 \times 7$ unitary matrix
$N$,
\begin{equation}
  \label{eq:79}
  N^*\,\mathbf{M_N}\,N^{-1}=\mathrm{diag}
  \left(
    m_{\tilde \chi_i^0},m_{\nu_j}
  \right)=\mathrm{diag}
  \left(
    m_{\tilde F_a^0}\qquad a=1,\ldots,7
  \right)
\end{equation}
where $i=1,\cdots,4$ for the neutralinos, and $j=1,\cdots,3$ for the neutrinos.

\section{The squark mass spectrum}
\label{sec:squark-slepton-mass}
The up and down-type squark mass matrices of our model have already
been given previously in Ref.~\cite{Diaz:1997xc}. Here we generalize
those to the three-generation case \cite{Diaz:1999ge}. The mass matrix
of the up squark sector follows from the quadratic terms in the scalar
potential
\begin{equation} 
  V_{\mathrm{quadratic}}=
  \begin{bmatrix}
    \tilde {\mathbf{u}}_L^\dagger&\tilde{\mathbf{ u}}_R^\dagger
  \end{bmatrix}
  \mathbf{{\tilde M_U^2}}
  \begin{bmatrix}
    \tilde{\mathbf{ u}}_L\\
    \tilde{\mathbf{ u}}_R
  \end{bmatrix}+\cdots
  \label{eq:80}
\end{equation} 
given by 

\begin{equation}
  \label{eq:81}
  \mathbf{\tilde M_U^2}=
  \begin{bmatrix}
    {\mathbf{M_Q^2}}+\frac12v_u^2{\mathbf{h_U}}{\mathbf{h_U}}^\dagger+\boldsymbol{\Delta}_{UL}&
    \frac1{\sqrt2}v_u{\mathbf{A_U^h}}-\frac1{\sqrt2}(\mu_0 v_0+\mu_3v_3)
    {\mathbf{h_U}}\\
    \frac1{\sqrt2}v_u{\mathbf{A_U^h}}^\dagger-\frac1{\sqrt2}(\mu_0 v_0+\mu_3v_3)
    {\mathbf{h_U}}^\dagger &
    {\mathbf{M_U^2}}+\frac12v_u^2{\mathbf{h_U}}^\dagger{\mathbf{h_U}}+\boldsymbol{\Delta}_{UR}
  \end{bmatrix}
\end{equation}

where $\boldsymbol{\Delta}_{UL}=\frac18\big(g^2-\frac13{g'}^2\big)\big(v_0^2-v_u^2
+v_3^2\big)\mathbf{1}$ and $\boldsymbol{\Delta}_{UR}=\frac16
{g'}^2(v_0^2-v_u^2+v_3^2)\mathbf{1}$ are the splitting in the squark
mass spectrum produced by electro-weak symmetry breaking, and
${A_U^h}_{ij}\equiv A_U^{ij}h_U^{ij}$. The eigenvalues of $\mathbf{\tilde
  M_U^2}$ are
\begin{equation}
  \textrm{diag}\,\{m_{\tilde u_1},m_{\tilde u_2},\ldots,m_{\tilde u_6}\}=
  \begin{bmatrix}
    \boldsymbol{\Gamma}_{UL}& 
    \boldsymbol{\Gamma}_{UR}
  \end{bmatrix}
  \mathbf{\tilde M_U^2}
  \begin{bmatrix}
    \boldsymbol{\Gamma}_{UL}^\dagger \\
    \boldsymbol{\Gamma}_{UR}^\dagger
  \end{bmatrix}
  \label{eq:82}
\end{equation}
This way the six weak-eigenstate fields $\tilde u_{iL}$ and $\tilde
u_{iR}$ ($i=1,2,3$) combine into six up-type mass eigenstate squarks
$\tilde u_k$ as follows: $\tilde u_{iL}=\Gamma^{\dagger ik}_{UL}\tilde
u_k=\Gamma^{* ki}_{UL}\tilde u_k$, $\tilde u_{iR}=\Gamma^{\dagger
  ik}_{UR}\tilde u_k=\Gamma^{* ki}_{UR}\tilde u_k$.

For completeness, we also give the mass matrix of the down squark
sector. The quadratic scalar potential includes
\begin{equation} 
  V_{\mathrm{quadratic}}=
  \begin{bmatrix}
    \tilde{\mathbf{d}}_L^*&\tilde{\mathbf{d}}_R^*
  \end{bmatrix}
  \mathbf{\tilde M_D^2}
  \begin{bmatrix}
    \tilde{\mathbf{d}}_L\\
    \tilde{\mathbf{d}}_R
  \end{bmatrix}+\cdots
  \label{eq:83}
\end{equation} 
given by 
\begin{equation}
  \label{eq:84}
  \mathbf{\tilde M_D^2}=
  \begin{bmatrix}
    \mathbf{M_Q^2}+\frac12v_0^2\mathbf{h_D}\mathbf{h_D}^\dagger+\boldsymbol{\Delta}_{DL}&
    \frac1{\sqrt2}v_0\mathbf{A_D^h}-\frac1{\sqrt2}\mu_0 v_u\mathbf{h_D}\\
    \frac1{\sqrt2}v_0\mathbf{A_D^h}^\dagger -\frac1{\sqrt2}\mu_0 v_u
    \mathbf{h_D}^\dagger  &
    \mathbf{M_D^2}+\frac12v_0^2\mathbf{h_D}^\dagger\mathbf{h_D}+\boldsymbol{\Delta}_{DR}
  \end{bmatrix}
\end{equation}
where $\boldsymbol{\Delta}_{DL}=-\frac18\big(g^2+\frac13{g'}^2\big)\big(v_0^2-v_u^2
+v_3^2\big)\mathbf{1}$, $\boldsymbol{\Delta}_{DR}=-\frac1{12}
{g'}^2(v_0^2-v_u^2+v_3^2)\mathbf{1}$, and ${A_D^h}_{ij}\equiv
A_D^{ij}h_D^{ij}$. The eigenvalues of $\mathbf{\tilde M_D^2}$ are
\begin{equation}
  \label{eq:85}
  \textrm{diag}\,\{m_{\tilde d_1},m_{\tilde d_2},\ldots,m_{d_6}\}=
  \begin{bmatrix}
      \boldsymbol{\Gamma}_{DL}&\boldsymbol{\Gamma}_{DR}
  \end{bmatrix}
  \mathbf{\tilde M_D^2}
  \begin{bmatrix}
    \boldsymbol{\Gamma}_{DL}^\dagger \\
    \boldsymbol{\Gamma}_{DR}^\dagger
  \end{bmatrix}
\end{equation}
One is left with six mass-eigenstate down squarks fields $\tilde
d_{k}$ related to $\tilde d_{iL}$ and $\tilde d_{iR}$ fields as
follows: $\tilde d_{iL}=\Gamma^{\dagger ik}_{DL}\tilde d_k=\Gamma^{*
  ki}_{DL}\tilde d_k$, $\tilde d_{iR}=\Gamma^{\dagger ik}_{DR}\tilde
d_k=\Gamma^{* ki}_{UR}\tilde d_k$.

\section{Scalar Mass Matrices}
\label{sec:scalar-mass-matrices}

The electroweak symmetry is broken when the Higgs and slepton fields
acquire non--zero vevs. These are calculated via the minimization of
the effective potential or, in the diagrammatic method, via the
tadpole equations.  The full scalar potential at tree level is
\begin{equation} 
  V_{\mathrm{total}}^0  = \sum_i \left| \frac{\partial W}{ \partial z_i} \right|^2 
  + V_D + V_{\mathrm{soft}}^\mathrm{MSSM} + V_{\mathrm{soft}}^\mathrm{SSSM}
  \label{eq:86}
\end{equation} 
where $z_i$ is any of the scalar fields in the superpotential in
eq.~(\ref{eq:35}), $V_D$ are the $D$-terms, and
$V_{\mathrm{soft}}^\mathrm{SSSM}$ is the $L$ violating part of eq.~(\ref{eq:25})

The tree level scalar potential contains the following linear terms 
\begin{equation}
  V_{\mathrm{linear}}^0=t_d^0\sigma^0_d+t_u^0\sigma^0_u+t_1^0\tilde\nu^R_1
  +t_2^0\tilde\nu^R_2+t_3^0\tilde\nu^R_3\,,
  \label{eq:87}
\end{equation}
where the different $t^0$ are the tadpoles at tree level. They are
given by
\begin{equation}
  \label{eq:88}
  \begin{split}
  t_u^0&=-\textstyle{\frac12}v_\beta(B_{\beta\gamma}+B_{\gamma\beta})\mu_\gamma+\Big[m_{H_u}^2+\mu_\beta\mu_\beta-
  \textstyle{\frac18}(g^2+{g'}^2)(v_\beta v_\beta-v_u^2)\Big]v_u\\
  t_\alpha^0&=\textstyle{\frac18}(g^2+{g'}^2)(v_\beta
  v_\beta-v_u^2)v_\alpha+\mu_\alpha\mu_\beta v_\beta-
   \textstyle{\frac12}(B_{\alpha\beta}+B_{\beta\alpha})\mu_\beta v_u+
  {\textstyle\frac12}(
  M^2_{L\alpha\beta}+M^2_{L \beta\alpha})v_\beta
  \end{split}
\end{equation}
Repeated indexes $\beta,\gamma$ in eq.~(\ref{eq:88}) implies summation over
$0,1,2,3$.  The five tree level tadpoles are equal to
zero at the minimum of the tree level potential.

We assume that non--diagonal $M_{L_{ij}}^2$ soft parameters are
negligible. Therefore we further denote $M_{L_{ii}}^2$ as $M_{L_i}^2$

The quadratic scalar potential includes~\cite{Akeroyd:1998iq,Porod:2000pw}.
\begin{multline} 
  V_{quadratic}=
  \begin{bmatrix}
    H_d^-&H_u^-&\tilde e_{iL}^-&\tilde e_{iR}^- 
  \end{bmatrix}\mathbf{M_{S^\pm}^2}
  \begin{bmatrix}
    H_d^+\\H_u^+\\\tilde e_{iL}^+\\\tilde e_{iR}^+\\ 
  \end{bmatrix}\\
 + {\textstyle\frac12}
  \left[
    \begin{array}{ccc}
      \sigma^0_d&\sigma^0_u&\tilde\nu_{i}^R 
    \end{array}
  \right]
  \mathbf{M^2_{S^0}}\left[
    \begin{array}{c}
      \sigma^0_d \\ \sigma^0_u \\\tilde\nu_{i}^R 
    \end{array}
  \right]+{\textstyle\frac12}
  \left[
    \begin{array}{ccc}
      \varphi^0_1&\varphi^0_2&\tilde\nu_{i}^I
    \end{array}
  \right]
  {\mathbf{M^2_{P^0}}}
  \left[
    \begin{array}{c}
      \varphi^0_1 \\ \varphi^0_2 \\ \tilde\nu_{i}^I 
    \end{array}
  \right] + \cdots
\label{eq:89} 
\end{multline} 

We will denote the physical scalar bosons by $S^0_i$, the
pseudo-scalar by $P^0_i$ and the charged bosons by $S^\pm_i$. Higgs
bosons mix with charged sleptons and the real (imaginary) parts of the
sneutrino mix with the scalar (pseudoscalar) Higgs bosons. The mass
matrices in eq.~(\ref{eq:89}), will be given in following subsections.

\subsection{Charged scalars}
\label{sec:charged-scalars}

The mass matrix of the charged scalar sector follows from the
quadratic terms in the scalar potential in eq.~\eqref{eq:89}. For
convenience reasons we will divide this $8\times8$ charged scalar mass
matrix into blocks in the following way:
\begin{equation} 
  \mathbf{M_{S^{\pm}}^2}=
  \left[
    \begin{array}{ccc}
      \mathbf{ M_{HH}^2} & \mathbf{ M_{H\tilde \ell}^{2}}^T \\ 
      \mathbf{ M_{H\tilde \ell}^2} & \mathbf{ M_{\tilde \ell\tilde \ell}^2} 
    \end{array}  
  \right] \, .
  \label{eq:90} 
\end{equation} 
Using the tadpole equations given in eqs.~(\ref{eq:88}) we obtain (sum
upon repeated indices):
\begin{equation*} 
  \begin{split}
    &\mathbf{ M_{HH}^2}= \\
   & \left[
      \begin{array}{ccc}
        \!\!B_0\mu_0{\frac{v_u}{v_0}}+{\textstyle\frac14} g^2(v_u^2-v_iv^i)-\mu_0\mu_i 
        {\frac{v^i}{v_0}}+{\textstyle\frac12} v^i(\mathbf{h^E}\mathbf{h^E}^{\dag})_{ij}v^j
        & B_0\mu_0+{\textstyle\frac14} g^2v_0v_u \\
        B_0\mu_0+{\textstyle\frac14} g^2v_0v_u 
        & B_0\mu_0{\frac{v_0}{v_u}}+{\textstyle\frac14} g^2(v_0^2+v_iv^j)+B_i\mu_i 
        {\frac{v^i}{v_u}}
      \end{array}
    \right] 
  \end{split}
\end{equation*} 
This matrix reduces to the usual charged Higgs mass matrix in the MSSM
when we set $v_i=\mu_i=0$. The slepton block is given by
\begin{equation*}
  \mathbf{ M_{\tilde \ell\tilde \ell}^2}=
  \begin{bmatrix}
    \mathbf{M_{LL}^2} & \mathbf{M_{LR}^2} \\
    \mathbf{M_{RL}^2} & \mathbf{M_{RR}^2} \\
  \end{bmatrix}
\end{equation*}
\begin{equation*}
  \begin{split}
    \left( \mathbf{ M_{LL}^2} \right)_{ij}
    =&\frac12\,v_o^2(\mathbf{h_E}^*\mathbf{h_E}^T)_{ij}+\frac14\,g^2
    \left( v_u^2- \textstyle{\sum_\alpha} v_\alpha^2
    \right)\delta_{ij}+\frac14\,g^2v_iv_j-\frac{v_u}{v_i}B_i\epsilon_i\delta_{ij}\\
    &-\mu_i\frac{\sum_\alpha v_\alpha\mu_\alpha}{v_i}\,\delta_{ij}+\mu_i\mu_j+{M_L^2}_{ji}-\frac12\sum_k\frac{v_k}{v_i}
    \left( {M^2_L}_{ik}+{M^2_L}_{ki}
    \right)\delta_{ij}\\
    \mathbf{M_{LR}^2}=&\frac{1}{\sqrt{2}} \left(
      v_0\mathbf{A_E^h}^*-\mu_0v_u\mathbf{h_E}^*
    \right)\\
    \mathbf{M_{RL}^2}=&\mathbf{M_{LR}^2}^{\dag}\\
    \left( \mathbf{M_{RR}^2} \right)_{ij}=&\frac14\,{g'}^2 \left(
      -\textstyle{\sum_\alpha} v_\alpha^2+v_u^2
    \right)\delta_{ij}+\frac12\,\textstyle{\sum_\alpha}v_\alpha^2 \left(
      \mathbf{h_E}^T\mathbf{h_E}^* \right)_{ij}+{M_E^2}_{ji}
  \end{split}
\end{equation*}
where ${A_E^h}^{ij}=A_E^{ij}h_E^{ij}$.

The mixing between the charged Higgs sector and the stau sector is 
given by the following $6\times2$ block: 
\begin{equation*} 
  \mathbf{ M_{H\tilde \ell}^2}=
  \left[
    \begin{array}{ccc}
      \mu_0\mu_i-\frac12v_0\sum_k(\mathbf{h_E}^T\mathbf{h_E}^*)_{ik}v_k+\frac14 g^2v_0v_i 
      & B_i\mu_i+\frac14 g^2v_uv_i 
      \\ {\frac1{\sqrt{2}}}v_u\sum_k(\mathbf{h_E}^T)_{ik}\mu_k-\frac1{\sqrt{2}}\textstyle{\sum}_k(\mathbf{A_E^L}^T)_{ik}v_k) 
      & -{\frac1{\sqrt{2}}}\sum_k(\mathbf{h_E}^T)_{ik}(\mu_0 v_k-\mu_kv_0) 
    \end{array}
  \right]
\end{equation*} 
The mass matrix in eqs.~(\ref{eq:90}), is diagonalized by rotation
matrices which define the eigenvectors
\begin{equation*} 
  \mathbf{S^+}=
  \mathbf{ R_{S^{\pm}}}\mathbf{{S'}^+}
\end{equation*} 
and the  eigenvalues in the one generation case are
\begin{equation*} 
  \rm{diag}(0,m_{S^{\pm}_2}^2,m_{S^{\pm}_3}^2, 
  m_{S^{\pm}_4}^2)=\mathbf{ R_{S^{\pm}}}\mathbf{ M_{S^{\pm}}^2} 
  \mathbf{ R_{S^{\pm}}^T} 
\end{equation*}

\subsection{CP--Even Neutral Scalars}
\label{sec:cp-even-neutral}
The neutral CP-even scalar sector mass matrix in
eq.~(\ref{eq:89}) is given by

\begin{equation}
  \label{eq:91}
    \mathbf{M_{S^0}^2} =
  \left[
    \begin{array}{cc}
      \mathbf{ M_{SS}^2} &  \mathbf{ M_{S\widetilde{\nu}_R}^2} \\
      \mathbf{ M_{S\widetilde{\nu}_R}^2}^T&
      \mathbf{ M_{\widetilde{\nu}_R \widetilde{\nu}_R}^2}  
    \end{array}
  \right]
\end{equation}

where
\begin{equation}
  \mathbf{ M_{SS}^2}= 
  \left[ 
    \begin{array}{cc}
      \displaystyle
      B_0\mu_0\frac{v_u}{v_0}+{\textstyle\frac14} g_Z^2v_0^2-\mu_0 
      \sum_{k=1}^3\mu_k 
      \frac{v_k}{v_0}+\frac{t_d}{v_0}  
      & -B_0\mu_0-{\textstyle\frac14} g^2_Zv_0v_u  
      \\
      -B_0\mu_0-{\textstyle\frac14} g^2_Zv_0v_u  
      &\displaystyle
      B_0\mu_0\frac{v_0}{v_u}+{\textstyle\frac14} g^2_Zv_u^2+\sum_{k=1}^3\, 
      B_k\mu_k \frac{v_k}{v_u}+\frac{t_u}{v_u}
    \end{array}
  \right] 
  \label{eq:92}
\end{equation}
\begin{equation*}
    \mathbf{ M_{S\widetilde{\nu}_R}^2}= 
  \left[
    \begin{array}{cc}
      \mu_0\mu_i+{\textstyle\frac14} g^2_Zv_0v_i  
      \\
      -B_i\mu_i-{\textstyle\frac14} g^2_Zv_uv_i  
    \end{array}
  \right] 
\end{equation*}
and
\begin{align}
  \left(\mathbf{ M_{\widetilde{\nu}_R \widetilde{\nu}_R}^2}\right)_{ij}&=
  \left(-\mu_0\mu_i\frac{v_0}{v_i}
    + B_i\mu_i\frac{v_u}{v_i}
    -\mu_i \sum_{k=1}^3 \mu_k \frac{v_k}{v_i}
    -  {\textstyle\frac12} \sum_{k=1}^3  \frac{v_k}{v_i} 
    \left(
      M^2_{L ik}+M^2_{L ki}
    \right)
    + \frac{t_i}{v_i}
  \right)\delta_{ij}  \nonumber \\
  & +{\textstyle\frac14} g^2_Zv_iv_j+ \mu_i \mu_j +{\textstyle\frac12} 
  \left(
    M^2_{L ij}+M^2_{L ji}
  \right)
  \label{eq:93}
\end{align}
where we have defined $g_Z^2\equiv g^2+{g'}^2$. In the upper--left
$2\times2$ block, in the limit $v_i=\mu_i=0$, the reader can
recognize the MSSM mass matrix corresponding to the CP--even neutral
Higgs sector.  To define the rotation matrices let us define the
unrotated fields by
\begin{equation}
   \mathbf{{S'}^0}=
  \left[
    \begin{array}{ccccc}
      \sigma_d^0 & \sigma_u^0 & \widetilde{\nu}_1^R &
      \widetilde{\nu}_2^R & \widetilde{\nu}_2^R
    \end{array}
  \right]
  \label{eq:94}
\end{equation}
Then the mass eigenstates are $S^0_i$ given by
\begin{equation} 
  S^0_i={R}^{S^0}_{ij} {S'}^0_j
  \label{eq:95}
\end{equation} 
with the eigenvalues $\mathrm{diag}(m^2_{S_1},\ldots,m^2_{S_5})= \mathbf{
  R^{S^0}}\, \mathbf{ M_{S^0}^2}\, \left(\mathbf{ R^{S^0}}\right)^T$.

\subsection{CP--Odd Neutral Scalars}

The quadratic scalar potential includes 
where the CP-odd neutral scalar mass matrix is 
\begin{equation}
  \mathbf{ M_{P^0}^2} =
  \left[
    \begin{array}{cc}
      \mathbf{ M_{PP}^2} &  \mathbf{ M_{P\widetilde{\nu}_I}^2} \\
      \mathbf{ M_{P\widetilde{\nu}_I}^2}^T&
      \mathbf{ M_{\widetilde{\nu}_I \widetilde{\nu}_I}^2}  
    \end{array}
  \right]
  \label{eq:96}
\end{equation}
where

\begin{equation}
  \mathbf{ M_{PP}^2}= 
  \left[
    \begin{array}{cc}
      \displaystyle B_0\mu_0\frac{v_u}{v_0}-\mu_0\sum_{k=1}^3
      \mu_k\frac{v_k}{v_0}+\frac{t_d}{v_0}
      & B_0\mu_0 \\
      B_0\mu_0 & \displaystyle
      B_0\mu_0\frac{v_0}{v_u}+\sum_{k=1}^3
      B_k\mu_k\frac{v_k}{v_u}+\frac{t_u}{v_u}
    \end{array}
  \right] 
  \label{eq:97}
\end{equation}
Note that in the limit $\mu_0\to0$ a Goldstone boson appear.
\begin{equation*}
  \mathbf{ M_{P\widetilde{\nu}_I}^2}= 
  \left[
    \begin{array}{c}
    \mu_0\mu_i  \\
    B_i\mu_i
    \end{array}
  \right] 
\end{equation*}
and
\begin{align}
  \left(
    \mathbf{ M_{\widetilde{\nu}_I \widetilde{\nu}_I}^2}
  \right)_{ij}&=  
  \left(-\mu_0\mu_i\frac{v_0}{v_i}
    + B_i\mu_i\frac{v_u}{v_i}
    -\mu_i \sum_{k=1}^3 \mu_k \frac{v_k}{v_i}
    - {\textstyle\frac12} \sum_{k=1}^3  \frac{v_k}{v_i} 
    \left(
      M^2_{L ik}+M^2_{L ki}\
    \right)
    + \frac{t_i}{v_i}
  \right)\delta_{ij}\nonumber  \\
  & + \mu_i \mu_j +{\textstyle\frac12}
  \left(
    M^2_{L ij}+M^2_{L ji}
  \right)
  \label{eq:98}
\end{align}
The neutral pseudo--scalar mass matrices are diagonalized by the 
following rotation matrices,
\begin{equation} 
  P_i={R}^{P^0}_{ij} P'_j
  \label{eq:99}
\end{equation} 
with the eigenvalues $\mathrm{diag}(m^2_{A_1},\ldots,m^2_{A_5})= \mathbf{
  R^{P^0}}\, \mathbf{ M_{P^0}^2}\, \left(\mathbf{ R^{P^0}}\right)^T$,
where the unrotated fields are
\begin{equation}
  \mathbf{{P'}^0}=
  \left[
    \begin{array}{ccccc}
      \varphi_d^0&\varphi_u^0&\widetilde{\nu}_1^I&
      \widetilde{\nu}_2^I& \widetilde{\nu}_2^I 
    \end{array}
  \right]
  \label{eq:100}
\end{equation}

May be convenient some times take the pseudoscalar mass eigenvalues as
input, it is worth briefly repeating here the discussion of the
pseudoscalar bosons masses.  The pseudo-scalar mass matrix in the
one generation case is given by
by:
\begin{equation} 
  \mathbf{ M^2_{P^0}}=
  \left[
    \begin{array}{ccc}
      B_0\mu_0{\frac{v_u}{v_0}}-\mu_0\mu_3{\frac{v_3}{v_0}}
      & B_0\mu_0 & +\mu_0\mu_3 \\ 
      B_0\mu_0 & B_0\mu_0{\frac{v_0}{v_u}}+B_3\mu_3{\frac{v_3}{v_u}}
      & +B_3\mu_3 \\ 
      +\mu_0\mu_3 & +B_3\mu_3 &  
      -\mu_0\mu_3{\frac{v_0}{v_3}}+B_3\mu_3{\frac{v_u}{v_3}} 
    \end{array}
  \right] \, .
\label{eq:101} 
\end{equation} 
As expected, this matrix has zero determinant, since the neutral
Goldstone boson eaten by the $Z$ is one of the corresponding states.
Therefore, the masses of the two physical states are given by the
formula:
\begin{equation} 
  m_{2,3}={\frac12}{\rm Tr}\mathbf{ M} 
  \pm {\frac12}\sqrt{\left({\rm Tr}\mathbf{ M}\right)^2 
    -4(M_{11}M_{22}-M_{12}^2+M_{11}M_{33}-M_{13}^2+M_{22}M_{33}-M_{23}^2)}.
  \label{eq:102}
\end{equation}
Therefore we can easily take one these masses as input and calculate
$B_0\mu_0$ from it using
\begin{equation*}
  B_0\mu_0=\frac{-m_{P^0_2}^4v_0v_uv_3+B_3\mu_3\mu_0 v_3(v_0^2+v_u^2+v_3^2)+
    \mu_3 m_{P^0_2}^2[\mu_0 v_u(v_0^2+v_3^2)]-B_3v_0(v_u^2+v_3^2)}
  {-m_{P^0_2}^2(v_0^2+v_u^2)v_3+\mu_{3}(\mu_0 v_0-B_3v_u)
    (v_0^2+v_u^2+v_3^2) }   
\end{equation*}
$B_3$ is obtained from the minimum equation for given $\mu_3$ and
$v_3$~\cite{Akeroyd:1998iq}.

\chapter{Sparticle decays}
\label{sec:sparticle-decays}

\section{Decays of neutralinos}
\label{sec:decays-neutr}
If unprotected by the ad hoc assumption of R--parity conservation the
LSP will decay as a result of gauge boson, squark, slepton and Higgs
boson exchanges.  

Although for the discussion of flavor--changing processes, such as
neutrino oscillations involving all three generations, it is important
to consider the full three-generation structure of the model, for the
following discussion of neutralino decay properties it will suffice to
assume $L$ violation only in the third generation, as a first
approximation.

\begin{figure}[h] 
\begin{center}
\begin{picture}(340,160)(0,0) 
\def\punto{2} 
\newcommand{\treebodyuno}[9]{
\def\xx{#1}
\def\largo{#2}           
\def\alto{#3}           
\def\altolabel{#4}      
\def\altotext{#5}       
\def\largophoton{#6}    
\def\altophoton{#7}     
\def\altophotontext{#8} 
\def\largofirst{#9}}      
\newcommand{\treebodydos}[5]{
\def\altofirst{#1}      
\def\largosecond{#2}    
\def\altosecond{#3}     
\def\largothird{#4}     
\def\altothird{#5}  
}      
\newcommand{\treebody}[6]{
\Text(\xx,\altolabel)[]{#1}
\Text(\xx,\altotext)[]{\hspace{45pt}#2}
\ArrowLine(\xx,\alto)(\largo,\alto)
\Vertex(\largo,\alto){\punto}
\Text(\largofirst,\altofirst)[]{\hspace{36pt}#3}
\ArrowLine(\largo,\alto)(\largofirst,\altofirst)
\Text(\largophoton,\altophotontext)[]{\hspace{-56pt}#4}
\Vertex(\largophoton,\altophoton){\punto}
\Text(\largosecond,\altosecond)[]{\hspace{33pt}#5}
\ArrowLine(\largophoton,\altophoton)(\largosecond,\altosecond)
\Text(\largothird,\altothird)[]{\hspace{30pt}#6}
\ArrowLine(\largophoton,\altophoton)(\largothird,\altothird)
}
\treebodyuno{0}{50}{110}{145}{120}{110}{100}{95}{80}
\treebodydos{130}{140}{120}{140}{80}
\treebody{a)}{${\tilde\chi_1^0}(p_1)$}{$\nu_3(p_2)$}{$Z^0$}{$f(p_3)$}{$\bar{f}(p_4)$}
\Photon(\largo,\alto)(\largophoton,\altophoton){2}{8}
\treebodyuno{190}{240}{110}{145}{120}{300}{100}{95}{270}
\treebodydos{130}{330}{120}{330}{80}
\treebody{b)}{${\tilde\chi_1^0}(p_1)$}{$\nu_3(p_2)$}{$S^0_i\,;P^0_i$}{$f(p_3)$}{$\bar{f}(p_4)$}
\DashArrowLine(\largo,\alto)(\largophoton,\altophoton){6}
\treebodyuno{0}{50}{30}{65}{40}{110}{20}{15}{80}
\treebodydos{50}{140}{40}{140}{0}
\treebody{c)}{${\tilde\chi_1^0}(p_1)$}{$f(p_3)$}{$\bar{\!\!\tilde f}_k$}{$\nu_3(p_2)$}{$\bar{f}(p_4)$}
\DashArrowLine(\largo,\alto)(\largophoton,\altophoton){6}
\treebodyuno{190}{240}{30}{65}{40}{300}{20}{15}{270}
\treebodydos{50}{330}{40}{330}{0}
\treebody{d)}{${\tilde\chi_1^0}(p_1)$}{$\bar{f}(p_4)$}{${\tilde f}_k$}{$\nu_3(p_2)$}{$f(p_3)$}
\DashArrowLine(\largo,\alto)(\largophoton,\altophoton){6}
\end{picture}
\end{center}
\caption{ Feynman graphs for the decay 
  $\tilde{\chi}^{0}_{1} \to \nu_3 \, f \, \bar{f}$ where $f \neq \tau$.}
\label{fig:4}
\end{figure}
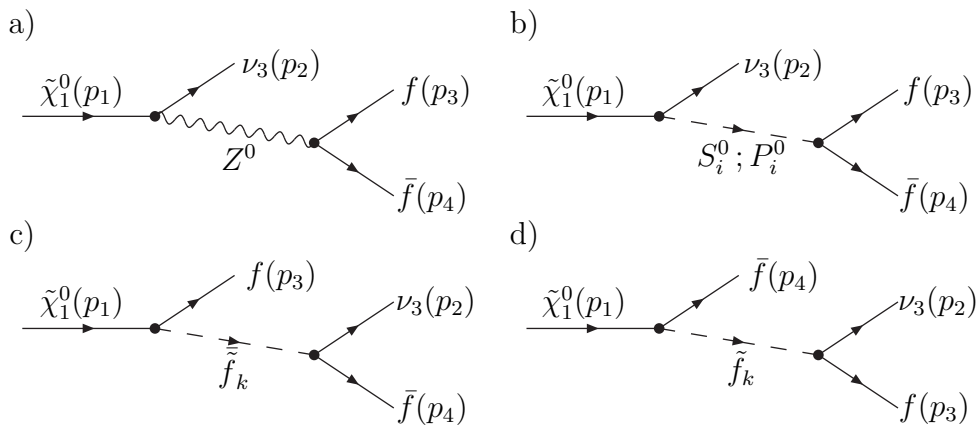

The relevant contributions to these decays are given
in Table 1.  The Feynman diagrams for the decays not involving taus,
i.e. $\tilde{\chi}^{0}_{1} \to \nu_3 f \bar{f}$ ($f=e$, $\nu_e$, $\mu$, $\nu_\mu$,
$u$, $d$, $c$, $s$, $b$) are shown explicitly in Fig.~\ref{fig:4}.

For this class of decays we
have $Z^0$, $P^0_i$, and $S^0_j$ exchange in the direct channel
(Fig.~\ref{fig:4}a and b) and $\tilde f$ exchange in the crossed
channels (Fig.~\ref{fig:4}c and d). In particular in the case $f=b$ the
$P^0_i$ and $S^0_j$ exchange contributions are significant.  This is
quite analogous to the results found in \cite{Bartl:1999iw} for $\tilde{\chi}^{0}_{2}
\to \tilde{\chi}^{0}_{1} f \bar{f}$ decays.  The particles exchanged in the $s$-,
$t$-, and $u$-channel for the decays $\tilde{\chi}^{0}_{1} \to \tau^\pm l^\mp
\nu_l$, ($l=e,\mu$), $\tilde{\chi}^{0}_{1} \to \tau^\pm q \bar{q}'$
($q,q'=u,d,s,c$), $\tilde{\chi}^{0}_{1} \to \tau^- \tau^+ \nu_l$, and $\tilde{\chi}^{0}_{1}
\to 3 \nu_3$ are given in Tab.~\ref{tab:6}.

\begin{table}
\begin{center}
\begin{tabular}{|l|c|c|}
\hline
Decay mode & exchanged particle & channel \\ \hline
$\tilde{\chi}^{0}_{1} \to 3 \, \nu_3$ & $Z$, $S^0_i$, $P^0_j$  & $s$ \\
                          & $Z$, $S^0_i$, $P^0_j$  & $t$ \\
                          & $Z$, $S^0_i$, $P^0_j$  & $u$ \\ \hline
$\tilde{\chi}^{0}_{1} \to \nu_3 \, \nu_l \, \bar{\nu}_l$ ($l=e,\mu$) & $Z$  & $s$ \\
                          & $\overline{\tilde \nu}_l$  & $t$ \\
                          & $\tilde \nu_l$ & $u$ \\ \hline
$\tilde{\chi}^{0}_{1} \to \nu_3 \, f \, \bar{f}$ ($f=e,\mu,u,d,s,c,b$)
                                         & $Z$, $S^0_i$, $P^0_j$ & $s$ \\
                          & $\overline{\tilde f}_{1,2}$  & $t$ \\
                          & $\tilde f_{1,2}$ & $u$ \\ \hline
$\tilde{\chi}^{0}_{1} \to \nu_3 \, \tau^+ \, \tau^-$&  $Z$, $S^0_i$, $P^0_j$ & $s$ \\
                          &  $W^-$, $S^-_k$ & $t$ \\
                          &  $W^+$, $S^+_k$ & $u$ \\ \hline
$\tilde{\chi}^{0}_{1} \to \nu_l \, \tau^{\pm} \, l^{\mp}$ ($l=e,\mu$) &
                             $W^\pm$, $S^\pm_k$ & $s$ \\
                          & $\overline{\tilde l}_{1,2}$  & $t$ \\
                          & $\tilde \nu_l$ & $u$ \\ \hline
$\tilde{\chi}^{0}_{1} \to \tau \, q \, \bar{q}'$ ($q=u,c$, $q'=d,s$) &
                             $W^\pm$, $S^\pm_k$ & $s$ \\
                          & $\overline{\tilde q}'_{1,2}$  & $t$ \\
                          & $\tilde q_{1,2}$ & $u$ \\ \hline
\end{tabular}
\end{center}
\caption[Contributions involved in the lightest neutralino 3-body 
decay modes]{Contributions involved in the lightest neutralino 3-body 
decay modes. The $s$-, $t$-, and  $u$-channels are defined by: 
$s=(p_1 - p_2)^2$, $t=(p_1 - p_3)^2$, and $u=(p_1 - p_4)^2$. 
See also Fig.~\ref{fig:4}.}
\label{tab:6}
\end{table}

In the calculations we have included all mixing effects, in particular
the standard MSSM ${\tilde f}_L - {\tilde f}_R$ mixing effects and
those induced by the bilinear $L$ violating terms, i.e.
$Re(\tilde \nu_3) - h^0 - H^0$, 
$Im(\tilde \nu_3) - A^0 - G^0$, 
\cite{deCampos:1995av}, ${\tilde \tau}_{L,R}^\pm - H^\pm - G^\pm$
\cite{Akeroyd:1998iq}, $\nu_3$ - $\tilde{\chi}^{0}_{i}$ \cite{mnutreeJ}, and
$\tau$ - $\tilde{\chi}^{-}_{j}$ mixings \cite{ROMA,Hirsch:2000xe,Barbieri:1990vb}. These mixing effects are
particularly important in the calculations of the various $L$
violating decay rates of $\tilde{\chi}^{0}_{1}$, which are discussed in
section \ref{sec:neutralinos}.

\section{Squark decays}
\label{sec:squark-decays}

\subsection{Two body decays}
\label{sec:two-body-decays-1}

In Appendix~\ref{sec:complete-formulas-2} we give the Feynman rules
for all vertices involving squarks, quarks, charginos and
neutralinos. The decays of the six squarks are given both in the MSSM and
the SSSM by

\begin{eqnarray}
  \Gamma(\tilde q_k \to q_i+F_j^0)&=&
  \frac{g^2\lambda^{1/2}(m_{\tilde q_k}^2,m_{q_i}^2,m_{F_j^0}^2)}{
    16\pi m_{\tilde q_k}^3}\bigg[-4h^{jki}_Qf^{jki}_Q
  m_{q_i}m_{F_j^0}\nonumber\\
  &&+\bigg((h^{jki}_Q)^2+(f^{jki}_Q)^2\bigg)\bigg(
  m_{\tilde q_k}^2-m_{q_i}^2-m_{F_j^0}^2\bigg)\bigg]
  \label{eq:103}\\
  \Gamma(\tilde q_k \to q'_i+F_j^\pm)&=&
  \frac{g^2\lambda^{1/2}(m_{\tilde q_k}^2,m_{q'_i}^2,m_{F_j^\pm}^2)}{
    16\pi m_{\tilde q_k}^3}\bigg[-4l^{jki}_{Q}H^{jki}_{QL}
  m_{q'_i}m_{F_j^\pm}
  \nonumber\\
  &&+\bigg((l^{jki}_{Q})^2+(H^{jki}_{QL})^2\bigg)\bigg(
  m_{\tilde q_k}^2-m_{q'_i}^2-m_{F_j^\pm}^2\bigg)\bigg]
  \label{eq:104}
\end{eqnarray}
where $Q=U,D$ refers to $\tilde q$ and
\begin{eqnarray}
  \label{eq:105}
  f^{jki}_Q&=&-(\sqrt2G_{0QL}^{jki}+H_{0QR}^{jki})\\
  \label{eq:106}
  h^{jki}_Q&=&\sqrt{2}G_{0QR}^{jki}-H_{0QL}^{jki}\\
  \label{eq:107}
  l^{jki}_Q&=&H_{QR}^{jki}-G_{QL}^{jki}
\end{eqnarray}
with  $G$ and $H$ being either the SSSM or the MSSM couplings defined in
Appendix~\ref{sec:complete-formulas-2}.

\subsection{Three body decays}
\label{sec:three-body-decays-1}

For definiteness and simplicity we assume only $L$ violation in the
third generation. A short discussion on ${\tilde t}_1 \to b \,
l^+$ in the three--generation model will be given at the end of
Subsect.~\ref{sec:three-body-decays}. In the one generation SSSM the
stop mass matrix is given by
\begin{equation*}
  \mathbf{M_{\tilde t}}^2=\left[ \begin{array}{cc}
      {M_Q^2}+\frac12v_u^2 {h_t}^2+\Delta_{UL}&
      \frac{h_t}{\sqrt2}
      \left( v_u{A_t}- \mu_0 v_0 +\mu_3 v_3 \right) \\
      \frac{h_t}{\sqrt2}
      \left( v_u{A_t}- \mu_0 v_0 +\mu_3 v_3 \right) &
      {M_U^2}+\frac12 v_u^2{h_t}^2+\Delta_{UR}
    \end{array} \right]
\end{equation*}
with $\Delta_{UL}=\frac18\big(g^2-\frac13{g'}^2\big)\big(v_0^2-v_u^2
+v_3^2\big)$ and $\Delta_{UR}=\frac16 {g'}^2(v_0^2-v_u^2+v_3^2)$.
The mass matrix for the sbottoms is given by
\begin{equation*}
  \mathbf{M_{\tilde b}}^2=
  \left[
    \begin{array}{cc}
      {M_Q^2}+\frac12v_0^2{h_b^2}+\Delta_{DL}&
      \frac{h_b}{\sqrt2}(v_0{A_D}-\mu_0 v_u)\\
      \frac{h_b}{\sqrt2}(v_0{A_D} -\mu_0 v_u)  &
      {M_D^2}+\frac12v_0^2{h_b^2}+\Delta_{DR}
    \end{array}
  \right]
\end{equation*}
where $\Delta_{DL}=-\frac18\big(g^2+\frac13{g'}^2\big)\big(v_0^2-v_u^2
+v_3^2\big)$, $\Delta_{DR}=-\frac1{12}
{g'}^2(v_0^2-v_u^2+v_3^2)$.
The  mass eigenstates are obtained by ($q=t,b$): 
\begin{equation*}
  \left[
    \begin{array}{c}
      \tilde q_1\\
      \tilde q_2
    \end{array}
  \right]=
  \left[
    \begin{array}{cc}
      \cos\theta_{\tilde q} & \sin\theta_{\tilde q}\\
      -\sin\theta_{\tilde q} & \cos\theta_{\tilde q}\\
    \end{array}
  \right] 
  \left[
    \begin{array}{c}
      \tilde q_L\\
      \tilde q_R
    \end{array}
  \right]=
  \mathcal{R}^{\tilde f}
  \left[
    \begin{array}{c}
      \tilde q_L\\
      \tilde q_R
    \end{array}
  \right]
\end{equation*}
with
\begin{equation*}
  \cos \theta_{\widetilde q} = \frac{- M^2_{\widetilde q_{12}}}{\sqrt{(M^2_{\widetilde q_{11}} -
      m^2_{\widetilde q_1})^2 + (M^2_{\widetilde q_{12}})^2}}, \qquad 
  \sin \theta_{\widetilde q} = \frac{M^2_{\widetilde q_{11}} - m^2_{\widetilde q_1}}
  {\sqrt{(M^2_{\widetilde q_{11}} - m^2_{\widetilde q_1})^2 + (M^2_{\widetilde q_{12}})^2}} \, .
\end{equation*}

The lighter stop three body channels are presented below. The complete
formulas for the several contributions to each channel are given in
\cite{Restrepo:2001me}
\begin{eqnarray}
 \Gamma({\tilde t}_1 \to W^+ \, b \, {\tilde \chi}^0_i)  &=& \nonumber \\
 & & \hspace{-30mm} 
   =  \frac{\alpha^2}{16 \, \pi m^3_{{\tilde t}_1} \sin^4 \theta_W}
  \int\limits^{(m_{{\tilde t}_1}-m_{\scriptscriptstyle{W}})^2}_{
           (m_b + m_{{\tilde \chi}^0_i})^2} \hspace{-8mm}
     d \, s \,
   \left( G^W_{{\tilde \chi}^+ {\tilde \chi}^+} +
   G^W_{{\tilde \chi}^+ t} +
   G^W_{{\tilde \chi}^+ {\tilde b}} +
   G^W_{t t} +
   G^W_{t {\tilde b}} +
   G^W_{{\tilde b} {\tilde b}} \right) \nonumber
\end{eqnarray}

\begin{eqnarray}
 \Gamma({\tilde t}_1 \to S_k^+ \, b \, {\tilde \chi}^0_i)  &=& \nonumber \\
 & & \hspace{-30mm} 
   =  \frac{\alpha^2}{16 \, \pi m^3_{{\tilde t}_1} \sin^4 \theta_W}
  \int\limits^{(m_{{\tilde t}_1}-m_{S_k^+})^2}_{
           (m_b + m_{{\tilde \chi}^0_i})^2} \hspace{-8mm}
     d \, s \,
   \left( G_{{\tilde \chi}^+ {\tilde \chi}^+} +
   G_{{\tilde \chi}^+ t} +
   G_{{\tilde \chi}^+ {\tilde b}} +
   G_{t t} +
   G_{t {\tilde b}} +
   G_{{\tilde b} {\tilde b}} \right) 
\nonumber
\end{eqnarray}

\begin{eqnarray}
 \Gamma({\tilde t}_1 \to S_k^0 \, b \, {\tilde \chi}^+_i)  &=& \nonumber \\
 & & \hspace{-30mm} 
   =  \frac{\alpha^2}{16 \, \pi m^3_{{\tilde t}_1} \sin^4 \theta_W}
  \int\limits^{(m_{{\tilde t}_1}-m_{S_k^0})^2}_{
           (m_b + m_{{\tilde \chi}^+_i})^2} \hspace{-8mm}
     d \, s \,
   \left( G^{S^0}_{{\tilde \chi}^+ {\tilde \chi}^+} +
   G^{S^0}_{{\tilde \chi}^+ t} +
   G^{S^0}_{{\tilde \chi}^+ {\tilde t}} +
   G^{S^0}_{t t} +
   G^{S^0}_{t {\tilde t}} +
   G^{S^0}_{{\tilde t} {\tilde t}} \right) 
\nonumber
\end{eqnarray}

\chapter{Numerical Results}
\label{sec:numerical-results}

\section{Sugra Case}
\label{sec:sugra-case}
For definiteness and simplicity we assume only $L$ violation in the
third generation case along this section. Note, in contrast, that in
order to describe Flavor Changing Neutral Current (FCNC) effects such
as the $L$ conserving process $\tilde t_1 \to c\,\tilde\chi_1^0$ we need
the three generations of quarks.

The soft SUSY breaking parameters at the electroweak scale needed for
the evaluation of the mass matrices and couplings are calculated by
solving the renormalization group equations (RGE's) of the SSSM and
imposing the radiative electroweak symmetry breaking condition. From
the measured quark masses, CKM matrix elements and $\tan\beta$ we first
solve one-loop RGE's for the gauge and Yukawa couplings to calculate
their corresponding values at the unification scale.  Assuming now
universal soft supersymmetry breaking boundary conditions, we evolve
downward the RGE's for all SSSM parameters, including full
three-generation mixing in the RGE's for Yukawa coupling constants, as
well as soft SUSY breaking parameters. Next, we evaluate the Higgs
potential at the $m_t$ scale including the one-loop corrections
induced by the Yukawa coupling constants of the third generation. The
radiative electroweak symmetry breaking requirement fixes the
magnitude of the SUSY Higgs mass parameter $\mu_0$ and the soft SUSY
breaking parameters $B_0$ and $B_3$.  Notice that due to the
minimization condition for $t_3$ in eq.~(\ref{eq:88}) one can solve
for $v_3$ as a function of $\mu_3$.  At this point, all $L$ violating
parameters at the electroweak scale are determined as functions of the
input parameters $(\tan\beta,\, m_0,\, A_0,m_{1/2}, \hbox{sign}(\mu_0),
m_{\nu_3})$, where $m_0$ the common scalar mass and $m_{1/2}$ is the common
gaugino mass. Iteration is required because $\mu_0$ and $\mu_3$ are
inputs to evaluate the loop-corrected minimum.  Having determined all
parameters at the electroweak scale, we obtain the masses and the
mixings of all the SUSY particles by diagonalizing the corresponding
mass matrices.  At this stage we also choose $\mu_3$ in order to get a
sufficiently light $\nu_3$.

\subsection{Neutralinos}
\label{sec:neutralinos}
In this section we present numerical predictions for the lightest and
second lightest neutralino production cross sections in $e^+ e^-$
collisions, namely, $e^+ e^-\to \tilde{\chi}^{0}_{1}
\tilde{\chi}^{0}_{1},\tilde{\chi}^{0}_{1} \tilde{\chi}^{0}_{2}$.  Moreover
we will characterize in detail all branching ratios for the lightest
neutralino decays in the SSSM. It will suffice to assume $L$ violation
only in the third generation, as a first approximation in this case.

The relevant parameters include the $L$ violating parameters and the
standard Minimal SUGRA (MSUGRA)\index{MSUGRA: Minimal SUGRA}
parameters $m_{1/2}$, $m_0$, $\tan \beta$.  The absolute
value of $\mu_0$ is fixed by radiative breaking of electroweak
symmetry.  We take $\mu_0$ positive to be in agreement with the $b \to s
\gamma$ decay \cite{Diaz:1999wq}.  As representative values of $\tan \beta$
we take $\tan \beta = 3$ and 50.  It is a feature of models with purely
spontaneous breaking of R--parity that neutrinos acquire a mass only
due to the violation of R-parity \cite{Ellis:1985gi,arca,mnutreeJ}.
This feature also applies to the SSSM.  As a result the $L$
violating parameters are directly related with $m_{\nu_3}$, the mass
of the neutrino $\nu_3$, which is generated due to the mixing implicit
in \eqref{eq:78}.

\subsubsection{Neutralino Production}

\begin{figure}
\setlength{\unitlength}{1mm}
\begin{picture}(150,90)
\put(-3,0){\includegraphics[height=8.7cm,width=7.cm]{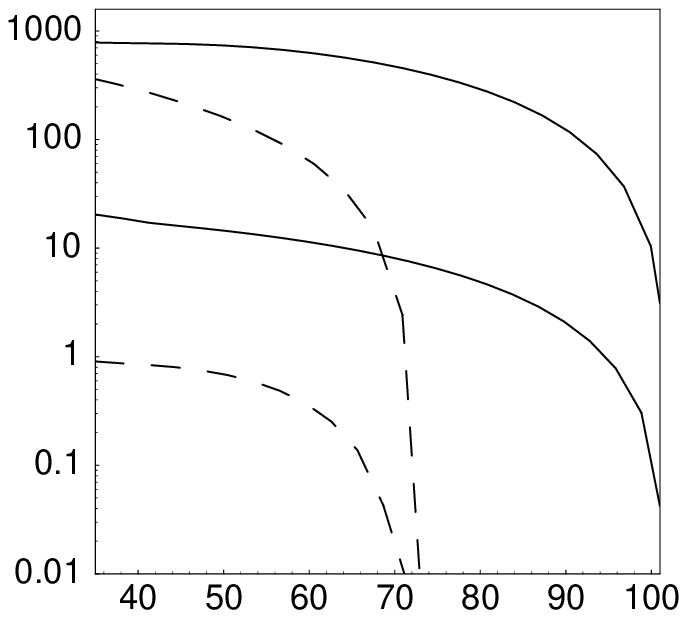}}
\put(-1,89){{\small \bf a)}}
\put(5,87){\makebox(0,0)[bl]{{\small $\sigma$~[fb]}}}
\put(45,16){\makebox(0,0)[bl]{{\small $\tan\beta =3$}}}
\put(10,44){\makebox(0,0)[bl]{{\small $e^+ e^- \to \tilde{\chi}^{0}_{1} \tilde{\chi}^{0}_{2}$}}}
\put(34,62){\makebox(0,0)[bl]{{\small $e^+ e^- \to \tilde{\chi}^{0}_{1} \tilde{\chi}^{0}_{1}$}}}
\put(69,-3){\makebox(0,0)[br]{{ $m_{\tilde{\chi}^{0}_{1}}$~[GeV]}}}
\put(73,0){\includegraphics[height=8.7cm,width=7.cm]{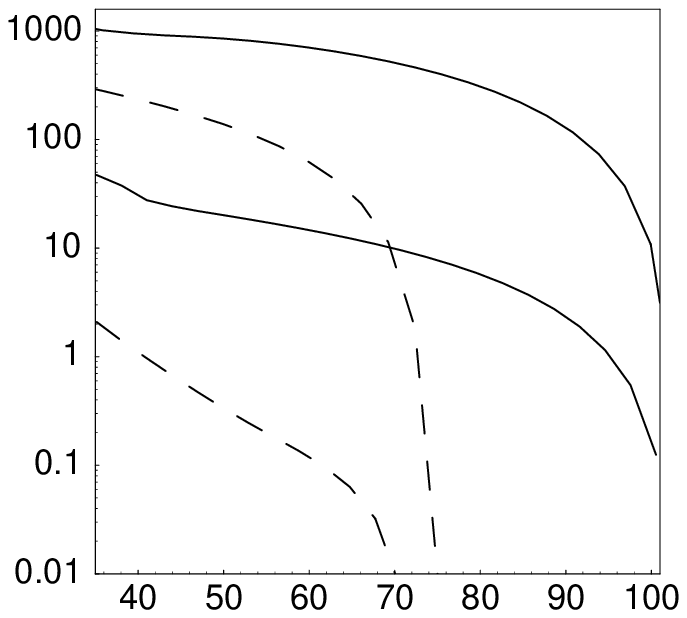}}
\put(75,89){{\small \bf b)}}
\put(81,87){\makebox(0,0)[bl]{{\small $\sigma$~[fb]}}}
\put(120,18){\makebox(0,0)[bl]{{\small $\tan\beta =50$}}}
\put(87,44){\makebox(0,0)[bl]{{\small $e^+ e^- \to \tilde{\chi}^{0}_{1} \tilde{\chi}^{0}_{2}$}}}
\put(108,62){\makebox(0,0)[bl]{{\small $e^+ e^- \to \tilde{\chi}^{0}_{1} \tilde{\chi}^{0}_{1}$}}}
\put(143,-3){\makebox(0,0)[br]{{ $m_{\tilde{\chi}^{0}_{1}}$~[GeV]}}}
\end{picture}
\caption[Neutralino cross section]{Maximum and minimum
  attainable values for the $e^+ e^- \to \tilde{\chi}^{0}_{1}
  \tilde{\chi}^{0}_{1}$ (full lines) and $e^+ e^- \to
  \tilde{\chi}^{0}_{1} \tilde{\chi}^{0}_{2}$ (dashed lines) production
  cross sections in fb as a function of $m_{\tilde{\chi}^{0}_{1}}$ for
  $\sqrt{s} = 205$~GeV, 50~GeV $<m_0<500$~GeV, 90~GeV
  $<m_{1/2}<270$~GeV, a) $\tan\beta=3$, and b) $\tan\beta=50$.  ISR
  corrections are included.}
\label{fig:5}
\end{figure}

While the $L$ violation would allow for the single production of
supersymmetric particles~\cite{ROMA,Hirsch:2000xe,Barbieri:1990vb},
for the assumed values of the $L$ violation parameters indicated
by the simplest interpretation of solar and atmospheric neutrino data
~\cite{Gonzalez-Garcia:2001sq,atm99,qva,MSW99},
these cross sections are typically too small to be observable.  As a
result neutralino production at LEP2 in our model typically occurs in
pairs with essentially the same cross sections as in the MSUGRA case.
In Fig.~\ref{fig:5}a and b we show the maximum and minimum
attainable values for the $e^+ e^- \to \tilde{\chi}^{0}_{1} \tilde{\chi}^{0}_{1}$ and $e^+ e^-
\to \tilde{\chi}^{0}_{1} \tilde{\chi}^{0}_{2}$ production cross sections as a function of
$m_{\tilde{\chi}^{0}_{1}}$ at $\sqrt{s}=205$~GeV. We compare the cases $\tan\beta=3$ and
$\tan\beta=50$, varying $m_{1/2}$ between 90~GeV and 260~GeV and $m_0$
between 50~GeV and 500~GeV. One can see that, indeed, these results
are identical to those obtained in the MSUGRA.  The $\tilde{\chi}^{0}_{1}
\tilde{\chi}^{0}_{1}$ production cross section can reach approximately 1~pb.
In our calculation we have used the formula as given in
\cite{Bartl:1999iw} and, in addition, we have included initial state
radiation (ISR) using the formula given in \cite{ISR}. Note that
$\tilde e_{L}$ and $\tilde e_{R}$ are exchanged in the $t$- and
$u$-channel implying that a large fraction of the neutralinos will be
produced in the forward and backward directions.

\begin{figure}
\setlength{\unitlength}{1mm}
\begin{picture}(150,160)
\put(-3,100){\includegraphics[height=8.7cm,width=7.cm]{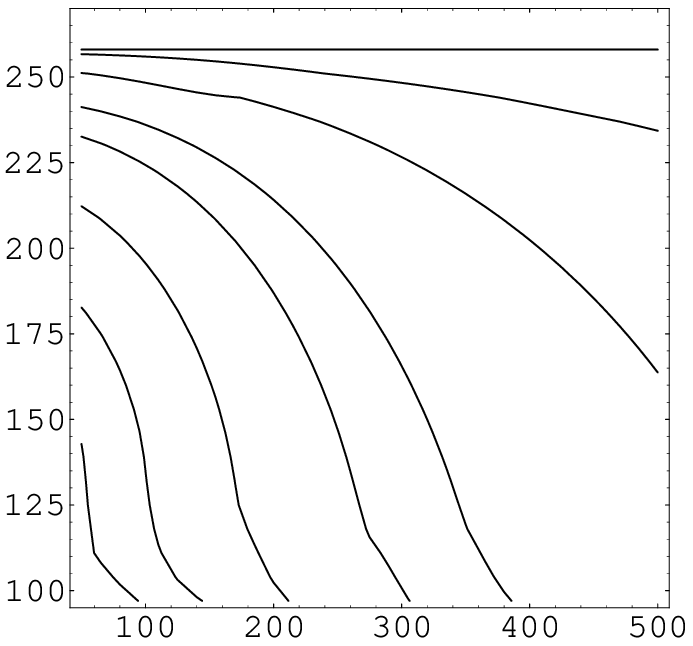}}
\put(-1,189){{\small \bf a)}}
\put(5,187){\makebox(0,0)[bl]{{\small $m_{1/2}$~[GeV]}}}
\put(29,180){\makebox(0,0)[bl]{{\small $0$}}}
\put(29,172.5){\makebox(0,0)[bl]{{\small $1$}}}
\put(29,165){\makebox(0,0)[bl]{{\small $10$}}}
\put(29,147){\makebox(0,0)[bl]{{\small $50$}}}
\put(25,129){\makebox(0,0)[bl]{{\small $100$}}}
\put(23,120){\makebox(0,0)[bl]{{\small $250$}}}
\put(15,116){\makebox(0,0)[bl]{{\small $500$}}}
\put(7,116){\makebox(0,0)[bl]{{\small $750$}}}
\put(47,130){\makebox(0,0)[bl]{{\small $\tan\beta =3$}}}
\put(69,97){\makebox(0,0)[br]{{ $m_0$~[GeV]}}}
\put(73,100){\includegraphics[height=8.7cm,width=7.cm]{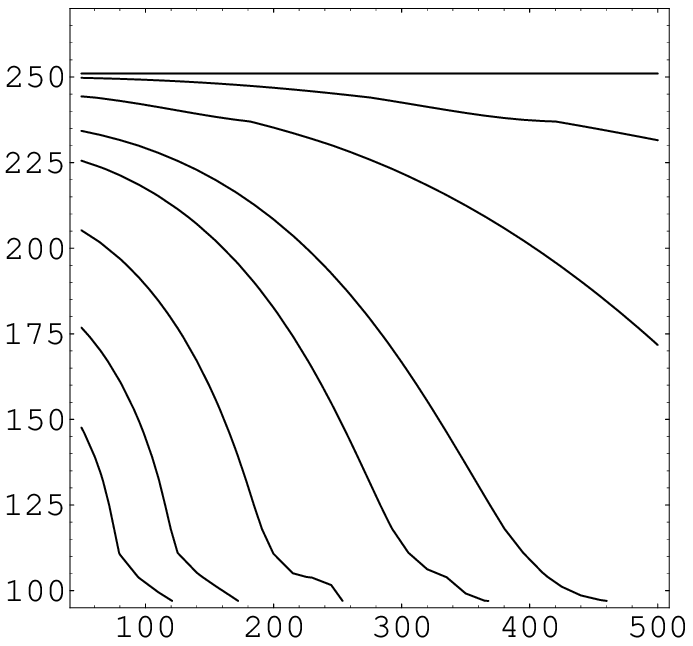}}
\put(75,189){{\small \bf b)}}
\put(81,187){\makebox(0,0)[bl]{{\small $m_{1/2}$~[GeV]}}}
\put(122,130){\makebox(0,0)[bl]{{\small $\tan\beta =50$}}}
\put(110,177){\makebox(0,0)[bl]{{\small $0$}}}
\put(110,169){\makebox(0,0)[bl]{{\small $1$}}}
\put(110,160){\makebox(0,0)[bl]{{\small $10$}}}
\put(110.5,146){\makebox(0,0)[bl]{{\small $50$}}}
\put(112,124){\makebox(0,0)[bl]{{\small $100$}}}
\put(100,121){\makebox(0,0)[bl]{{\small $250$}}}
\put(91.5,117){\makebox(0,0)[bl]{{\small $500$}}}
\put(86,113){\makebox(0,0)[bl]{{\small $750$}}}
\put(143,97){\makebox(0,0)[br]{{ $m_0$~[GeV]}}}
\put(-3,0){\includegraphics[height=8.7cm,width=7.cm]{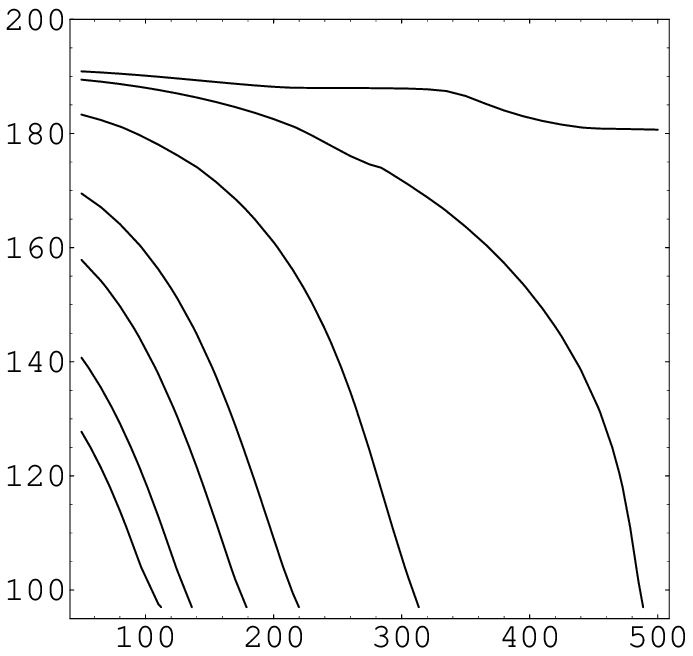}}
\put(-1,89){{\small \bf c)}}
\put(5,87){\makebox(0,0)[bl]{{\small $m_{1/2}$~[GeV]}}}
\put(29,70){\makebox(0,0)[bl]{{\small $10^{-3}$}}}
\put(27,65){\makebox(0,0)[bl]{{\small $1$}}}
\put(26,55){\makebox(0,0)[bl]{{\small $10$}}}
\put(11,55){\makebox(0,0)[bl]{{\small $50$}}}
\put(12,39){\makebox(0,0)[bl]{{\small $100$}}}
\put(11,25){\makebox(0,0)[bl]{{\small $200$}}}
\put(5,9){\makebox(0,0)[bl]{{\small $300$}}}
\put(25,77){\makebox(0,0)[bl]{{\small $\tan\beta =3$}}}
\put(69,-3){\makebox(0,0)[br]{{ $m_0$~[GeV]}}}
\put(73,0){\includegraphics[height=8.7cm,width=7.cm]{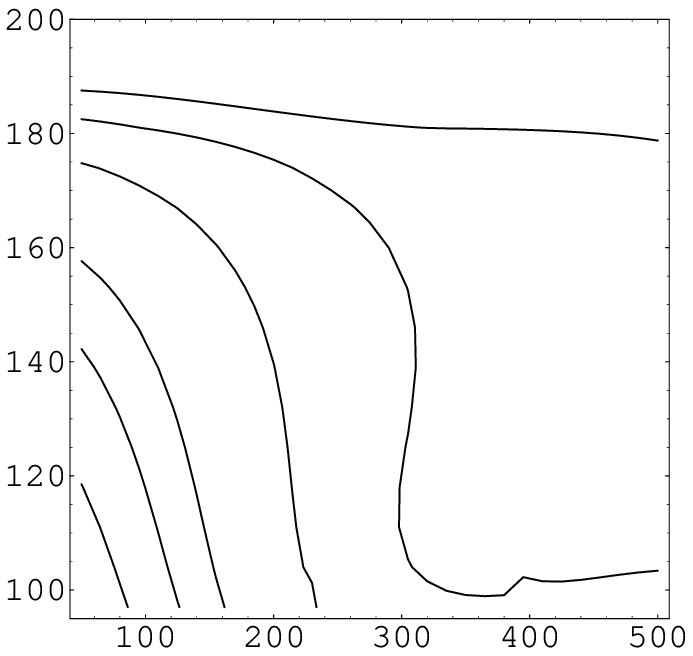}}
\put(75,89){{\small \bf d)}}
\put(81,87){\makebox(0,0)[bl]{{\small $m_{1/2}$~[GeV]}}}
\put(105,72){\makebox(0,0)[bl]{{\small $\tan\beta =50$}}}
\put(115,66){\makebox(0,0)[bl]{{\small $10^{-3}$}}}
\put(105,64){\makebox(0,0)[bl]{{\small $1$}}}
\put(98,51.5){\makebox(0,0)[bl]{{\small $10$}}}
\put(89,42){\makebox(0,0)[bl]{{\small $50$}}}
\put(81.5,41.5){\makebox(0,0)[bl]{{\small $100$}}}
\put(85,13){\makebox(0,0)[bl]{{\small $200$}}}
\put(143,-3){\makebox(0,0)[br]{{ $m_0$~[GeV]}}}
\end{picture}
\caption[Contour lines of the neutralino production cross sections]{
  Contour lines of the production cross sections in fb, in the
  $m_0$--$m_{1/2}$ plane for $\sqrt{s} = 205$~GeV, a) $e^+ e^- \to
  \tilde{\chi}^{0}_{1} \tilde{\chi}^{0}_{1}$, $\tan\beta=3$, b) $e^+
  e^- \to \tilde{\chi}^{0}_{1} \tilde{\chi}^{0}_{1}$, $\tan\beta=50$,
  c) $e^+ e^- \to \tilde{\chi}^{0}_{1} \tilde{\chi}^{0}_{2}$,
  $\tan\beta=3$, and d) $e^+ e^- \to \tilde{\chi}^{0}_{1}
  \tilde{\chi}^{0}_{2}$, $\tan\beta=50$.  ISR corrections are
  included.}
\label{fig:6}
\end{figure}

In order to show more explicitly the dependence of the cross sections
on the parameters $m_0$ and $m_{1/2}$ we plot in
Fig.~\ref{fig:6}a and b the contour lines of $\sigma(e^+ e^- \to
\tilde{\chi}^{0}_{1} \tilde{\chi}^{0}_{1})$ in the $m_0$-$m_{1/2}$ plane at $\sqrt{s} =
205$~GeV for $\tan\beta =3$ and $\tan\beta = 50$. The contour lines for
$\sigma(e^+ e^- \to \tilde{\chi}^{0}_{1} \tilde{\chi}^{0}_{2})$ are given in
Fig.~\ref{fig:6}c and d.

\subsubsection{Neutralino Decay Length}

For the observability of the $L$ violating effects it is crucial
that with this choice of parameters the LSP will decay most of the
time inside the detector.
The neutralino decay path expected at LEP2 depends crucially on the
values of $L$ violating parameters or, equivalently, on the value of
the heaviest neutrino mass, $m_{\nu_3}$.
We fix the value of $m_{\nu_3}$ as indicated by the analysis of the
atmospheric neutrino data \cite{Gonzalez-Garcia:2001sq,atm99}. It is important to note that,
as explained in \cite{Romao:2000up}, due to the projective nature of the
neutrino mass matrix \cite{arca}, only one of the
three neutrinos picks up a mass in tree approximation. This means
that, neglecting radiative corrections which give small masses to the
first two neutrinos in order to account for the solar neutrino data,
the neutralino decay length scale is set mainly by the tree--level
value of $m_{\nu_3}$. In ref.~\cite{Romao:2000up} we have explicitly shown
that this is a good approximation for most points in parameter space.

\begin{figure}
\setlength{\unitlength}{1mm}
\begin{picture}(150,90)
\put(-1,-4){\includegraphics[height=9.0cm,width=7.6cm]{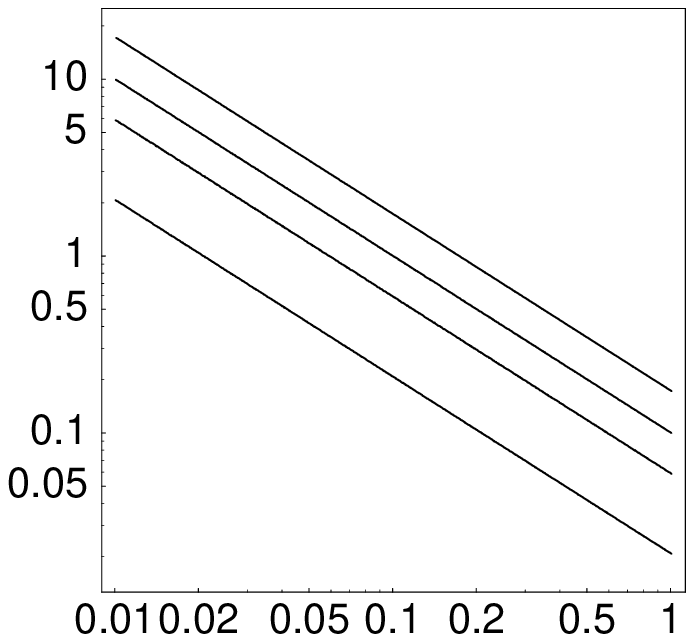}}
\put(-1,87){{\small \bf a)}}
\put(5,85){\makebox(0,0)[bl]{{\small $L(\tilde{\chi}^{0}_{1})$~[cm]}}}
\put(55,48){\makebox(0,0)[bl]{{\small 60}}}
\put(55,41){\makebox(0,0)[bl]{{\small 70}}}
\put(55,36){\makebox(0,0)[bl]{{\small 80}}}
\put(55,26){\makebox(0,0)[bl]{{\small 90}}}
\put(77,-3){\makebox(0,0)[br]{{ $m_{\nu_3}$~[eV]}}}
\put(82,-6){\includegraphics[height=9.4cm,width=7.6cm]{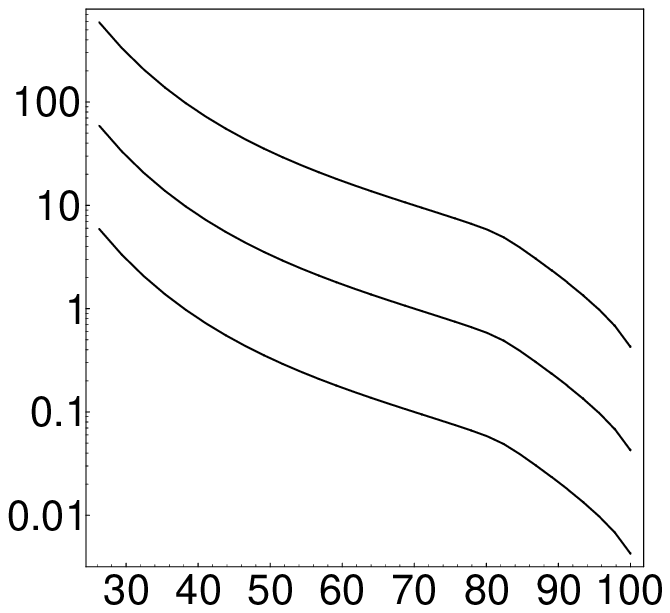}}
\put(81,87){{\small \bf b)}}
\put(89,85){\makebox(0,0)[bl]{{\small $L(\tilde{\chi}^{0}_{1})$~[cm]}}}
\put(125,60){\makebox(0,0)[bl]{{\small 0.01}}}
\put(125,46){\makebox(0,0)[bl]{{\small 0.1}}}
\put(125,32){\makebox(0,0)[bl]{{\small 1}}}
\put(158,-3){\makebox(0,0)[br]{{ $m_{\tilde{\chi}^{0}_{1}}$~[GeV]}}}
\end{picture}
\caption[Decay length of the lightest neutralino in cm]{Decay 
  length of the lightest neutralino in cm for $\sqrt{s} = 205$~GeV, a)
  as a function of $m_{\nu_3}$ for $m_{\tilde{\chi}^{0}_{1}} = 60, 70,
  80$, and 90 GeV, b) as a function of $m_{\tilde{\chi}^{0}_{1}}$ for
  $m_{\nu_3} = 0.01, 0.1$, and 1 eV.}
\label{fig:7}
\end{figure}

\begin{figure}
\setlength{\unitlength}{1mm}
\begin{picture}(150,95)
\put(-3,0){\includegraphics[height=8.7cm,width=7.cm]{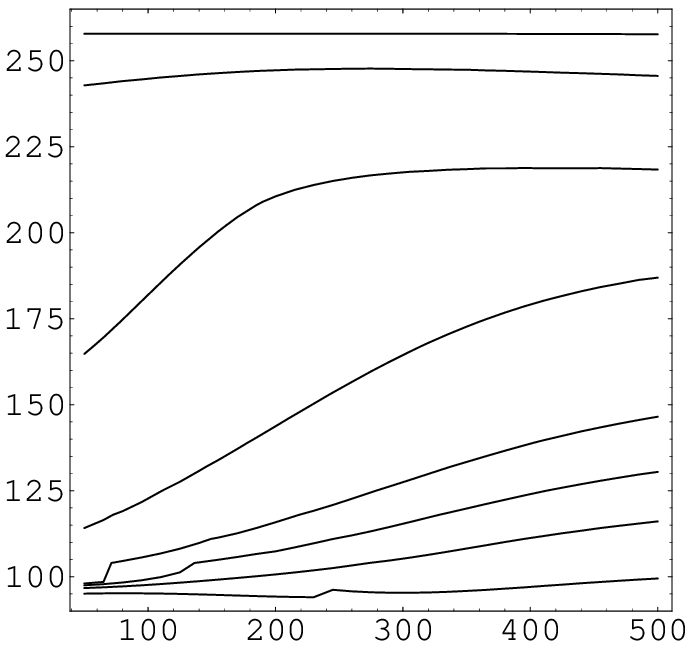}}
\put(-1,89){{\small \bf a)}}
\put(5,87){\makebox(0,0)[bl]{{\small $m_{1/2}$~[GeV]}}}
\put(38,78){\makebox(0,0)[bl]{{\small $0.001$}}}
\put(26,73){\makebox(0,0)[bl]{{\small $0.1$}}}
\put(26,63){\makebox(0,0)[bl]{{\small $1$}}}
\put(26,36){\makebox(0,0)[bl]{{\small $10$}}}
\put(57,33){\makebox(0,0)[bl]{{\small $50$}}}
\put(57,27){\makebox(0,0)[bl]{{\small $100$}}}
\put(57,20){\makebox(0,0)[bl]{{\small $200$}}}
\put(57,13){\makebox(0,0)[bl]{{\small $500$}}}
\put(30,53){\makebox(0,0)[bl]{{\small $\tan\beta =3$}}}
\put(69,-3){\makebox(0,0)[br]{{ $m_0$~[GeV]}}}
\put(73,0){\includegraphics[height=8.7cm,width=7.cm]{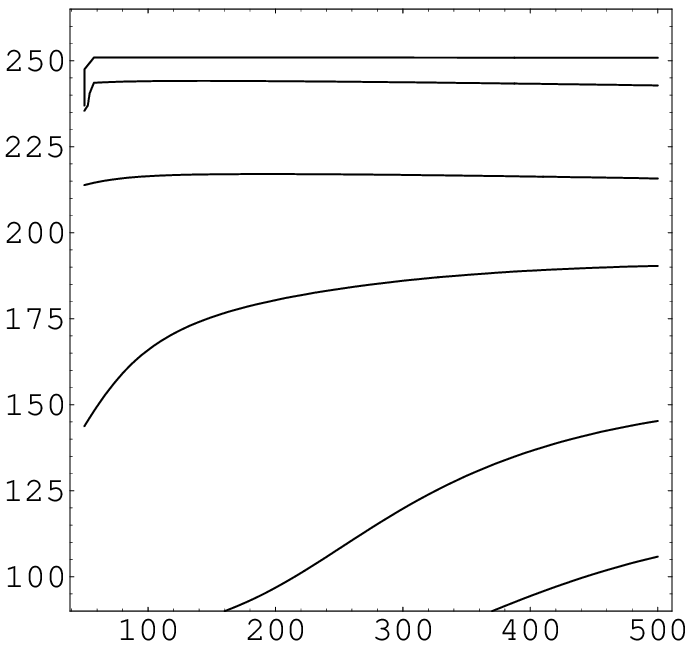}}
\put(75,89){{\small \bf b)}}
\put(81,87){\makebox(0,0)[bl]{{\small $m_{1/2}$~[GeV]}}}
\put(91,53){\makebox(0,0)[bl]{{\small $\tan\beta =50$}}}
\put(98,79){\makebox(0,0)[bl]{{\small $0.001$}}}
\put(96,72){\makebox(0,0)[bl]{{\small $0.1$}}}
\put(96,64){\makebox(0,0)[bl]{{\small $1$}}}
\put(130,53){\makebox(0,0)[bl]{{\small $10$}}}
\put(130,31){\makebox(0,0)[bl]{{\small $50$}}}
\put(130,14.5){\makebox(0,0)[bl]{{\small $100$}}}
\put(143,-3){\makebox(0,0)[br]{{ $m_0$~[GeV]}}}
\end{picture}
\caption[Decay length of the lightest neutralino in cm]{Decay 
  length of the lightest neutralino in cm in the $m_0$--$m_{1/2}$
  plane for $\sqrt{s} = 205$~GeV, a) $\tan\beta=3$, and b)
  $\tan\beta=50$. The $L$ violating parameters are fixed such that
  $m_{\nu_3} = 0.06$~GeV.}
\label{fig:8}
\end{figure}

In Fig.~\ref{fig:7} we plot the $\tilde{\chi}^{0}_{1}$ decay length in cm
expected at LEP2 for $\sqrt{s} = 205$~GeV. Here and later on
we consider the neutralinos
stemming from the process $e^+ e^- \to \tilde{\chi}^{0}_{1} \tilde{\chi}^{0}_{1}$ when discussing
the decay length.
In Fig.~\ref{fig:7}a we plot the $\tilde{\chi}^{0}_{1}$ decay length in cm
as a function of neutrino
mass $m_{\nu_3}$, for different $m_{\tilde{\chi}^{0}_{1}}$ between 60 and 90 GeV,
with $m_0 = 100$ ~GeV, and $\tan \beta = 3$. As can be seen the
expected neutralino decay length is typically such that the decays
occur inside the detector, leading to a drastic modification of the
MSUGRA signals.  An equivalent way of presenting the neutralino decay
path at LEP2 is displayed in Fig.~\ref{fig:7}b, which gives the
decay length of $\tilde{\chi}^{0}_{1}$ as a function of $m_{\tilde{\chi}^{0}_{1}}$ for $m_{\nu_3}
= 0.01$, 0.1, and 1 eV.
Finally, we show the dependence of the neutralino decay path on the
supergravity parameters fixing the magnitude of $L$ violating
parameters or, equivalently, the magnitude of the heaviest neutrino
mass, $m_{\nu_3}$.
In Fig.~\ref{fig:8}a and b we plot the contour lines of the decay
length of $\tilde{\chi}^{0}_{1}$ in the $m_0$-$m_{1/2}$ plane for $m_{\nu_3} =
0.06$~eV, $\tan\beta =3$ and 50. Note that the decay length is short
enough that it may happen inside typical high energy collider
detectors even for the small neutrino mass values $\sim 0.06$~eV
indicated by the atmospheric neutrino data~\cite{Gonzalez-Garcia:2001sq,atm99}.
For large values of $\tan\beta$ the total decay width increases and,
correspondingly, the decay path decreases due to the tau Yukawa
coupling and the bottom Yukawa coupling.

\subsubsection{Neutralino Branching Ratios}

As discussed in the beginning of this section, the lightest neutralino
$\tilde{\chi}^{0}_{1}$ will typically decay in the detector. In the
following we present our results for the branching ratios of all $L$
violating 3-body decay of $\tilde{\chi}^{0}_{1}$, and of the radiative
decay $\tilde{\chi}^{0}_{1} \to \nu_3 \gamma$.  The Feynman diagrams
for the decays $\tilde{\chi}^{0}_{1} \to \nu_3 f \bar{f}$ ($f=e$,
$\nu_e$, $\mu$, $\nu_\mu$, $u$, $d$, $c$, $s$, $b$) are shown in
Fig.~\ref{fig:4}.

\begin{figure}
\setlength{\unitlength}{1mm}
\begin{picture}(150,100)
\put(-3,0){\includegraphics[height=8.7cm,width=7.cm]{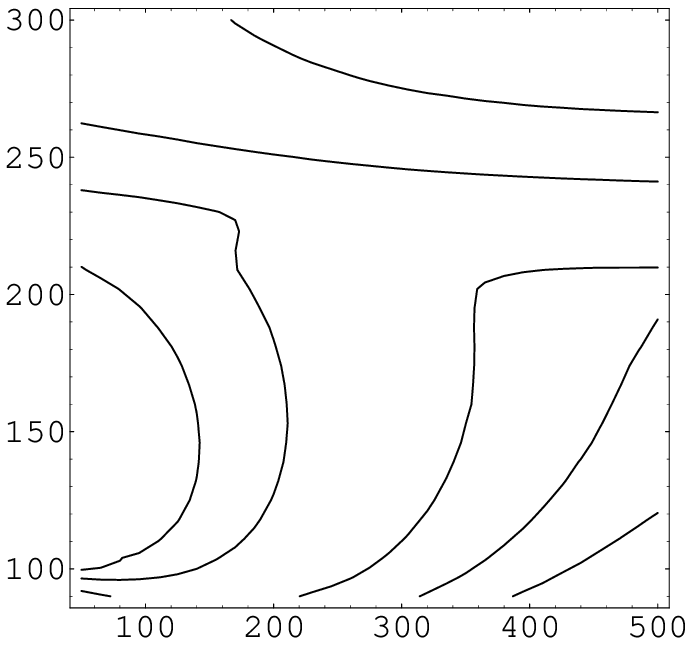}}
\put(-1,89){{\small \bf a)}}
\put(5,87){\makebox(0,0)[bl]{{\small $m_{1/2}$~[GeV]}}}
\put(39,74){\makebox(0,0)[bl]{{\small $5$}}}
\put(26,66){\makebox(0,0)[bl]{{\small $3$}}}
\put(58,12){\makebox(0,0)[bl]{{\small $7$}}}
\put(57,23){\makebox(0,0)[bl]{{\small $5$}}}
\put(47,42){\makebox(0,0)[bl]{{\small $3$}}}
\put(26,42){\makebox(0,0)[bl]{{\small $1$}}}
\put(18,25){\makebox(0,0)[bl]{{\small $0.5$}}}
\put(7,75){\makebox(0,0)[bl]{{\small $\tan\beta =3$}}}
\put(69,-3){\makebox(0,0)[br]{{ $m_0$~[GeV]}}}
\put(73,0){\includegraphics[height=8.7cm,width=7.cm]{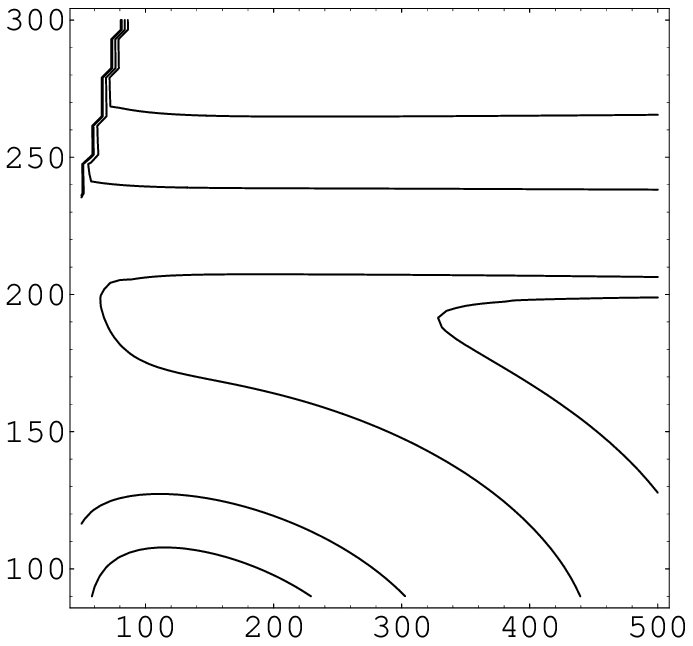}}
\put(75,89){{\small \bf b)}}
\put(81,87){\makebox(0,0)[bl]{{\small $m_{1/2}$~[GeV]}}}
\put(105,75){\makebox(0,0)[bl]{{\small $\tan\beta =50$}}}
\put(94,71){\makebox(0,0)[bl]{{\small $5$}}}
\put(94,62.5){\makebox(0,0)[bl]{{\small $3$}}}
\put(117,40){\makebox(0,0)[bl]{{\small $5$}}}
\put(94,51){\makebox(0,0)[bl]{{\small $3$}}}
\put(92,23){\makebox(0,0)[bl]{{\small $1$}}}
\put(92,15){\makebox(0,0)[bl]{{\small $0.5$}}}
\put(143,-3){\makebox(0,0)[br]{{ $m_0$~[GeV]}}}
\end{picture}
\caption[Branching ratios for  $\tilde{\chi}^{0}_{1} \to 3 \, \nu$]{Branching ratios for  $\tilde{\chi}^{0}_{1} \to 3 \, \nu$ in \%
  in the $m_0$--$m_{1/2}$ plane for a) $\tan\beta=3$, and b)
  $\tan\beta=50$.  The $L$ violating parameters are fixed such that
  $m_{\nu_3} = 0.06$~GeV.}
\label{fig:9}
\end{figure}

\begin{figure}
\setlength{\unitlength}{1mm}
\begin{picture}(150,100)
\put(-3,0){\includegraphics[height=8.7cm,width=7.cm]{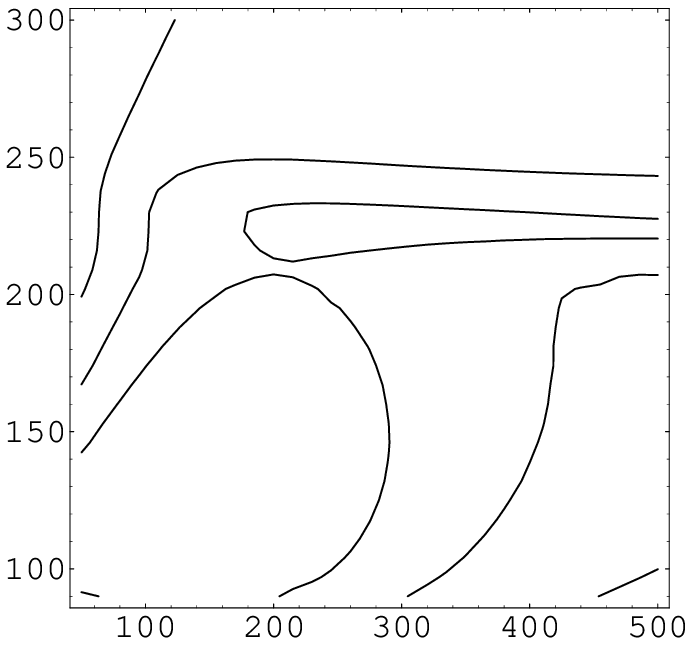}}
\put(-1,89){{\small \bf a)}}
\put(5,87){\makebox(0,0)[bl]{{\small $m_{1/2}$~[GeV]}}}
\put(14,76){\makebox(0,0)[bl]{{\small $2$}}}
\put(21,66){\makebox(0,0)[bl]{{\small $1$}}}
\put(27,54){\makebox(0,0)[bl]{{\small $0.5$}}}
\put(56,10){\makebox(0,0)[bl]{{\small $2$}}}
\put(50,41){\makebox(0,0)[bl]{{\small $1$}}}
\put(29.5,26){\makebox(0,0)[bl]{{\small $0.5$}}}
\put(25,75){\makebox(0,0)[bl]{{\small $\tan\beta =3$}}}
\put(69,-3){\makebox(0,0)[br]{{ $m_0$~[GeV]}}}
\put(73,0){\includegraphics[height=8.7cm,width=7.cm]{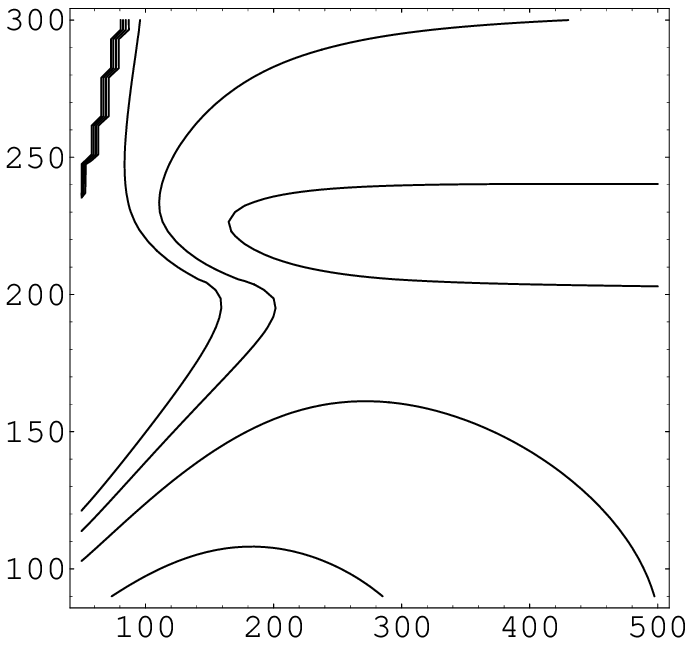}}
\put(75,89){{\small \bf b)}}
\put(81,87){\makebox(0,0)[bl]{{\small $m_{1/2}$~[GeV]}}}
\put(107,40){\makebox(0,0)[bl]{{\small $\tan\beta =50$}}}
\put(83,62){\makebox(0,0)[bl]{{\small $3$}}}
\put(100,78){\makebox(0,0)[bl]{{\small $2$}}}
\put(100,61){\makebox(0,0)[bl]{{\small $1$}}}
\put(100,33.5){\makebox(0,0)[bl]{{\small $1$}}}
\put(100,16){\makebox(0,0)[bl]{{\small $0.2$}}}
\put(143,-3){\makebox(0,0)[br]{{ $m_0$~[GeV]}}}
\end{picture}
\caption[Branching ratios for $\tilde{\chi}^{0}_{1} \to \nu_3 \, l^+ \, l^-$]{
  Branching ratios for $\tilde{\chi}^{0}_{1} \to \nu_3 \, l^+ \, l^-$
  in \% in the $m_0$--$m_{1/2}$ plane for a) $\tan\beta=3$, and b)
  $\tan\beta=50$. Here $l$ is the sum of $e$ and $\mu$.  The $L$
  violating parameters are fixed such that $m_{\nu_3} = 0.06$~GeV.}
\label{fig:10}
\end{figure}

\begin{figure}
\setlength{\unitlength}{1mm}
\begin{picture}(150,100)
\put(-3,0){\includegraphics[height=8.7cm,width=7.cm]{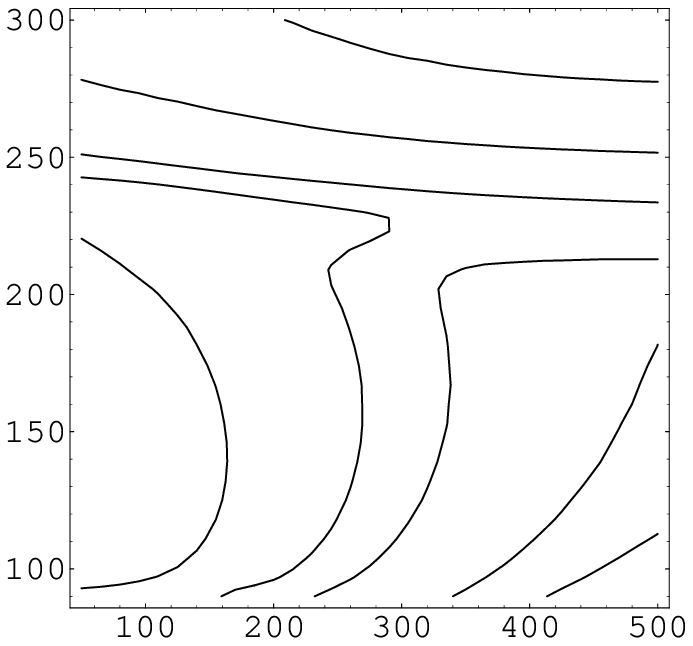}}
\put(-1,89){{\small \bf a)}}
\put(5,87){\makebox(0,0)[bl]{{\small $m_{1/2}$~[GeV]}}}
\put(39,79){\makebox(0,0)[bl]{{\small $15$}}}
\put(39,68){\makebox(0,0)[bl]{{\small $10$}}}
\put(26,64){\makebox(0,0)[bl]{{\small $5$}}}
\put(57,16){\makebox(0,0)[bl]{{\small $15$}}}
\put(57,22){\makebox(0,0)[bl]{{\small $10$}}}
\put(45,42){\makebox(0,0)[bl]{{\small $5$}}}
\put(34,42){\makebox(0,0)[bl]{{\small $3$}}}
\put(21,25){\makebox(0,0)[bl]{{\small $1$}}}
\put(7,76){\makebox(0,0)[bl]{{\small $\tan\beta =3$}}}
\put(69,-3){\makebox(0,0)[br]{{ $m_0$~[GeV]}}}
\put(73,0){\includegraphics[height=8.7cm,width=7.cm]{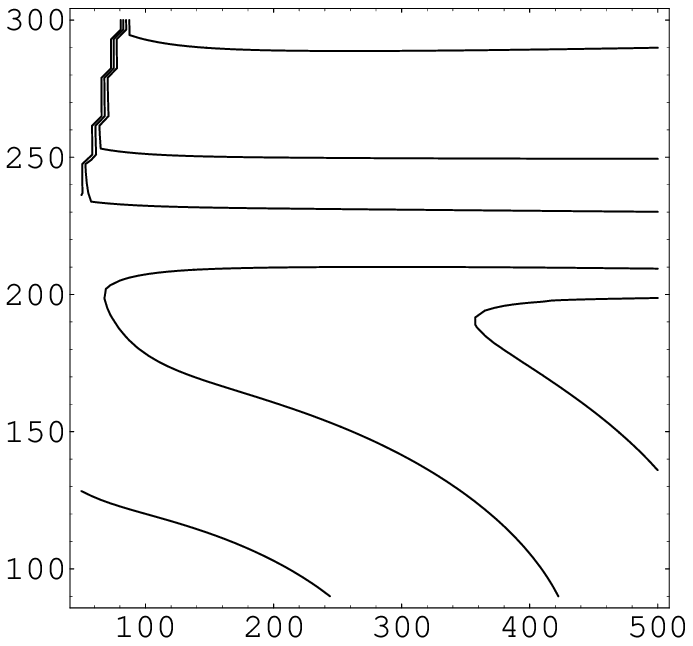}}
\put(75,89){{\small \bf b)}}
\put(81,87){\makebox(0,0)[bl]{{\small $m_{1/2}$~[GeV]}}}
\put(100,40){\makebox(0,0)[bl]{{\small $\tan\beta =50$}}}
\put(94,79.5){\makebox(0,0)[bl]{{\small $15$}}}
\put(94,66){\makebox(0,0)[bl]{{\small $10$}}}
\put(94,59){\makebox(0,0)[bl]{{\small $5$}}}
\put(94,52){\makebox(0,0)[bl]{{\small $5$}}}
\put(128,40){\makebox(0,0)[bl]{{\small $10$}}}
\put(92,19){\makebox(0,0)[bl]{{\small $1$}}}
\put(143,-3){\makebox(0,0)[br]{{ $m_0$~[GeV]}}}
\end{picture}
\caption[Branching ratios for $\tilde{\chi}^{0}_{1} \to \nu_3 \, q \, \bar{q}$]{Branching ratios for $\tilde{\chi}^{0}_{1} \to \nu_3 \, q \, \bar{q}$ in \%
  in the $m_0$--$m_{1/2}$ plane for a) $\tan\beta=3$, and b)
  $\tan\beta=50$.  Here $q$ is the sum over $u$, $d$, $s$, and $c$.
  The $L$ violating parameters are fixed such that $m_{\nu_3} =
  0.06$~GeV.}
\label{fig:11}
\end{figure}

\begin{figure}
\setlength{\unitlength}{1mm}
\begin{picture}(150,100)
\put(-3,0){\includegraphics[height=8.7cm,width=7.cm]{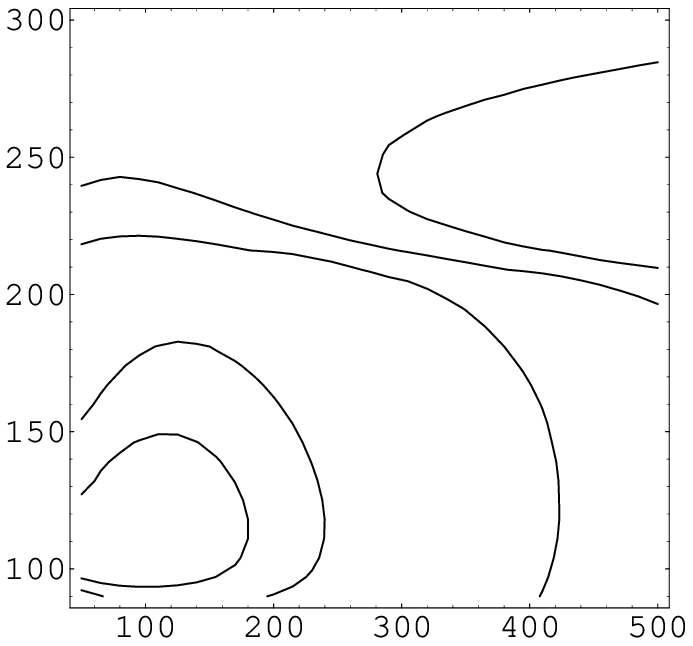}}
\put(-1,89){{\small \bf a)}}
\put(5,87){\makebox(0,0)[bl]{{\small $m_{1/2}$~[GeV]}}}
\put(39,65){\makebox(0,0)[bl]{{\small $15$}}}
\put(18,63){\makebox(0,0)[bl]{{\small $10$}}}
\put(43,42){\makebox(0,0)[bl]{{\small $5$}}}
\put(19,36){\makebox(0,0)[bl]{{\small $1$}}}
\put(7,20){\makebox(0,0)[bl]{{\small $0.1$}}}
\put(7,75){\makebox(0,0)[bl]{{\small $\tan\beta =3$}}}
\put(69,-3){\makebox(0,0)[br]{{ $m_0$~[GeV]}}}
\put(73,0){\includegraphics[height=8.7cm,width=7.cm]{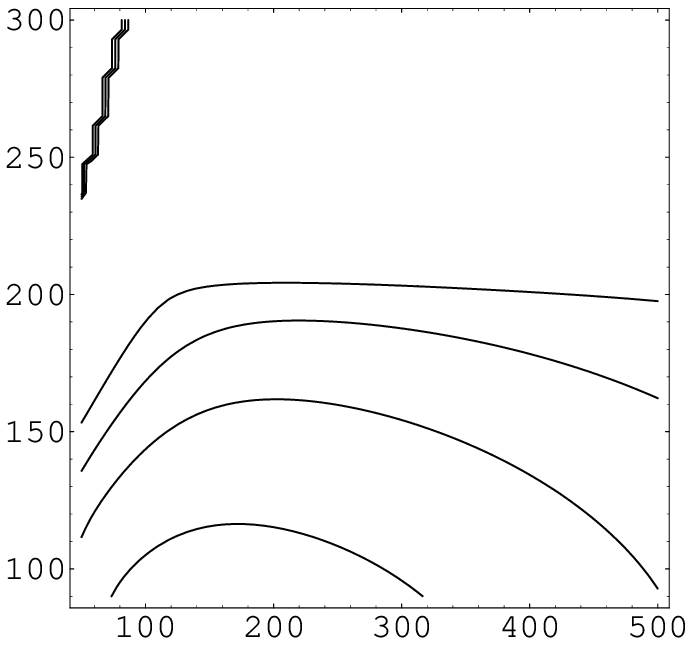}}
\put(75,89){{\small \bf b)}}
\put(81,87){\makebox(0,0)[bl]{{\small $m_{1/2}$~[GeV]}}}
\put(103,75){\makebox(0,0)[bl]{{\small $\tan\beta =50$}}}
\put(86,77){\vector(-1,0){4}}
\put(86,77){\makebox(0,0)[bl]{{\small $2.6$}}}
\put(85,70){\makebox(0,0)[bl]{{\small $15$}}}
\put(95,50){\makebox(0,0)[bl]{{\small $15$}}}
\put(95,45){\makebox(0,0)[bl]{{\small $10$}}}
\put(95,36){\makebox(0,0)[bl]{{\small $5$}}}
\put(95,20){\makebox(0,0)[bl]{{\small $1$}}}
\put(143,-3){\makebox(0,0)[br]{{ $m_0$~[GeV]}}}
\end{picture}
\caption[Branching ratios for $\tilde{\chi}^{0}_{1} \to \nu_l \, 
\tau^{\pm} \, l^{\mp}$]{Branching ratios for $\tilde{\chi}^{0}_{1} \to
  \nu_l \, \tau^{\pm} \, l^{\mp}$ in \% in the $m_0$--$m_{1/2}$ plane
  for a) $\tan\beta=3$, and b) $\tan\beta=50$. Here $l$ is the sum of
  $e$ and $\mu$.  The $L$ violating parameters are fixed such that
  $m_{\nu_3} = 0.06$~GeV.}
\label{fig:12}
\end{figure}

\begin{figure}
\setlength{\unitlength}{1mm}
\begin{picture}(150,100)
\put(-3,0){\includegraphics[height=8.7cm,width=7.cm]{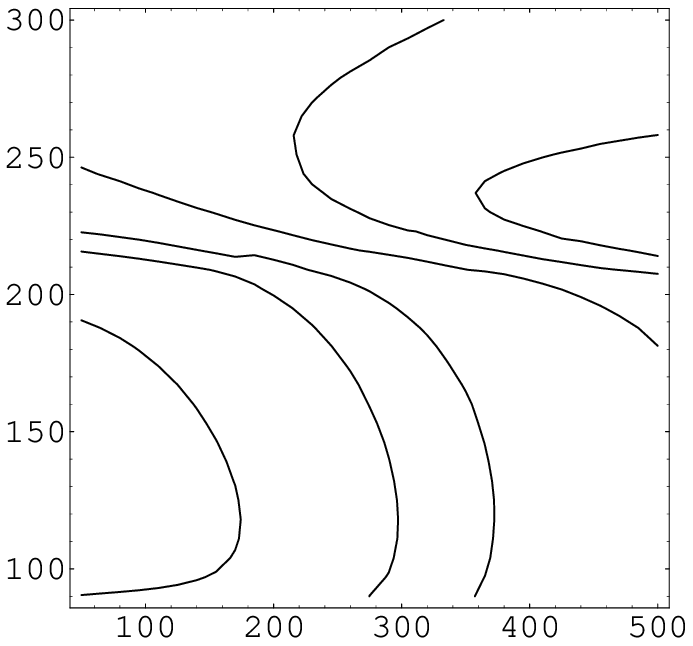}}
\put(-1,89){{\small \bf a)}}
\put(5,87){\makebox(0,0)[bl]{{\small $m_{1/2}$~[GeV]}}}
\put(42,63){\makebox(0,0)[bl]{{\small $50$}}}
\put(22,63){\makebox(0,0)[bl]{{\small $40$}}}
\put(55,42){\makebox(0,0)[bl]{{\small $25$}}}
\put(34,42){\makebox(0,0)[bl]{{\small $10$}}}
\put(24,42){\makebox(0,0)[bl]{{\small $5$}}}
\put(16,25){\makebox(0,0)[bl]{{\small $1$}}}
\put(12,75){\makebox(0,0)[bl]{{\small $\tan\beta =3$}}}
\put(69,-3){\makebox(0,0)[br]{{ $m_0$~[GeV]}}}
\put(73,0){\includegraphics[height=8.7cm,width=7.cm]{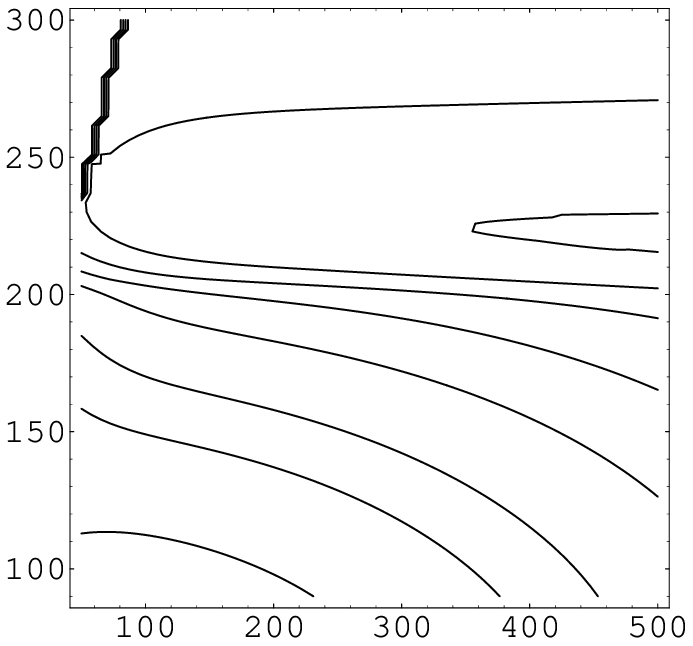}}
\put(75,89){{\small \bf b)}}
\put(81,87){\makebox(0,0)[bl]{{\small $m_{1/2}$~[GeV]}}}
\put(103,65){\makebox(0,0)[bl]{{\small $\tan\beta =50$}}}
\put(82,77){\makebox(0,0)[bl]{{\small $7$}}}
\put(85,72){\makebox(0,0)[bl]{{\small $50$}}}
\put(100,51.5){\makebox(0,0)[bl]{{\small $50$}}}
\put(130,59){\makebox(0,0)[bl]{{\small $60$}}}
\put(129,46){\makebox(0,0)[bl]{{\small $40$}}}
\put(129,41){\makebox(0,0)[bl]{{\small $30$}}}
\put(100,43){\makebox(0,0)[bl]{{\small $20$}}}
\put(100,34){\makebox(0,0)[bl]{{\small $10$}}}
\put(100,27){\makebox(0,0)[bl]{{\small $5$}}}
\put(100,14){\makebox(0,0)[bl]{{\small $1$}}}
\put(143,-3){\makebox(0,0)[br]{{ $m_0$~[GeV]}}}
\end{picture}
\caption[Branching ratios for $\tilde{\chi}^{0}_{1} \to \tau^{\pm} 
\, q \, \bar{q}'$]{Branching ratios for $\tilde{\chi}^{0}_{1} \to
  \tau^{\pm} \, q \, \bar{q}'$ in \% in the $m_0$--$m_{1/2}$ plane for
  a) $\tan\beta=3$, and b) $\tan\beta=50$.  Here $q$ is the sum over
  $u$, $d$, $s$, and $c$.  The $L$ violating parameters are fixed such
  that $m_{\nu_3} = 0.06$~GeV.}
\label{fig:13}
\end{figure}

\begin{figure}
\setlength{\unitlength}{1mm}
\begin{picture}(150,100)
\put(-3,0){\includegraphics[height=8.7cm,width=7.cm]{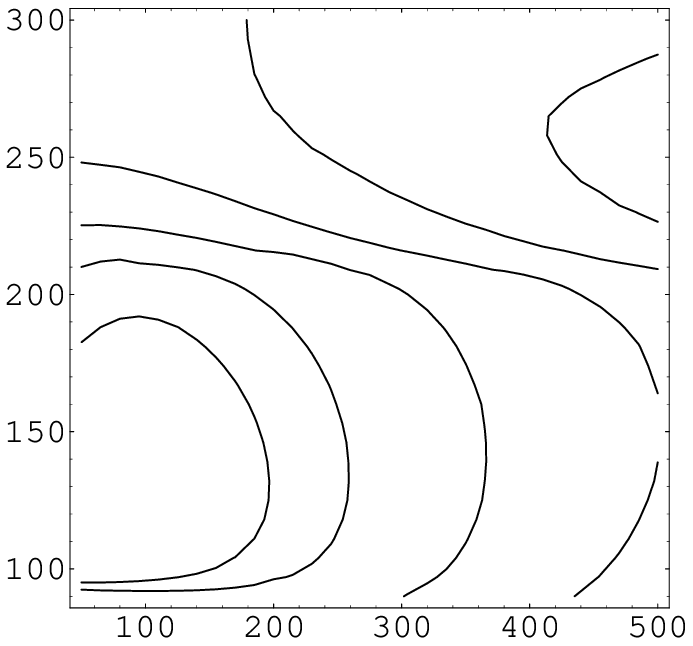}}
\put(-1,89){{\small \bf a)}}
\put(5,87){\makebox(0,0)[bl]{{\small $m_{1/2}$~[GeV]}}}
\put(48,67){\makebox(0,0)[bl]{{\small $10$}}}
\put(26,62){\makebox(0,0)[bl]{{\small $25$}}}
\put(55,42){\makebox(0,0)[bl]{{\small $50$}}}
\put(36,42){\makebox(0,0)[bl]{{\small $75$}}}
\put(21,42){\makebox(0,0)[bl]{{\small $90$}}}
\put(19,25){\makebox(0,0)[bl]{{\small $95$}}}
\put(27,75){\makebox(0,0)[bl]{{\small $\tan\beta =3$}}}
\put(69,-3){\makebox(0,0)[br]{{ $m_0$~[GeV]}}}
\put(73,0){\includegraphics[height=8.7cm,width=7.cm]{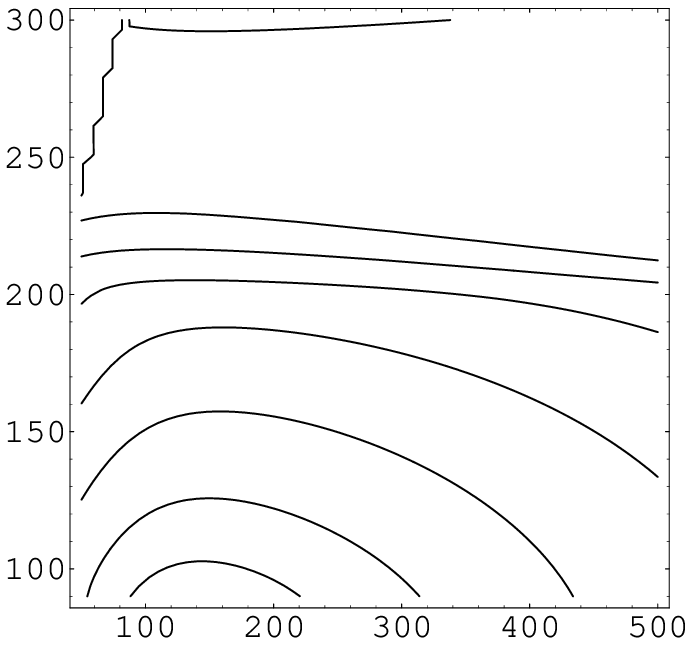}}
\put(75,89){{\small \bf b)}}
\put(81,87){\makebox(0,0)[bl]{{\small $m_{1/2}$~[GeV]}}}
\put(110,72){\makebox(0,0)[bl]{{\small $\tan\beta =50$}}}
\put(84,72){\makebox(0,0)[bl]{{\small $1$}}}
\put(92,77){\makebox(0,0)[bl]{{\small $5$}}}
\put(94,59){\makebox(0,0)[bl]{{\small $5$}}}
\put(92,54.5){\makebox(0,0)[bl]{{\small $10$}}}
\put(92,50){\makebox(0,0)[bl]{{\small $25$}}}
\put(92,45){\makebox(0,0)[bl]{{\small $50$}}}
\put(92,34){\makebox(0,0)[bl]{{\small $75$}}}
\put(92,23){\makebox(0,0)[bl]{{\small $90$}}}
\put(92,15){\makebox(0,0)[bl]{{\small $95$}}}
\put(143,-3){\makebox(0,0)[br]{{ $m_0$~[GeV]}}}
\end{picture}
\caption[Branching ratios for $\tilde{\chi}^{0}_{1} \to \nu_3 \, b 
\, \bar{b}$]{Branching ratios for $\tilde{\chi}^{0}_{1} \to \nu_3 \, b
  \, \bar{b}$ in \% in the $m_0$--$m_{1/2}$ plane for a)
  $\tan\beta=3$, and b) $\tan\beta=50$.  The $L$ violating parameters
  are fixed such that $m_{\nu_3} = 0.06$~GeV.}
\label{fig:14}
\end{figure}

\begin{figure}
\setlength{\unitlength}{1mm}
\begin{picture}(150,100)
\put(-3,0){\includegraphics[height=8.7cm,width=7.cm]{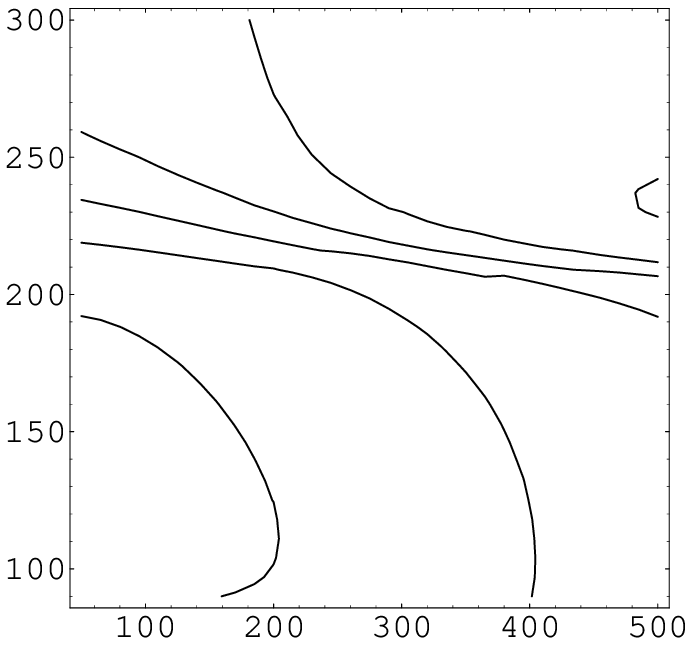}}
\put(-1,89){{\small \bf a)}}
\put(5,87){\makebox(0,0)[bl]{{\small $m_{1/2}$~[GeV]}}}
\put(59,62){\makebox(0,0)[bl]{{\small $9$}}}
\put(32,63){\makebox(0,0)[bl]{{\small $7$}}}
\put(21,62){\makebox(0,0)[bl]{{\small $5$}}}
\put(54,44){\makebox(0,0)[bl]{{\small $3$}}}
\put(34,42){\makebox(0,0)[bl]{{\small $1$}}}
\put(24,26){\makebox(0,0)[bl]{{\small $0.1$}}}
\put(35,75){\makebox(0,0)[bl]{{\small $\tan\beta =3$}}}
\put(69,-3){\makebox(0,0)[br]{{ $m_0$~[GeV]}}}
\put(73,0){\includegraphics[height=8.7cm,width=7.cm]{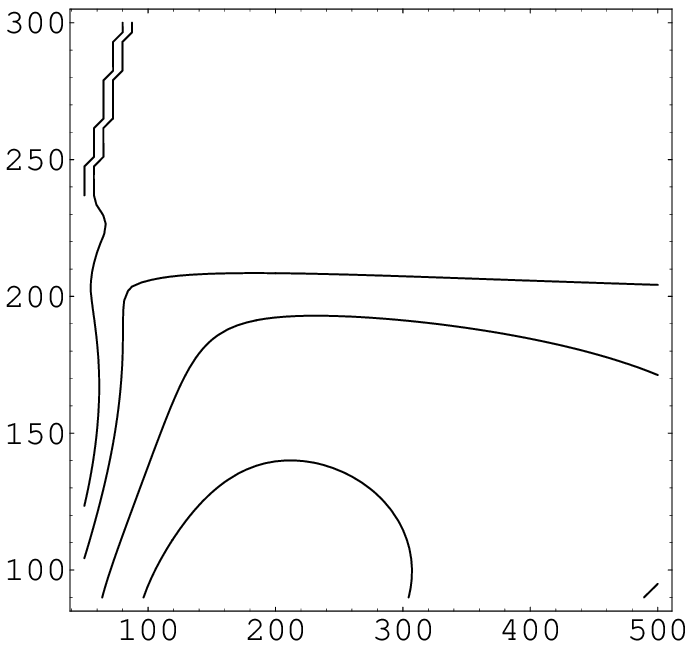}}
\put(75,89){{\small \bf b)}}
\put(81,87){\makebox(0,0)[bl]{{\small $m_{1/2}$~[GeV]}}}
\put(105,72){\makebox(0,0)[bl]{{\small $\tan\beta =50$}}}
\put(87,77){\vector(-1,0){5}}
\put(87,77){\makebox(0,0)[bl]{{\small $90.3$}}}
\put(85,69){\makebox(0,0)[bl]{{\small $10$}}}
\put(100,52){\makebox(0,0)[bl]{{\small $8$}}}
\put(100,46){\makebox(0,0)[bl]{{\small $5$}}}
\put(100,28){\makebox(0,0)[bl]{{\small $3$}}}
\put(143,-3){\makebox(0,0)[br]{{ $m_0$~[GeV]}}}
\end{picture}
\caption[Branching ratios for $\tilde{\chi}^{0}_{1} \to \nu_3 
\, \tau^+ \, \tau^-$]{Branching ratios for $\tilde{\chi}^{0}_{1} \to
  \nu_3 \, \tau^+ \, \tau^-$ in \% in the $m_0$--$m_{1/2}$ plane for
  a) $\tan\beta=3$, and b) $\tan\beta=50$.  The $L$ violating
  parameters are fixed such that $m_{\nu_3} = 0.06$~GeV.}
\label{fig:15}
\end{figure}

\begin{figure}
\setlength{\unitlength}{1mm}
\begin{picture}(150,100)
\put(-3,0){\includegraphics[height=8.7cm,width=7.cm]{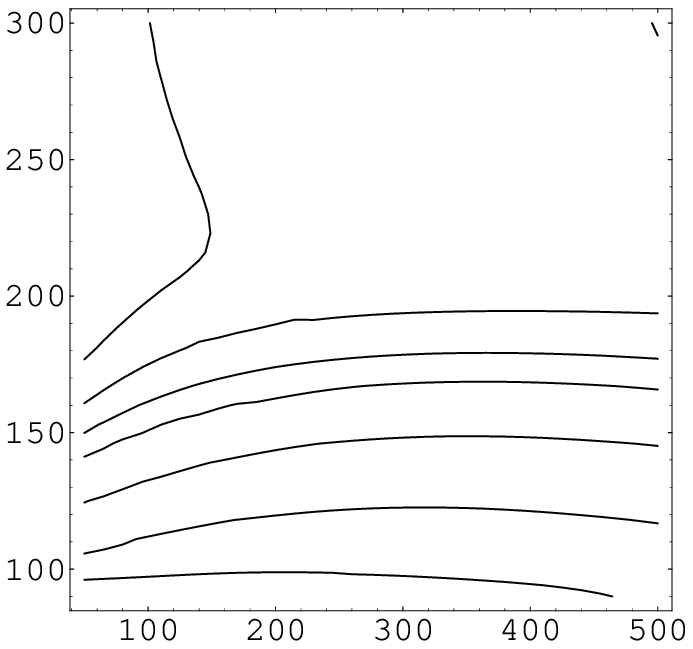}}
\put(-1,89){{\small \bf a)}}
\put(5,87){\makebox(0,0)[bl]{{\small $m_{1/2}$~[GeV]}}}
\put(18,61.5){\makebox(0,0)[bl]{{\small $10^{-3}$}}}
\put(30,46){\makebox(0,0)[bl]{{\small $10^{-4}$}}}
\put(30,41){\makebox(0,0)[bl]{{\small $10^{-4}$}}}
\put(21,35.5){\makebox(0,0)[bl]{{\small $10^{-3}$}}}
\put(30,31){\makebox(0,0)[bl]{{\small $0.01$}}}
\put(30,22){\makebox(0,0)[bl]{{\small $0.1$}}}
\put(30,14){\makebox(0,0)[bl]{{\small $1$}}}
\put(30,75){\makebox(0,0)[bl]{{\small $\tan\beta =3$}}}
\put(69,-3){\makebox(0,0)[br]{{ $m_0$~[GeV]}}}
\put(73,0){\includegraphics[height=8.7cm,width=7.cm]{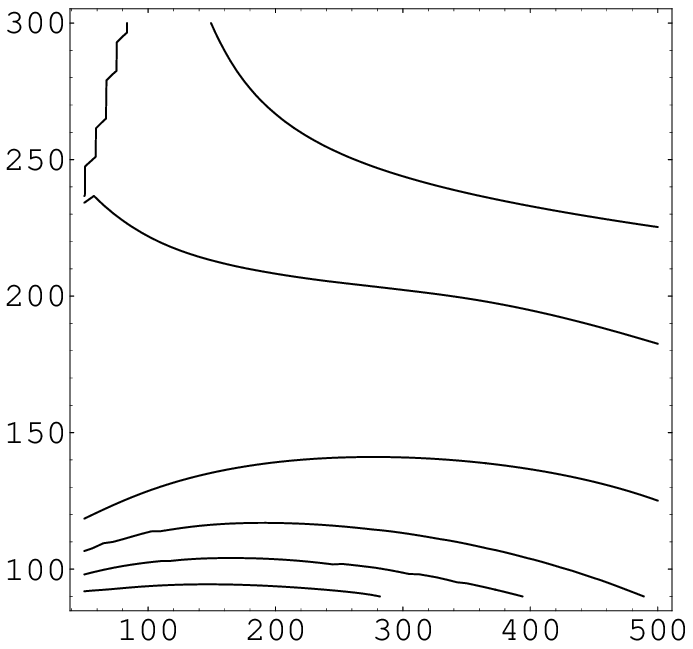}}
\put(75,89){{\small \bf b)}}
\put(81,87){\makebox(0,0)[bl]{{\small $m_{1/2}$~[GeV]}}}
\put(87,35){\makebox(0,0)[bl]{{\small $\tan\beta =50$}}}
\put(85,72){\makebox(0,0)[bl]{{\small $10^{-3}$}}}
\put(105,68){\makebox(0,0)[bl]{{\small $10^{-2}$}}}
\put(105,51){\makebox(0,0)[bl]{{\small $10^{-3}$}}}
\put(105,28){\makebox(0,0)[bl]{{\small $10^{-3}$}}}
\put(105,14.5){\makebox(0,0)[bl]{{\small $10^{-2}$}}}
\put(105,10){\makebox(0,0)[bl]{{\small $10^{-3}$}}}
\put(143,-3){\makebox(0,0)[br]{{ $m_0$~[GeV]}}}
\end{picture}
\caption[Branching ratios for $\tilde{\chi}^{0}_{1} \to \nu_3 \, \gamma$]{
  Branching ratios for $\tilde{\chi}^{0}_{1} \to \nu_3 \, \gamma$ in
  \% in the $m_0$--$m_{1/2}$ plane for a) $\tan\beta=3$, and b)
  $\tan\beta=50$.  The $L$ violating parameters are fixed such that
  $m_{\nu_3} = 0.06$~GeV.}
\label{fig:16}
\end{figure}

In the following plots Fig.~\ref{fig:9} - \ref{fig:16} we
show contour lines in the $m_0$-$m_{1/2}$ plane for the branching
ratios in \% of the various $\tilde{\chi}^{0}_{1}$ decays, in (a) for $\tan\beta = 3$
and in (b) for $\tan\beta=50$. We have fixed the mass of the heaviest
neutrino to $m_{\nu_3} = 0.06 $~eV \cite{Gonzalez-Garcia:2001sq,atm99}. It turns out, that in
the range $10^{-2}$~eV $\leq m_{\nu_3} \leq 1$~keV all the $\tilde{\chi}^{0}_{1}$
decay branching ratios are rather insensitive to the actual value of
$m_{\nu_3}$.  This is an important feature of our supergravity--type
SSSM. It is a consequence of the fact that, as a
result of the universal supergravity boundary conditions on the soft
breaking terms, all $L$ violating couplings are proportional to a
unique common parameter which may be taken as $\mu_3 / \mu_0$. For
a more detailed discussion on this proportionality the reader is
referred to ref.~\cite{Romao:2000up}.
Also note that for $m_{1/2} \gtrsim$ 220~GeV the neutralino mass becomes
larger than $m_W$ and $m_Z$ so that $\tilde{\chi}^{0}_{1}$ decays into real $W$
and $Z$ are possible. The effects of these real decays can be seen for
$m_{1/2} \gtrsim$ 220~GeV in most of the following plots.
For the large $\tan\beta$ case ($\tan\beta = 50$) and $m_{1/2} \gg m_0$ the
mass of the lighter charged boson $S^\pm_1$ is smaller than
$m_{\tilde{\chi}^{0}_{1}}$ (upper left corner of Fig.~\ref{fig:9}b -
\ref{fig:16}b). In this region of the parameter space the two 2-body
decays $\tilde{\chi}^{0}_{1} \to W^\pm \tau^\pm$ and $\tilde{\chi}^{0}_{1} \to
S^\pm_1 \tau^\pm$ compete.  The first one is $L$ violating, but has more
phase space than the second one which is $L$ conserving, since $S^\pm_1$
is mainly a stau. For this reason, the most import final state is
$\tau^+ \tau^- \nu_3$, followed by $\tau^\pm q \bar{q}'$ and $\tau^\pm l^\mp \nu_i$
($l=e,\mu$) as shown in Figs.~\ref{fig:15}, \ref{fig:13}, and
\ref{fig:12}, respectively.  All other final states have nearly
vanishing branching ratios in this corner of the parameter space.

Fig.~\ref{fig:9}a and b exhibit the contour lines for the
branching ratio of the invisible decay $\tilde{\chi}^{0}_{1} \to 3 \, \nu$. This
branching ratio can reach 7\% for the parameters chosen. In
Figs.~\ref{fig:10}, \ref{fig:11} we show the branching
ratio for the decays $\tilde{\chi}^{0}_{1} \to \nu_3 \, l^+ \, l^-$ and
$\tilde{\chi}^{0}_{1} \to \nu_3 \, q \, \bar{q}$ where $l$ and $q$ denote the
leptons and quarks of the first two generations, summed over all
flavors. These branching ratios can go up to 3\% and 15\%,
respectively. Notice that the sneutrino, slepton, and squark exchange
contributions to the $\tilde{\chi}^{0}_{1}$ decays become larger with increasing
$m_0$, despite the fact that the increase of the scalar masses
$m_{\tilde \nu}$, $m_{\tilde l}$, $m_{\tilde q}$ suppresses these
exchange contributions. 
This trend can also be observed in Fig.~\ref{fig:11},
\ref{fig:12} and \ref{fig:13}.
This happens because the tadpole equations correlate $\mu_0$ to $m_0$.
Increasing $\mu_0$ while keeping $M_1$ and $M_2$ fixed implies
increasing the gaugino content of $\tilde{\chi}^{0}_{1}$ and, hence, enhancing the
$\tilde{\chi}^{0}_{1}$-$f$-$\tilde f$ couplings.

In Figs.~\ref{fig:12}, \ref{fig:13} we show the contour
lines for the branching ratios of the LSP decays involving a single
tau, namely $\tilde{\chi}^{0}_{1} \to \nu_l \, \tau^{\pm} \, l^{\mp}$ and
$\tilde{\chi}^{0}_{1} \to \tau^{\pm} \, q \, \bar{q}'$, where $l$, $q$, and $q'$
are summed over the first two generations.  The branching for these
decay modes can reach up to 20\% and 60\% respectively. For
$m_{1/2}\gtrsim 220$~GeV decays into real $W^\pm$ dominate. If this is
the case and if both $\tilde{\chi}^{0}_{1}$ produced in $e^+ e^- \to \tilde{\chi}^{0}_{1}
\tilde{\chi}^{0}_{1}$ decay according to these modes this would lead to very
distinctive final states, such as $4 j \tau^+ \tau^+$, $\tau^+ \tau^+
l^- l^-$ ($l=e,\mu$), or $\tau^+ \tau^+ e^- \mu^-$. The full list of
expected signals is given in Tab.~\ref{tab:7}.  The first column in this table
specifies the two pairs of $\tilde{\chi}^{0}_{1}$ decay modes, while the second
one gives the corresponding signature. In the last column we state
whether the corresponding signature exists for $e^+ e^- \to \tilde{\chi}^{0}_{1}
\tilde{\chi}^{0}_{2}$ production within MSUGRA.

The LSP decays involving only third generation fermions, namely,
$\tilde{\chi}^{0}_{1} \to \nu_3 \, b \, \bar{b}$ and $\tilde{\chi}^{0}_{1} \to \nu_3 \, \tau^+
\, \tau^-$ are different from those into the first and second
generation fermion pairs, because the Higgs boson exchanges and the
Yukawa terms play a very important role.  This can be seen in
Figs.~\ref{fig:14}, \ref{fig:15} , where we plot the
contour lines for these decays.  The branching ratio of $\tilde{\chi}^{0}_{1} \to
\nu_3 \, b \, \bar{b}$ can reach up to 97\%. The decay rate is large
because the scalar exchange contributions ($S^0_j,P^0_j,\tilde b_k$)
are large for $m_{1/2} \lesssim 200$~GeV. Note that this is also the case
for $\tan\beta =3$, because not only the neutrino-neutralino mixing
proportional to $m_{\nu_3}$ is important but also the
neutrino-higgsino mixing proportional to $\mu_3 / \mu_0$. The
decrease of the branching ratio with increasing $m_0$ is due to the
decrease of the higgsino component of $\tilde{\chi}^{0}_{1}$ and the increase of
the Higgs boson masses.  For $m_{1/2} \gtrsim 200$~GeV the decays into
real $W^+$ and $Z^0$ are possible, reducing the branching ratio of
$\tilde{\chi}^{0}_{1} \to \nu_3 \, b \, \bar{b}$.
As shown in Fig.~\ref{fig:15} the branching ratio for $\tilde{\chi}^{0}_{1} \to
\nu_3 \, \tau^+ \, \tau^-$ is very small for $\tan\beta = 3$ and $m_{1/2}
\lesssim 200$~GeV.  This is due to the destructive interference between
$Z^0$ contribution and the contributions of the exchanged charged
scalar particles (mainly due to the stau components of $S^\pm_k$).

Finally we have also considered the radiative LSP decay mode $\tilde{\chi}^{0}_{1}
\to \nu_3 \, \gamma$ \cite{Mukhopadhyaya:1999gy}.  In Fig.~\ref{fig:16} the
branching ratio for this mode is shown. This decay proceeds only at
one-loop level and therefore is in general suppressed compared to the
three-body decay modes.  However, for $m_{1/2} \lesssim 125$~GeV and
large $\tan\beta$ it exceeds 1\%, leading to interesting signatures like
$e^+ e^- \to \tilde{\chi}^{0}_{1} \tilde{\chi}^{0}_{1} \to \tau^\pm \, \mu^\mp \,\gamma$ +
$\not\!\!P_T$.  Due to initial state radiation it can easily happen that a
second photon is observed in the same event.

\subsubsection{Signals at $e^+\,e^-$ colliders}
\label{sec:exper-sign-supersymm}

The complete list of possible signatures stemming from LSP decays in
SSSM is shown in Tab.~\ref{tab:7}. In this table we also indicate
whether the same signatures could also arise in MSUGRA as a result of
$e^+ e^- \to \tilde{\chi}^{0}_{1} \tilde{\chi}^{0}_{2}$ followed by the
MSSM decay modes of $\tilde{\chi}^{0}_{2}$ if its production is
kinematically allowed.  The final states 4 jets + $\not\!\!P_T$
\cite{Diaz:1998zg}, $\tau$ + 2 jets + $\not\!\!P_T$, and $\tau$ + ($e$ or
$\mu$) + {$\not\!\!P_T$} would also occur in MSUGRA via the decay of
$\tilde{\chi}^{0}_{2}$ into $\tilde{\chi}^{\pm}_{1}$.  However one expects
in general that these decay modes are suppressed within MSUGRA.
In contrast in the SSSM these signatures can be rather large as can be
seen from Figs.~\ref{fig:12}, \ref{fig:13}.  Note moreover, that some
of the $L$ signatures are practically background free.  For
example, due to the Majorana nature of $\tilde{\chi}^{0}_{1}$, one can
have two same--sign $\tau$ leptons + 4 jets + {$\not\!\!P_T$}.  Other
interesting signals are: $\tau$ + 3 ($e$ and/or $\mu$) +
$\not\!\!P_T$, 3 $\tau$ + ($e$ or $\mu$) + $\not\!\!P_T$, $\tau$ +
($e$ or $\mu$) + 2 jets + $\not\!\!P_T$, $\tau$ + 4 jets +
$\not\!\!P_T$, $\tau^\pm \tau^\pm$ + ($e$ or $\mu$) + 2 jets +
$\not\!\!P_T$, or $\tau^\pm \tau^\pm$ + $l^\mp {l'}^\mp$ +
{$\not\!\!P_T$} with $l=e,\mu$.
In Table~\ref{tab:8} we give masses and branching ratios for
typical examples.

\begin{table}
\begin{center}
\begin{tabular}{|l|l|l|}
\hline
Combination of $\tilde{\chi}^{0}_{1}$ decay modes & signature & MSUGRA-like \\ \hline
$(3 \, \nu)$ $(3 \, \nu)$ & $\not\!\!P_T$  & yes \\ \hline
$(3 \, \nu)$ $(\nu_3 l^+ l^-)$ & 2 leptons + $\not\!\!P_T$  & yes \\ \hline
$(3 \, \nu)$ $(\nu_3 q \bar{q})$ & 2 jets + $\not\!\!P_T$  & yes \\
$(3 \, \nu)$ $(\nu_3 b \bar{b})$ &                   &     \\ \hline
$(3 \, \nu)$ $(\nu_l \tau^\pm l^\mp)$ with $l=e,\mu$ &
      $\tau$ + ($e$ or $\mu$) + $\not\!\!P_T$  & yes, but suppressed \\ \hline
$(3 \, \nu)$ $(\tau^\pm q \bar{q}')$ &
      $\tau$ + 2 jets + $\not\!\!P_T$  & yes, but suppressed \\ \hline
$(3 \, \nu)$ $(\nu_3 \, \gamma)$ & $\gamma$ + $\not\!\!P_T$  & yes \\ \hline
$(\nu_3 l^+ l^-)$ $(\nu_3 {l'}^+ {l'}^-)$ & 4 leptons + $\not\!\!P_T$  & no \\ \hline
$(\nu_3 l^+ l^-)$ $(\nu_3 q \bar{q})$ & 2 leptons + 2jets + $\not\!\!P_T$ 
                                                                 & no \\
$(\nu_3 l^+ l^-)$ $(\nu_3 b \bar{b})$ & & \\ \hline
$(\nu_3 l^+ l^-)$ $(\nu_l \tau^\pm l^\mp)$ with $l=e,\mu$ &
      $\tau$ + 3 ($e$ and/or $\mu$) + $\not\!\!P_T$  & no \\ \hline
$(\nu_3 \tau^+ \tau^-)$ $(\nu_l \tau^\pm l^\mp)$ with $l=e,\mu$ &
      3 $\tau$ +  ($e$ or $\mu$) + $\not\!\!P_T$  & no \\ \hline
$(\nu_3 l^+ l^-)$ $(\tau^\pm q \bar{q}')$ &
      $\tau$ + 2 leptons + 2 jets + $\not\!\!P_T$  & no \\ \hline
$(\nu_3 \tau^+ \tau^-)$ $(\tau^\pm q \bar{q}')$ &
      3 $\tau$  + 2 jets + $\not\!\!P_T$  & no \\ \hline
$(\nu_3 l^+ l^-)$ $(\nu_3 \gamma)$ & 2 leptons + $\gamma$ + $\not\!\!P_T$
                                                          & no \\ \hline
$(\nu_3 q \bar{q})$ $(\nu_3 q \bar{q})$ & & \\
$(\nu_3 q \bar{q})$ $(\nu_3 b \bar{b})$ & 4 jets + $\not\!\!P_T$ & yes, but
                                                          suppressed\\
$(\nu_3 b \bar{b})$ $(\nu_3 b \bar{b})$ & & \\ \hline
$(\nu_3 q \bar{q})$ $(\nu_l \tau^\pm l^\mp)$ with $l=e,\mu$ & 
          $\tau$ + ($e$ or $\mu$) + 2 jets + $\not\!\!P_T$ & no \\
$(\nu_3 b \bar{b})$ $(\nu_l \tau^\pm l^\mp)$ with $l=e,\mu$ & & \\ \hline
$(\nu_3 q \bar{q})$  $(\tau^\pm q \bar{q}')$ & $\tau$ + 4 jets + $\not\!\!P_T$
                                                     & no \\
$(\nu_3 b \bar{b})$  $(\tau^\pm q \bar{q}')$ & & \\ \hline
$(\nu_3 q \bar{q})$ $(\nu_3 \gamma)$ & 2 jets + $\gamma$ + $\not\!\!P_T$
                                                          & no \\
$(\nu_3 b \bar{b})$ $(\nu_3 \gamma)$ &  &  \\ \hline
$(\nu_l \tau^\pm l^\mp)$ $(\nu_l \tau^\pm {l'}^\mp)$ &
     $\tau^\pm \tau^\pm$ + $l^\mp {l'}^\mp$ + $\not\!\!P_T$ & no \\
   &  $\tau^\pm \tau^\mp$ + $l^\mp {l'}^\pm$ + $\not\!\!P_T$ & no \\ \hline
$(\nu_l \tau^\pm l^\mp)$ $(\tau^\pm q \bar{q}')$ &
     $\tau^\pm \tau^\pm$ + ($e$ or $\mu$) + 2 jets + $\not\!\!P_T$ & no \\
   &  $\tau^\pm \tau^\mp$ + ($e$ or $\mu$) + 2 jets + $\not\!\!P_T$ & no \\ \hline
$(\nu_l \tau^\pm l^\mp)$  $(\nu_3 \gamma)$ & 
     $\tau$ + ($e$ or $\mu$) + $\gamma$ + $\not\!\!P_T$ & no \\ \hline
$(\tau^\pm q \bar{q}')$ $(\tau^\pm q \bar{q}')$ &
     $\tau^\pm \tau^\pm$ + 4 jets + $\not\!\!P_T$ & no \\
   &  $\tau^\pm \tau^\mp$ + 4 jets + $\not\!\!P_T$ & no \\ \hline
$(\tau^\pm q \bar{q}')$  $(\nu_3 \gamma)$ & 
     $\tau$ + 2 jets + $\gamma$ + $\not\!\!P_T$ & no \\ \hline
$(\nu_3 \gamma)$ $(\nu_3 \gamma)$ & 2 $\gamma$ + $\not\!\!P_T$ & no \\ \hline
\end{tabular}
\end{center}
\caption{The signatures  expected from the process 
$e^+ e^- \to \tilde{\chi}^{0}_{1} \tilde{\chi}^{0}_{1}$ in the SSSM. }
\label{tab:7}
\end{table}
\begin{table}[h]
\begin{center}
\begin{tabular}{|c||c|c|c||c|c|c|} \hline
 & \multicolumn{3}{c||}{$\tan \beta = 3$}
 & \multicolumn{3}{c|}{$\tan \beta = 50$} \\ 
 & A & B & C & A & B & C \\ \hline
$m_{\tilde{\chi}^{0}_{1}}$ & 54.6 & 59.0 & 92.5 & 60.0 & 61.5 & 94.4 \\ 
$m_{S^0_1}$ & 91.0 & 96.8 & 102.9 & 107.2 & 111.1 & 116.4 \\
$m_{\tilde \nu}$ & 180.5 & 449.6 & 466.0 & 178.2 & 448.7 & 465.1 \\
$m_{\tilde e_R}$ & 170.5 & 445.7 & 450.6 & 171.6 & 446.1 & 451.0 \\
$m_{\tilde e_L}$ & 194.2 & 455.2 & 471.4 & 195.3 & 455.8 & 471.9 \\
$m_{\tilde q}$ & 398.1 & 572.8 & 705.4 &  398.1 & 572.8 & 705.4 \\ 
$m_{\tilde t_1}$ & 261.4 & 328.5 & 442.2 & 279.9 & 355.2 & 466.3 \\
$m_{\tilde b_1}$ & 361.3 & 479.1 & 612.1 & 243.0 & 343.0 & 470.1 \\ \hline
$BR(\tilde{\chi}^{0}_{1} \to 3 \nu$)               &  0.5 &  4.5 &  1.8 &  0.3 &  1.2 &
  1.9 \\
$BR(\tilde{\chi}^{0}_{1} \to l^- l^+ \nu_3)$       &  0.2 &  1.1 &  0.5 &  0.2 &  0.3 &
  0.6 \\
$BR(\tilde{\chi}^{0}_{1} \to q \bar{q} \nu_3)$     &  1.0 &  8.6 &  4.0 &  0.5 &  2.2 &
  4.4 \\
$BR(\tilde{\chi}^{0}_{1} \to l^\pm \tau^\mp \nu)$  &  0.6 &  5.6 & 18.0 &  0.5 &  1.8 &
 17.8 \\
$BR(\tilde{\chi}^{0}_{1} \to q \bar{q}' \tau^\pm)$ &  1.1 & 16.1 & 53.7 &  0.9 &  5.1 &
 53.2 \\
$BR(\tilde{\chi}^{0}_{1} \to b \bar{b} \nu_3)$     & 96.5 & 62.6 & 13.4 & 97.1 & 88.4 &
 13.3 \\
$BR(\tilde{\chi}^{0}_{1} \to \tau^- \tau^+ \nu_3)$ &  0.1 &  1.5 &  8.6 &  0.5 &  1.0 &
  8.8 \\
\hline
\end{tabular}
\end{center}
\caption[Neutralino masses and branching ratios]{Masses and branching 
ratios for the points:
 A $(m_{1/2},m_0)$ = (153,155),
 B  $(m_{1/2},m_0)$ = (153,440), and  C $(m_{1/2},m_0)$ = (251,440) for
both $\tan\beta =3$ and 50.
The masses are given in GeV and the branching ratios in \% and we only
give those larger than 0.1\%. Here the same
summations of the final states are performed as in the figures. 
$m_{\tilde q}$ is the averaged squark mass for the first two generations.}
\label{tab:8}
\end{table}

As it is well known, also in gauge mediated supersymmetry breaking
models (GMSB) \cite{Giudice:1999bp} the neutralino can decay inside
the detector, because the gravitino $\tilde G$ is the LSP. It is
therefore an interesting question if the SSSM can be confused with
GMSB. To answer this question let us have a look at the dominant decay
modes of the lightest neutralino in GMSB.
If the lightest neutralino is the NLSP, its main decay mode in GMSB is
\begin{eqnarray}
\tilde{\chi}^{0}_{1} \to \gamma \, \tilde G \, \, , \nonumber
\end{eqnarray}
where $\tilde G$ is the gravitino.  For the case where at least one of
the sleptons is lighter than the lightest neutralino the latter has
the following decay chain $ \tilde{\chi}^{0}_{1} \to {\tilde l}^\pm \,
l^\mp \to l^\pm \, l^\mp \, \tilde G \, \, , $.  In principle
three-body decay modes mediated by virtual photon, virtual Z-boson and
virtual sfermions also exist. However, in the neutralino mass range
considered here these decays are phase--space--suppressed
\cite{Giudice:1999bp,Bagger:1997bt}. This implies that the SSSM can
not be confused with GMSB, because (i) in GMSB the final states
containing quarks are strongly suppressed, and (ii) GMSB have lepton
flavor conservation, and therefore there are no final states like $e^+
e^+ \tau^- \tau^-$ + $\not\!\!P_T$. A further interesting question
would be how the neutralino phenomenology changes in a GMSB scenario
with $L$ violation. The main consequence would be an enhancement of
final states containing photons and/or leptons. A detailed study of
this question is, however, beyond the scope of the present work.

\subsection{Lightest Stop Two-Body Decays in SUGRA}
\label{sec:lightest-stop-two}
For definiteness and simplicity we assume only $L$ violation in the
third generation case. The discussion on ${\tilde t}_1 \to b \, l^+$ in
the three generation model will be given at the end of
Subsect.~\ref{sec:three-body-decays}

Note, in contrast, that in order to describe Flavor Changing Neutral
Current (FCNC) effects such as the $L$ conserving process $\tilde t_1
\to c\,\tilde\chi_1^0$ we need the three generations of quarks.

We scan the soft SUSY breaking parameter space in the range 
\begin{eqnarray}
  &m_0&\leq 700 \,\mathrm{GeV}\nonumber \\
  100 \:\mathrm{GeV}<&m_{1/2}&\leq 400 \,\mathrm{GeV}\nonumber \\ 
  \label{eq:108}
  &|A_0|&\leq 1000 \,\mathrm{GeV}, \\
  &m_{\nu_3}&<18\,\mathrm{MeV}\nonumber \\
  1.8<&\tan\beta&<60 \nonumber
\end{eqnarray}
the previous range on $\tan\beta$ guarantee that both $h_t$ and $h_b$
will be perturbative. For the CKM matrix, we use the Particle Data
Group convention~\cite{Groom:2000in}, taking $K_{us}=0.2205$,
$K_{cb}=0.041$, $|K_{ub}/K_{cb}|=0.08$ and neglecting CP violation,
i.e. $\delta=0$. Notice that here we scan over a much larger range for
$\mu_3$ than used in ref.~\cite{Bartl:1996gz}.

\begin{figure}[htp]
  \begin{center}
    \begin{picture}(232,203)
      \put(0,0){\includegraphics[height=8.0cm]{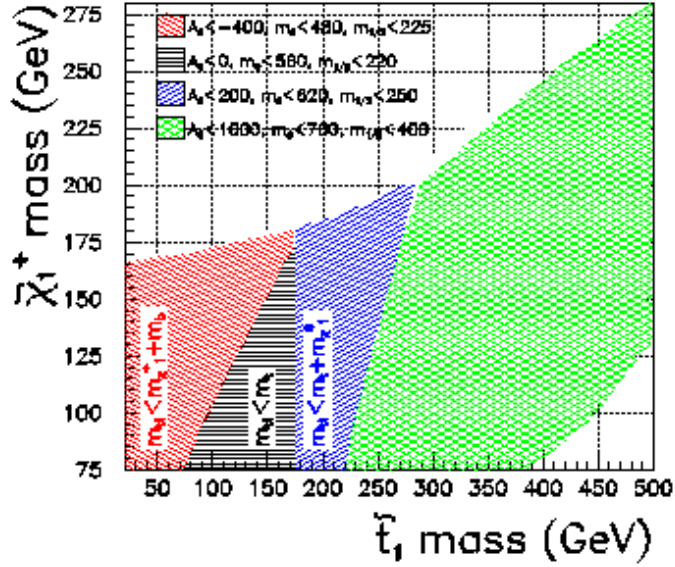}}
    \end{picture}
    \caption[Kinematical regions in the $m_{{\tilde
        t}_1}$--$m_{{\tilde\chi}_1^+}$ plane] {\small Kinematical
      regions in the $m_{{\tilde t}_1}$--$m_{{\tilde\chi}_1^+}$ plane.
      From left to right: Region I $m_{\tilde t_1} <
      m_{{\tilde\chi}_1^+}+m_b$; Region II
      $m_{{\tilde\chi}_1^+}+m_b<m_{{\tilde t}_1}<m_t$; Region III
      $m_t<m_{{\tilde t}_1}<m_{{\tilde\chi}_1^0}+m_t$; and region IV
      $m_{{\tilde t}_1}>m_{{\tilde\chi}_1^0}+m_t$ }
    \label{fig:17}
  \end{center}
\end{figure}

The resulting region of lightest stop and chargino masses is displayed
in Fig.~\ref{fig:17}. Neglecting the three-body decays, we find that in
Region I of the $m_{{\tilde t}_1}$--$m_{{\tilde\chi}_1^+}$ plane,
$BR(\tilde t_1 \to c\,\tilde\chi_1^0)+ BR(\tilde t_1 \to
b\,\tau)\approx 1$. In Region II $BR(\tilde t_1 \to b\,\tau)+BR(\tilde
t_1 \to b\,\tilde\chi_i^+)\approx 1$ (i=1,2).  In Region III
$BR(\tilde t_1 \to b\,\tau)+BR(\tilde t_1 \to
b+\tilde\chi_i^+)+BR(\tilde t_1 \to t\,\nu_3)\approx 1$ (i=1,2),
while in region IV $BR(\tilde t_1 \to b\,\tau)+BR(\tilde t_1 \to
b\,\tilde\chi_i^+)+BR(\tilde t_1 \to t\,\nu_3)+BR(\tilde t_1 \to
t\,\tilde\chi_j^0) \approx 1$ $(j=1,\ldots,4)$. Note that in each
region the exact equality to 1 is reached when the FCNC processes are
fully included.

In Sec.~\ref{sec:squark-decays} we give the two--body squark
decay-widths, for squarks of all three generations.  These equations
reduce to the expressions found in Ref.~\cite{Bartl97a} provided one
identifies $\Gamma_{UL}^{33}=\cos\theta_{\tilde t}$ and
$\Gamma_{UR}^{33}=\sin\theta_{\tilde t}$.  They also generalize the
results for the SSSM to the three-generation case.

\begin{figure}[htp]
  \center{\includegraphics[height=7cm]{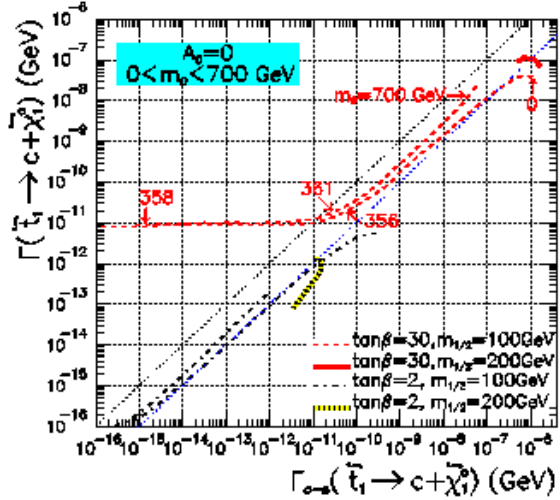}}
  \caption[Comparison between the exact numerical calculation  and
  the one--step approximation for the $\tilde t_1 \to
  c\,\tilde\chi_1^0$ decay width.]{\small Comparison between the exact
    numerical calculation (ordinate) and the one--step approximation
    (abscissa) for the $\tilde t_1 \to c\,\tilde\chi_1^0$ decay width
    for various values of $\tan\beta$ and $m_{1/2}$ with $A_0=0$ and
    $m_0$ varying in the indicated range. The dotted left diagonal
    line would signify the equality between the estimates, while the
    right diagonal line would indicate one order of magnitude
    difference. Results of both estimates indicated in the lower right
    legend. More details are found in the text.}
  \label{fig:18}
\end{figure}

In the MSSM the main $\tilde t_1$ decay channel expected in region I
of Fig.~\ref{fig:17} is the loop--induced and flavor--changing $\tilde
t_1 \to c\,\tilde\chi_1^0$ ~\cite{Hikasa:1987db,baer,Porod:1999yp,Porod:1997at,Porod:1998kk}.  As is
well-known, the FCNC processes in the MSSM in general involve a very
large number of input parameters.  For this reason, following common
practice, we prefer to perform the phenomenological study of flavor
changing processes in the framework of a supergravity theory with
universal supersymmetry breaking.  The simplest description of FCNC
processes in SUGRA models uses the
so-called one-step approximation.  Here we start by reproducing the
standard calculation for $\tilde t_1 \to c\,\tilde\chi_1^0$ as
in~\cite{Hikasa:1987db}. To do this consider only the effect of the
third generation Yukawa coupling. From our general eq.~(\ref{eq:103})
we have for $\tilde t_1= \tilde u_l$
\begin{equation}
  \Gamma(\tilde t_1 \to c\,\tilde\chi_1^0)
  \approx \frac{g^2}{8\pi} \left(\Gamma_{UL13}\right)^2
  \left[
    {\textstyle\frac23} \sin\theta_W
    N'_{11}+\left({\textstyle\frac12}-{\textstyle\frac23}
      \sin^2\theta_W\right)
    \frac{N'_{12}}{\cos\theta_W}
  \right]^2
  m_{\tilde t_1}
  \left(1-\frac{m_{\tilde \chi_1^0}^2}{m_{\tilde t_1}^2}
  \right)^2
  \label{eq:109}
\end{equation}
with
\begin{equation}
  \Gamma_{UL13} =
  \frac{\Delta_L\cos\theta_{\tilde t}-\Delta_R\sin\theta_{\tilde t}}
  {m_{\tilde c_L}^2-m_{\tilde t_L}^2}
  \label{eq:110}
\end{equation}
In the one--step approximation $\Delta_L, \Delta_R$ are given by
\begin{eqnarray}
  \label{eq:111}
  \Delta_L&=&(\tilde M_U^2)_{23}\approx (M_Q^2)_{23}\approx-
  \frac{t_U}{16\pi^2}K_{cb}K_{tb}h_b^2(M_Q^2+M_D^2+m_{L_0}^2+A_b^2)\\
  \label{eq:112}
  \Delta_R&=&(\tilde M_U^2)_{26}\approx (A_U^h)_{23}\approx-
  \frac{t_U}{16\pi^2}K_{cb}K_{tb}h_b^2m_t(A_b+\frac12A_t)
\end{eqnarray}
with $t_U=\ln(M_G/m_t)$. So, in the one--step approximation we have
\begin{equation}
  \label{eq:113}
  \Gamma(\tilde t_1 \to c\,\tilde\chi_1^0)\approx F h_b^4(
  \delta_{m_0^2}\cos\theta_{\tilde t}-\delta_A\sin\theta_{\tilde t})^2
  \left[
    \frac{\sqrt2}6(\tan\theta_W\,N_{11}+3N_{12})
  \right]^2
  m_{\tilde t_1} 
  \left(1-\frac{m_{\tilde \chi_1^0}^2}{m_{\tilde t_1}^2}
  \right)^2
\end{equation}
where the pre--factor $F = \frac{g^2}{16\pi}\left(
  \frac{t_U}{16\pi^2}K_{cb}K_{tb}\right)^2 \sim 6 \times 10^{-7}$ and
the parameter $\delta_{m_0^2}$ is given by
\begin{equation}
  \delta_{m_0^2}=\frac{M_Q^2+M_D^2+m_{L_0}^2+A_b^2}{m_{\tilde c_L}^2-
    m_{\tilde t_L}^2}\sim 1
  \label{eq:114}
\end{equation}
is basically independent of the initial conditions due to the $m_0$
dependence both in the numerator as in the denominator and
\begin{equation}
  \delta_A=\frac{m_t(A_b+\frac12A_t)}{m_{\tilde c_L}^2-m_{\tilde t_L}^2}
  \label{eq:115}
\end{equation}
Note however, that the one-step approximation includes only the third
generation Yukawa couplings and neglects the running of the soft
breaking
terms~\cite{Hikasa:1987db,baer,Porod:1999yp,Porod:1997at,Porod:1998kk,Boehm:1999tr}.
Such an approximation is rather poor for our purposes, since we will
be interested in comparing with $L$ violating decay modes (see section
\ref{sec:two-body-decays}). In order to have an accurate calculation of the respective
branching ratios we need to go beyond the one-step approximation.  We
therefore use a exact numerical calculation for the FCNC process
$\tilde t \to c\, \tilde\chi_1^0$ in which the running of the Yukawa
couplings and soft breaking terms is taken into account.  First we
have checked that indeed the effect of the Yukawas from the two first
generations is negligible. However the same is not true for the
running of the soft breaking terms.  As can be seen from
Figure~\ref{fig:18} the range of variation that we obtain from the
numerical solution is
\begin{equation}
  \label{eq:116}
  \Gamma(\tilde t_1 \to c\,\tilde\chi_1^0)\sim(10^{-16}\hbox{ --
    }10^{-6})\hbox{GeV}
\end{equation}
depending on the assumed value of $m_{1/2}$ and $\tan\beta$. In this
figure we have compared the decay width obtained from
eq.~(\ref{eq:103}) with the approximate formula in eq.~(\ref{eq:109})
for two fixed values of $m_{1/2}$, $\tan\beta$ and taking $A_0=0$. The
approximate formula only reproduce well the numerical result for the
academic case of no SUSY breaking gaugino mass, $m_{1/2}=0$. For the
realistic case $m_{1/2}>100\,$GeV, the exact solution is usually one
decade smaller than the approximate one. In the one-step approximation
$\Gamma(\tilde t_1 \to c\,\tilde\chi_1^0)$ can be arbitrarily small if
the two terms $\delta_{m_0^2} \cos\theta_{\tilde t}$ and
$\delta_A\sin\theta_{\tilde t}$ in eq.~(\ref{eq:113}) cancel.  This
behavior can be illustrated in Figure~\ref{fig:18} by the dashed line
labeled 358, which corresponds to $m_0 = 358$ GeV.  One sees clearly
that while the approximate solution goes to zero, the numerical one
reaches a minimum value around $10^{-11}$ GeV.  The wrong behavior of
the approximate solution indicates that the $\delta_A$ depends
strongly on the scale.  For example, the RGE for $A_b$ is very
sensitive on $m_{1/2}$ and $\tan\beta$ and in the one-step
approximation there is no explicit dependence on $m_{1/2}$, which is
crucial. Both solutions increase with $\tan\beta$, as expected by the
bottom Yukawa dependence explicit in eq.~(\ref{eq:113}) and remain
practically constant for large $m_{1/2}$ values.

\subsection{Two-Body Decays of the Lightest Stop: the SSSM  case}
\label{sec:two-body-decays}
In contrast to the case of an $L$ conserving supergravity theory
(MSUGRA), in the $L$ violating through Superrenormalizable terms 
(SSUGRA)\index{SSUGRA: Superrenormalizable SUGRA} case, one can have a
competing $L$ violating stop decay mode in region I of
Fig.~\ref{fig:17}.  From eq.~(\ref{eq:104}) with $\tau=F_3^+$ one can
easily compute the $L$ violating stop decay width $\tilde t_1 \to
b\,\tau$,
\begin{eqnarray}
  \Gamma(\tilde t_1 \to b\,\tau)&=&
  \frac{g^2\lambda^{1/2}(m_{\tilde t_1}^2,m_{b}^2,m_{\tau}^2)}{
    16\pi m_{\tilde t_1}^3}\{-4U_{32}^*\hat h_bc_{\theta_{\tilde t}}
  (V_{32}^*\hat h_ts_{\theta_{\tilde t}}-V_{31}^*
  c_{\theta_{\tilde t}})m_{b}m_{\tau}\nonumber\\
  &&+[(V_{32}^*\hat h_ts_{\theta_{\tilde t}}-
  V_{31}^*c_{\theta_{\tilde t}})^2+U_{32}^{*2}
  \hat h_b^2c^2_{\theta_{\tilde t}}](m_{\tilde t_1}^2-
  m_{b}^2-m_{\tau}^2)\}
  \label{eq:117}
\end{eqnarray}
which coincides with the result found in Ref.~\cite{Bartl:1996gz}.
In~\cite{Akeroyd:1998sv} it was shown that, except for $U_{32}$ which
determines the SU(2)-conserving mixing of the Higgsino with the
left-handed $\tau$, all other mixing matrix elements $V_{3i}$ and
$U_{3i}$ are proportional to $v_3^{II}$ (defined in eq.~(\ref{eq:18})
)and therefore to the $\nu_3$ mass.  Neglecting these terms we have
from eq.~(\ref{eq:117})
\begin{equation}
  \Gamma(\tilde t_1 \to b\,\tau)\approx
  \frac{g^2\lambda^{1/2}(m_{\tilde t_1}^2,m_{b}^2,m_{\tau}^2)}{
    16\pi m_{\tilde t_1}^3}\sin^2\xi'
  \hat h_b^2c^2_{\theta_{\tilde t}}(m_{\tilde t_1}^2-m_{b}^2-m_{\tau}^2)
  \label{eq:118}
\end{equation}
noting that, to a good approximation,
\begin{equation}
  |U_{32}|\approx\left|\frac{\mu_3}{\mu}\right|=|\sin\xi'|
  \label{eq:119}
\end{equation}
where $\mu_3$ corresponds to the superrenormalizable parameter in basis
I defined in section~\ref{sec:one-gener-supersymm}.  The lesson here
is that the $L$ violating decay rate $\Gamma(\tilde t_1 \to b\,\tau)$ is
proportional to $\mu_3$ or, equivalently, to $\sin^2\xi'$, instead of
$\sin^2\xi$, and thus not necessarily small, since it is \emph{not directly
controlled by the neutrino mass}. In other words, there can be
cancellations in the latter but not in the $L$ violating branching
ratio.

The meaning of the factor $\sin^2\xi' h_b^2$ may also be seen 
in basis II, where $\epsilon^{II}_3=0$.  In this case $v_3^{II}$
is proportional to the $\nu_3$ mass so that, as already
mentioned, in this basis all the elements $U_{3i}$ and $V_{3i}$ are
small~\cite{javi}.  Neglecting these terms, $\Gamma(\tilde t_1 \to
b\,\tau)$ may be written directly from the interaction term $\tilde
t_Lb_R\tau_L$, which is induced by the trilinear term in the
$\epsilon^{II}_3=0$--basis given in eq.~\eqref{eq:20} as
\begin{equation}
  \label{eq:120}
  \lambda_3^{II}=\left(\mu_3/\mu\right)h_b=h_b\sin\xi'
\end{equation}
which is the factor in eq.~(\ref{eq:118}). Note, however, that in our
numerical calculation to be described below we have used
for $\Gamma(\tilde t_1 \to b\,\tau)$ the full expression given in
eq.~(\ref{eq:103}).

Next section we will determine the conditions under which the
$L$ violating decay width $\Gamma(\tilde t_1 \to b\,\tau)$ can be dominant
over the $L$ conserving ones, $\Gamma(\tilde t_1 \to c\,\tilde\chi_1^0)$ and
$\Gamma(\tilde t_1 \to b\,\tilde\chi_1^+)$.

\subsubsection{Region I}
\label{sec:region-i}

\begin{figure}[htp]
  \begin{minipage}[t]{7.7cm}
    \begin{picture}(232,190) 
      \put(0,0){\includegraphics[height=6.0cm]{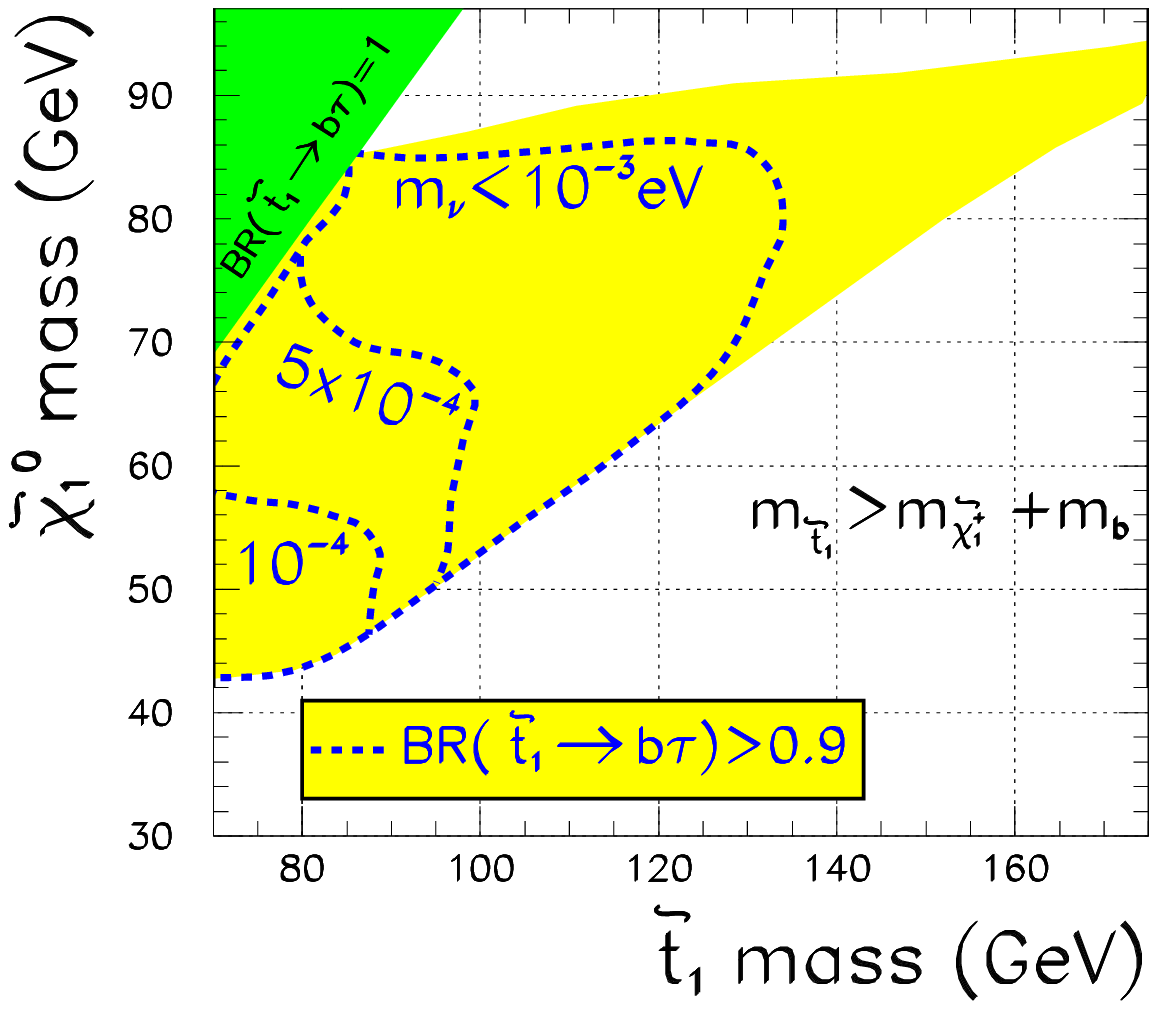}}
    \end{picture}
    \caption[Regions where the ${\tilde t}_1 \to b\,\tau$ decay branching ratio
    exceeds 90\%]{\footnotesize Regions where the ${\tilde t}_1 \to b\,\tau$
      decay branching ratio exceeds 90\% in the $m_{{\tilde
          t}_1}$--$m_{{\tilde\chi}_1^0}$ plane for different
      $m_{\nu_3}$ values.  The SSSM parameters are randomly varied
      as indicated in the text under the restriction $m_{\tilde
        t_1}<m_{\tilde\chi^{\pm}_1}+m_b$.  The upper--left triangular
      region corresponds to $m_{\tilde t_1}<m_{\tilde\chi^0_1}+m_c$ so
      that only the $\tilde t_1 \to b\,\tau$ decay channel is open.
      The lower--right
      unshaded 
      region corresponds to $m_{\tilde t_1}>m_{\tilde\chi^+_1}+m_b$.}
    \label{fig:19}
    \hfill
  \end{minipage}
  \parbox[t]{0.6cm}{}\hfill
  \begin{minipage}[t]{7.7cm}
    \hfill
    \begin{picture}(232,190) 
      \put(0,0){\includegraphics[height=6.0cm]{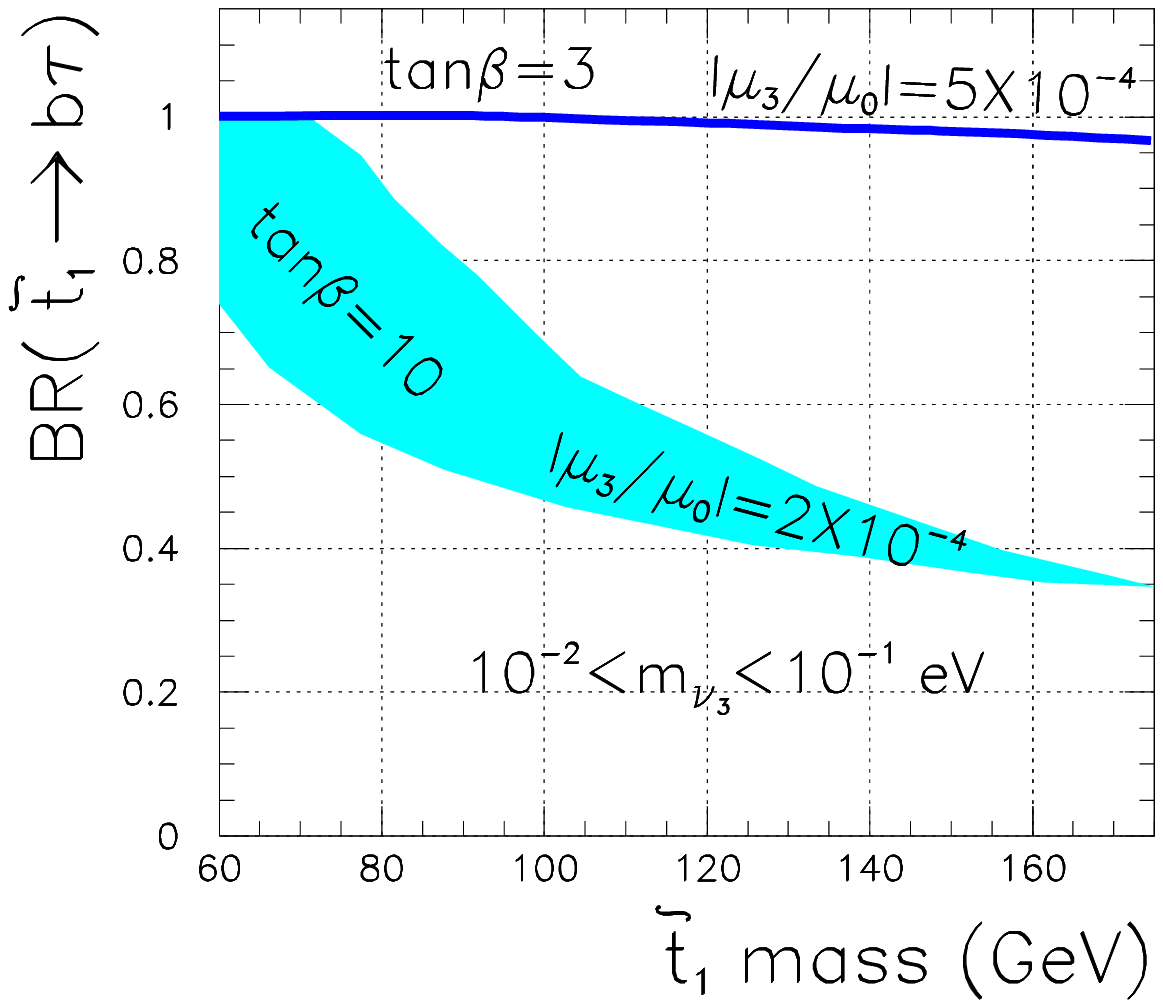}}
    \end{picture}
    \caption[$BR(\tilde t_1 \to b\,\tau)$ as function of the lighter stop
    mass]{\footnotesize $BR(\tilde t_1 \to b\,\tau)$ as function of the
      lighter stop mass for $\nu_3$ mass in the sub--eV range
      and two different values of $\tan\beta$ and $\mu_3/\mu_0$.  This
      prediction is natural in the sense that we have allowed only up
      to one order of magnitude of cancellation between the two terms
      that contribute to $\sin\xi$.}
    \label{fig:20}
  \end{minipage}
\end{figure}

Using the one-step approximation for $\Gamma(\tilde t_1 \to c\,\tilde\chi_1^0)$
one finds from eq.~(\ref{eq:109})
\begin{equation}
  \label{eq:121}
  \Gamma(\tilde t_1 \to c\,\tilde\chi_1^0)\sim 10^{-6}h_b^4m_{\tilde t_1} 
  \left(1-\frac{m_{\tilde \chi_1^0}^2}{m_{\tilde t_1}^2}\right)^2
\end{equation}
Using the eq.~(\ref{eq:118}) and neglecting charm, tau and bottom
masses we get
\begin{equation}
  \label{eq:122}
  \frac{\Gamma(\tilde t_1 \to c\,\tilde\chi_1^0)}{\Gamma(\tilde t_1 \to b\,\tau)}\sim
  10^{-5} \frac{h_b^2}{\sin^2\xi'}
  \left(1-\frac{m_{\tilde \chi_1^0}^2}{m_{\tilde t_1}^2}\right)^2
\end{equation}
Therefore $\Gamma(\tilde t_1 \to c\,\tilde\chi_1^0)$ will start to
compete with $\Gamma(\tilde t_1 \to b\,\tau)$ from $\sin\xi'\lesssim
5\times 10^{-3}$ ($10^{-4}$) for $\tan\beta$ large (small). In 
Fig.~\ref{fig:19} we compare $BR({\tilde t}_1 \to c\,{\tilde{\chi}_1^0})$
[calculated numerically from their exact formula in~(\ref{eq:103})]
with $BR({\tilde t}_1 \to b\,\tau)$ within the restricted region of
the $m_{{\tilde t}_1}$--$m_{{\tilde\chi}_1^0}$ plane where only those
two decay modes are open. We consider different $m_{\nu_3}$ values
(these correspond to relatively small values of the $L$ violating
parameters $|\mu_3|, |v_3| \lesssim 1$ GeV). We vary the SSSM
parameters randomly obeying the condition $m_{\tilde
  t_1}<m_{\tilde\chi^{\pm}_1}+m_b$ and depict the corresponding region
in light grey. The upper--left triangular region is defined by
kinematics and corresponds to $m_{\tilde t_1}<m_{\tilde\chi^0_1}+m_c$,
so that $BR(\tilde t_1 \to b\,\tau)=100 \: \%$. The lower--right grey
corresponds to $m_{\tilde t_1}>m_{\tilde\chi^0_1}+m_c$ when the
sampling is done over the region defined by eq.~\eqref{eq:108}.  One notices
from Fig.~\ref{fig:19} that in the central region the dominant stop
decay mode is $\tilde t_1 \to b\,\tau$ with branching ratio $BR(\tilde
t_1 \to b\, \tau)>0.9$.  The dotted lines in the light grey region
indicate maximum $\nu_3$ mass values obtained in the scan. In the
calculation of the $\nu_3$ mass, we have allowed only up to one order of
magnitude of cancellation between the two terms which contribute to
$\sin\xi$.  Therefore if the lightest stop only decays into the two
modes considered here, the processes $\tilde t_1 \to b\,\tau$, will be
important even for the case of very light $\nu_3$ masses.

We note however that we can use the limits obtained from leptoquark
searches~\cite{Abe:1997dn} in order to derive limits on the top-squark
in the SSSM. In particular, if $BR({\tilde t}_1\to b\tau)=1$ stop
masses less than 99 GeV are excluded at 95\% of CL., under the
assumption that the three--body decays of the stops are negligible.
Therefore, the dark region in Fig.~\ref{fig:19} would be ruled out.
In ref.~\cite{deCampos:1998kw} we have determined the corresponding
restrictions on the SUGRA parameter space.

The dependence on the $\nu_3$ mass may be seen in
Fig.~\ref{fig:20} where the role played by $\tan\beta$ is manifest.
In this figure we have shown $BR(\tilde t_1 \to b\,\tau)$ as function of
the lighter stop mass for $\nu_3$ mass in the sub--eV range,
indicated by the simplest oscillation interpretation of the
Super-Kamiokande atmospheric neutrino data.
We have obtained such $\nu_3$ mass values numerically, allowing
only one decade of cancellation between the two terms that contribute
to $\sin\xi$ in eq.~\eqref{eq:30}. The degree of suppression for
$\mu_3/\mu_0$ obtained numerically agrees very well with the
expectations from the approximate formula for the minimal
$\nu_3$ mass in eq.~\eqref{eq:33}. In contrast with
Ref.~\cite{Bartl:1996gz}, in our case $BR({\tilde t}_1\to b\tau)$ decreases
with $\tan\beta$. The reason for this difference is that here we take
into account the fact that the mixing parameter $\Gamma_{UL13}$
obtained from the RGE depends on $h_b^2$ in eq.~(\ref{eq:109}), while
in ref.~\cite{Bartl:1996gz} was simply regarded as a phenomenological input
parameter (called $\delta$ there).

The message from this subsection is that in our SSUGRA the $L$
violating decay mode $\tilde t_1 \to b\,\tau$ can very easily dominate
the $L$ conserving decay mode $\tilde t_1 \to c\,\tilde\chi_1^0$, even
for very small neutrino masses.

\subsubsection{Region II}
\label{sec:region-ii}
In region II the $L$ conserving decay mode $\tilde t_1 \to b
\tilde\chi^+_1$ is open (but not $\tilde t_1 \to t \nu$), and competes
with the $L$ violating mode $\tilde t_1 \to b\,\tau$.  Replacing
the subindex 3 by 1 on the diagonalization matrices $U$ and $V$ in
eq.~(\ref{eq:117}) we get the corresponding expression for
$\Gamma(\tilde t_1 \to b\,\tilde\chi_1^+)$. In order to get an
approximate expression for the ratio of the two main decay rates in
this region, we note that in SUGRA with universality at the
unification scale, the lightest chargino is usually gaugino-like,
implying that $V_{11}^2\sim 1$. In addition, the lightest stop is
usually right-handed, hence $\sin^2\theta_{\tilde
  t}\gtrsim\cos^2\theta_{\tilde t}$.  This way we find
\begin{equation}
  \frac{\Gamma(\tau)}{\Gamma(\tilde\chi_1^+)} \equiv
  \frac{\Gamma(\tilde t_1 \to b\,\tau)}{\Gamma(\tilde t_1 \to
    b\,\tilde\chi_1^+)}\approx
  \frac{\sin^2\xi'\hat h_b^2\cos^2{\theta_{\tilde t}}}
  {\left[(V_{11}^{*}\cos{\theta_{\tilde t}}-
      V_{12}^{*}\hat h_t\sin{\theta_{\tilde t}})^2+U_{12}^{*2}\hat h_b^2
      \cos^2{\theta_{\tilde t}}\right]}\,K
\label{eq:123}
\end{equation}
where $K$ is a kinematical factor depending on the lightest stop and
chargino masses, and here we have defined $\hat h_{t,b} \equiv
h_{t,b}/g$. The presence of the bottom quark Yukawa coupling indicates
that large values of $\tan\beta$ are necessary to have large $L$
violating branching ratios in this region. In fact, we have checked
numerically with the exact expressions that in Region II (RII)
${\Gamma(\tau)}/{\Gamma(\tilde\chi_1^+)}\gtrsim1$ only for large $\tan\beta$
as we will see in the next figures. 

In Fig.~\ref{fig:21} we show the regions in the
$m_{\chi^{\pm}_1}-m_{\tilde t_1}$ plane where $BR(\tilde t_1 \to
b\,\tau)$ dominates over $BR(\tilde t_1 \to b\widetilde\chi^+_1)$. In
the upper--left region the decay mode $\tilde t_1 \to
b\widetilde\chi^+_1$ is not allowed and corresponds to Region I. Below
and to the right of this zone, and above and to the left of three
rising lines, lies region RII where
${\Gamma(\tau)}/{\Gamma(\tilde\chi_1^+)}>1$. The three lines
correspond to $|\mu_3|<80$ GeV (dashed), $|\mu_3|<60$ GeV (dotted),
and $|\mu_3|<40$ GeV (dot--dashed), respectively. The proximity to the
upper-left zone indicates that the $L$ violating decay dominates only
close to the threshold where there is a high kinematical suppression
of the $L$ conserving one, through the factor $K$. Unlike the case of
region I this requires large values of the $L$ violating parameters.
Note, moreover, that if the stops have a small mixing
($\cos{\theta_{\tilde t}}\approx 0$), then
${\Gamma(\tau)}/{\Gamma(\tilde\chi_1^+)} \ll 1$ in RII.

\begin{figure}[htp]
  \center{\includegraphics[height=7cm]{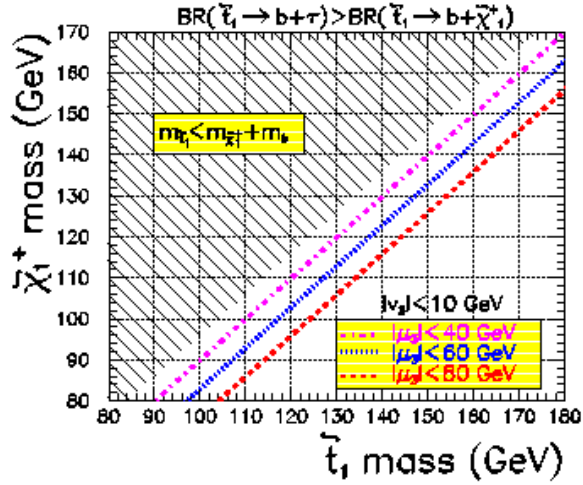}}
  \caption[Contours of $BR({\tilde t}_1 \to b\,\tau)>BR({\tilde t}_1 \to
  b\tilde\chi_1^+)$]{\small Contours of $BR({\tilde t}_1 \to
    b\,\tau)>BR({\tilde t}_1 \to b\tilde\chi_1^+)$ in the $m_{{\tilde
        t}_1}$--$m_{{\tilde\chi}_1^+}$ plane for $|v_3|<10$ GeV.
    Three different maximum values for $|\mu_3|$ are considered:
    $|\mu_3|<40$ GeV (dot-dash), $|\mu_3|<60$ GeV (dots), and
    $|\mu_3|<80$ GeV (dashes).  The region where $m_{\tilde
      t_1}<m_b+m_{\widetilde\chi^+_1}$ corresponds to the previously
    studied Region I.}
  \label{fig:21}
\end{figure}

A simpler expression for the ratio of decay rates in eq.~(\ref{eq:123})
is obtained if we take $V_{11}\approx1$ and assume no kinematical
suppression in eq.~(\ref{eq:123}) through the factor $K$:
\begin{equation}
\frac{\Gamma(\tau)}{\Gamma(\tilde\chi_1^+)}\sim
\sin^2\xi'\hat h_b^2\,.
\label{eq:124}
\end{equation}
Note that the presence of the parameter $\sin\xi'=\mu_3/\mu$
indicates that the $L$ violating decay mode is \emph{not}
strictly proportional to the neutrino mass, but proportional to the
SSSM parameter $\mu_3^2$. 

\begin{figure}[htp]
  \begin{minipage}[t]{7.7cm}
    \begin{picture}(232,190)
      \put(0,0){\includegraphics[height=7cm,width=7.5cm]{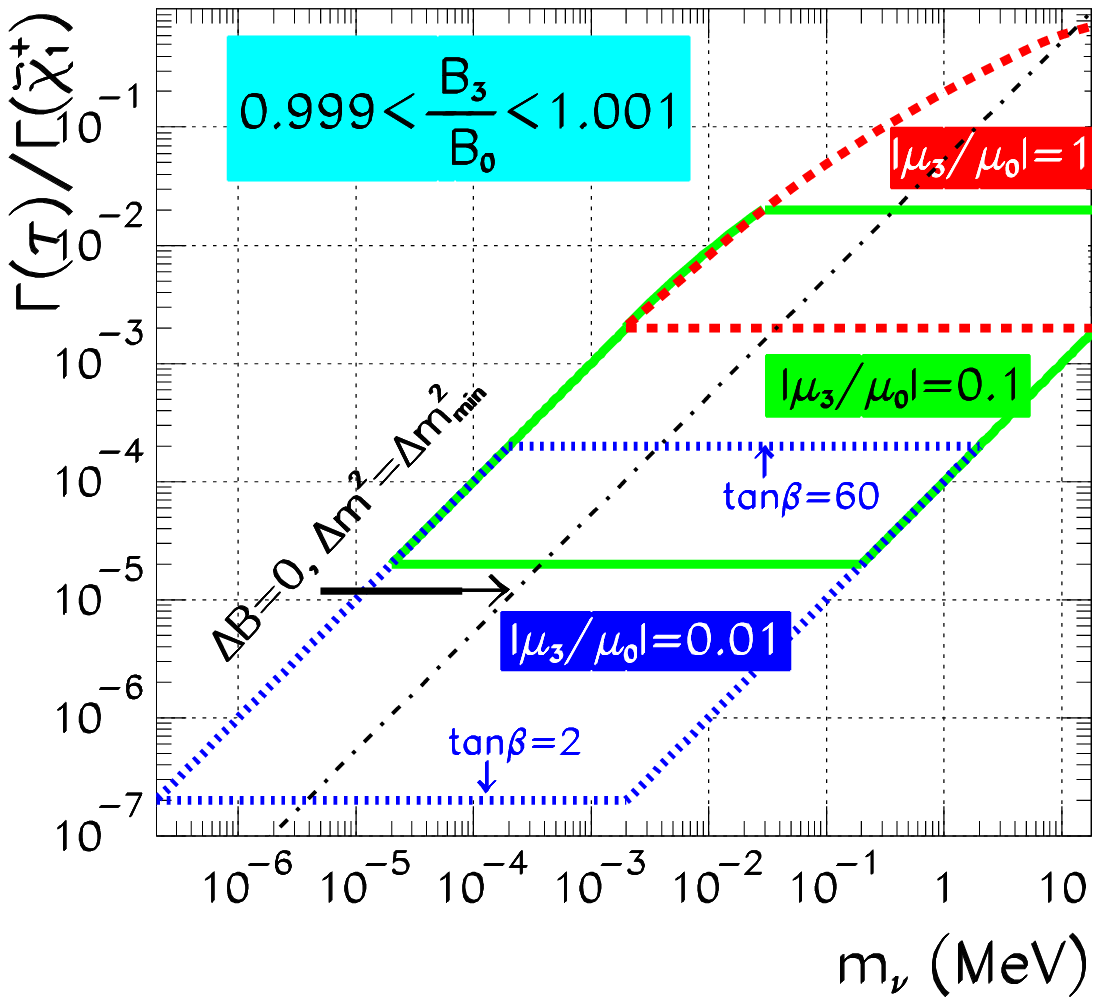}}
    \end{picture}
    \caption[Regions for $\Gamma({\tilde t}_1 \to b\,\tau)/\Gamma({\tilde
      t}_1 \to b\, {\tilde{\chi}_1^+})$ as a function of the
    $\nu_3$ mass]{\small Regions for $\Gamma({\tilde t}_1 \to
      b\,\tau)/\Gamma({\tilde t}_1 \to b\, {\tilde{\chi}_1^+})$ as a
      function of the $\nu_3$ mass with the universality
      condition $B_0=B_3$ at the unification scale imposed at the 0.1\%
      level as indicated.  Its effect is to alter the maximum
      attainable $\nu_3$ mass. The dot-dashed line corresponds
      to the case where $\Delta B=0$ at the weak scale.}
    \label{fig:22}
    \hfill
  \end{minipage}
  \parbox[t]{0.6cm}{}\hfill
  \begin{minipage}[t]{7.7cm}
    \begin{picture}(232,190)
      \put(0,0){\includegraphics[height=7.0cm,width=7.5cm]{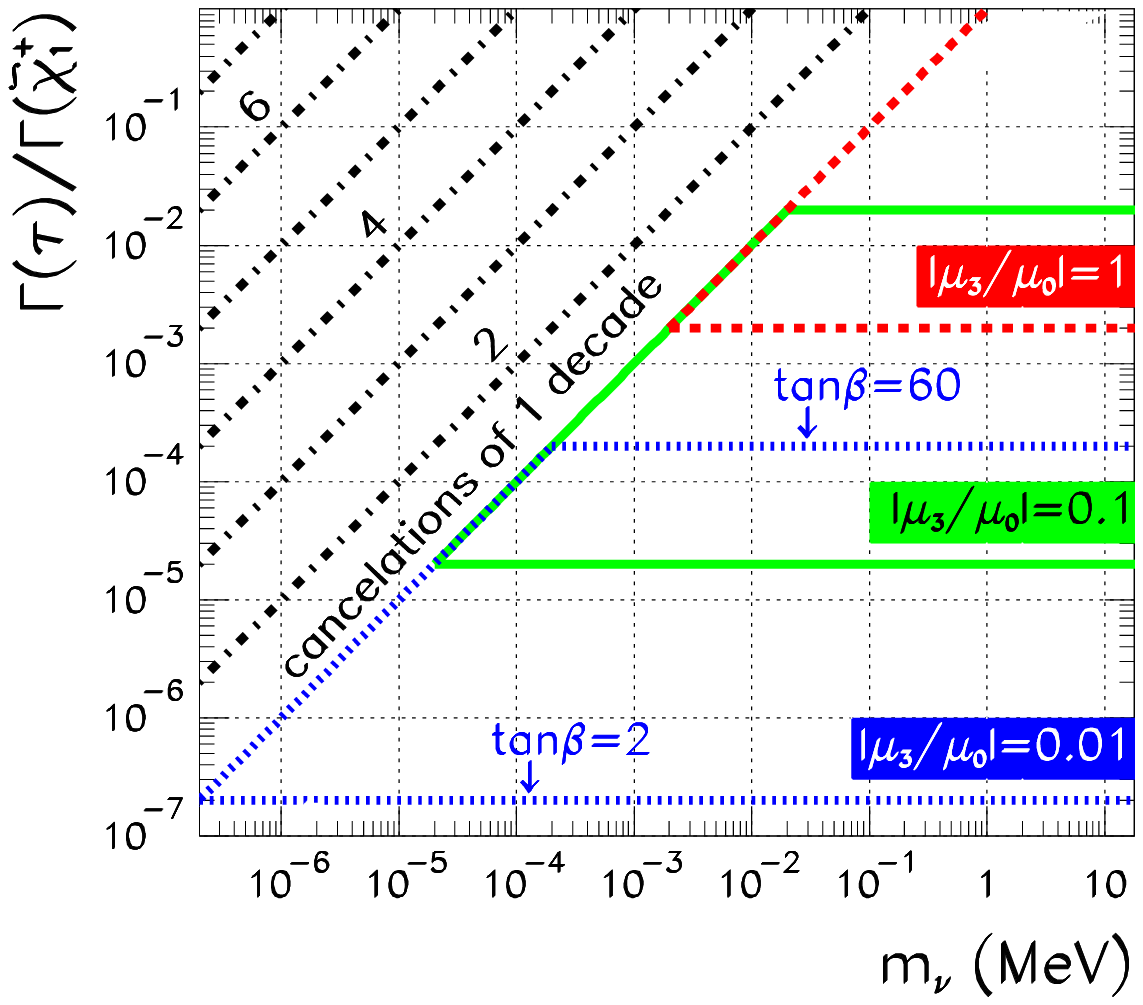}}
    \end{picture}
    \caption[Regions for $\Gamma({\tilde t}_1 \to b\,\tau)/\Gamma({\tilde
      t}_1 \to b\, {\tilde{\chi}_1^+})$ as a function of the
    $\nu_3$ mass]{\small Regions for $\Gamma({\tilde t}_1 \to
      b\,\tau)/\Gamma({\tilde t}_1 \to b\, {\tilde{\chi}_1^+})$ as a
      function of the $\nu_3$ mass for different levels of
      cancellation between the two terms that contribute to the
      neutrino mass. We impose the universality condition
      $m_{L_0}^2=M_L^2$ at the unification scale, but $B_0$ and $B_3$
      are not universal. We take $|\mu_3/\mu_0|=1$ (inside the dashed
      lines), $|\mu_3/\mu_0|=0.1$ (solid lines), and
      $|\mu_3/\mu_0|=0.01$ (dotted lines). }
    \label{fig:23}
  \end{minipage}
\end{figure}

However generically we expect some correlation with the $\nu_3$
mass, especially in the case where the boundary conditions in the RGE
are universal and there are no strong cancellations between two terms
that contribute to $\sin\xi$ as shown in Fig.~\ref{fig:22}.  In this
figure we plot the ratio $\Gamma(\tau)/\Gamma(\tilde\chi_1^+)$ in RII
as a function of the $\nu_3$ mass. Both decay rates have been
calculated numerically from the exact formulas. In this figure we have
imposed both $m_{L_0}^2=M_L^2$ and $B_0=B_3$ within 0.1\% at the GUT
scale.  Cancellation between the $\Delta m^2$ and $\Delta B$ terms in
the neutrino mass formula of eq.~\eqref{eq:30} are accepted only
within 1 decade. As a reference we have drawn the line corresponding
to $\Delta B=0$ and $\Delta m^2=\Delta m^2_{min}$ ($\Delta m^2$ is
negative and its magnitude is bounded from below by $\Delta
m^2_{min}$) at the weak scale, which gives an idea of the value of the
neutrino mass when there is no cancellation between the $\Delta B$ and
$\Delta m^2$ terms.

We have imposed an upper bound on $m_{\nu_3}$ at the collider 
experimental limit of the $\nu_3$ mass, and have chosen fixed
values 
of $\mu_3/\mu_0=1$, 0.1, and 0.01.  The allowed region for
$\mu_3/\mu_0=1$ is above the dashed line.  In the case of
$\mu_3/\mu_0=0.1$ (0.01) the allowed region lies enclosed between
the solid (dotted) lines. The effect of $\tan\beta$ is to increase the
ratio $\Gamma(\tau)/\Gamma(\tilde\chi_1^+)$: the minimum value of the
ratio is obtained for $\tan\beta\approx 2$ and the maximum corresponds
to $\tan\beta \approx60$.  The extreme values of $\tan\beta$ are
dictated by perturbativity.

%

A number of statistically less significant points appear outside the
drawn regions in Fig.~\ref{fig:22} and are not depicted. They
correspond to points with $m_{\tilde
  t_1}-m_b-m_{{\tilde\chi}_1^\pm}<10\,$ GeV which appear above the
diagonal line, and points with $\cos\theta_{\tilde t}<0.1$ which
appear below the horizontal line corresponding to the lowest values of
$\tan\beta$. In the last case, our approximation in eq.~(\ref{eq:124})
does not work any more. On the other hand, eq.~(\ref{eq:124}) predicts
very well the behavior of $\Gamma(\tau)/\Gamma(\tilde\chi_1^+)$ if
$\cos\theta_{\tilde t}>0.1$. For example for $\mu_3/\mu_0=1$, or
equivalently $\sin\xi'=1/\sqrt2$, we expect from eq.~(\ref{eq:124}) a
maximum value of order 1 for large $\tan\beta$ ($h_b\approx1$) and a
minimum value of order $10^{-3}$ for small $\tan\beta$ ($h_b^2\approx
10^{-3}$), and this is confirmed by Fig.~\ref{fig:22}. High values of
the $L$ violating branching ratio for large $\mu_3$ values
are highly restricted for large $\tan\beta$. This can be understood as
follows.  In the case of $\mu_3/\mu_0=1$ and $\tan\beta=60$
acceptable neutrino masses are obtained only if $\sin\xi\sim 1$. On
the other hand, in this regime we find from eq.~\eqref{eq:30} that the
$\Delta B$ term is large because of the high value of $\tan\beta$, and
that the $\Delta m^2$ term is large because $m_{L_0}^2$ becomes
negative and $\Delta m^2=m_{L_0}^2-M_L^2$ grows in magnitude. This
way, acceptable neutrino masses are achieved only with cancellation
within more than one decade. In any case, we think that
Fig.~\ref{fig:22} is very conservative considering that in SSSM--SUGRA
with unification of top-bottom-tau Yukawa couplings, the large value
of $\tan\beta$ implies that a cancellation of four decades among vev's
is needed.

The width of the band in Fig.~\ref{fig:22} reflects the degree of
correlation between the ratio $\Gamma({\tilde t}_1 \to b\,\tau)/\Gamma ({\tilde
  t}_1 \to b\, {\tilde{\chi}_1^+})$ and the neutrino mass under the
mentioned conditions. Note that one would have an indirect measurement
of the neutrino mass if this ratio were determined independently. The
band will open to the left if one allows a stronger cancellation
between the terms in $\Delta B$ and $\Delta m^2$. On the other hand it will
open to the right if the universality between $B_0$ and $B_3$ is
relaxed. This is shown in Fig.~\ref{fig:23} where we plot the ratio
$\Gamma(\tau)/\Gamma(\tilde\chi_1^+)$ in RII as a function of the $\nu_3$ mass, but
without imposing universality between $B_3$ and $B_0$. If we accept
cancellation within one decade between the $\Delta B$ and $\Delta m^2$ terms,
then the allowed region is at the right and below the corresponding
dashed tilted line.  If a larger degree of cancellation is accepted,
the left boundary of the allowed region moves to the left as indicated
in the figure, enhancing the $L$ violating channel.  In addition if we
accept only a decade of cancellation between the two terms that
contribute to the $\nu_3$ mass, then our approximate formula which
predicts the minimum $\nu_3$ mass in eq.~\eqref{eq:33} works very well.

\begin{figure}[htp]
  \center{\includegraphics[height=7cm]{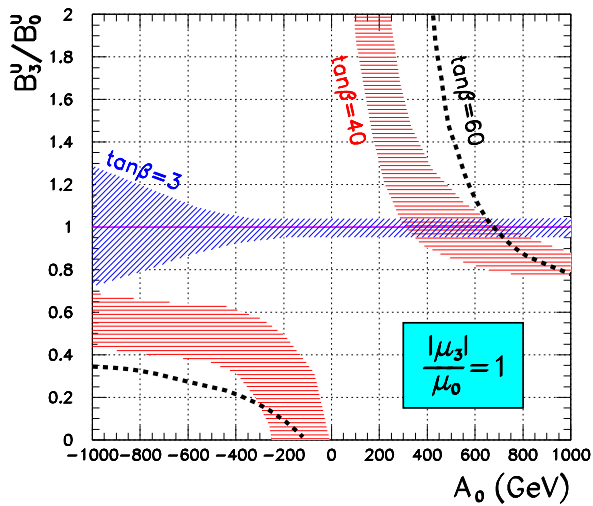}}
  \caption[Universality condition $B_0=B_3$ at the unification scale as a
  function of $A_0$.]{\small Universality condition $B_0=B_3$ at the
    unification scale as a function of $A_0$. As $\tan\beta$
    increases, the allowed values of $A_0$ are more constrained.}
  \label{fig:24}
\end{figure}

In summary, in this subsection we have shown that even in region II,
where the $L$ conserving decay mode ${\tilde t}_1 \to
b\,{\tilde{\chi}_1^+}$ is also open, the $L$ violating decay mode
${\tilde t}_1\to b\,\tau$ can be comparable to ${\tilde t}_1\to
b\,{\tilde{\chi}_1^+}$ for large $\tan\beta$ and $\mu_3$, and
relatively close to the chargino production threshold. In general,
this implies a large neutrino mass unless a cancellation is accepted
between the two terms contributing to the tree level neutrino mass. In
addition, the non-universality of the $B_0$ and $B_3$ terms at the GUT
scale does not increase appreciably the allowed parameter space,
except at large $\tan\beta$. The main consequence of this
non-universality is to restrict the allowed values of $A_0$ at large
$\tan\beta$.  In the next subsection we study the effects introduced
by the non-universality of $m_{L_0}^2$ and $m_{L_3}^2$.

\subsubsection{Effects of non--Universality}
\label{sec:effects-non-univ}
We now study the effect of possible non-universality of soft-breaking
SUSY parameters on our previous results. In particular, the
non-universality between $m_{L_0}^2$ and $m_{L_3}^2$ at the GUT scale.
The SUGRA model with universality at the unification scale, while
highly predictive, rests upon a number of simplifying assumptions
which do not necessarily hold in specific models due to the possible
evolution of the physical parameters in the range from $M_{Planck}$ to
$M_{\mathrm{GUT}}$. Specifically, there are several models in the literature
with non-universal soft SUSY breaking mass parameters at high scales.
A recent survey can be found in \cite{nusug}, where several models
such as based on string theory, M-theory, and anomaly mediated
supersymmetry are analyzed. For this reason we find interesting to
explore here the effects of non-universal soft terms.

The SUGRA spectra are typically found for given values of $m_{1/2}$,
$m_0$, $A_0$, $\tan\beta$ and Sgn($\mu_0$). In our case we have in
addition $\sin\xi'$ (or equivalently, $\mu_3$). The value of
$v_3$ is determined by the previous parameters through the
minimization conditions.  In addition, a relation between $A_0$ and
the ratio $B_3/B_0$ at the GUT scale (which indicates the degree of
universality) emerges. This relation can be seen in Fig.~\ref{fig:24}
for $\mu_3/\mu_0=1$ and the values $\tan\beta=3$, 40, and 60, for
$m_{L_0}^2=M_L^2$. The relation becomes more restrictive as
$\tan\beta$ is increased, starting from $-1000<A_0<1000$ GeV allowed
for $\tan\beta=3$, to a single $A_0$ value compatible with unification
for $\tan\beta=60$.
\begin{figure}[htp]
  \begin{minipage}[t]{7.7cm}
    \begin{picture}(232,220)
      \put(0,0){\includegraphics[height=8.0cm,width=8.7cm]{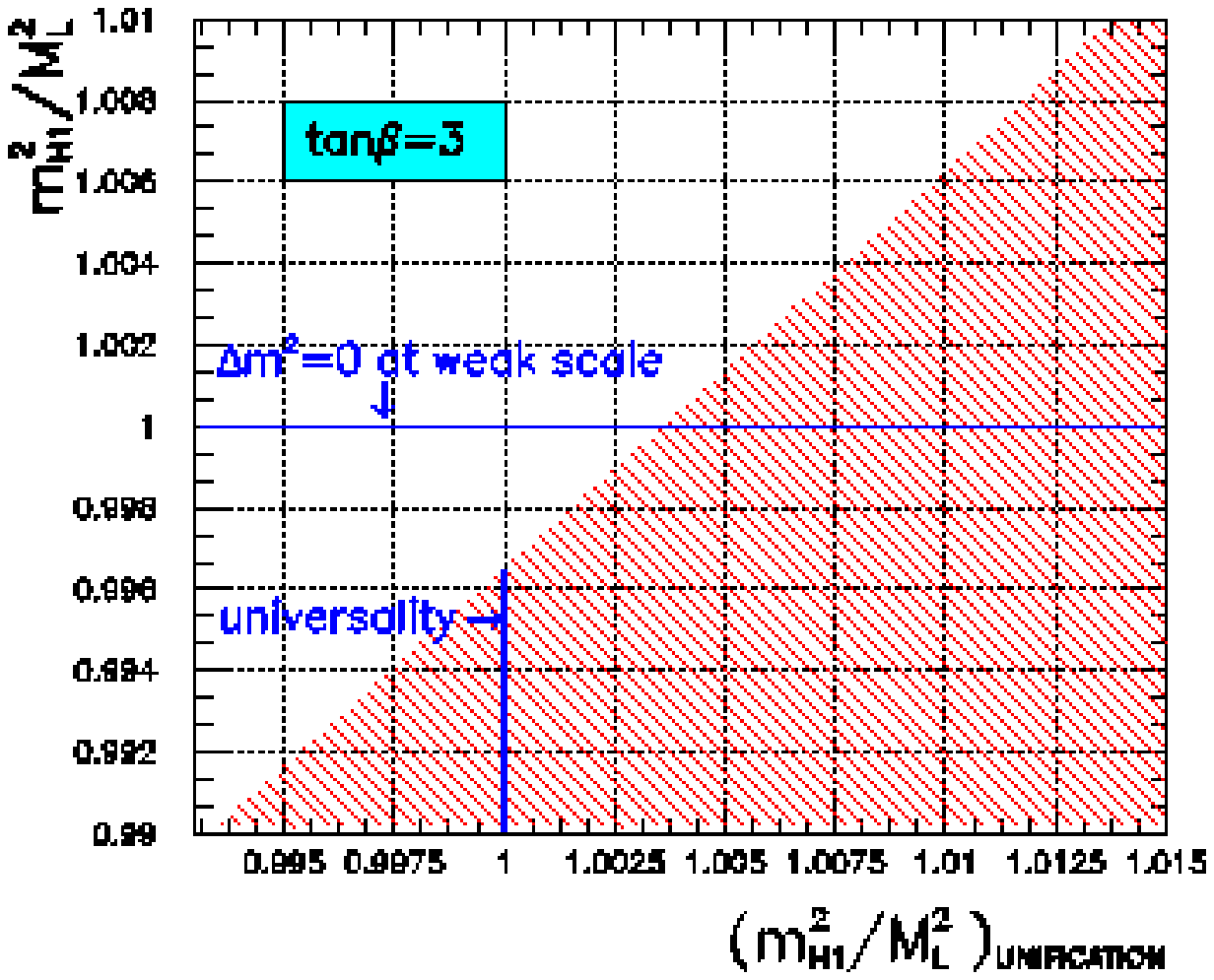}}
    \end{picture}
    \caption[Comparison between the ratio $m_{L_0}^2/M_L^2$ at the
    weak and the unification scales]{\small Comparison between the
      ratio $m_{L_0}^2/M_L^2$ at the weak and the unification scales
      for $\tan\beta=3$. Universality at the unification scale,
      $m_{L_0}^2/M_L^2=1$, implies a maximum value for this ratio at
      the weak scale.}
    \label{fig:25}
    \hfill
  \end{minipage}
  \parbox[t]{0.6cm}{}\hfill
  \begin{minipage}[t]{7.7cm}
    \begin{picture}(232,220)
      \put(0,0){\includegraphics[height=8.0cm,width=8.7cm]{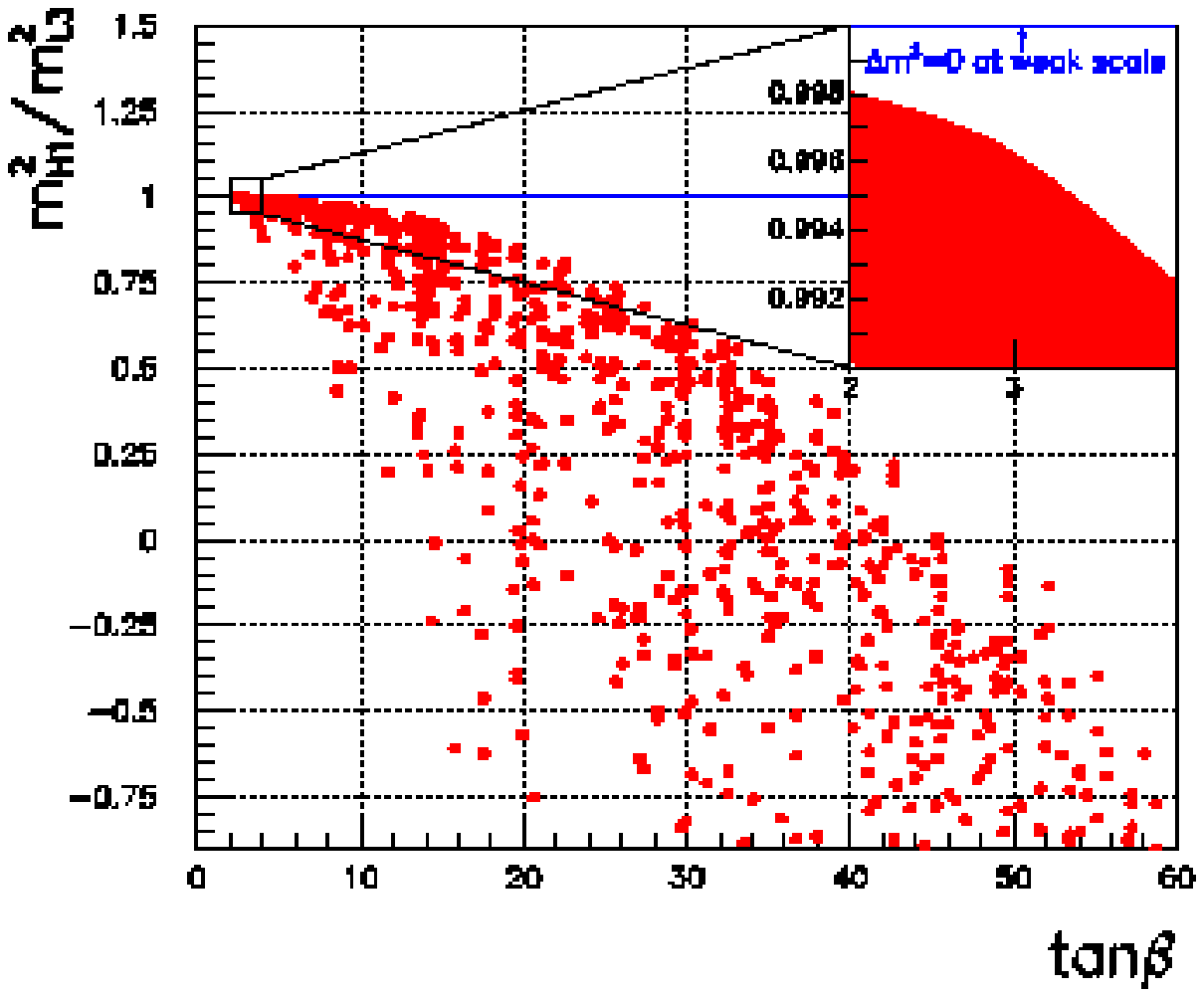}}
    \end{picture}
    \caption[$(m_{L_0}^2/M_L^2)$ evaluated at the weak scale as a function of
    $\tan\beta$.]{\small $(m_{L_0}^2/M_L^2)$ evaluated at the weak
      scale as a function of $\tan\beta$. This ratio is always less
      than one and decreases with $\tan\beta$.}
    \label{fig:26}
  \end{minipage}
\end{figure}
Another way to enhance the $L$ violating channel, enlarging the band
toward the left in Fig.~\ref{fig:22}, is by relaxing the universality
between $m_{L_0}^2$ and $M_L^2$ at the GUT scale. In Fig.~\ref{fig:25}
we plot the ratio $m_{L_0}^2/M_L^2$ at the weak scale as a function of
the same ratio at the unification scale $M_{\mathrm{GUT}}$ for $\tan\beta=3$.
The shaded region is allowed, implying a maximum value for the ratio
$m_{L_0}^2/M_L^2$ at the weak scale for a given value of the ratio at
the GUT scale. We see from Fig.~\ref{fig:25} that a relaxation of
universality of 0.5\% or more is enough to make
$(m_{L_0}^2/M_L^2)_{weak}=1$ possible, meaning that smaller neutrino
masses are attainable without having to rely on a cancellation between
the $\Delta m^2$ and $\Delta B$ terms or small values of
$\Gamma({\tilde t}_1\to b\tau)$.

However as we increase $\tan\beta$ the maximum value of
$m_{L_0}^2/M_L^2$ decreases, and thus, the required non-universality
between $m_{L_0}$ and $M_L$ at unification scale grows drastically. In
Fig.~\ref{fig:26} we show the ratio $m_{L_0}^2/M_L^2$ at the weak
scale as a function of $\tan\beta$. We appreciate clearly the growing
of $|\Delta m^2|_{min}$ with $\tan\beta$. We remind the reader that
this kind of non--universality in the soft terms is not uncommon in
string models~\cite{Brignole:1997dp}, or GUT models based on
$SU(5)$~\cite{Polonsky:1994sr} or $SO(10)$~\cite{so10} for example.  
There are in fact some $SO(10)$ models for non-universality of the GUT
scale scalar masses which naturally favor light neutrino mass
\cite{so10}.

\begin{figure}[htp]
  \centerline{\includegraphics[height=7cm]{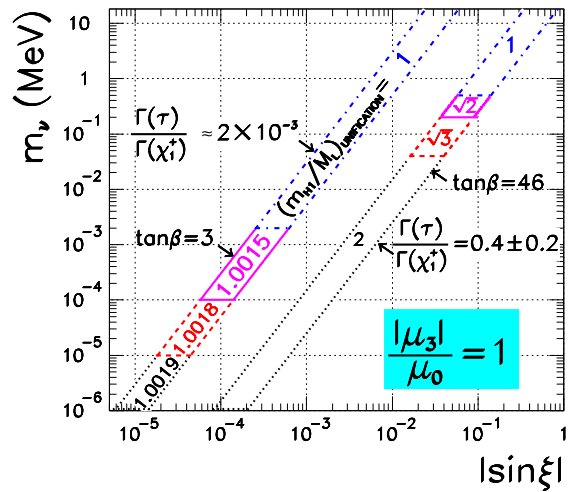}}
  \caption[Minimum value of the $\nu_3$ mass as a function of 
  $\sin\xi$]{\small Minimum value of the $\nu_3$ mass as a
    function of $\sin\xi$ for different values of $m_{L_0}/M_L$ at the
    GUT scale and two values of $\tan\beta$. The ratio $\mu_3/\mu_0$
    is fixed to the indicated value, leading to a nearly constant
    value for $\Gamma({\tilde t}_1 \to b\, \tau)/\Gamma({\tilde t}_1
    \to b\,{\tilde{\chi}_1^+})$. Here we assume that the two terms
    contributing to the $\nu_3$ mass cancel to within an order
    of magnitude.}
\label{fig:27}
\end{figure}
The effect of non--universality it is also explored in
Fig.~\ref{fig:27} where it is shown the relation between the neutrino
mass and the parameter $\sin\xi$ for $\mu_3/\mu_0=1$. Two different
bands are shown: one for $\tan\beta=3$ and
$\Gamma(\tau)/\Gamma(\tilde\chi_1^+)=2\times10^{-3}$, and a second one
for $\tan\beta=46$ and
$\Gamma(\tau)/\Gamma(\tilde\chi_1^+)=0.4\pm0.2$. The required degree
of universality at the GUT scale is indicated inside the bands.
For example, in order to have neutrino masses of the order of eV for
$\tan\beta=3$, $m_{L_0}^2$ needs to be at least 0.2\% larger than
$M_L^2$. Similarly, for $\tan\beta=46$ we need a $m_{L_0}^2$ twice as
large as $M_L^2$ at the GUT scale in order to have neutrino masses of
1 eV. We stress the fact that for Fig.~\ref{fig:27} we have
conservatively accepted cancellation at the level of one
order-of-magnitude only.

In summary, the lesson to learn here is that non-universal soft SUSY
breaking terms at the GUT scale have the potential of making it easier
to reconcile sizeable $L$ violating effects in the stop sector
with very small neutrino masses, without resorting to cancellations.

\section{Three Body Decays of stop}
\label{sec:three-body-decays}
We discuss the phenomenology of the lightest stop in the SSSM. In this
class of models we consider scenarios where the $L$ violating two-body
decay ${\tilde t}_1 \to \tau^+ \, b $ competes with the leading
three-body decays such as ${\tilde t}_1 \to W^+ \, b \, {\tilde
  \chi}^0_1$.  We demonstrate that the $L$ violating decay can be
sizable and in some parts of the parameter space even the dominant
one.  Moreover we discuss the expectations for ${\tilde t}_1 \to \mu^+
\, b $ and ${\tilde t}_1 \to e^+ \, b $. The recent results from solar
and atmospheric neutrinos suggest that these are as important as the
$\tau^+ b$ mode.  The ${\tilde t}_1 \to l^+ \, b$ decays are of
particular interest for hadron colliders, as they may allow a full
mass reconstruction of the lighter stop.  Moreover these decay modes
allow cross checks on the neutrino mixing angle involved in the solar
neutrino puzzle complementary to those possible using neutralino
decays \cite{Porod:2000hv}. For the so--called small mixing angle or
SMA solution ${\tilde t}_1 \to e^+ \, b$ should be negligible, while
for the large mixing angle type solutions \emph{all} ${\tilde t}_1 \to
l^+ \, b$ decays should have comparable magnitude. We first explore
the extent to which the decay $\tilde t_1\to b\,\tau$ can be sizeable
when compared with the 3--body decay modes.  Moreover, we discuss the
connections between the decay modes ${\tilde t}_1 \to b \, l^+$ and
neutrino physics, in particular we discuss a possible test of the
solution to the solar neutrino puzzle. The appendixes
in~\cite{Restrepo:2001me} contain complete formulas for the total
widths of the three-body decay modes as well as for the couplings.

Notice that in SSUGRA models~\cite{Diaz:1999ge} $m_{\nu_3}$ is
calculable through the RGE evolution and one finds in this case
cancellations up to two orders of magnitude for the combination
$\Lambda_3=-\mu_3 v_0 +\mu_0 v_3$. In general the smallness of $m_{\nu_3}$
requires relatively small $\mu_3$ as in ref.~\cite{Mira:2000gg}.
The remaining two neutrinos acquire mass radiatively.
Rigorous quantitative results were given in the second paper in
ref.~\cite{Romao:2000up}. Typically they are hierarchically lighter
than the heaviest neutrino, whose mass arises at the tree-level.  This
way one accounts for the observed hierarchy between the solar and the
atmospheric neutrino mass scales.

In this section we present our numerical results for the branching
ratios of the lighter stop $\tilde t_1$. Here we consider scenarios
where all two-body decays induced at tree-level are kinematically
forbidden except the $b \, l^+$ decays.  Before going into detail it
is useful to have some approximate formulas at
hand~\cite{Diaz:1999ge}:
\begin{align}
  \Gamma(\tilde t_1\to b\,\tau) \approx\,&
  \frac{g^2 |U_{32}|^2 h_b^2 \cos^2_{\theta_{\tilde t}} m_{\tilde t_1}}{16\pi} 
  \approx
  \frac{g^2 |\mu_3|^2 h_b^2 \cos^2_{\theta_{\tilde t}} m_{\tilde t_1}}{16\pi |\mu_0|^2} 
  \label{eq:125} \\
  \Gamma(\tilde t_1 \to c\tilde\chi_1^0)\approx\,& F h_b^4(
  \delta_{m_0^2}\cos\theta_{\tilde t}-\delta_A\sin\theta_{\tilde t})^2
  f_L^2 
   m_{\tilde t_1} 
  \left(
    1-\frac{m_{\tilde \chi_1^0}^2}{m_{\tilde t_1}^2}
  \right)^2\,
\label{eq:126} ,
\end{align}
where $F = \frac{g^2}{16\pi}\left(
 \log(m_{\mathrm{GUT}}/m_Z) K_{cb}K_{tb}/16 \pi^2 \right)^2
 \sim 6 \times 10^{-7}$,
$f_L = \sqrt2(\tan\theta_W\,N_{11}+3N_{12}) / 6$ and
the parameter $\delta_{m_0^2}$ is given by
\begin{equation*}
\delta_{m_0^2}=\frac{M_Q^2+M_D^2+m_{L_0}^2+A_b^2}{m_{\tilde c_L}^2-m_{\tilde
t_L}^2}
\end{equation*}
For the minimal SUGRA models one finds $\delta_{m_0^2}={\cal O}(1)$ which is
basically independent of the initial conditions due to the $m_0$
dependence both in the numerator and in the denominator. Finally we
have
\begin{equation*}
\delta_A=\frac{m_t(A_b+\frac12A_t)}{m_{\tilde c_L}^2-m_{\tilde t_L}^2}
\end{equation*}
The complete formulas are given in \cite{Diaz:1999ge,Bartl:1996gz},
while for the three--body decays they are given in
Section~\ref{sec:three-body-decays-1} . They reduce to the ones given
in~\cite{Porod:1999yp,Porod:1997at} for vanishing $L$ violating
parameters.

We have fixed the parameters as in \cite{Porod:1999yp} to avoid color
breaking minima, while in the top squark sector we have used
$m_{\tilde t_1}$, $\cos \theta_{\tilde t}$, $\tan \beta$, and $\mu_0$ as
input parameters.  For the sbottom sector we have fixed $M_{\tilde Q},
M_{\tilde D}$ and $A_b$ as input parameters whereas for the charged
scalars we took $m_{P^0_2}$, $M_{\widetilde E}, M_{\tilde L}$, and
$A_\tau$ as input~\footnote{ $m_{P^0_2}$ plays here the same role as
  $m_{A^0}$ in the MSSM case \cite{Porod:1999yp}.}.  In addition we
have chosen the $L$ violating parameters $\mu_3$ and $v_3$
in such a way that the heaviest neutrino mass is fixed with the help
of eq.~\eqref{eq:28}.
For simplicity, we have also assumed that the soft SUSY breaking
parameters are equal for all generations.

In order to get a feeling for minimal size of branching ratios that
can be measured let us first shortly discuss the expected size for the
direct production of light stops at future colliders.
One expects for example at the LHC a production cross section of $\sim
35$~pb for 220 GeV stop mass.  Therefore, once the full luminosity has
been reached, one has to expect approximately 3.5 $10^6$ events per
year.
The corresponding stop production cross section at a future $e^+ e^-$
linear collider of $800$ c.m.s.~energy is of ${\cal O}(10-100 \mathrm{fb})$
\cite{Bartl:1997wt}.  For an integrated luminosity of 500 fb$^{-1}$
per year one can expect ${\cal O}(10^4)$ stop pairs per year.  This implies
that branching ratio as low as $10^{-3}$ can in principle be measured.

We consider first the simplest case of one generation model which, as
already mentioned, is sufficient to describe the relative importance
of the ${\tilde t}_1 \to \tau^+ \, b $ decay mode relative to the
possible 3-body decay modes
\begin{eqnarray}
{\tilde t}_1 &\to& W^+ \, b \, {\tilde \chi}^0_1 \nonumber\\
{\tilde t}_1 &\to& S^+_k \, b \, {\tilde \chi}^0_1 \nonumber\\
{\tilde t}_1 &\to& S^+_k \, b \, \nu_3 \nonumber\\
\label{eq:127}
{\tilde t}_1 &\to& b \, S^0_i \, \tau^+ \,,  \\
{\tilde t}_1 &\to& b \, P^0_j \, \tau^+ \, \nonumber\\
{\tilde t}_1 &\to& b \, {\tilde l}^+_i \, \nu_l \,, \nonumber\\
{\tilde t}_1 &\to& b \, {\tilde \nu}_l \, l^+ \hspace{1cm} (l=e,\mu) \, .
\nonumber
\end{eqnarray}
In general the important final states are those that conserve $L$. For
example, for the case of decays involving $S^+_k$ the most important
are those in which the scalars will be mainly a stau.  Due to the fact
that the existing bounds on the MSSM sneutrinos are below 100 GeV
there exists the possibility that the sneutrino has nearly the same
mass as one of the Higgs boson.  Similarly it could be that the
charged MSSM boson has nearly the same mass as one of the staus. This
implies large mixing effects even for small $L$ violating parameters
\cite{deCampos:1995av,Akeroyd:1998iq,Porod:2000pw}.  We therefore have
used the complete formulas for the 3-body decay modes which are
presented in the Appendix of \cite{Restrepo:2001me}. The latter
include $L$ violating decays such as $\tilde t_1 \to W^+ b \nu_3$. In
addition to the above mentioned decays there is also ${\tilde t}_1 \to
b \, Z^0 \, \tau^+$.  This decay mode is kinematically suppressed
compared to ${\tilde t}_1 \to b \, \tau^+$ and there is no possible
enhancement due to a mixing with an $L$ conserving final state.
Therefore it can be safely neglected.

\begin{figure}[htp]
 \setlength{\unitlength}{1mm}
 \begin{picture}(142,180)
 \put(5,175){\mbox{{\bf a)}}}
 \put(0,90){\includegraphics[height=8.5cm,width=8.5cm]{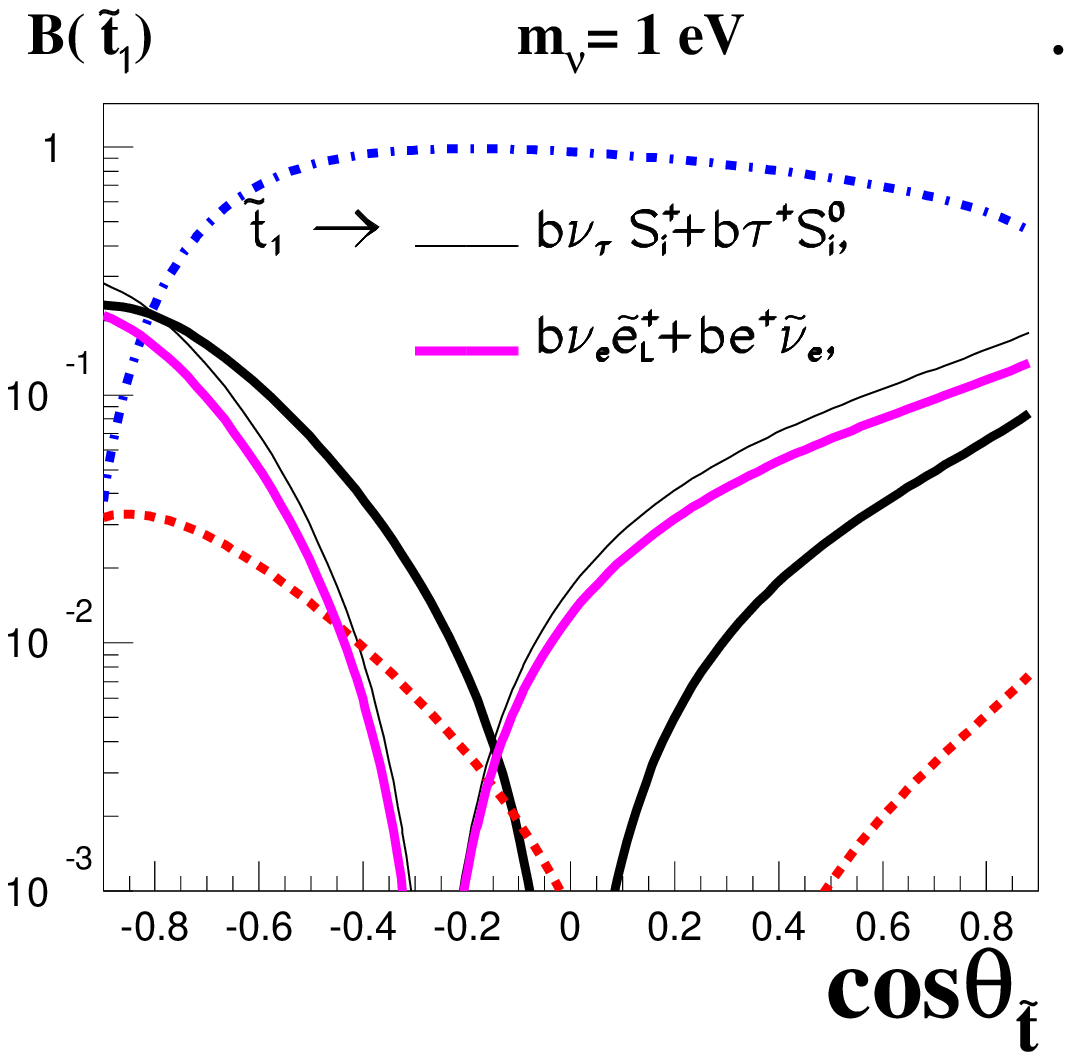}}
 \put(75,175){\mbox{{\bf b)}}}
 \put(70,90){\includegraphics[height=8.5cm,width=8.5cm]{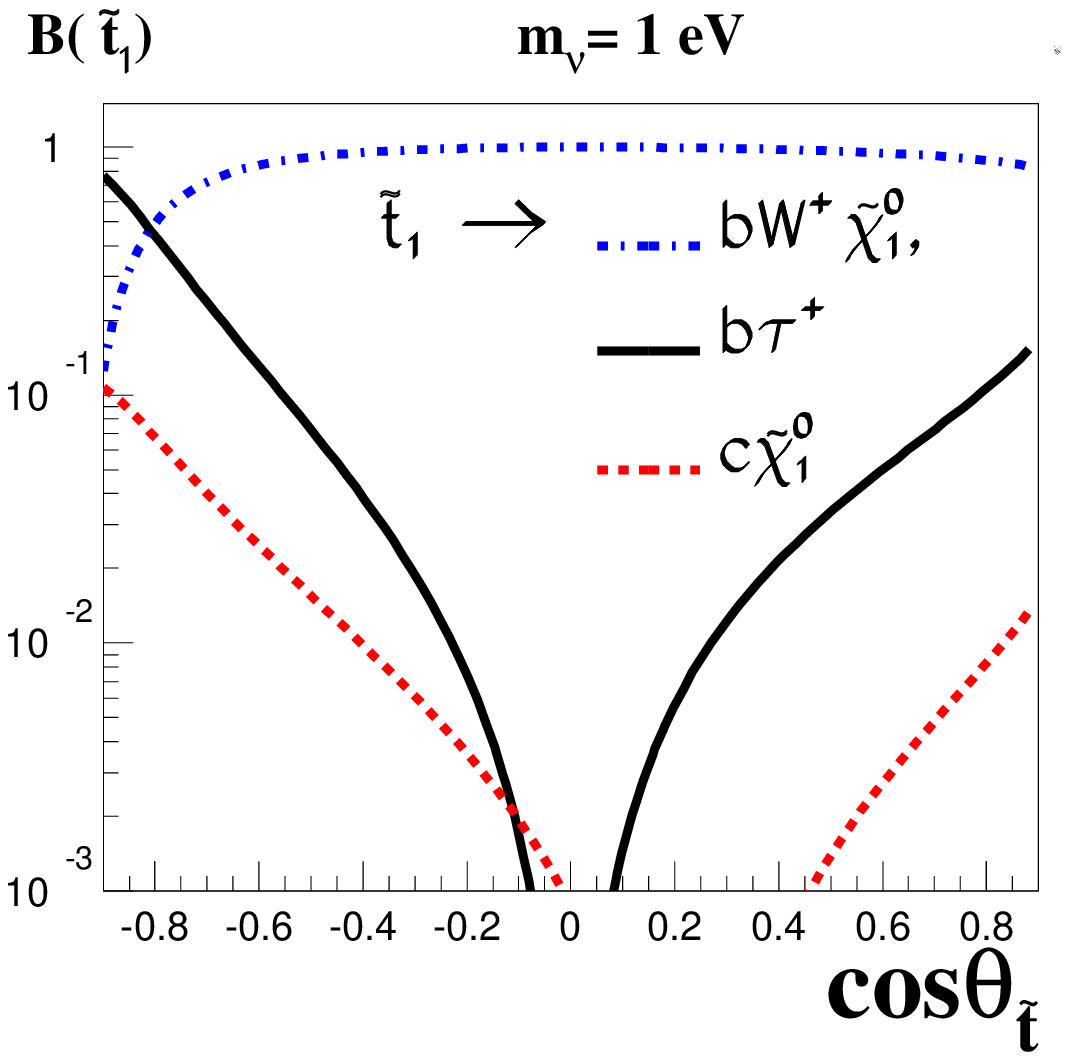}}
 \put(5,85){\mbox{{\bf c)}}}
 \put(0,0){\includegraphics[height=8.5cm,width=8.5cm]{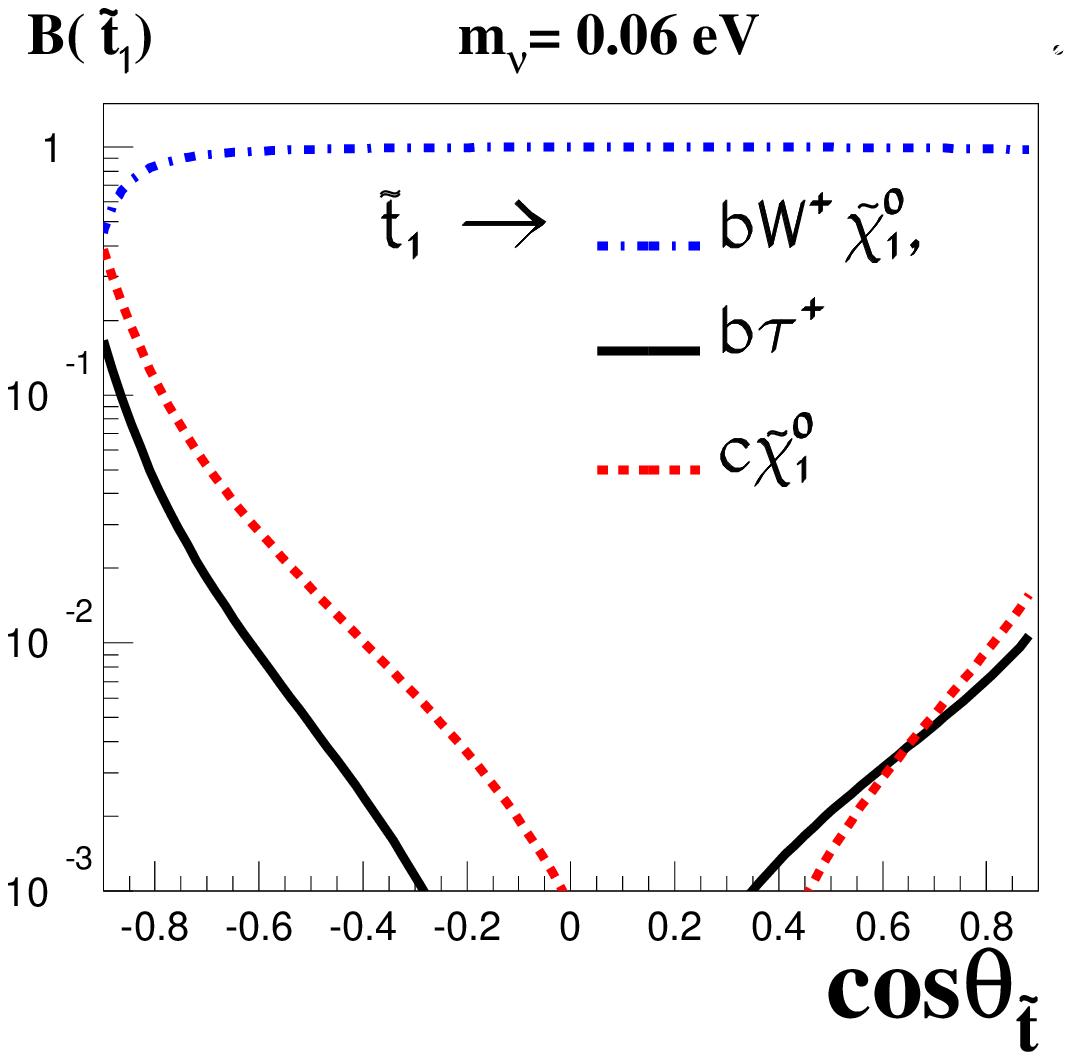}}
 \put(75,85){\mbox{{\bf d)}}}
 \put(70,0){\includegraphics[height=8.5cm,width=8.5cm]{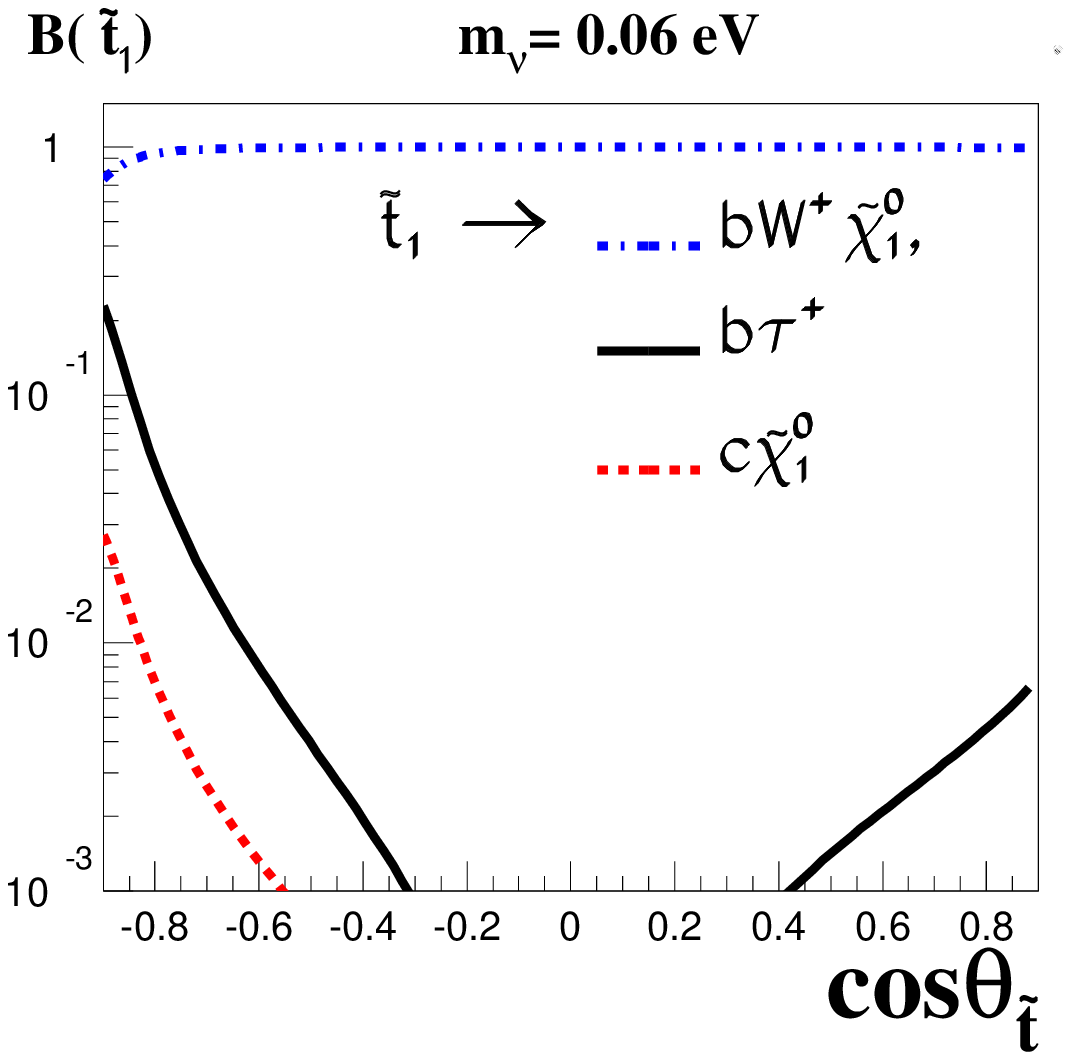}}
\end{picture}
    \caption[Branching ratios for the $\tilde t_1$ as a function of 
    $\cos \theta_{\tilde t}$]{Branching ratios for the $\tilde t_1$ as
      a function of $\cos \theta_{\tilde t}$ for different scenarios.
      We have fixed in a) $m_{\nu_3}=1\,$eV, b) $m_{\nu_3}=1\,$eV,
      $M_{\tilde E}>225\,$GeV, $M_{\tilde L}>225\,$GeV, c)
      $m_{\nu_3}=0.06\,$eV, $M_{\tilde E}>225\,$GeV, $M_{\tilde
        L}>225\,$GeV, d) Branching ratios for the $\tilde t_1$ as a
      function of $\cos \theta_{\tilde t}$ for $\tan \beta = 3$.
      $m_{\nu_3}=0.06\,$eV, $M_{\tilde E}>225\,$GeV, $M_{\tilde
        L}>225\,$GeV.  All the other inputs are given in
      Table~\protect{\ref{tab:9}}.  }
\label{fig:28}
\end{figure}

In Fig.~\ref{fig:28} we show the branching ratios for the $\tilde t_1$
as a function of $\cos \theta_{\tilde t}$ in different scenarios.  In
order to calculate the partial width for the decay $\tilde t_1\to
c\tilde\chi_1^0$ we have taken the formula given in ref.
\cite{Hikasa:1987db}. According to the analysis performed
in~\cite{Diaz:1999ge}, where the full calculation was done in the
SUGRA scenario with universality at the unification scale, the result
obtained with the present approximation should be taken as one upper
bound. This implies also that the shown branching ratio for $\tilde
t_1 \to b \tau^+$ can be viewed as a lower bound.
The parameters and physical quantities used in Fig.~\ref{fig:28} are
given in Tab.~\ref{tab:9}. For the case of Fig.~\ref{fig:28}(a) we have
fixed in addition the $L$ violating parameters such that $m_{\nu}
\approx 1$~eV. With this choice of parameters $S^0_1$, $S^0_3$,
$P^0_3$, and $S^-_4$ are mainly the MSSM Higgs-bosons whereas $S^0_2$,
$P^0_2$, $S^-_2$, and $S^-_3$ are mainly the MSSM sleptons of the
third generation.
In the plot we show the various branching ratios of the lighter stop
summing up those branching ratios for the decays into sleptons that
give the same final state, for example:
\begin{equation*}
{\tilde t}_1 \to b \, \nu_e \, {\tilde e}^+_L \,
          \to \, b \, e^+ \, \nu_e \, {\tilde \chi}^0_1 \;, \hspace{5mm}
{\tilde t}_1 \to b \, e^+ \, {\tilde \nu}_e \,
          \to \, b \, e^+ \, \nu_e \, {\tilde \chi}^0_1 \, .
\end{equation*}
The branching ratios for decays into $\tilde{\mu}_{L}$ or
$\tilde{\nu}_{\mu}$ are practically the same as those into
$\tilde{e}_{L}$ or $\tilde{\nu}_{e}$.  Note that the energy spectrum
of the leptons in the final will be somewhat different depending on
whether the scalar in the intermediate step is charged or neutral.
This offers in principle the possibility of determining the branching
ratios of the different decay chains even if the final state topology
is common.  Note, that states containing scalars or neutralinos will
lead to additional jet and/or lepton multiplicities absent in the
MSSM.

In Fig.~\ref{fig:28}(b) the slepton mass parameters are chosen such
that decays into scalars are kinematically forbidden.  Here we display
the channels ${\tilde t}_1 \to b \, W^+ {\tilde \chi}^0_1$, ${\tilde
  t}_1 \to b\, \tau^+$ and ${\tilde t}_1 \to c \, {\tilde \chi}^0_1$.
The remaining modes, such as ${\tilde t}_1 \to b \, S^0_1 \, \tau^+$,
turn out to be completely negligible.  In both cases, with and without
sleptons in the final state, one can see that in general the three
body mode ${\tilde t}_1 \to b \, W^+ {\tilde \chi}^0_1$, dominates
except for a somewhat narrow range of negative $\cos \theta_{\tilde
  t}$.  However, the branching ratio for $\tilde t_1\to b\,\tau^+$ is
above 0.1\% for most values of $|\cos \theta_{\tilde t}|$ implying the
observability of this mode.  Most importantly, note that even in the
parameter ranges where the three-body decay mode is dominant, its
resulting signature is rather different from that of the MSSM due to
the fact the lightest neutralino decays into SM-fermions, leading to
enhanced jet and/or lepton multiplicities, as discussed in detail in
\cite{Bartl2000rp,Porod:2000hv}. In the remaining part of this section
we assume that 3-body decays into scalars are kinematically forbidden.

\begin{table}
  \begin{center}
    \begin{tabular}{|l|lll|}\hline
      Input: & $\tan \beta = 6$ & $\mu_0 = 500$ GeV & $M_2 = 250$ GeV \\
      & $M_{\tilde D}=370$ GeV & $M_{\tilde Q}=340$ GeV & $A_b=150$ GeV \\
      & $M_{\tilde E}=210$ GeV & $M_{\tilde L}= 210$ GeV & $A_\tau=150$ GeV \\ 
      & $m_{{\tilde t}_1}=220$ GeV & $\cos \theta_{\tilde t}=-0.8$ &
      $m_{P^0_3}=300$ GeV \\ \hline
      Calculated & $m_{{\tilde \chi}^0_1}=122$ GeV &
      $m_{{\tilde \chi}^+_1}=234$ GeV & $m_{{\tilde \chi}^+_2}=519$ GeV \\
      &  $m_{{\tilde b}_1}=334$ GeV & $m_{{\tilde b}_2}=381$ GeV & 
      $\cos \theta_{\tilde b}=0.879$ \\
      & $m_{S^0_1}=107$ GeV & $m_{S^0_2}=200$ GeV & $m_{S^0_3}=302$ GeV \\
      &  $m_{P^0_2}=200$ GeV & $m_{P^0_3}=300$ GeV  & \\
      & $m_{S^-_2}=203$ GeV & $m_{S^-_3}=226$ GeV & $m_{S^-_4}=311$ GeV  \\
      & $m_{{\tilde e}_L}=215$ GeV &
      $m_{{\tilde \nu}_e}=m_{{\tilde \nu}_\mu }=200$ GeV  & \\ \hline
    \end{tabular}
  \end{center}
  \caption[Input parameters used]{Input parameters and resulting 
quantities used in Fig.~\protect{~\ref{fig:28}}.}
\label{tab:9}
\end{table}

In Fig.~\ref{fig:28}(c) the $L$ violating parameters are fixed in
such a way that the heaviest neutrino mass is in the range suggested
by the oscillation interpretation of the atmospheric neutrino anomaly
\cite{Gonzalez-Garcia:2001sq}.  

\begin{figure}[htp]
 \begin{center}
    \begin{picture}(232,190) 
      \put(0,-25){\includegraphics[height=8.0cm,width=8.5cm]{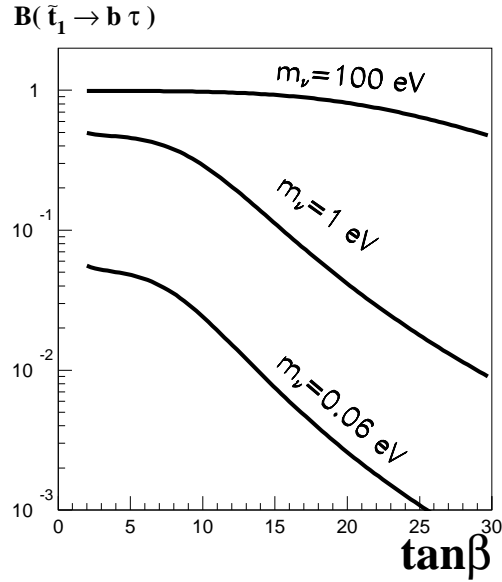}}
    \end{picture}
 \end{center}
    \caption[Branching ratios for ${\tilde t}_1$ decays]{\small Branching 
      ratios for ${\tilde t}_1$ decays for $m_{{\tilde t}_1} =
      220$~GeV, $\mu_0 = 500$~GeV, $M_2 = 240$~GeV, and $m_\nu = 100$,
      1 and $0.06\,$eV.  The branching ratios are shown as a function
      of $\tan \beta$.  ($\cos\theta_{\tilde t}=-0.8$)}
   \label{fig:29}
\end{figure}

In Fig.~\ref{fig:28}(d) we show the same scenario as in
Fig.~\ref{fig:28}(c) but for $\tan \beta = 3$. The branching ratio into
$b \tau$ now increases, whereas the branching ratio into $c \tilde
\chi^0_1$ decreases.  This is easily understood by inspecting
eqs.~(\ref{eq:125}) and (\ref{eq:126}). Indeed for the $b \, \tau$ case
the partial width is proportional to $h_b^2$, whereas for
$c\tilde\chi_1^0$ it is proportional to $h_b^4$.  This implies that
the partial width for $\tilde t_1\to c\tilde\chi_1^0$ grows faster
with $\tan \beta$ than the width for $\tilde t_1\to b \, \tau$.
This is also demonstrated in Fig.~\ref{fig:29} where we show the $\tan
\beta$ dependence of the branching ratio for the decay of $\tilde t_1$
into $b\tau^+$ for several values of the neutrino mass. For
$m_{\nu_3}=0.06\,$eV the $BR(\tilde t_1\to b\,\tau)$ is still above
$0.1\%$ if $\tan\beta$ is not too large, as favored by the explanation
of the neutrino anomalies in this model~\cite{Romao:2000up}. As seen
from the figure, the the $\tilde t_1 \to b\tau^+$ branching ratio is
also somewhat correlated to the $\nu_3$ mass.  Should one add a
sterile neutrino to the model~\cite{Hirsch:2000xe}, then the neutrino
state $\nu_3$ could in principle be heavier than assumed above,
favoring ${\tilde t}_1 \to \tau^+ \, b $ decay mode.

Let us now turn to the general three neutrinos case. There are new
features that arise in this case, as opposed to the 1-generation case
considered so far.  In this model the solution to the present neutrino
anomalies implies that all the $\mu_i$ are of the same order of
magnitude~\cite{Romao:2000up}.  

Two further important results of \cite{Romao:2000up} are that the
atmospheric neutrino angle is controlled by the ratio $(-\mu_2 v_0
+\mu_0 v_2)/(-\mu_3 v_0 +\mu_0 v_3)$ and that the solar mixing angle is
controlled by $(\mu_1/\mu_2)^2$.  One can get approximate formulas for
the decay widths ${\tilde t}_1 \to b \, e^+$ and ${\tilde t}_1 \to b \,
\mu^+$ similar to eq.~(\ref{eq:125}) by replacing $\mu_3$ by $\mu_{1,2}$.
This implies that {\it (i)} The decays into $ b \, e^+$ and $ b \,
\mu^+$ are as important as the decay into $ b \, \tau^+$.  {\it (ii)} The
decays ${\tilde t}_1 \to b \, e^+$ and ${\tilde t}_1 \to b \, \mu^+$ are
related with the solar mixing angle.  Moreover, we find that
$\sum_{l=e,\mu,\tau} \Gamma(\tilde t_1 \to b \, l^+)$ in the 3-generation model
is nearly equal to $\Gamma(\tilde t_1 \to b \, \tau^+)$ in the 1-generation
model provided that $\sum_{i=1}^3 \mu_i^2$ is identified to $\mu_3^2$ in
the 1-generation model.

\begin{figure}
\setlength{\unitlength}{1mm}
\begin{center}
\begin{picture}(70,70)
\put(-5,-5){\includegraphics[height=7.0cm,width=7.cm]{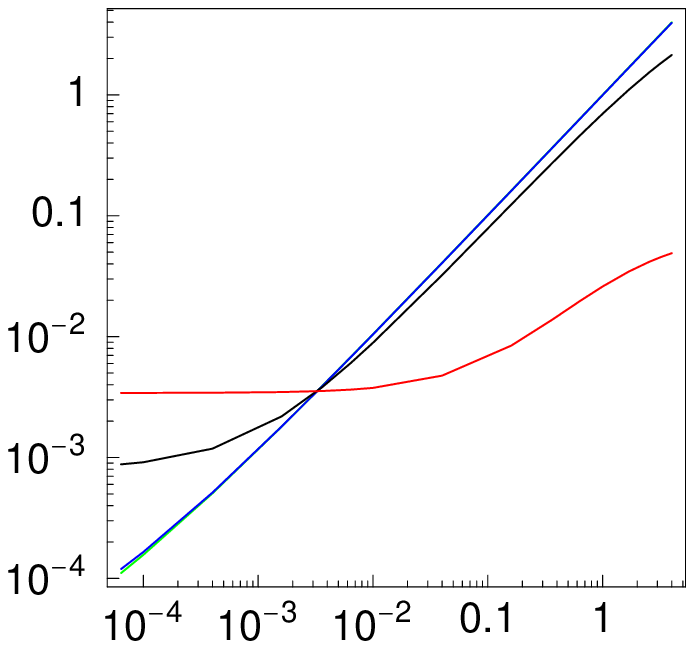}}
\put(0,63){\makebox(0,0)[bl]{{\small $BR({\tilde t}_1 \to b \, e^+)/
                                    BR({\tilde t}_1 \to b \, \mu^+) $}}}
\put(68,-8){\makebox(0,0)[br]{{\small $(\mu_1 / \mu_2)^2$}}}
\put(42,50){{\small $\geq 0.1$}}
\put(49,38){{\small $ 10^{-2}$}}
\put(52,41){\vector(0,1){6}}
\put(8,18){{\small $10^{-2}$}}
\put(52,28){{\small $10^{-3}$}}
\put(15,25){{\small $10^{-3}$}}
\put(12,7){{\small $\geq 0.1$}}
\end{picture}
\end{center}
\caption[Ratio of branching ratios: $BR({\tilde t}_1 \to b e^+)/
BR({\tilde t}_1 \to b \mu^+) $ as a function of $(\mu_1 /
\mu_2)^2$]{Ratio of branching ratios: $BR({\tilde t}_1 \to b
  e^+)/ BR({\tilde t}_1 \to b \mu^+) $ as a function of $(\mu_1
  / \mu_2)^2$ for $m_{{\tilde t}_1} = 220$~GeV, $\mu_0 = 500$~GeV,
  $M_2 = 240$~GeV; $|\cos \theta_{\tilde t}| \geq 0.1, 0.01, 10^{-3}$,
  $m_{\nu_3} = 0.6$~eV.}
\label{fig:30}
\end{figure}

In Fig.~\ref{fig:30} we show the ratio of $BR({\tilde t}_1 \to b \,
e^+)/BR({\tilde t}_1 \to b \, \mu^+)$ versus
$(\mu_1/\mu_2)^2$ for different values of $\cos
\theta_{\tilde t}$. For definiteness we have fixed the heaviest
neutrino mass at the best-fit value indicated by the atmospheric
neutrino anomaly.
One can see that the dependence is nearly linear even for rather small
$\cos \theta_{\tilde t}$.  For $|\cos \theta_{\tilde t}| \lesssim 10^{-2}$ the
approximation in eq.~(\ref{eq:125}) breaks down and additional pieces
dependent on $\sin \theta_{\tilde t}$ \cite{Diaz:1999ge,Bartl:1996gz}
become important, leading to the non-linear dependence.  One sees from
the figure that, as long as $\cos \theta_{\tilde t} \gtrsim 10^{-2}$ there is a
good degree of correlation between the branching ratios into
$BR(\tilde t_1 \to b\,e^+)$ and $BR(\tilde t_1 \to b\,\mu^+)$ and the
ratio $(\mu_1/\mu_2)^2$. Thus by measuring these branchings one will get
information on the solar neutrino mixing, since $\tan^2 \theta_{sol}$ is
proportional to $(\mu_1/\mu_2)^2$~\cite{Romao:2000up} which makes it a
rather important quantity.
For the so--called small mixing angle or SMA solution of the solar
neutrino problem we expect ${\tilde t}_1 \to e^+ \, b$ to be
negligible. In contrast, for the large mixing angle type solutions
(LMA, LOW and QVAC, see ref.~\cite{Gonzalez-Garcia:2001sq} and
references therein) we expect \emph{all} ${\tilde t}_1 \to l^+ \, b$
decays to have comparable rates.
As a result in this model one can directly test the solution to the
solar neutrino problem against the lighter stop decay pattern. 
This is also complementary to the case of neutralino decays considered
in \cite{Porod:2000hv}. In that case the sensitivity is mainly to
atmospheric mixing, as opposed to solar mixing. Testing the latter in
neutralino decays at a collider experiment requires more detailed
information on the complete spectrum to test the solar angle
\cite{Porod:2000hv}. In contrast we have obtained here a rather neat
connection of stop decays with the solar neutrino physics.

Note, that this result is much more general than the scenarios
discussed in this work. It is of particular importance in scenarios
where only the $L$ violating decays and the decay into $\tilde
\chi^0_1 \, c$ are present \cite{Diaz:1999ge,Bartl:1996gz}. Similarly,
the other ratios of the final states $b \, l^+$ are proportional to
the square of the ratio of corresponding $\mu_i$ provided that
$\cos \theta_{\tilde t}$ is not too small.

\chapter{Conclusions}
\label{sec:conclusions}
We have seen in this work that the SSSM is a well motived model which
also provides a solid explanation of the neutrino anomalies
\cite{Romao:2000up,Hempfling:1996wj} which can be tested at
accelerators experiments. It reproduces the same signals that general
$B_p$--SSM but with a minimum set of $L$ violating terms and
automatically satisfies all the experimental bounds on induced
dimension--4 $L$ violating operators.  We have explicitly studied the
phenomenology of neutralino and stop decays.

Concerning the Neutralinos, we have studied the production of the
lightest neutralino $\tilde{\chi}^{0}_{1}$ at LEP2 and the resulting
phenomenology in models where an effective superrenormalizable term
parameterizes the explicit breaking of $L$. We have considered
supergravity scenarios which can be explored at LEP2 in which the
lightest neutralino is also the lightest supersymmetric particle. We
have presented a detailed study of the LSP $\tilde{\chi}^{0}_{1}$
decay properties and studied the general features of the corresponding
signals expected at LEP2.  A detailed investigation of the possible
detectability of the signals discussed in Tab.~\ref{tab:7} taking into
account realistic detector features is beyond the scope of this thesis.
Clearly, existing LEP2 data are already probing the part of the
parameter region which corresponds to approximately
$m_{\tilde{\chi}^{0}_{1}} \lesssim 40$~GeV.  Finally, we note that, in
addition to important modifications in the $\tilde{\chi}^{0}_{1}$
decay properties, the SSSM lead also to new interesting features in
other decays, such as charged \cite{Akeroyd:1998iq} and neutral
\cite{deCampos:1995av} Higgs boson and slepton decays, stop
decays~\cite{Diaz:1999ge,Bartl:1996gz}, and gluino cascade decays
\cite{Bartl97a,gluino}. In addition we have shown that the SSSM can not be
confused with gauge mediated supersymmetry breaking and conserved
R-parity due to the absence of several final states in the GMSB case.

Concerning the stops we have studied the two--body decays of the
lightest top squark in SUGRA models with and without $L$ violating
terms. We have improved the calculation for the decay $\tilde t_1 \to
c\,\tilde\chi^0$ by numerically solving the renormalization group
equations (RGE's) of the SSSM including full generation mixing in the
RGE's for Yukawa couplings as well as soft SUSY breaking parameters.
The decay-width is in general one order of magnitude smaller than the
one obtained in the usual one--step approximation. This result will
therefore enlarge the regions of parameter space where the four--body
decays of lightest stop dominate over the decay into a charm quark and
the lightest neutralino. As a result it will affect the present
experimental lower bound on the $\tilde t_1$ mass even in the $L$
conserving case~\cite{Boehm:1999tr}. In the SSSM of course new decay
modes appear and, as we have shown, they can be sizeable. In fact we
have shown that the lightest stop can be the LSP, decaying with 100\%
rate into a bottom quark and a tau lepton. We have shown that the
decay mode $\tilde t_1 \to b\, \tau$ dominates over $\tilde t_1 \to
c\,\tilde\chi^0$ even for neutrino masses in the range suggested by the
simplest oscillation interpretation of the Super-Kamiokande
atmospheric neutrino data. This result would have a strong impact on
the top squark search strategies at LEP~\cite{Abbiendi:1999yz} and
TEVATRON~\cite{Holck:1999tp}, where it is usually assumed that the
$\tilde t_1 \to c\,\tilde\chi^0$ decay mode is the main channel. In
addition to the signal of two jets and two taus present when the two
produced stops decays through the $L$ violating channel, one expects a
plethora of exotic high--multiplicity fermion events arising from
neutralino decay, since such decay can happen inside the detector even
for the small neutrino masses in the range suggested by the $\nu_\mu$ to
$\nu_\tau$ oscillation interpretation of the atmospheric neutrino
anomaly~\cite{Romao:2000up}.

In addition the $L$ violating decay ${\tilde t}_1 \to b \, \tau^+$ may
compete with three--body decays. Specifically, we have found that for
$m_{{\tilde t}_1} \lesssim 250$~GeV there are regions of parameter where
${\tilde t}_1 \to b \, \tau^+$ is an important decay mode if not the most
important one. This implies that there exists the possibility of full
stop mass reconstruction from $\tau^+ \, \tau^- \, b \, \bar{b}$ final
states, favoring the prospects for its discovery. In contrast, in the
MSSM the discovery of the lightest stop might not be possible at the
LHC within this mass range.
This implies that it is important to take into account this new decay
mode when designing the stop search strategies at a future $e^+ e^-$
Linear Collider.  The SSSM also imply additional leptons and/or jets
in stop cascade decays.  Looking at the three generation model the
decays into ${\tilde t}_1 \to b \, l^+$ imply the possibility of
probing $\mu^2_1 /\mu^2_2$ and thus the solar mixing angle.
This complements information which can be obtained using neutralino
decays. In the latter case the sensitivity is mainly to the
atmospheric mixing, as opposed to solar mixing \cite{Porod:2000hv}. In
SSSM neutralino decays is ideal to test the atmospheric anomaly
at a collider experiment, while stop decays provide neat complementary
information on the solar mixing angle.
Obtaining solar mixing information from neutralino decays would
require more detailed knowledge on the supersymmetric spectrum, since
it would be involved in the relevant loop calculations of the solar
neutrino mass scale and mixing angle.
By combining the two one can probe the parameters associated with both
solar and atmospheric neutrino anomalies at collider experiments.

\appendix

\chapter{Constraints on horizontal charges}
\label{cha:constr-horiz-charg}
In this Appendix we derive the general expressions for the individual
field charges satisfying the set of 13 phenomenological and
theoretical constraints corresponding to the six mass ratios for the
quarks and the charged leptons plus the two quark mixing angles
summarized in Table~\ref{tab:4}; the two relations provided by the
absolute value of the masses of the third generation fermions; one
phenomenological assumption about the charge of the $\mu_0$ term
(\ref{eq:66}); one theoretical constraint corresponding to the
consistency conditions for the coefficients of the mixed linear
anomalies and one additional constraint from the vanishing of the
mixed anomaly quadratic in the horizontal charges.  As discussed in
section \ref{sec:flavour-symmetries}, this leaves us with four free
parameters that we choose to be $n_i$ ($i=1,2,3$) and $x$.  We obtain
\begin{eqnarray}
  Q_3&=& \frac1{15\,\left( 7 + x  \right) }
  \Big[-180 - 45x - 3x^2 + Q_{13}(41 + 5x)
  -7L_{23}  + L_{23}^2
  \nonumber\\
  &&\hspace{1.5cm} 
  + n_1( 2 + x + L_{23} )
  + n_2( 9 + x - L_{23} )  
  + n_3(9 + x) 
  \Big],  \nonumber \\  
  L_3&=&\frac1{15
    \left( 7 + x  \right) }
  \Big[20  + 50x + 6x^2+ 18Q_{13} - 21 L_{23} + 3 L_{23}^2
  \nonumber\\
  &&\hspace{1.5cm}
  - n_1(  29 + 2x - 3 L_{23}) 
  -n_2( 8 + 2x +3 L_{23}) 
  +n_3(97+13x)
  \Big],   \nonumber
\end{eqnarray}
where $L_{23}=L_{23}+l_{23}$ and 
$Q_{13}$ parametrized the two different possibilities  
for the quark and lepton charge differences given 
in Tab.~(\ref{tab:4}).
In terms of $Q_3$ and $L_3$ and of our  
four free parameters we have 
\begin{equation*}
  \begin{array}{lll}
    H_u&=&n_3-L_3\\
    L_0&=&-1+H_u
  \end{array}
  \qquad\qquad\qquad
  \begin{array}{lll}
    u_3&=&-Q_3-H_u\\
    d_3&=&-Q_3+x-L_0\\
    l_3&=&-L_3+x-L_0
  \end{array}
\end{equation*}
and from these all the other individual charges can be 
straightforwardly determined from
the charge differences in Tab.~(\ref{tab:4}).
The solution for the charges in model MQ1+ML1 
for the preferred values $n_1=n_2=n_3=-8$ and $x=1$ 
is given  in Table~\ref{tab:10}. For the $n_0=0$ case and the same
value for the neutrino mass which implies in this case $n_i=-6-x$, we
present in Table~\ref{tab:11} the individual charges in in model MQ1+ML1 
for  $x=0$ 
\begin{table}[ht]
  \begin{center}
    \begin{footnotesize}
      \begin{tabular}{|ccccccccccccccccc|}\hline
        &&&&&&&&&&&&&&&&\\
        $Q_1$&$Q_2$&$Q_3$&$u_1$&$u_2$&$u_3$&$d_1$&$d_2$&$d_3$&$L_1$&
        $L_2$&$L_3$&$l_1$&$l_2$&$l_3$&$H_u$&$L_0$\\ 
        $\frac{161}{30}$&$\frac{131}{30}$&$\frac{71}{30}$
        &$\frac{103}{15}$&$\frac{58}{15}$&$\frac{28}{15}$
        &$-\frac{18}{5}$&$-\frac{23}{5}$&$-\frac{23}{5}$
        &$-\frac{113}{30}$&$-\frac{113}{30}$&$-\frac{113}{30}$
        &$\frac{98}{15}$&$\frac{53}{15}$&$\frac{23}{15}$
        &$-\frac{127}{30}$&$\frac{97}{30}$\\
         &&&&&&&&&&&&&&&&\\\hline
      \end{tabular}
    \end{footnotesize}
    \caption{The anomaly free set of charges 
 of  model MQ1+ML1 for $x=1$ and  $n_1=n_2=n_3=-8$} 
    \label{tab:10}
  \end{center}
\end{table}

\begin{table}[ht]
  \begin{center}
    \begin{footnotesize}
      \begin{tabular}{|ccccccccccccccccc|}\hline
        &&&&&&&&&&&&&&&&\\
        $Q_1$&$Q_2$&$Q_3$&$u_1$&$u_2$&$u_3$&$d_1$&$d_2$&$d_3$&$L_1$&
        $L_2$&$L_3$&$l_1$&$l_2$&$l_3$&$H_u$&$L_0$\\ 
        $\frac{23}{5}$&$\frac{18}{5}$&$\frac{8}{5}$
        &$\frac{28}{5}$&$\frac{13}{5}$&$\frac{3}{5}$
        &$-\frac{14}{5}$&$-\frac{19}{5}$&$-\frac{19}{5}$
        &$-\frac{19}{5}$&$-\frac{19}{5}$&$-\frac{19}{5}$
        &$\frac{28}{5}$&$\frac{18}{5}$&$\frac{8}{5}$
        &$-\frac{11}{5}$&$\frac{11}{5}$\\
         &&&&&&&&&&&&&&&&\\\hline
      \end{tabular}
    \end{footnotesize}
    \caption{The anomaly free set of charges 
 of  model MQ1+ML1 for $n_0=0$, $x=0$ and  $n_1=n_2=n_3=-6$} 
    \label{tab:11}
  \end{center}
\end{table}

\chapter{Complete formulas for 2 body decays}
\label{sec:complete-formulas-2}

In this Appendix we derive the Feynman rules $F_j^0q_i\tilde q_k$
(involving a neutralino/tau-neutrino, a quark, and a squark) and
$F_j^{\pm}q_i\tilde q'_k$ (involving a chargino/tau, a quark, and a
squark of different electric charge) in the case of three generations
and $L$ violating in the third generation. This is a generalization of
the Feynman rules contained in~\cite{Gunion:1986yn}, which hold for the MSSM
and for one generation of quarks and squarks.

\section{An explicit calculation}
\label{sec:an-expl-calc}
Before the general case, we will calculate here $\tilde
t_1\to\tilde\chi_1^0\,t$.

From eq.~(\ref{eq:8}), the lagrangian in terms of the two body decays
are
\begin{multline}
  \label{eq:137}
  \mathcal{L}_{t\tilde t\tilde\chi_1^0}=
  \left\{ 
    \frac{1}{\sqrt{2}}\tilde t_L^*
    \left[
      g(i\lambda^3)t_L+g'h_t(i\lambda')t_L +\mathrm{h.c}
    \right]+\frac{1}{\sqrt{2}}g'h_t
    \left[
      (i\lambda')t_L\tilde t_L^*+\bar{t}_L(-i\bar{\lambda'})\tilde t_L
    \right]  
  \right.\\
  -h_t\left.
    \left(
      \tilde t_R^* t_L\widetilde H_u^0+\widetilde t_L\bar{t}_R\widetilde H_u^0
    \right)
  \right\} 
\end{multline}
In terms of the four component spinors
\begin{equation}
  \label{eq:138}
  \widetilde B=
  \begin{bmatrix}
    -i\lambda'\\
    i\bar{\lambda}'
  \end{bmatrix},\qquad
  \widetilde W_3=
  \begin{bmatrix}
    -i\lambda^3\\
    i\bar{\lambda}^3
  \end{bmatrix},\qquad
  \widetilde H_u=
  \begin{bmatrix}
    \widetilde H_u^0\\
    \bar{\widetilde H}_u^0
  \end{bmatrix}
\end{equation}
we have
\begin{multline}
  \label{eq:139}
  \mathcal{L}_{t\tilde t\tilde\chi_1^0}=
  -\frac{1}{\sqrt{2}}\,\tilde t_L^*
  \left[
    g\overline{\widetilde
    W}_3P_Lt+g'h_t\overline{\widetilde B}P_L t
  \right]
  -\frac{1}{\sqrt{2}}\,g'
  \left[
    h_t\bar{t}P_L\widetilde B\tilde t_R+h_t^*\overline{\widetilde
      B}P_Rt\tilde t_R^*
  \right]\\
  -\frac{g\,m_t}{\sqrt{2}m_W\sin\beta}\,
  \left[
    \overline{\widetilde H}_uP_L t \tilde t_R^*+\bar tP_L\widetilde
    H_u\tilde t_L
  \right]
\end{multline}
The result in the mass basis is finally
\begin{equation}
  \label{eq:140}
    \mathcal{L}_{t\tilde t\tilde\chi_1^0}=g\,
    \left\{ 
      \bar{t}\,
      \left[
        h^t_{ij}P_L+f^t_{1j}P_R
      \right]\tilde \chi_j^0\tilde t_1+\mathrm{h.c}
    \right\} 
\end{equation}
where
\begin{align}
  h^t_{1j}=&\frac{m_t}{\sqrt{2}\,m_W\sin\beta}\,
  \left(
    \sin\beta {N'_{j3}}^*-\cos\beta {N'_{j4}}^*
  \right)\cos\theta_{\tilde t}
  \nonumber\\
  \label{eq:141}
  &\quad+\left[
    \frac{-2\sqrt{2}}{3}\,\sin\theta_W
    \left(
      \tan\theta_W {N'_{j2}}^*-{N'_{j1}}^*
    \right)
  \right]\sin\theta_{\tilde t}\\
  f^t_{1j}=&
  \left[
    \frac{-2\sqrt{2}}{3}\,\sin\theta_W N'_{j1}-\sqrt{2}
    \left(
      \frac12-\frac{-2}{3}\sin\theta_W 
    \right)\frac{N'_{j2}}{\cos\theta_W}
  \right]\cos\theta_{\tilde t}
  \nonumber\\
  \label{eq:142}
  &\quad+\left[
    \frac{m_t}{\sqrt{2}\,m_W\sin\beta}
    \left(
      \sin\beta N'_{j3}-\cos\beta N_{j4}^*
    \right)
  \right]\sin\theta_{\tilde t}
\end{align}
where we have redefined the neutralino diagonalization matrix as
\begin{equation}
  \label{eq:143}
  \begin{bmatrix}
    N'_{j1}\\
    N'_{j2}
  \end{bmatrix}=
  \begin{bmatrix}
    \cos\theta_W  & \sin\theta_W\\
    -\sin\theta_W & \cos\theta_W
  \end{bmatrix}
  \begin{bmatrix}
    N_{j1}\\
    N_{j2}
  \end{bmatrix},\qquad
  \begin{bmatrix}
    N'_{j3}\\
    N'_{j4}
  \end{bmatrix}=
  \begin{bmatrix}
    \cos\beta & -\sin\beta\\
    \sin\beta & \cos\beta
  \end{bmatrix}
  \begin{bmatrix}
    N_{j3}\\
    N_{j4}
  \end{bmatrix}
\end{equation}

We now calculate the two body width decay $\Gamma(\tilde
t_1\to t+\tilde\chi_1^0)$
\begin{equation}
  \label{eq:144}
  \Gamma(\tilde t_1\to t+\tilde\chi_1^0)=\frac{\Sigma_S|\mathcal{M}|^2}{16\pi\,m_{\tilde
      t_1}^3}\lambda^3(m_{\tilde t_1}^2,m_t^2,m_{\tilde\chi_1^0}^2)
\end{equation}
where $\lambda(a,b,c)=(a - b - c)^2 - 4ac$.  The corresponding Feynman
diagram is

\noindent
\begin{picture}(120,70)(0,25) 
\ArrowLine(110,0)(60,25)
\Vertex(60,25){3}
\DashArrowLine(60,25)(10,25){6}
\ArrowLine(110,60)(60,25)
\Text(35,35)[]{${\tilde t}_1$}
\Text(85,55)[]{\hspace{-0.3cm}$\tilde\chi_j^0(p_1)$}
\Text(85,0)[]{$\bar{t}(p_2)$}
\end{picture}
\hspace{1cm}$-ig[h^t_{ij}P_L+f^t_{1j}P_R]$
\vspace{30pt} \hfill \\

\noindent
Consequently we can write down the amplitude as
\begin{equation}
  \label{eq:145}
  \mathcal{M}=g\,
  \left[
    h^t_{1j}\,\bar{u}(p_1)P_Lv(p_2)+f^t_{1j}\,\bar{u}(p_1)P_Rv(p_2)
  \right]
\end{equation}
The final result is
\begin{multline}
  \label{eq:146}
  \Gamma(\tilde t_1\to t+\tilde\chi_1^0)=\frac{g^2\lambda^{1/2}(m_{\tilde
      t_1}^2,m_t^2,m_{\tilde\chi_1^0}^2)}{16\pi\,m_{\tilde t_1}^3}
  \left\{ 
    \left[
      (h^t_{1j})^2+(f^t_{1j})^2
    \right]
    \left(
      m_{\tilde t_1}^2-m_t^2-m_{\tilde\chi_1^0}^2
    \right)
  \right.\\
  -\left.
    4h^t_{ij}f^t_{1j}m_tm_{\tilde \chi_1^0}
  \right\} 
\end{multline}

\section{Couplings Neutralino--Squark-Quark}
\label{sec:coupl-neutr-squark}
Following~\cite{Bertolini:1991if} we work in a quark interaction basis where 
$d_{L,R}=d_{L,R}^0$, $u_{L}=K u_{L}^0$, and $u_{R}=u_{R}^0$ (we 
denote $q$ and $q^0$ the mass and current eigenstates respectively), 
as opposed to Ref.~\cite{Rosiek:1990rs} where a more general basis is used. 
In addition, we implement the notation $\tilde q_{L,R}\equiv \tilde 
q^0_{L,R}$ for the interaction basis.

The starting point is the following piece of the Lagrangian
\begin{eqnarray}
  {\cal L}_{u \tilde u F^0}&=& 
  -g{\bar u}_i^0
  \Bigg\{
  \sqrt2
  \left[
    \sin\theta_W e_U {N'}_{J1}+\frac 1{\cos\theta_W}
    ({\textstyle\frac12}-e_U\sin^2\theta_W){N'}_{J2}
  \right]
  \tilde u_{iL}\nonumber\\
  &&+
  \frac{m^{0U}_{ij}}{\sqrt2m_W\sin\beta\sin\theta}{N'}_{J4}\tilde u_{jR}
  \Bigg\}
  P_R F_J^0\nonumber\\
  &&+g{\bar u}^0_i
  \Bigg\{
  \sqrt2
  \left[
    \sin\theta_W e_U {N'}_{J1}^*+\frac 1{\cos\theta_W}(-e_U\sin^2\theta_W)
    {N'}_{J2}^*
  \right]
  \tilde u_{iR}\nonumber\\
  &&-
  \frac{m^{0U\dagger}_{ij}}{\sqrt2m_W\sin\beta\sin\theta}{N'}_{J4}^*\tilde 
  u_{jL}\Bigg\}P_L F_J^0+\textrm{h.c.}
  \label{eq:128}
\end{eqnarray}
written in the quark interaction basis. The $5\times5$ matrix $N'$ 
diagonalizes the neutralino/neutrino mass matrix in the 
$(\tilde\gamma,\tilde Z,\tilde L_0^0,\tilde H_u^0,\nu_{\tau})$ basis 
as defined in~\cite{Akeroyd:1998iq}, with the index $J=1...5$. The $3\times3$ 
up--type quark mass matrix $m^{0U}$ is not diagonal, with the indexes 
$i,j=1,2,3$. 

In order to write the above Lagrangian with mass eigenstates we use
the basic relations mentioned before, in particular, $u_{iL}^0=
(K^\dagger)^{ij}u_{jL}$, which implies that ${\bar u}_{iL}^0={\bar
u}_{jL}K^{ji}$. We need the following relations:
\begin{equation}
  \label{eq:129}
  \begin{split}
    {\bar u}_{iL}^0\tilde u_{iL}=&
    {\bar u}_{iL}\left(\Gamma_{UL}^{*}{(K^\dagger)}^*\right)^{ki}\tilde
    u_{k}\\
    {\bar u}_{iL}^0m^{0U}_{ij}\tilde u_{jR}
    =&{\bar u}_{iL}(\Gamma_{UR}^{*}m^U)^{ki}\tilde u_{k}\\
    {\bar u}_{iR}\tilde u_{iR}=&{\bar u}_{iR}\Gamma_{UR}^{*ki}\tilde u_{k}
    \\
    {\bar u}_{iR}m^{0U}_{ij}\tilde u_{jL}=& 
    {\bar u}_{iR}(\Gamma_{UL}^{*}K^*{m}^U)^{ki}\tilde u_{k}
  \end{split}
\end{equation}
where $i,j=1,2,3$ label the quark flavors, $k=1...6$ labels the
squarks, and $m^U\equiv{\mathrm{diag}}\,\{m_u,m_c,m_t\}$ is the
diagonal up--type quark mass matrix. In this way, the Lagrangian in
eq.~(\ref{eq:128}) can be written as
\begin{equation}
  {\cal L}_{u \tilde u F^0}= 
  -g{\bar u}_i[(\sqrt2G_{0UL}^{*jki}+H_{0UR}^{*jki})P_R-
  (\sqrt{2}G_{0UR}^{*jki}-H_{0UL}^{*jki})P_L]F_j^0\tilde u_k+\textrm{h.c.}
  \label{eq:130}
\end{equation}
where the different couplings are
\begin{eqnarray}
G_{0UL}^{jki}&=&\left[
\sin\theta_W e_U {N'}^*_{j1}+\frac 1{\cos\theta_W}
({\textstyle\frac12}-e_U\sin^2\theta_W){N'}^*_{j2}
\right]\left(\Gamma_{UL}K^\dagger\right)^{ki}\nonumber\\
G_{0UR}^{jki}&=&\left[
\sin\theta_W e_U {N'}_{j1}+\frac 1{\cos\theta_W}(-e_U\sin^2\theta_W)
{N'}_{j2}
\right]\Gamma_{UR}^{ki}\label{eq:131}\\
H_{0UL}^{jki}&=&{N'}_{j4}(\Gamma_{UL}K^{\dagger}{\hat h}_U)^{ki}
\nonumber\\
H_{0UR}^{jki}&=&{N'}^*_{j4}(\Gamma_{UR}\hat h_U)^{ki}\nonumber
\end{eqnarray}
and $\hat
h_U\equiv$diag$\,(m_u,m_c,m_t)/(\sqrt2m_W\sin\beta\sin\theta)$.
Graphically, the $F_j^0u_i\tilde u_k$ Feynman rules are given by

\begin{picture}(120,60)(0,25) 
\DashArrowLine(60,25)(110,0){6}
\Vertex(60,25){3}
\ArrowLine(10,25)(60,25)
\ArrowLine(110,60)(60,25)
\Text(35,35)[]{$F_j^0$}
\Text(85,55)[]{$u_i$}
\Text(85,0)[]{${\tilde u}_k$}
\end{picture}
$
-ig[(\sqrt2G_{0UL}^{jki}+H_{0UR}^{jki})P_L
-(\sqrt{2}G_{0UR}^{jki}-H_{0UL}^{jki})P_R]
$

\begin{picture}(120,70)(0,25) 
\DashArrowLine(110,0)(60,25){6}
\Vertex(60,25){3}
\ArrowLine(60,25)(10,25)
\ArrowLine(60,25)(110,60)
\Text(35,35)[]{$F_j^0$}
\Text(85,55)[]{$u_i$}
\Text(85,0)[]{${\tilde u}_k$}
\end{picture}
$
-ig[(\sqrt2G_{0UL}^{*jki}+H_{0UR}^{*jki})P_R
-(\sqrt{2}G_{0UR}^{*jki}-H_{0UL}^{*jki})P_L]
$
\vspace{30pt} \hfill \\
The analogous Feynman rules in the MSSM are obtained by replacing
$F_i^0 \to \tilde\chi^0_i$, by interpreting the matrix $N$
as the usual $4\times4$ neutralino mixing matrix, and by setting
$\theta=\pi/2$ in the formula for the Yukawa couplings.

Similarly, replacing all $u(\tilde u)$ by $d(\tilde d)$ in 
eq.~(\ref{eq:128}) and
starting from 
\begin{equation}
  \label{eq:132}
  \begin{split}
    {\cal L}_{q \tilde q' F^+}=&g\bar{d}_i\left[
      \frac{m^{0U\dagger}_{ij}}{\sqrt2m_W\sin\beta\sin\theta}V_{J2}\tilde
      u_{jR}- V_{J1}\tilde
      u_{iL}\right]P_RF^c_J
    \\
    &+g\bar{d}_i\frac{m^{D}_{ij}}{\sqrt2m_W\cos\beta\sin\theta}
    U^*_{J2}\tilde u_{jL}P_LF^c_J\\
    &+g\bar{u}^0_i\left[
      \frac{m^{D\dagger}_{ij}}{\sqrt2m_W\cos\beta\sin\theta}U_{J2}\tilde
      d_{jR}- U_{J1}\tilde
      d_{iL}\right]P_RF^+_J
    \\
    &+g\bar{u}^0_i\frac{m^{0U}_{ij}}{\sqrt2m_W\cos\beta\sin\theta}
    V^*_{J2}\tilde b_{jL}P_LF^+_J+\hbox{h.c}
  \end{split}
\end{equation}
we can obtain the complete Feynman
rules for the neutralino/tau--neutrino and chargino/tau with quarks and 
squarks. The results, that complements the obtained in~\cite{Bertolini:1991if}, are

Neutralino--(d)quark--(d)squark

\begin{picture}(120,60)(0,25) 
\DashArrowLine(60,25)(110,0){6}
\Vertex(60,25){3}
\ArrowLine(10,25)(60,25)
\ArrowLine(110,60)(60,25)
\Text(35,35)[]{$F_j^0$}
\Text(85,55)[]{$d_i$}
\Text(85,0)[]{${\tilde d}_k$}
\end{picture}
$
-ig[(\sqrt2G_{0DL}^{jki}+H_{0DR}^{jki})P_L
-(\sqrt{2}G_{0DR}^{jki}-H_{0DL}^{jki})P_R]
$

\begin{picture}(120,70)(0,25) 
\DashArrowLine(110,0)(60,25){6}
\Vertex(60,25){3}
\ArrowLine(60,25)(10,25)
\ArrowLine(60,25)(110,60)
\Text(35,35)[]{$F_j^0$}
\Text(85,55)[]{$d_i$}
\Text(85,0)[]{${\tilde d}_k$}
\end{picture}
$
-ig[(\sqrt2G_{0DL}^{*jki}+H_{0DR}^{*jki})P_R
-(\sqrt{2}G_{0DR}^{*jki}-H_{0DL}^{*jki})P_L]
$
\vspace{30pt} \hfill \\

\noindent
The mixing matrices $G_{0D}$ and $H_{0D}$ are defined as 
\begin{equation}
  \label{eq:133}
  \begin{split}
    G_{0DL}^{jki}=&\left[
      \sin\theta_W e_D {N'}_{j1}^*+\frac 1{\cos\theta_W}
      (T_{3D}-e_D\sin^2\theta_W){N'}^*_{j2}
    \right]\Gamma_{DL}^{ki}\\
  G_{0DR}^{jki}=&\left[
    \sin\theta_W e_D {N'}_{j1}+\frac 1{\cos\theta_W}(-e_D\sin^2\theta_W)
    {N'}_{j2}
  \right]\Gamma_{DR}^{ki}\\
  H_{0DL}^{jki}=&{N'}_{j3}(\Gamma_{DL}{\hat h}_D)^{ki}\\
  H_{0DR}^{*jki}=&{N'}_{j3}^*(\Gamma_{DR}\hat h_D)^{ki}
\end{split}
\end{equation}

\section{Couplings Chargino-Squark-Quark}
\label{sec:coupl-charg-squark}

Here we give the couplings that were used in
Sections \ref{sec:coupl-neutr-squark} and \ref{sec:three-body-decays-1}

Chargino/tau--(d)quark--(u)squark


\begin{picture}(120,60)(0,25) 
\DashArrowLine(60,25)(110,0){6}
\Vertex(60,25){3}
\ArrowLine(10,25)(60,25)
\ArrowLine(110,60)(60,25)
\Text(35,35)[]{$F_j^+$}
\Text(85,55)[]{$d_i$}
\Text(85,0)[]{${\tilde u}_k$}
\end{picture}
$
-ig(-C^{-1})[(G_{UL}^{jki}-H_{UR}^{jki})P_L-H_{UL}^{jki}P_R]
$

\vglue0.5truecm

\begin{picture}(120,70)(0,25) 
\DashArrowLine(110,0)(60,25){6}
\Vertex(60,25){3}
\ArrowLine(60,25)(10,25)
\ArrowLine(60,25)(110,60)
\Text(35,35)[]{$F_j^+$}
\Text(85,55)[]{$d_i$}
\Text(85,0)[]{${\tilde u}_k$}
\end{picture}
$
-ig[(G_{UL}^{*jki}-H_{UR}^{*jki})P_R-H_{UL}^{*jki}P_L]C
$
\vspace{30pt} \hfill \\

\noindent
where $C$ is the charge conjugation matrix (in spinor space) and the
mixing matrices $G_U$ and $H_U$ are defined as
\begin{equation}
  \label{eq:134}
  \begin{split}
    G_{UL}^{jki}&\equiv V_{j1}^*\Gamma_{UL}^{ki},\qquad
    H_{UL}^{jki}\equiv U_{j2}^*(\Gamma_{UL}\hat h_D)^{ki},\\
    H_{UR}^{jki}&\equiv V_{j2}^*(\Gamma_{UR}\hat h_UK)^{ki},
\end{split}
\end{equation}
Chargino/tau--(u)quark--(d)squark

\begin{picture}(120,60)(0,25) 
\DashArrowLine(60,25)(110,0){6}
\Vertex(60,25){3}
\ArrowLine(10,25)(60,25)
\ArrowLine(110,60)(60,25)
\Text(35,35)[]{$F_j^+$}
\Text(85,55)[]{$u_i$}
\Text(85,0)[]{${\tilde d}_k$}
\end{picture}
$
-ig(-C^{-1})[(G_{DL}^{jki}-H_{DR}^{jki})P_L-H_{DL}^{jki}P_R]
$

\begin{picture}(120,70)(0,25) 
\DashArrowLine(110,0)(60,25){6}
\Vertex(60,25){3}
\ArrowLine(60,25)(10,25)
\ArrowLine(60,25)(110,60)
\Text(35,35)[]{$F_j^+$}
\Text(85,55)[]{$u_i$}
\Text(85,0)[]{${\tilde d}_k$}
\end{picture}
$
-ig[(G_{DL}^{*jki}-H_{DR}^{*jki})P_R-H_{DL}^{*jki}P_L]C
$
\vspace{30pt} \hfill \\

\noindent
where the mixing matrices $G_D$ and $H_D$ are defined as
\begin{equation}
  \label{eq:135}
  \begin{split}
    G_{DL}^{jki}\equiv& U_{j1}^*(\Gamma_{DL}K^\dagger)^{ki},\qquad
    H_{DL}^{jki}\equiv V_{j2}^*(\Gamma_{DL}K^\dagger\hat
    h_U)^{ki},\\
    H_{DR}^{jki}\equiv&U_{j2}^*(\Gamma_{DR}\hat h_DK^\dagger)^{ki},
  \end{split}
\end{equation}
In order to derive the decays widths we write, for example 
eq.~(\ref{eq:130}) as
\begin{equation}
  \label{eq:136}
  {\cal L}_{u \tilde u F^0}= g{\bar u}_i^0(f^{*jki}_UP_R
  +h^{*jki}_UP_L)F_j^0\tilde u_k+\textrm{h.c}
\end{equation}
The result is presented in Sec.~\ref{sec:two-body-decays-1}.

\def\jhep{} \def\pr{}

\begin{theindex}
\addcontentsline{toc}{chapter}{Index of abbreviations}

  \item \textbf{Abbreviations:}

  \indexspace

  \item $B$: Baryon Number, 1, 5, 16
  \item $B_p$--SSM: Baryon   parity Supersymmetric Standard Model, 19
  \item $D=4$: Dimension--4, 16
  \item $D=5$:     Dimension--5, 16
  \item $L$: Lepton Number, 1, 5, 16
  \item $L_p$--SSM: Lepton parity Supersymmetric   Standard Model, 19
  \item $R$--parity, 1

  \indexspace

  \item Dimension--4, 16
  \item Dimension--5, 16

  \indexspace

  \item FCNC: Flavor Changing   Neutral Current, 3
  \item FN, Froggatt and Nielsen, 24

  \indexspace

  \item GMSB: Gauge Mediated Supersymmetry Breaking, 3
  \item GS: Green-Schwartz, 21
  \item GUT: Grand Unification Theory, 17

  \indexspace

  \item LSP: Lightest   Supersymmetric Particle, 2

  \indexspace

  \item MSSM: Minimal   Supersymmetric Standard Model, 1
  \item MSUGRA: Minimal SUGRA, 46

  \indexspace

  \item RGE: Renormalization Group   Equations, 3

  \indexspace

  \item SM: Standard Model, 1
  \item SSM: Supersymmetric Standard Model, 1, 5, 15
  \item SSSM: Superrenormalizable Supersymmetric Standard Model, 2, 8
  \item SSUGRA: Superrenormalizable SUGRA, 62
  \item SUGRA:   Supergravity, 10
  \item Superrenormalizable, 16
  \item SUSY: Supersymmetric, 1

\end{theindex}

\chapter*{Agradecimientos\markboth{Agradecimientos}{Agradecimientos}}
\label{sec:agradecimientos}
\addcontentsline{toc}{chapter}{Agradecimientos}
Una ventaja de haber realizado el doctorado fuera de mi pa{\'\i}s es que
ahora puedo agradecer a todas las personas que he conocido durante los
cuatro a{\~n}os que m{\'a}s r{\'a}pido han transcurrido en mi vida.

En primer lugar me gustar{\'\i}a agradecer la paciencia de mi familia para
afrontar mis prolongadas ausencias sobre todo en la fase final de la
escritura de este trabajo: mi esposa Tatiana, mi hija Sara y su
abuela Margarita. 

A Jos{\'e} por su constante apoyo en lo acad{\'e}mico y en lo
personal. A los dem{\'a}s que hicieron posible esta Tesis: Andrew
Akeroyd, Alfred Bartl, Alexander Belayev, Fernando de Campos, Marco A.
D{\'\i}az, Oscar Eboli, Miguel {\'A}ngel Garc{\'\i}a Jare{\~n}o,
Javier Ferrandis, Mart{\'\i}n Hirsch, Mauricio Magro, Jes{\'u}s Maria
Mira, Enrico Nardi, William Ponce, Werner Porod, Diego Francisco
Rodrigez, Jorge Crispim Romao y Emilio Torrente. Mis especiales
agradecimientos al grupo de los Profesores Bartl y Majerotto en Viena,
de Oscar Eboli en Brasil, y Enrico Nardi en Colombia por la cordial
invitaci{\'o}n, la amabilidad dispensada y el buen ambiente de trabajo
generado en mis respectivas visitas a sus grupos.  A mi familia en
Colombia (mis padres y Wilson, Liliana, M{\'o}nica y Jorge) y,
finalmente a COLCIENCIAS\footnote{This work was also supported by
  Spanish grant PB98-0693, by the European Commission RTN network
  HPRN-CT-2000-00148 and by the European Science Foundation network grant
  N.~86}

A la amistad y/o ayuda de las personas relacionadas con el grupo
durante este tiempo: Orlando, Sergio, Omar, Nicolao, Hiroshi, Nuria,
Concha, Ricard, Ver{\'o}nica, Carlos, Isabel, Sabina, Florian, Cecilia,
Victor, Kostas, Michele, Ara, Alexander Res, Timur, Maria Amparo,
Thomas,~$\ldots$.

A las personas que conoc{\'\i} a trav{\'e}s de mi esposa y con quien
departimos tantas veladas m{\'u}sico--gastron{\'o}micas. En especial a
{\'A}ngeles y su familia por su innumerables atenciones, a Nuria a
quien nucna olvidar{\'e}, a Virgilio (sus incontables ayudas
inform{\'a}ticas) y su familia en Albacete, a Alejandro y su familia
en Denia, a los de la casa de todos: Gerardo, Julio, Celine, Oscar (y
los paseos al campo), Pilar. A Daniela, Ana, Amparo, Pi{\~n}ero,
Marcos, Sonia, Maria Angeles, Tony y Rosana, Rub{\'e}n, Carlos, Kico y
M{\'o}nica, Manolo,~$\ldots$

A los que pasaron por nuestra casa dejando huella: Ricardo, William,
Enrico, Maria Fernanda, Jorge, Diego, Wilson, Sonia, Jes{\'u}s Maria,
Maria Eugenia, Sara Maria Mira Puerta, In{\'e}s Dalila, Orlando, Omar,
Antonio y Parm{\'e}nides.

Finalmente a los que desarrollaron
xemacs($+$Auc\TeX$+$ref$+$X-symbol$+$flyspell) por facilitarme de
manera sustancial la escritura de {\'e}ste trabajo.

\begin{thebibliography}{99}
\addcontentsline{toc}{chapter}{Bibliography }

\bibitem{Nilles:1984ge} \footnote{Bash scripts to automatically
    produce a new \LaTeX\ file with the \texttt{$\backslash$cite}'s
    and \texttt{$\backslash$bibitem's} ordered, or to convert
    ``\texttt{$\backslash$def$\backslash$journal-name$\#$1$\#$2$\#$3...}''
    citations into the spires format
    (\texttt{wget} is used for the automatic download), are available
 at \texttt{http://ific.uv.es/$\sim$restrepo/autoeditlatex/}}
H.~P.~Nilles,
Phys.\ Rept.\  {\bf 110} (1984) 1.

\bibitem{Haber:1985rc}
H.~E.~Haber and G.~L.~Kane,
Phys.\ Rept.\  {\bf 117} (1985) 75.

\bibitem{Tata:1995zj}
X.~Tata,
hep-ph/9510287,
  Talk given at Theoretical Advanced Study Institute in Elementary Particle
  Physics (TASI 95): QCD and Beyond, Boulder, CO. In ''Boulder
  1995, QCD and beyond'' 163-219.

\bibitem{Carone:1996nd}
C.~D.~Carone, L.~J.~Hall and H.~Murayama,
Phys.\ Rev.\ D {\bf 54} (1996) 2328
[hep-ph/9602364].
 
\bibitem{Farrar:1978xj}
G.~R.~Farrar and P.~Fayet,
Phys.\ Lett.\ B {\bf 76}, 575 (1978).

\bibitem{Dimopoulos:1981zb}
S.~Dimopoulos and H.~Georgi,
Nucl.\ Phys.\ B {\bf 193}, 150 (1981).

\bibitem{Ibanez:1992pr}
L.~E.~Ibanez and G.~G.~Ross,
Nucl.\ Phys.\ B {\bf 368}, 3 (1992).

\bibitem{Bento:1987mu}
M.~C.~Bento, L.~Hall and G.~G.~Ross,
Nucl.\ Phys.\ B {\bf 292}, 400 (1987).

\bibitem{Ibanez:1991hv}
L.~E.~Ibanez and G.~G.~Ross,
Phys.\ Lett.\ B {\bf 260}, 291 (1991).

\bibitem{Ibanez:1993ji}
L.~E.~Ibanez,
Nucl.\ Phys.\ B {\bf 398}, 301 (1993)
[hep-ph/9210211].

\bibitem{Hinchliffe:1993ad}
I.~Hinchliffe and T.~Kaeding,
Phys.\ Rev.\ D {\bf 47}, 279 (1993).

\bibitem{Eyal:1999gq}
G.~Eyal and Y.~Nir,
JHEP {\bf 9906} (1999) 024
[hep-ph/9904473]. and ref 1-16 there

\bibitem{Choi:1999}
K.~Choi, K.~Hwang and E.~J.~Chun,
Phys.\ Rev.\  D {\bf 60} (1999) 031301
[hep-ph/9811363];

\bibitem{aul} 
C.~S.~Aulakh and R.~N.~Mohapatra,
Phys.\ Lett.\  B {\bf 121} (1983) 147.

\bibitem{ross:1985}
G.~G.~Ross and J.~W.~F.~Valle,
Phys.\ Lett.\  B {\bf 151} (1985) 375;

\bibitem{Hall:1984id}
L.~J.~Hall and M.~Suzuki,
Nucl.\ Phys.\  B {\bf 231} (1984) 419.

\bibitem{Ellis:1985gi} J.~Ellis, G.~Gelmini, C.~Jarlskog, G.~G.~Ross
  and J.~W.~F.~Valle, Phys.\ Lett.\ B {\bf 150}, 142 (1985)

\bibitem{Lee:1984kr}
I.~Lee,
Phys.\ Lett.\ B {\bf 138}, 121 (1984).
I.~Lee,
Nucl.\ Phys.\ B {\bf 246}, 120 (1984).

\bibitem{Dawson:1985vr}
S.~Dawson,
Nucl.\ Phys.\ B {\bf 261}, 297 (1985).

\bibitem{eigthies}
M.~J.~Hayashi and A.~Murayama,
Phys.\ Lett.\ B {\bf 153}, 251 (1985);
%
R.~N.~Mohapatra,
Phys.\ Rev.\ D {\bf 34}, 3457 (1986);
%
R.~N.~Mohapatra,
Phys.\ Rev.\ Lett.\  {\bf 56}, 561 (1986);
%
H.~Konig,
Z.\ Phys.\ C {\bf 44}, 401 (1989).

\bibitem{Dimopoulos:1988jw}
S.~Dimopoulos and L.~J.~Hall,
Phys.\ Lett.\  B {\bf 207} (1988) 210;

\bibitem{arca}
A.~Santamaria and J.~W.~Valle,
Phys.\ Lett.\  B {\bf 195} (1987) 423;
%
Phys.\ Rev.\  D {\bf 39} (1989) 1780;
%
and Phys.\ Rev.\ Lett.\  {\bf 60} (1988) 397.

\bibitem{beyond}
For a recent review see J.~W.~Valle,
Proceedings of PASCOS98, ed. P. Nath,
W. Scientific, hep-ph/9808292;
%
J.~W.~Valle,
lectures given at the {\sl VIII Jorge Andre Swieca
Summer School} (Rio de Janeiro, February 1995) and at {\sl V Taller
Latinoamericano de Fenomenologia de las Interacciones Fundamentales}
(Puebla, Mexico, October 1995),
hep-ph/9603307.

\bibitem{Dreiner:1997uz}
H.~Dreiner,
hep-ph/9707435.
 
\bibitem{Barbier:1998fe}
R.~Barbier {\it et al.},
hep-ph/9810232.
 
\bibitem{Gonzalez-Garcia:2001sq}
M.~C.~Gonzalez-Garcia, M.~Maltoni, C.~Pena-Garay and J.~W.~F.~Valle,
Phys.\ Rev.\ D {\bf 63} (2001) 033005 [hep-ph/0009350] and references
therein.

\bibitem{Abreu:2001nc}
P.~Abreu {\it et al.}  [DELPHI Collaboration],
Phys.\ Lett.\ B {\bf 502}, 24 (2001)
[hep-ex/0102045].

\bibitem{Abreu:2001mm}
P.~Abreu {\it et al.}  [DELPHI Collaboration],
Phys.\ Lett.\ B {\bf 502}, 24 (2001).

\bibitem{Abreu:2001ne}
P.~Abreu {\it et al.}  [DELPHI Collaboration],
Phys.\ Lett.\ B {\bf 500}, 22 (2001)
[hep-ex/0103015];
%
D.~Buskulic {\it et al.}  [ALEPH Collaboration],
Phys.\ Lett.\ B {\bf 384}, 461 (1996);
%
D.~Buskulic {\it et al.}  [ALEPH Collaboration],
Phys.\ Lett.\ B {\bf 349}, 238 (1995);
%
J.~F.~Cavaignac, B.~Vignon and R.~Wilson,
Phys.\ Lett.\ B {\bf 67}, 148 (1977).

\bibitem{Barger:1989rk}
V.~Barger, G.~F.~Giudice and T.~Han,
Phys.\ Rev.\ D {\bf 40}, 2987 (1989).

\bibitem{Diaz:1999ge}
M.~A.~Diaz, D.~A.~Restrepo and J.~W.~F.~Valle,
Nucl.\ Phys.\ B {\bf 583} (2000) 182
[hep-ph/9908286].

\bibitem{Allanach:1999bf}
B.~Allanach {\it et al.},
``Searching for R-parity violation at Run-II of the Tevatron,''
hep-ph/9906224.

\bibitem{Datta:2000xq}
A.~Datta, B.~Mukhopadhyaya and F.~Vissani,
Phys.\ Lett.\ B {\bf 492} (2000) 324
[hep-ph/9910296].
  
\bibitem{Romao:2000up}
J.~C.~Romao, M.~A.~Diaz, M.~Hirsch, W.~Porod and J.~W.~Valle,
Phys.\ Rev.\ D {\bf 61} (2000) 071703 [hep-ph/9907499];
%
M.~Hirsch, M.~A.~Diaz, W.~Porod, J.~C.~Romao and J.~W.~Valle,
Phys.\ Rev.\ D {\bf 62} (2000) 113008
[hep-ph/0004115].

\bibitem{lastdiego} A.~Belyaev, O.J.P.~Eboli, W.~Porod D.A.~Restrepo,
  and J.W.F.~Valle, work in progress

\bibitem{Martin:1997ns} 
S.~P.~Martin,
``A supersymmetry primer,''hep-ph/9709356;

\bibitem{Smirnov:1996ey}
A.~Y.~Smirnov and F.~Vissani,
Nucl.\ Phys.\ B {\bf 460}, 37 (1996)
[hep-ph/9506416].

\bibitem{Hempfling:1996wj}
R.~Hempfling,
Nucl.\ Phys.\ B {\bf 478} (1996) 3
[hep-ph/9511288].
 
\bibitem{Tamvakis:1996dk}
K.~Tamvakis,
Phys.\ Lett.\ B {\bf 383} (1996) 307
[hep-ph/9602389].
 
\bibitem{Choi:1997se}
K.~Choi, E.~J.~Chun and H.~Kim,
Phys.\ Lett.\ B {\bf 394} (1997) 89
[hep-ph/9611293].
 
\bibitem{Nilles:1997ij}
H.~Nilles and N.~Polonsky,
Nucl.\ Phys.\ B {\bf 484} (1997) 33
[hep-ph/9606388].
 
\bibitem{Benakli:1997iu}
K.~Benakli and A.~Y.~Smirnov,
Phys.\ Rev.\ Lett.\  {\bf 79} (1997) 4314
[hep-ph/9703465].
 
\bibitem{Giudice:1997wb}
G.~F.~Giudice and R.~Rattazzi,
Phys.\ Lett.\ B {\bf 406} (1997) 321
[hep-ph/9704339].
 
\bibitem{Binetruy:1998sm}
P.~Binetruy, E.~Dudas, S.~Lavignac and C.~A.~Savoy,
Phys.\ Lett.\ B {\bf 422} (1998) 171
[hep-ph/9711517].

\bibitem{Joshipura:1998sp}
A.~S.~Joshipura and A.~Y.~Smirnov,
Phys.\ Lett.\ B {\bf 439} (1998) 103
[hep-ph/9806376].
 
\bibitem{Chun:1999cq}
E.~J.~Chun and H.~B.~Kim,
Phys.\ Rev.\ D {\bf 60} (1999) 095006
[hep-ph/9906392].

\bibitem{kaplan}
D.~E.~Kaplan and A.~E.~Nelson,
JHEP {\bf 0001} (2000) 033
[arXiv:hep-ph/9901254].

\bibitem{Chkareuli:2000at}
J.~L.~Chkareuli, I.~G.~Gogoladze, A.~B.~Kobakhidze, M.~G.~Green and D.~E.~Hutchcroft,
Phys.\ Rev.\ D {\bf 62} (2000) 015014
[hep-ph/9908451].
  
\bibitem{Chkareuli:2000tp}
J.~L.~Chkareuli and C.~D.~Froggatt,
Phys.\ Lett.\ B {\bf 484} (2000) 87
[hep-ph/0004090].
 
\bibitem{Mira:2000gg}
J.~M.~Mira, E.~Nardi, D.~A.~Restrepo and J.~W.~Valle,
Phys.\ Lett.\ B {\bf 492} (2000) 81
[hep-ph/0007266].

\bibitem{Valle:1987sq}
J.~W.~Valle,
Phys.\ Lett.\ B {\bf 196} (1987) 157.

\bibitem{MASIpot3}
A.~Masiero and J.~W.~Valle,
Phys.\ Lett.\ B {\bf 251} (1990) 273;

\bibitem{Romao:1992vu}
J.~C.~Romao, C.~A.~Santos and J.~W.~Valle,
Phys.\ Lett.\ B {\bf 288} (1992) 311;
%
M.~Chaichian and A.~V.~Smilga,
Phys.\ Rev.\ Lett.\  {\bf 68} (1992) 1455;
%
J.~C.~Romao, A.~Ioannisian and J.~W.~Valle,
Phys.\ Rev.\ D {\bf 55} (1997) 427
[hep-ph/9607401].

\bibitem{MASI}
G.~F.~Giudice, A.~Masiero, M.~Pietroni and A.~Riotto,
Nucl.\ Phys.\ B {\bf 396} (1993) 243;
[hep-ph/9209296];
%
M.~Shiraishi, I.~Umemura and K.~Yamamoto,
Phys.\ Lett.\ B {\bf 313} (1993) 89; see also
%
I.~Umemura and K.~Yamamoto,
Nucl.\ Phys.\ B {\bf 423} (1994) 405;
%
K.~Enqvist, K.~Huitu and P.~N.~Pandita,
Phys.\ Lett.\ B {\bf 366} (1996) 181
[hep-ph/9507227];
%
J.~R.~Espinosa,
Phys.\ Lett.\ B {\bf 353} (1995) 243
[hep-ph/9503255].

\bibitem{Chaichian:1993ra} 
M.~Chaichian and R.~Gonzalez Felipe,
Phys.\ Rev.\ D {\bf 47} (1993) 4723.

\bibitem{Adhikari:1996bm} 
R.~Adhikari and B.~Mukhopadhyaya,
Phys.\ Lett.\ B {\bf 378} (1996) 342
[Erratum-ibid.\ B {\bf 384} (1996) 492]
[hep-ph/9601382].
 
\bibitem{Frank:2001tr}
M.~Frank and K.~Huitu,
hep-ph/0106004.

\bibitem{ROMA}
P.~Nogueira, J.~C.~Romao and J.~W.~Valle,
Phys.\ Lett.\ B {\bf 251} (1990) 142;
J.~C.~Romao, J.~Rosiek and J.~W.~Valle,
Phys.\ Lett.\ B {\bf 351} (1995) 497
[hep-ph/9502211];
%
J.~C.~Romao, N.~Rius and J.~W.~Valle,
Nucl.\ Phys.\ B {\bf 363} (1991) 369.

\bibitem{Hirsch:2000xe}
M.~Hirsch and J.~W.~F.~Valle,
Phys.\ Lett.\ B {\bf 495} (2000) 121
[hep-ph/0009066].

\bibitem{Barbieri:1990vb}
R.~Barbieri, D.~E.~Brahm, L.~J.~Hall and S.~D.~Hsu,
Phys.\ Lett.\ B {\bf 238} (1990) 86.

\bibitem{Mukhopadhyaya:1998xj}
B.~Mukhopadhyaya, S.~Roy and F.~Vissani,
Phys.\ Lett.\ B {\bf 443} (1998) 191
[hep-ph/9808265].
 
\bibitem{Navarro:1999tz}
L.~Navarro, W.~Porod and J.~W.~Valle,
Phys.\ Lett.\ B {\bf 459} (1999) 615
[hep-ph/9903474].

\bibitem{otros} 
T.~Feng,
Commun.\ Theor.\ Phys.\  {\bf 33} (2000) 421
[hep-ph/9806505];
%
hep-ph/9808379;
%
C.~Chang and T.~Feng,
Eur.\ Phys.\ J.\ C {\bf 12} (2000) 137
[hep-ph/9901260].

\bibitem{Choi:1999tq}
S.~Y.~Choi, E.~J.~Chun, S.~K.~Kang and J.~S.~Lee,
Phys.\ Rev.\ D {\bf 60} (1999) 075002
[hep-ph/9903465].
 
\bibitem{GMSB1}
The signals relevant for LEP2 are discussed e.g.~in 
S.~Ambrosanio and B.~Mele,
Phys.\ Rev.\  D {\bf 52} (1995) 3900.

\bibitem{Giudice:1999bp}
For a general review see of GMSB: G.~F.~Giudice and R.~Rattazzi,
Phys.\ Rept.\  {\bf 322} (1999) 419.

\bibitem{LEPSEARCH}
P.~Abreu {\it et al.}
[DELPHI Collaboration],
CERN-EP-99-049; J.-F. Grivaz, Rapporteur Talk, International Europhysics
Conference on High Energy Physics, Brussels, 1995;
%
D.~Buskulic {\it et al.}  [ALEPH Collaboration],
Phys.\ Lett.\ B {\bf 373} (1996) 246.

\bibitem{D0} 
B.~Abbott {\it et al.}  [D0 Collaboration],
Phys.\ Rev.\ Lett.\  {\bf 83} (1999) 4937
[hep-ex/9902013];
%
S.~Abachi {\it et al.}  [D0 Collaboration],
Phys.\ Rev.\ Lett.\  {\bf 75} (1995) 618;
%
F.~Abe {\it et al.}  [CDF Collaboration],
Phys.\ Rev.\ Lett.\  {\bf 69} (1992) 3439;
%
J.~Alitti {\it et al.}  [UA2 Collaboration],
Phys.\ Lett.\ B {\bf 235} (1990) 363.

\bibitem{Abbott:2000yu}
B.~Abbott {\it et al.}  [D0 Collaboration],
Phys.\ Rev.\ D {\bf 62} (2000) 071701
[hep-ex/0005034].

\bibitem{Bartl:1996gz}
A.~Bartl, W.~Porod, M.~A.~Garcia-Jareno, M.~B.~Magro, J.~W.~F.~Valle and W.~Majerotto,
Phys.\ Lett.\ B {\bf 384} (1996) 151
[hep-ph/9606256].

\bibitem{Hikasa:1987db}
K.~Hikasa and M.~Kobayashi,
Phys.\ Rev.\ D {\bf 36} (1987) 724.

\bibitem{baer} 
H.~Baer, M.~Drees, R.~Godbole, J.~F.~Gunion and X.~Tata,
Phys.\ Rev.\ D {\bf 44} (1991) 725;
%
T.~Kon and T.~Nonaka,
hep-ph/9404230;
%
T.~Kon and T.~Nonaka,
Phys.\ Lett.\ B {\bf 319} (1993) 355
[hep-ph/9309286];
%
T.~Kon, T.~Kobayashi, S.~Kitamura, K.~Nakamura and S.~Adachi,
Z.\ Phys.\ C {\bf 61} (1994) 239
[hep-ph/9307312];

\bibitem{Porod:1999yp}
W.~Porod,
Phys.\ Rev.\  D {\bf 59} (1999) 095009
[hep-ph/9812230].

\bibitem{Porod:1997at}
W.~Porod and T.~Wohrmann,
Phys.\ Rev.\  D {\bf 55} (1997) 2907
[hep-ph/9608472].

\bibitem{Porod:1998kk}
W.~Porod,
Ph.D thesis, hep-ph/9804208, and references therein.

\bibitem{Boehm:1999tr} 
C.~Boehm, A.~Djouadi and Y.~Mambrini,
Phys.\ Rev.\ D {\bf 61} (2000) 095006
[arXiv:hep-ph/9907428].

\bibitem{Bartl97a}
A.~Bartl, W.~Majerotto and W.~Porod,
Z.\ Phys.\ C {\bf 64} (1994) 499; erratum ibid. C~{\bf 68}, 518  (1995); 

\bibitem{Bartl:1997wt}
A.~Bartl, H.~Eberl, S.~Kraml, W.~Majerotto and W.~Porod,
Z.\ Phys.\ C {\bf 73} (1997) 469 
[hep-ph/9603410];
%
A.~Bartl, H.~Eberl, S.~Kraml, W.~Majerotto, W.~Porod and A.~Sopczak,
Z.\ Phys.\ C {\bf 76} (1997) 549
[hep-ph/9701336].

\bibitem{Ellis83}
J.~Ellis and S.~Rudaz,
Phys.\ Lett.\ B {\bf 128} (1983) 248;
%
G.~Altarelli and R.~Ruckl,
Phys.\ Lett.\ B {\bf 144} (1984) 126;
%
I.~I.~Bigi and S.~Rudaz,
Phys.\ Lett.\ B {\bf 153} (1985) 335.

\bibitem{Djouadi:2000bx}
A.~Djouadi and Y.~Mambrini,
Phys.\ Rev.\ D {\bf 63} (2001) 115005
[hep-ph/0011364].

\bibitem{Boehm:2000tr}
C.~Boehm, A.~Djouadi and Y.~Mambrini,
Phys.\ Rev.\ D {\bf 61} (2000) 095006
[hep-ph/9907428].

\bibitem{Bartl:2000kw}
A.~Bartl, H.~Eberl, S.~Kraml, W.~Majerotto and W.~Porod,
Eur.\ Phys.\ J.\ directC {\bf 6} (2000) 1
[hep-ph/0002115].


\bibitem{Datta:2000yc}
A.~Datta and B.~Mukhopadhyaya,
Phys.~Rev.~Lett.~{\bf 85} (2000) 248,  [hep-ph/0003174].

\bibitem{deCampos:1995av} 
F.~de Campos, M.~A.~Garcia-Jareno, A.~S.~Joshipura, J.~Rosiek and J.~W.~Valle,
Nucl.\ Phys.\ B {\bf 451} (1995) 3
[hep-ph/9502237].

\bibitem{Akeroyd:1998iq}
A.~Akeroyd, M.~A.~Diaz, J.~Ferrandis, M.~A.~Garcia-Jareno and J.~W.~Valle,
Nucl.\ Phys.\ B {\bf 529} (1998) 3
[hep-ph/9707395].

\bibitem{ulrike} U.~Dydak, Search for the supersymmetric Scalar top
  Quark with the CMS Detector at the LHC, Diploma Thesis, University
  of Vienna, 1996; 
  F.~Gianotti, ATLAS Internal Note PHYS-No-110 (1997);
  L.~Poggioli, G.~Polesello, E.~Richter-Was, and
  J.~Soderqvist, ATL-PHYS-97-111.

\bibitem{Roy:1997bu}
S.~Roy and B.~Mukhopadhyaya,
Phys.\ Rev.\ D {\bf 55} (1997) 7020
[hep-ph/9612447].

\bibitem{Joshipura:1995ib}
A.~S.~Joshipura and M.~Nowakowski,
Phys.\ Rev.\ D {\bf 51} (1995) 2421
[hep-ph/9408224].

\bibitem{Borzumati:1996hd}
F.~M.~Borzumati, Y.~Grossman, E.~Nardi and Y.~Nir,
Phys.\ Lett.\ B {\bf 384} (1996) 123
[hep-ph/9606251].
 
\bibitem{Diaz:1998xc}
M.~A.~Diaz, J.~Ferrandis, J.~C.~Romao and J.~W.~Valle,
Phys.\ Lett.\ B {\bf 453} (1999) 263
[hep-ph/9801391].

\bibitem{Diaz:1999wq} 
M.~A.~Diaz, E.~Torrente-Lujan and J.~W.~Valle,
Nucl.\ Phys.\ B {\bf 551} (1999) 78
[hep-ph/9808412].

\bibitem{javi} 
J.~Ferrandis,
Phys.\ Rev.\ D {\bf 60} (1999) 095012
[hep-ph/9810371];
%
M.~A.~Diaz, J.~Ferrandis, J.~C.~Romao and J.~W.~Valle,
Nucl.\ Phys.\ B {\bf 590} (2000) 3
[hep-ph/9906343].

\bibitem{Mukhopadhyaya:1999gy}
B.~Mukhopadhyaya and S.~Roy,
Phys.\ Rev.\ D {\bf 60} (1999) 115012
[hep-ph/9903418].


\bibitem{Datta:2000yd}
A.~Datta, B.~Mukhopadhyaya and S.~Roy,
Phys.\ Rev.\ D {\bf 61} (2000) 055006
[hep-ph/9905549].
 
\bibitem{Diaz:1997xc}
M.~A.~Diaz, J.~C.~Romao and J.~W.~Valle,
Nucl.\ Phys.\ B {\bf 524} (1998) 23
[hep-ph/9706315].

\bibitem{Nardi:1997iy}
E.~Nardi,
Phys.\ Rev.\  D {\bf 55}, 5772 (1997).

\bibitem{Joshipura:1995wm}
A.~S.~Joshipura and M.~Nowakowski,
Phys.\ Rev.\ D {\bf 51} (1995) 5271
[hep-ph/9403349].

\bibitem{deCarlos:1996du}
B.~de Carlos and P.~L.~White,
Phys.\ Rev.\ D {\bf 54} (1996) 3427
[hep-ph/9602381].

\bibitem{Nowakowski:1996dx}
M.~Nowakowski and A.~Pilaftsis,
Nucl.\ Phys.\ B {\bf 461} (1996) 19
[hep-ph/9508271].
 
\bibitem{Banks:1995by}
T.~Banks, Y.~Grossman, E.~Nardi and Y.~Nir,
Phys.\ Rev.\  D {\bf 52} (1995) 5319
[hep-ph/9505248].

\bibitem{Diaz:1998vf}
M.~A.~Diaz,
in {\it Beyond the Standard Model: From Theory to Experiment},
Proceedings of Valencia 97, ed. I. Antoniadis, L. E. Ibanez and J. W.
F. Valle, W. Scientific, p~188, hep-ph/9802407.

\bibitem{Comelli:1994nt}
D.~Comelli, A.~Masiero, M.~Pietroni and A.~Riotto,
Phys.\ Lett.\ B {\bf 324} (1994) 397
[hep-ph/9310374].

\bibitem{Bartl2000rp}
A.~Bartl, W.~Porod, D.~Restrepo, J.~Romao and J.~W.~F.~Valle,
Nucl. Phys. B {\bf 600} (2001) 39 
[hep-ph/0007157.]

\bibitem{Hirsch:1999kc}
M.~Hirsch and J.~W.~Valle,
Nucl.\ Phys.\ B {\bf 557} (1999) 60
[hep-ph/9812463].

\bibitem{Porod:2000pw}
W.~Porod, D.~Restrepo and J.~W.~F.~Valle,
hep-ph/0001033.

\bibitem{Restrepo:2001me}
D.~Restrepo, W.~Porod and J.~W.~Valle,
Phys.\ Rev.\ D {\bf 64} (2001) 055011
[hep-ph/0104040].


\bibitem{Akeroyd:1998sv}
A.~G.~Akeroyd, M.~A.~Diaz and J.~W.~Valle,
Phys.\ Lett.\ B {\bf 441} (1998) 224
[hep-ph/9806382].

\bibitem{Diaz:2000wm}
M.~A.~Diaz, J.~Ferrandis and J.~W.~Valle,
Nucl.\ Phys.\ B {\bf 573} (2000) 75
[hep-ph/9909212].
 
\bibitem{Borzumati:1999th}
F.~Borzumati, J.~Kneur and N.~Polonsky,
Phys.\ Rev.\  D {\bf 60} (1999) 115011
[hep-ph/9905443].

\bibitem{Grossman:1998py}
Y.~Grossman and H.~E.~Haber,
Phys.\ Rev.\ D {\bf 59} (1999) 093008
[hep-ph/9810536];
%
Y.~Grossman and H.~E.~Haber,
Phys.\ Rev.\ Lett.\  {\bf 78} (1997) 3438
[hep-ph/9702421].

\bibitem{Bisset:1999nw} 
M.~Bisset, O.~C.~Kong, C.~Macesanu and L.~H.~Orr,
hep-ph/9907359;
%
M.~Bisset, O.~C.~Kong, C.~Macesanu and L.~H.~Orr,
Phys.\ Lett.\ B {\bf 430} (1998) 274
[hep-ph/9804282].

\bibitem{sacha}
S.~Davidson and J.~Ellis,
Phys.\ Rev.\ D {\bf 56} (1997) 4182
[hep-ph/9702247];
%
S.~Davidson and J.~Ellis,
Phys.\ Lett.\ B {\bf 390} (1997) 210
[hep-ph/9609451];
%
S.~Davidson,
Phys.\ Lett.\ B {\bf 439} (1998) 63
[hep-ph/9808425].

\bibitem{expl}
G.~Bhattacharyya, D.~Choudhury and K.~Sridhar,
Phys.\ Lett.\ B {\bf 349} (1995) 118
[hep-ph/9412259];
%
E.~Ma and D.~Ng,
Phys.\ Rev.\ D {\bf 41} (1990) 1005.

\bibitem{bartl:1996snow}
A. Bartl, {\it et al}, Published in the
Proceedings of the 1996 DPF/DPB Summer Study on New Directions For
High-Energy Physics, Edited by D.G. Cassel, L. Trindle Gennari,
R.H. Siemann, 1997, p.~693-705. See also
http://www.slac.stanford.edu/pubs/snowmass96/.

\bibitem{Barate:1998zg}
R.~Barate {\it et al.}  [ALEPH Collaboration],
Eur.\ Phys.\ J.\ C {\bf 2} (1998) 395.

\bibitem{Binetruy:1996xk}
P.~Binetruy, S.~Lavignac and P.~Ramond,
Nucl.\ Phys.\ B {\bf 477} (1996) 353
[hep-ph/9601243].
 
\bibitem{Weinberg:1982wj}
S.~Weinberg,
Phys.\ Rev.\ D {\bf 26} (1982) 287.

\bibitem{Murayama:1994tc}
H.~Murayama and D.~B.~Kaplan,
Phys.\ Lett.\ B {\bf 336}, 221 (1994)
[hep-ph/9406423].

\bibitem{Ben-Hamo:1994bq}
V.~Ben-Hamo and Y.~Nir,
Phys.\ Lett.\ B {\bf 339}, 77 (1994)
[hep-ph/9408315].

\bibitem{Murayama:1996qe}
H.~Murayama,
hep-ph/9610419.
 
\bibitem{Moreau:2000hz}
See G.~Moreau,
hep-ph/0012156 and references therein.

\bibitem{Pati:1996fn}
J.~C.~Pati,
Phys.\ Lett.\ B {\bf 388} (1996) 532
[hep-ph/9607446].

\bibitem{Choi:1997fr}
K.~Choi, E.~J.~Chun and H.~Kim,
Phys.\ Rev.\ D {\bf 55} (1997) 7010
[hep-ph/9610504].
 
\bibitem{Kurosawa:2001iq}
K.~Kurosawa, N.~Maru and T.~Yanagida,
Phys.\ Lett.\ B {\bf 512} (2001) 203
[hep-ph/0105136].

\bibitem{Chamseddine:1996gb}
A.~H.~Chamseddine and H.~Dreiner,
Nucl.\ Phys.\ B {\bf 458} (1996) 65
[hep-ph/9504337].
 
\bibitem{Chun:1996xv}
E.~J.~Chun and A.~Lukas,
Phys.\ Lett.\  B {\bf 387} (1996) 99
[hep-ph/9605377];

\bibitem{Berezinsky:1998pb}
V.~Berezinsky, A.~S.~Joshipura and J.~W.~Valle,
Phys.\ Rev.\ D {\bf 57} (1998) 147
[hep-ph/9608307].

\bibitem{Castano:1994ec}
D.~J.~Castano and S.~P.~Martin,
Phys.\ Lett.\ B {\bf 340}, 67 (1994)
[hep-ph/9408230].

\bibitem{Green:1984sg}
M.~B.~Green and J.~H.~Schwarz,
Phys.\ Lett.\  B {\bf 149} (1984) 117.

\bibitem{Brahm:1989iy}
D.~E.~Brahm and L.~J.~Hall,
Phys.\ Rev.\ D {\bf 40}, 2449 (1989).

\bibitem{Gato:1985jq}
B.~Gato, J.~Leon, J.~Perez-Mercader and M.~Quiros,
Nucl.\ Phys.\ B {\bf 260}, 203 (1985).

\bibitem{Chaichian:1996wr}
M.~Chaichian and K.~Huitu,
Phys.\ Lett.\ B {\bf 384} (1996) 157
[hep-ph/9603412].

\bibitem{Gonzalez-Garcia:1991qf}
M.~C.~Gonzalez-Garcia and J.~W.~Valle,
Nucl.\ Phys.\ B {\bf 355} (1991) 330;


\bibitem{Hall:1990dg}
L.~J.~Hall,
Mod.\ Phys.\ Lett.\ A {\bf 5} (1990) 467.

\bibitem{Smirnov:1996jt}
A.~Y.~Smirnov,
hep-ph/9611465.

\bibitem{Suematsu:2001sp}
D.~Suematsu,
hep-ph/0104187.

\bibitem{Shafi:2000gw}
Q.~Shafi and Z.~Tavartkiladze,
Nucl.\ Phys.\ B {\bf 580} (2000) 83
[hep-ph/9909238].
 
\bibitem{Chun:1999kd}
E.~J.~Chun,
Phys.\ Lett.\ B {\bf 454} (1999) 304
[hep-ph/9901220].

\bibitem{Froggatt:1979nt}
C.~D.~Froggatt and H.~B.~Nielsen,
Nucl.\ Phys.\  B {\bf 147} (1979) 277.

\bibitem{Joshipura:2000sn}
A.~S.~Joshipura, R.~D.~Vaidya and S.~K.~Vempati,
Phys.\ Rev.\ D {\bf 62} (2000) 093020
[hep-ph/0006138].

\bibitem{Mira:2000fx}
J.~M.~Mira, E.~Nardi and D.~A.~Restrepo,
Phys.\ Rev.\  D {\bf 62} (2000) 016002
[hep-ph/9911212].

\bibitem{Giudice:1988yz}
G.~F.~Giudice and A.~Masiero,
Phys.\ Lett.\  B {\bf 206} (1988) 480.;

\bibitem{Fukuda:1998mi}
Y.~Fukuda {\it et al.}  [Super-Kamiokande Collaboration],
Phys.\ Rev.\ Lett.\ {\bf 81} (1998) 1562 [hep-ex/9807003]; 
%
H. Sobel's
talk at Neutrino 2000 Conference, Sudbury, 16-21 June.

  
\bibitem{atm99} For recent global fits of atmospheric neutrino data
  see, for example, 
%
N.~Fornengo, M.~C.~Gonzalez-Garcia and J.~W.~Valle,
Nucl.\ Phys.\ B {\bf 580} (2000) 58 [hep-ph/0002147];
%
M.~C.~Gonzalez-Garcia, H.~Nunokawa, O.~L.~Peres and J.~W.~Valle,
Nucl.\ Phys.\ B {\bf 543} (1999) 3
[hep-ph/9807305];
R.~Foot, R.~R.~Volkas and O.~Yasuda,
Phys.\ Rev.\ D {\bf 58} (1998) 013006
[hep-ph/9801431];
M.~C.~Gonzalez-Garcia, M.~Maltoni, C.~Pena-Garay and J.~W.~Valle,
Phys.\ Rev.\ D {\bf 63} (2001) 033005
[hep-ph/0009350].

\bibitem{qva}
A.~Friedland,
Phys.\ Rev.\ Lett.\  {\bf 85} (2000) 936
[hep-ph/0002063];
%
A.~de Gouvea, A.~Friedland and H.~Murayama,
Phys.\ Lett.\ B {\bf 490} (2000) 125
[hep-ph/0002064];
%
G.~L.~Fogli, E.~Lisi, D.~Montanino and A.~Palazzo,
Phys.\ Rev.\ D {\bf 62} (2000) 113004
[hep-ph/0005261].
  
\bibitem{MSW99} 
J.~N.~Bahcall, P.~I.~Krastev and A.~Y.~Smirnov,
Phys.\ Rev.\ D {\bf 62} (2000) 093004
[hep-ph/0002293];
%
M.~C.~Gonzalez-Garcia, P.~C.~de Holanda, C.~Pena-Garay and J.~W.~Valle,
Nucl.\ Phys.\ B {\bf 573} (2000) 3
[hep-ph/9906469].

\bibitem{Nir:1995bu}
Y.~Nir,
Phys.\ Lett.\ B {\bf 354}, 107 (1995)
[hep-ph/9504312].

\bibitem{Fusaoka:1998vc}
H.~Fusaoka and Y.~Koide,
Phys.\ Rev.\  D {\bf 57} (1998) 3986
[hep-ph/9712201].

\bibitem{Leurer:1993wg}
M.~Leurer, Y.~Nir and N.~Seiberg,
Nucl.\ Phys.\  B {\bf 398} (1993) 319
[hep-ph/9212278];
%
Nucl.\ Phys.\  B {\bf 420} (1994) 468
[hep-ph/9310320].

\bibitem{Binetruy:1995ru}
P.~Binetruy and P.~Ramond,
Phys.\ Lett.\  B {\bf 350} (1995) 49
[hep-ph/9412385].

\bibitem{Dudas:1995yu}
E.~Dudas, S.~Pokorski and C.~A.~Savoy,
Phys.\ Lett.\ B {\bf 356}, 45 (1995)
[hep-ph/9504292].

\bibitem{Barger:1993ac}
V.~Barger, M.~S.~Berger and P.~Ohmann,
Phys.\ Rev.\  D {\bf 47} (1993) 1093
[hep-ph/9209232].

\bibitem{Grossman:1994ax}
Y.~Grossman and Z.~Ligeti,
Phys.\ Lett.\  B {\bf 332} (1994) 373;
%
P.~Abreu {\it et al.}  [DELPHI Collaboration],
CERN-EP-99-162.

\bibitem{Barbieri:1990qj}
R.~Barbieri, M.~M.~Guzzo, A.~Masiero and D.~Tommasini,
Phys.\ Lett.\  B {\bf 252} (1990) 251;
%
E.~Roulet and D.~Tommasini,
Phys.\ Lett.\  B {\bf 256} (1991) 218;
%
K.~Enqvist, A.~Masiero and A.~Riotto,
Nucl.\ Phys.\  B {\bf 373} (1992) 95;

\bibitem{Davidson:2000uc}
S.~Davidson and M.~Losada,
JHEP {\bf 0005} (2000) 021
[hep-ph/0005080].
 
\bibitem{Takayama:2000pc}
F.~Takayama and M.~Yamaguchi,
Phys.\ Lett.\ B {\bf 476} (2000) 116
[hep-ph/9910320].
 
\bibitem{Bednyakov:1998cx}
V.~Bednyakov, A.~Faessler and S.~Kovalenko,
Phys.\ Lett.\ B {\bf 442} (1998) 203
[hep-ph/9808224];
%
O.~C.~Kong,
Mod.\ Phys.\ Lett.\ A {\bf 14} (1999) 903
[hep-ph/9808304];
%
R.~Adhikari and G.~Omanovic,
hep-ph/9802390;
%
S.~Rakshit, G.~Bhattacharyya and A.~Raychaudhuri,
Phys.\ Rev.\ D {\bf 59} (1999) 091701
[hep-ph/9811500].

\bibitem{Chun:1999gp}
E.~J.~Chun, S.~K.~Kang, C.~W.~Kim and U.~W.~Lee,
Nucl.\ Phys.\ B {\bf 544} (1999) 89
[hep-ph/9807327].
 
\bibitem{Bhattacharyya:1999tv}
G.~Bhattacharyya, H.~V.~Klapdor-Kleingrothaus and H.~Pas,
Phys.\ Lett.\ B {\bf 463} (1999) 77
[hep-ph/9907432].

\bibitem{Haug:2000kr}
O.~Haug, J.~D.~Vergados, A.~Faessler and S.~Kovalenko,
Nucl.\ Phys.\ B {\bf 565}, 38 (2000)
[hep-ph/9909318];
%
A.~Abada and M.~Losada,
Nucl.\ Phys.\ B {\bf 585} (2000) 45
[hep-ph/9908352];
%
R.~Kitano and K.~Oda,
Phys.\ Rev.\ D {\bf 61} (2000) 113001
[hep-ph/9911327].

\bibitem{Mukhopadhyaya:2000wn}
B.~Mukhopadhyaya,
Pramana {\bf 54} (2000) 147
[hep-ph/9907275].

\bibitem{Chun:2000bq}
E.~J.~Chun and S.~K.~Kang,
Phys.\ Rev.\ D {\bf 61} (2000) 075012
[hep-ph/9909429].
 
\bibitem{Davidson:2000ne}
S.~Davidson and M.~Losada,
hep-ph/0010325.

\bibitem{Abada:2000xr}
A.~Abada and M.~Losada,
Phys.\ Lett.\ B {\bf 492} (2000) 310
[hep-ph/0007041].
 
\bibitem{Abada:2001ma}
A.~Abada and G.~Bhattacharyya,
Phys.\ Rev.\ D {\bf 63} (2001) 017701
[hep-ph/0007016].
 
\bibitem{Hempfling:1997je}
R.~Hempfling,
hep-ph/9702412.

\bibitem{Porod:2000hv}
W.~Porod, M.~Hirsch, J.~Romao and J.~W.~Valle,
Phys.\ Rev.\ D {\bf 63} (2001) 115004
[hep-ph/0011248].

\bibitem{Godbole:1993fb}
R.~M.~Godbole, P.~Roy and X.~Tata,
Nucl.\ Phys.\  B {\bf 401} (1993) 67
[hep-ph/9209251].

\bibitem{Hirsch:2000jt}
M.~Hirsch, J.~C.~Romao and J.~W.~Valle,
Phys.\ Lett.\ B {\bf 486} (2000) 255
[hep-ph/0002264].

\bibitem{Chun:1999ub}
E.~J.~Chun and J.~S.~Lee,
Phys.\ Rev.\ D {\bf 60} (1999) 075006
[hep-ph/9811201].
 
\bibitem{Chun:2001mm}
E.~J.~Chun,
hep-ph/0105157.


\bibitem{Grossman:1999hc}
Y.~Grossman and H.~E.~Haber,
hep-ph/9906310.
 
\bibitem{Grossman:2001ex}
Y.~Grossman and H.~E.~Haber,
Phys.\ Rev.\ D {\bf 63} (2001) 075011
[hep-ph/0005276].

\bibitem{Bar-Shalom:2001ew}
S.~Bar-Shalom, G.~Eilam and B.~Mele,
hep-ph/0106053.

\bibitem{Faessler:1998db}
A.~Faessler, S.~Kovalenko and F.~Simkovic,
Phys.\ Rev.\ D {\bf 58} (1998) 055004
[hep-ph/9712535].

\bibitem{Bhattacharyya:1998dt}
G.~Bhattacharyya and P.~B.~Pal,
Phys.\ Lett.\ B {\bf 439} (1998) 81
[hep-ph/9806214].

\bibitem{Bhattacharyya:1999bx}
G.~Bhattacharyya and P.~B.~Pal,
Phys.\ Rev.\ D {\bf 59} (1999) 097701
[hep-ph/9809493].

\bibitem{Frank:2000gs}
M.~Frank,
Phys.\ Rev.\ D {\bf 62} (2000) 015006.

\bibitem{Bisset:1999hz}
M.~Bisset, O.~C.~Kong, C.~Macesanu and L.~H.~Orr,
Nucl.\ Phys.\ Proc.\ Suppl.\  {\bf 76} (1999) 201
[hep-ph/9811499].
 
\bibitem{Joshipura:1999fn}
A.~S.~Joshipura and S.~K.~Vempati,
Phys.\ Rev.\ D {\bf 60} (1999) 095009
[hep-ph/9808232].
 
\bibitem{Choi:2001bg}
K.~Choi, E.~J.~Chun and K.~Hwang,
Phys.\ Rev.\ D {\bf 63} (2001) 013002
[hep-ph/0004101].
 
\bibitem{Babu:1990px}
K.~S.~Babu and R.~N.~Mohapatra,
Phys.\ Rev.\ Lett.\  {\bf 64} (1990) 1705;

\bibitem{Huitu:1999qu}
K.~Huitu, J.~Maalampi and K.~Puolamaki,
Eur.\ Phys.\ J.\ C {\bf 6} (1999) 159
[hep-ph/9705406].

\bibitem{deCampos:1999mf}
F.~de Campos, O.~J.~Eboli, M.~A.~Garcia-Jareno and J.~W.~Valle,
Nucl.\ Phys.\ B {\bf 546} (1999) 33
[hep-ph/9710545].

\bibitem{Bartl:1999iw}
A.~Bartl, W.~Majerotto and W.~Porod,
Phys.\ Lett.\ B {\bf 465} (1999) 187
[hep-ph/9907377].
%
A.~Bartl, H.~Fraas and W.~Majerotto,
Nucl.\ Phys.\ B {\bf 278} (1986) 1.

\bibitem{mnutreeJ}
J.~C.~Romao and J.~W.~Valle,
Phys.\ Lett.\ B {\bf 272} (1991) 436;
J.~C.~Romao and J.~W.~Valle,
Nucl.\ Phys.\ B {\bf 381} (1992) 87.

\bibitem{ISR}
E.~A.~Kuraev and V.~S.~Fadin,
Sov.\ J.\ Nucl.\ Phys.\  {\bf 41} (1985) 466
[Yad.\ Fiz.\  {\bf 41} (1985) 733];
%
M.E.~Peskin, in {\em Physics at the 100 GeV Mass Scale}, 
             17th SLAC Summer Institute, 1989.

\bibitem{Diaz:1998zg}
M.~A.~Diaz, M.~A.~Garcia-Jareno, D.~A.~Restrepo and J.~W.~Valle,
Nucl.\ Phys.\ B {\bf 527} (1998) 44
[hep-ph/9803362].


\bibitem{Bagger:1997bt}
J.~A.~Bagger, K.~Matchev, D.~M.~Pierce and R.~Zhang,
Phys.\ Rev.\  D {\bf 55} (1997) 3188
[hep-ph/9609444].

\bibitem{Groom:2000in}
D.~E.~Groom {\it et al.}  [Particle Data Group Collaboration],
Eur.\ Phys.\ J.\ C {\bf 15} (2000) 1.

\bibitem{Abe:1997dn}
F.~Abe {\it et al.}  [CDF Collaboration],
Phys.\ Rev.\ Lett.\  {\bf 78} (1997) 2906.

\bibitem{deCampos:1998kw} F.~de Campos, M.~A.~Diaz, O.~J.~P. Eboli,
  M.~B.~Magro, L.~Navarro, W.~Porod, D.~A.~Restrepo, J.~W.~F. Valle,
  Physics at Run II: QCD and Weak Boson Physics Workshop Batavia, IL ;
  4 - 6 Mar 1999 Publ. in: Proceedings G Landsberg,
  hep-ph/9903245.

\bibitem{nusug}
H.~Baer, M.~A.~Diaz, P.~Quintana and X.~Tata,
JHEP {\bf 0004} (2000) 016
[arXiv:hep-ph/0002245].

\bibitem{Brignole:1997dp}
A.~Brignole, L.~E.~Ibanez and C.~Munoz,
hep-ph/9707209.

\bibitem{Polonsky:1994sr}
N.~Polonsky and A.~Pomarol,
Phys.\ Rev.\ Lett.\  {\bf 73} (1994) 2292
[hep-ph/9406224].

\bibitem{so10}
H.~Murayama, M.~Olechowski and S.~Pokorski,
Phys.\ Lett.\ B {\bf 371} (1996) 57
[hep-ph/9510327];
%
H.~Baer, M.~A.~Diaz, J.~Ferrandis and X.~Tata,
Phys.\ Rev.\ D {\bf 61} (2000) 111701
[hep-ph/9907211].

\bibitem{gluino} 
H.~Dreiner, M.~Guchait and D.~P.~Roy,
Phys.\ Rev.\ D {\bf 49} (1994) 3270
[hep-ph/9310291];
%
H.~Dreiner and G.~G.~Ross,
Nucl.\ Phys.\ B {\bf 365} (1991) 597;
%
A.~Bartl, W.~Porod, F.~de Campos, M.~A.~Garcia-Jareno, M.~B.~Magro, J.~W.~Valle and W.~Majerotto,
Nucl.\ Phys.\ B {\bf 502} (1997) 19
[hep-ph/9612436];
A.~Bartl, F.~De Campos, M.~A.~Garcia-Jareno, M.~B.~Magro, W.~Majerotto, W.~Porod and J.~W.~Valle,
Proceedings of the ''International Workshop on Physics Beyond the
Standard Model: From Theory to Experiment'', Valencia, Spain, 13-17
Okt, 1997, Eds.~I.~Antoniadis, L.E.~Ib{\'a}{\~n}ez, J.W.F.~Valle, p.~200-204,
hep-ph/9712484.

\bibitem{Abbiendi:1999yz}
G.~Abbiendi {\it et al.}  [OPAL Collaboration],
Phys.\ Lett.\ B {\bf 456} (1999) 95
[hep-ex/9903070].

\bibitem{Holck:1999tp}
C.~Holck  [CDF Collaboration],
Talk given at American Physical Society (APS) Meeting of the Division
of Particles and Fields (DPF 99), Los Angeles, CA, 5-9 Jan 1999,
hep-ex/9903060.

\bibitem{Gunion:1986yn}
J.~F.~Gunion and H.~E.~Haber,
Nucl.\ Phys.\ B {\bf 272} (1986) 1
[Erratum-ibid.\ B {\bf 402} (1986) 567].


\bibitem{Bertolini:1991if}
S.~Bertolini, F.~Borzumati, A.~Masiero and G.~Ridolfi,
Nucl.\ Phys.\ B {\bf 353} (1991) 591.


\bibitem{Rosiek:1990rs}
J.~Rosiek,
Phys.\ Rev.\ D {\bf 41} (1990) 3464.


\end{thebibliography}
\end{document}